%% file: main.tex
\documentclass[10pt]{article}

\usepackage[T1]{fontenc}
\usepackage[utf8]{inputenc}
\usepackage{lmodern}
\usepackage{microtype}
\usepackage{geometry}
\usepackage{amsmath,amssymb,bm}
\usepackage{mathtools}
\usepackage{booktabs,array,tabularx,multirow}
\usepackage{graphicx}
\graphicspath{{figures/}}
\usepackage{placeins}
\usepackage{xcolor}
\usepackage{enumitem}
\usepackage{listings}
\usepackage{tikz}
\usetikzlibrary{arrows.meta,positioning,fit,calc}
\usepackage{cite}
\usepackage{hyperref}
\usepackage[nameinlink,noabbrev]{cleveref}
\hypersetup{
  colorlinks=true,
  hypertexnames=false,
  citecolor=blue!55!black,
  linkcolor=blue!55!black,
  urlcolor=blue!55!black,
  pdftitle={Hadronic tau decays at higher orders in QCD},
  pdfauthor={Gauhar Abbas, Debabina Biswas, and Vartika Singh}
}

\setlist{topsep=0.35em,itemsep=0.2em,parsep=0pt}
\allowdisplaybreaks[3]
\numberwithin{equation}{section}

\newcommand{\as}{\alpha_s}
\newcommand{\MSbar}{\overline{\mathrm{MS}}}
\newcommand{\Dhat}{\widehat D}

\newcommand{\cO}{\mathcal O}
\newcommand{\arxiv}[1]{\href{https://arxiv.org/abs/#1}{arXiv:#1}}
\newcommand{\doi}[1]{\href{https://doi.org/#1}{doi:#1}}

\newcommand{\parttitle}[1]{\par\bigskip\phantomsection\section*{#1}%
  \addcontentsline{toc}{section}{#1}\nopagebreak}

\title{\textbf{Hadronic $\tau$ decays at higher orders in QCD}}
\author{Gauhar Abbas\footnote{gauhar.phy@itbhu.ac.in},
Debabina Biswas\footnote{debabinabiswas.rs.phy24@itbhu.ac.in}, Pratham Dadhich\footnote{pratham.rs.phy24@itbhu.ac.in }, and Vartika Singh\footnote{vartikasingh.rs.phy19@itbhu.ac.in}\\[3pt]
\normalsize\itshape Department of Physics, Indian Institute of Technology
(BHU), Varanasi 221005, India}

\date{November 4, 2026}

\begin{document}

\maketitle

\begin{abstract}
We ask what the four known coefficients of the Adler function in massless QCD can support about the
unknown coefficient $c_{5,1}$ of $a^5$.  In fixed-order perturbation theory the
Adler-to-moment map is triangular with unit diagonal; hence a multi-weight
relative test of a common forecast reduces exactly to one absolute coefficient
tolerance.  For three complementary weights, a ten-percent withheld-order test
requires $|\widehat c_{4,1}-c_{4,1}|\leq2.1509$ and resolves $c_{5,1}$ only to
about $\pm24$.  Direct Shanks/Pad\'e continuation cannot reach the usual
literature range without reversing the sign of its next defect.  We therefore
apply a cross-validated renormalon-closure (CVRC) protocol to a pure-power
Beneke--Jamin-type Borel ansatz.  CVRC calibrates and tests this declared
structure; it is not a new renormalon model.  The calibrated construction gives
the conditional estimate
$c_{5,1}=248\pm30\,(\mathrm{struct})\pm74\,(\mathrm{method})=248\pm80$.
The band is nonstatistical, and a recursive test misses the known $c_{4,1}$ by
$29.9\%$.  Thus the result is a model-qualified estimate consistent with
existing determinations, not evidence that four coefficients determine the
fifth.
\end{abstract}

\setcounter{tocdepth}{1}
{\small\setlength{\parskip}{0pt}\tableofcontents}
\clearpage

\section{Introduction}
Hadronic $\tau$ decay provides a precise low-energy determination of the QCD
coupling~\cite{Davier:2005xq,Pich:2020gzz,ParticleDataGroup:2024cfk}.  Its
massless perturbative correction is governed by the Adler function, known
through $\mathcal O(\alpha_s^4)$~\cite{Baikov:2008jh}, and is combined with the
operator product expansion (OPE)~\cite{Wilson:1969zs} through finite-energy sum
rules~\cite{Narison:1988ni,Braaten:1991qm}.  The four-loop calculation
underpins a series of determinations of
$\alpha_s(m_\tau)$~\cite{Davier:2008sk,Maltman:2008nf,Beneke:2008ad,
Bethke:2009jm,Pich:2010xb,Pich:2011cj,Bethke:2011tr,Boito:2011qt,Boito:2014sta,
Pich:2016bdg,Boito:2020xli,Ayala:2022cxo,Boito:2025pwg}.  The same analyses must
control higher-dimensional power corrections~\cite{Braaten:1991qm,
Davier:2005xq,Davier:2008sk,Beneke:2008ad,Pich:2010xb} and possible duality
violations near the timelike axis~\cite{Cata:2008ye,Boito:2011qt,
Boito:2014sta,Pich:2016bdg,Pich:2022tca,Boito:2024gtb}.

The principal perturbative limitations are the unknown higher-order Adler
coefficients and the different organization of the contour expansion in
fixed-order (FOPT) and contour-improved perturbation theory
(CIPT)~\cite{Pivovarov:1991rh,LeDiberder:1992jjr}, whose difference is the
dominant theoretical uncertainty in present $\alpha_s$
extractions~\cite{Jamin:2005ip,Beneke:2008ad,Menke:2009vg,Cvetic:2010ut,
Groote:2012jq,Beneke:2012vb,Hoang:2020mkw,Hoang:2021nlz,Golterman:2023oml,
Gracia:2023qdy}.  Subtracting the gluon-condensate renormalon reduces this
difference substantially~\cite{Benitez-Rathgeb:2022yqb,
Benitez-Rathgeb:2022hfj,Beneke:2023wkq,Beneke:2025hlg}.  Renormalons link
these issues to the large-order structure of the series, but four known
nontrivial coefficients cannot by themselves reveal that structure.  A successful interpolation of a
short series is therefore not automatically a successful reconstruction of
its Borel singularities~\cite{Beneke:1998ui}.

We compare Pad\'e approximants and nonlinear sequence transformations with
renormalon-informed constructions under a common validation protocol.  The
methods include Shanks/Wynn transformations~\cite{Shanks1955,Wynn1956},
Borel-normalized variants, the $\theta$~\cite{Brezinski1971},
$\rho$~\cite{Wynn1956rho}, and Levin-type~\cite{Levin:1973,Weniger:1989}
algorithms, D-log Pad\'es~\cite{BakerGravesMorris}, and transformations supplied with external
renormalon information.  The objective is not to enlarge the catalogue of
extrapolants, but to determine which statements remain supported after
renormalon-sensitive moment tests, a blind $\beta$-function benchmark, the
pole--$\MSbar$ mass relation, and the exactly solvable large-$\beta_0$ limit.

The obstacle is structural rather than technical.  Perturbative QCD series are
asymptotic, and their factorial large-order growth is encoded by ultraviolet
(UV) and infrared (IR) renormalon singularities in the Borel
plane~\cite{Beneke:1998ui}.  The known massless Adler function contains only four
nontrivial coefficients,
\begin{equation}
 \Dhat(Q^2)=\sum_{n\geq1}c_{n,1}a^n(Q^2),\qquad
 a\equiv\frac{\as}{\pi},
 \label{eq:adler-series}
\end{equation}
with, for $n_f=3$ in the $\MSbar$ scheme~\cite{Bardeen:1978yd},
\begin{equation}
 \begin{aligned}
 c_{1,1}&=1, & c_{2,1}&=1.639821204,\\
 c_{3,1}&=6.371014, & c_{4,1}&=49.07570
 \end{aligned}
 \label{eq:known-adler}
\end{equation}
(Ref.~\cite{Baikov:2008jh} and references therein).
Estimating additional coefficients is therefore useful but intrinsically model
dependent~\cite{Pich:2020gzz,David:2013gaa}.  Scheme-invariant
(effective-charge and minimal-sensitivity) estimates~\cite{Kataev:1995vh},
renormalon models of the Adler function~\cite{Beneke:2008ad,Caprini:2009vf,
DescotesGenon:2010cr,Jamin:2010cv,Caprini:2010tf,Caprini:2011ya,Beneke:2012vb},
Pad\'e and Borel--Pad\'e approximants~\cite{Boito:2018rwt}, and conformal
mappings of the Borel plane~\cite{Caprini:2019kwp} have all been applied to
this problem.  More broadly, the series has been reorganized by renormalon-chain
resummation~\cite{LovettTurner:1994hx,Ball:1995ni,LovettTurner:1995ti,
Neubert:1995gd,Maxwell:1996he,Maxwell:1997yw,Maxwell:2001he}, modified Borel
and ultraviolet-renormalon studies~\cite{Altarelli:1994vz,Cvetic:2001ws},
renormalization-group-summed and conformally mapped
expansions~\cite{Abbas:2012py,Abbas:2012fi,Abbas:2013usa,Caprini:2020lff,
Caprini:2023tfa,Caprini:2024unl}, series acceleration~\cite{Schilcher:2024fjx},
and scheme or scale variations~\cite{Boito:2016pwf,Wu:2018cmb,Caprini:2018agy,
Wu:2019mky}.  Sequence transformations with explicit remainder estimates have a
broader mathematical history~\cite{Brezinski1971,Levin:1973,Smith:1979,
Smith:1982,Weniger:1989} and have been used to predict perturbative
coefficients in quantum field theory~\cite{Jentschura:1999vn,
Jentschura:2000iw}; Levin-type transformations of the FOPT series for
$\delta^{(0)}$ gave $c_{5,1}=278^{+27}_{-19}$ in Ref.~\cite{Abbas:2025dpm}.  We distinguish throughout
between data-driven transformations, which infer a remainder pattern from the
available terms, and renormalon-informed transformations, which test a
predeclared Borel structure.  Only the latter can encode a sign-alternating
ultraviolet contribution that is not visible in four positive coefficients.

The renormalon ansatz used for the physical forecast must also be attributed
precisely.  Its leading singularity positions and renormalization-group
exponents follow Beneke and Jamin~\cite{Beneke:2008ad}, but its singularities
are retained as pure powers: the subleading cut terms and the gluon-condensate
Wilson-coefficient correction of the full Beneke--Jamin (BJ) model are omitted.
We therefore call it a \emph{pure-power Beneke--Jamin-type ansatz}.  The term
\emph{cross-validated renormalon closure} (CVRC) denotes only the calibration,
acceptance, and rejection protocol applied to that ansatz.  For brevity,
``BJ--CVRC'' refers to their combination; it does not claim a new Borel-model
class.

All coefficient forecasts in this paper are constructed in fixed-order
perturbation theory (FOPT).  The contour-improved calculation is used only as
a coupling-dependent diagnostic of how the tolerance changes; no CIPT
coefficient forecast or FOPT--CIPT average is formed.  Numerical spreads are
method or model envelopes, not confidence intervals.  The abbreviations used
throughout, and the status labels attached to every forecast, are collected in
Table~\ref{tab:acronyms}.

The presentation follows the validation logic.  Part~I defines the observable
and moment map.  Part~II states the forecast families, the predeclared
selection rules, and the single Adler tolerance to which the three-weight
criterion reduces.  Part~III summarizes the external validation before the
unknown coefficient is forecast.  Part~IV proves the defect-sign obstruction,
calibrates the pure-power BJ-type ansatz, and derives the one headline result.
Part~V compares that result with every other continuation constructed here,
in one convention, and states which combinations are legitimate.
All arguments needed to assess the transfer identity, the one-tolerance
collapse, the defect-sign obstruction, and the conditional BJ--CVRC estimate
are contained in the main text.  The appendices contain the extended
physical-order survey, complete algorithmic derivations, benchmark tables,
and stability scans on which the summaries in the main text rest.

A shorter, self-contained account of the main results, the transfer
identity and the one-tolerance collapse in FOPT and CIPT, the defect-sign
obstruction, and the conditional BJ--CVRC estimate, is given in the
companion letter~\cite{Abbas:2026letter}.  The present paper supplies the full
algorithm families, external benchmarks, stability and misspecification scans, and 
numerical tables on which that account rests.

\parttitle{Part I. Observable, moments, and the renormalon problem}
This part fixes the observable, coupling convention, contour weight, and the
distinction between Adler and moment coefficients.  It then derives the
selection rule that determines which moments retain the $u=2$ singularity.
Nothing in this part depends on an extrapolation algorithm.

\section{QCD framework for inclusive hadronic \texorpdfstring{$\tau$}{tau} decay}
\label{sec:qcd-framework}
A coefficient extrapolation is meaningful only after fixing the observable,
the coupling convention, the contour weight, and the distinction between the
Adler coefficients and the integrated moment coefficients.  This section
therefore discusses the massless perturbative quantity to which every sequence
transformation below is applied and states what the transformation can and
cannot say about the OPE and the FOPT--CIPT separation.

\begin{table*}[t]
\centering
\small
\caption{Abbreviations and status labels used throughout the paper, including its
appendices.  The status labels describe logical role, not statistical quality.}
\label{tab:acronyms}
\begin{tabularx}{\textwidth}{@{}lXlX@{}}
\toprule
symbol & meaning & symbol & meaning\\
\midrule
BJ & Beneke--Jamin & FOPT & fixed-order perturbation theory\\
CIPT & contour-improved perturbation theory & OPE & operator product expansion\\
CVRC & cross-validated renormalon closure & DSO & defect-sign obstruction\\
ARTS & Adler-projected renormalon-transported Shanks
 & PORT & position-only ratio transport\\
OAMDT & order-adaptive multi-mode defect transport
 & ATO & accuracy through order\\
\addlinespace
\textsc{base} & predeclared reference continuation, retained only for comparison
 & \textsc{cond} & conditional one-order-ahead continuation\\
\textsc{roll} & model rolled beyond one validated step
 & \textsc{indet} & rolled interval includes zero\\
\textsc{diag} & diagnostic only; fails a declared test or lacks ATO
 & \textsc{abst} & method declines to forecast\\
\bottomrule
\end{tabularx}
\end{table*}

\subsection{Inclusive rate and current correlators}

For the nonstrange charged currents
\begin{equation}
 J_{V/A}^{\mu}(x)=\bar u(x)\gamma^\mu(1,\gamma_5)d(x),
\end{equation}
we define
\begin{align}
 \Pi_{V/A}^{\mu\nu}(q)
 &=i\!\int d^4x\,e^{iqx}
 \langle0|T\{J_{V/A}^{\mu}(x)J_{V/A}^{\nu\dagger}(0)\}|0\rangle
 \nonumber\\
 &=(q^\mu q^\nu-q^2g^{\mu\nu})\Pi_{V/A}^{(1)}(q^2)
   +q^\mu q^\nu\Pi_{V/A}^{(0)}(q^2).
 \label{eq:correlator}
\end{align}
The optical theorem~\cite{Braaten:1991qm} expresses the inclusive spectral
distribution through
$\rho^{(J)}(s)=\pi^{-1}\operatorname{Im}\Pi^{(J)}(s+i0)$.  Up to the
well-known electroweak factor, the nonstrange rate is
\begin{align}
 R_{\tau,V/A}={}&12\pi S_{\rm EW}|V_{ud}|^2
 \int_0^{m_\tau^2}\frac{ds}{m_\tau^2}
 \left(1-\frac{s}{m_\tau^2}\right)^2
 \nonumber\\[-1mm]
 &\times\left[\left(1+2\frac{s}{m_\tau^2}\right)
 \operatorname{Im}\Pi_{V/A}^{(1)}(s)
 +\operatorname{Im}\Pi_{V/A}^{(0)}(s)\right].
 \label{eq:Rtau-spectral}
\end{align}
The transverse factor in Eq.~\eqref{eq:Rtau-spectral} is the fixed physical
spectral weight
\begin{equation}
 w_\tau(x)=(1-x)^2(1+2x)=1-3x^2+2x^3,
 \qquad x=\frac{s}{m_\tau^2}.
 \label{eq:physical-spectral-weight}
\end{equation}
It is fixed by the phase space and spin structure of the total inclusive
width; it is not selected to improve perturbation theory.  The experiment
also provides a binned differential spectrum and its covariance matrix.  They
permit additional observables
\begin{equation}
 I_{\rm exp}^{(w)}(s_0)=\frac{1}{s_0}\int_0^{s_0}ds\,
 w\!\left(\frac{s}{s_0}\right)\rho_{\rm exp}(s),
 \qquad s_0\leq m_\tau^2,
 \label{eq:experimental-moment-main}
\end{equation}
with weights chosen for a stated theoretical or experimental purpose.  For
binned data, $I_w=\sum_i q_i[w]\rho_i$ and
$\operatorname{Cov}(I_w,I_v)=\sum_{ij}q_i[w]C_{ij}q_j[v]$, with
$q_i[w]\simeq(\Delta s_i/s_0)w(s_i/s_0)$ (or the corresponding bin-averaged
weight) and $C_{ij}$ the published spectral covariance matrix.  It contains
statistical, normalization, efficiency, background, bin-migration,
and unfolding correlations.  Thus many moments can be formed, but they are
not independent measurements merely because their polynomial weights differ.
The unmodified total width is the special case $s_0=m_\tau^2$ with
$w=w_\tau$; every other weight, or a lower $s_0$, requires the differential
spectrum and its covariance.

In the massless limit the longitudinal nonstrange contribution vanishes,
apart from the separately treated pion pole.  The inclusive $V+A$ rate is
conventionally organized as
\begin{equation}
 R_{\tau,V+A}=N_c|V_{ud}|^2S_{\rm EW}
 \left[1+\delta'_{\rm EW}+\delta^{(0)}
 +\sum_{D\geq2}\delta^{(D)}\right].
 \label{eq:Rtau-decomposition}
\end{equation}
The electroweak factors $S_{\rm EW}$ and $\delta'_{\rm EW}$ are given in
Refs.~\cite{Marciano:1988vm,Braaten:1990ef,Davier:2002dy}; mass and power
corrections $\delta^{(D)}$ are discussed in
Refs.~\cite{Braaten:1991qm,Pich:1999hc}.  The spectral functions that enter
Eq.~\eqref{eq:Rtau-spectral} are those measured by ALEPH~\cite{ALEPH:2005qgp,
Davier:2013sfa}, with more recent combinations including other
data~\cite{Boito:2020xli,Boito:2025pwg}.  The present work concerns only the massless
perturbative term $\delta^{(0)}$; it does not refit electroweak factors,
quark-mass corrections, condensates, or duality violations.

\subsection{Analyticity, the Adler function, and the contour moment}

Analyticity converts the weighted spectral integral into a contour integral
on $|s|=s_0$~\cite{Narison:1988ni,Braaten:1991qm}.  We normalize the reduced Adler function by
\begin{equation}
 \Dhat(Q^2)=D(Q^2)-1,
 \qquad D(Q^2)=-4\pi^2Q^2\frac{d\Pi^{(1+0)}(Q^2)}{dQ^2},
 \label{eq:adler-definition}
\end{equation}
so that the first perturbative coefficient is unity.  After integration by
parts a polynomial Adler weight $W$ gives
\begin{equation}
 \delta_W^{(0)}(s_0)=\frac{1}{2\pi i}
 \oint_{|x|=1}\frac{dx}{x}\,W(x)\Dhat(-s_0x),
 \qquad x=\frac{s}{s_0}.
 \label{eq:general-contour-moment}
\end{equation}
For a polynomial spectral weight normalized by
$2\int_0^1\!dz\,w(z)=1$, our Adler-weight convention is
\begin{equation}
 W(x)=2\int_x^1 dz\,w(z).
 \label{eq:spectral-to-adler-weight}
\end{equation}
Consequently Eq.~\eqref{eq:physical-spectral-weight} gives
$W_\tau(x)=(1-x)^3(1+x)=1-2x+2x^3-x^4$.  The distinction between $w_\tau$
and $W_\tau$ is essential: the former multiplies the measured spectral
function, whereas the latter multiplies the Adler function on the contour.

\subsection{Renormalization group, FOPT, and CIPT}

With $a=\alpha_s/\pi$ we use
\begin{equation}
 \begin{aligned}
 \frac{da}{d\ln\mu^2}&=-\sum_{j\geq0}\beta_j a^{j+2},\\
 \beta_0&=\frac94,\qquad \beta_1=4,\qquad
 \beta_2=\frac{3863}{384}\quad(n_f=3).
 \end{aligned}
 \label{eq:beta-convention}
\end{equation}
The logarithmic Adler expansion can be written
\begin{equation}
 \Dhat(a,L)=\sum_{n\geq1}a^n\sum_{k=1}^{n}k c_{n,k}L^{k-1},
 \qquad L=\ln\frac{Q^2}{\mu^2}.
 \label{eq:adler-log-expansion}
\end{equation}
Only $c_{n,1}$ is independent at each order; RG invariance fixes the
$k>1$ coefficients from lower orders and the beta function.  Where needed
below, the four-loop coefficient $\beta_3$ and the five-loop coefficient
$\beta_4$ are taken from the standard $\MSbar$ results, in particular
Refs.~\cite{Baikov:2016tgj,Herzog:2017ohr}.  Where a numerical coupling is required we take
$\alpha_s(M_Z)=0.11873$~\cite{Brida:2025gii}, as in
Ref.~\cite{Abbas:2025dpm}, and evolve it to the $\tau$ mass with
RunDec/CRunDec~\cite{Chetyrkin:2000yt,Herren:2017osy} using five-loop running
and $\MSbar$ decoupling of the bottom and charm quarks, which gives
\begin{equation}
 \alpha_s(m_\tau)=0.3196,\qquad a(m_\tau^2)=0.10173\quad(n_f=3).
 \label{eq:alphas-input}
\end{equation}
Only the contour-improved tolerances of Sec.~\ref{sec:tolerance-cipt} and the illustrative moment values
quoted there depend on this input; every statement about Adler coefficients
is coupling independent.  For scale, $a^5=1.090\times10^{-5}$, so a change
$\Delta c_{5,1}=80$, the total uncertainty of Eq.~\eqref{eq:cvrc-final},
shifts every FOPT moment by $\Delta c_{5,1}a^5\simeq8.7\times10^{-4}$ (the
unit diagonal makes the shift weight independent), about $0.4\%$ of
$\delta^{\rm FOPT}_{W_\tau}\simeq0.203$ evaluated through fifth order at
$c_{5,1}=277$, the central value of the independent Borel--Pad\'e estimate of
Ref.~\cite{Boito:2018rwt}.  This value is used only as a common fifth-order
sensitivity benchmark in the contour-comparison tables; it is neither an
input to CVRC nor the central value adopted later in this paper.

Both prescriptions below are truncations of the moment $\delta^{(0)}_W$ of
Eq.~\eqref{eq:general-contour-moment} at order $N$.  FOPT sets $\mu^2=s_0$,
expands the running coupling and logarithms, and integrates term by term:
\begin{equation}
 \delta^{\rm FOPT}_{W,N}(s_0)=
 \sum_{n=1}^{N}d_n[W]a^n(s_0).
 \label{eq:fopt-series}
\end{equation}
CIPT~\cite{Pivovarov:1991rh,LeDiberder:1992jjr} instead uses the running
coupling along the contour and truncates the Adler series before integration,
\begin{equation}
 \delta^{\rm CIPT}_{W,N}(s_0)=\frac{1}{2\pi i}
 \sum_{n=1}^{N}c_{n,1}\oint_{|x|=1}\frac{dx}{x}
 W(x)a^n(-s_0x).
 \label{eq:cipt-series}
\end{equation}
These operations need not commute at finite or asymptotically large
order~\cite{Beneke:2008ad,Hoang:2020mkw,Hoang:2021nlz,Gracia:2023qdy}.
Moreover,
the forecasts in this paper concern $c_{n,1}$ or the derived FOPT
coefficients $d_n[W]$.

\subsection{OPE, renormalons, and the scope of resummation}

Away from the timelike axis the correlator admits the schematic
OPE~\cite{Wilson:1969zs}
\begin{equation}
 D(Q^2)=D_{\rm pert}(Q^2)+
 \sum_D\frac{C_D(Q^2,\mu)\langle O_D(\mu)\rangle}{Q^D}.
 \label{eq:ope}
\end{equation}
An IR Borel singularity at $u=p$ is associated parametrically with a
$D=2p$ power correction, while negative-$u$ singularities are UV
renormalons~\cite{Beneke:1998ui}.  The ambiguity of a lateral Borel sum is cancelled, in a
consistent scheme and prescription, by the corresponding ambiguity in the
operator matrix element; renormalon-free definitions of the gluon
condensate~\cite{Benitez-Rathgeb:2022yqb,Benitez-Rathgeb:2022hfj} and
gradient-flow formulations of the OPE~\cite{Beneke:2023wkq,Beneke:2025hlg}
make this compensation explicit.  Therefore neither Pad\'e approximation nor any
sequence transformation removes the physical OPE prescription dependence.
The transformations can accelerate or model a declared perturbative
prescription; they cannot manufacture the nonperturbative matrix elements.

This distinction also explains why moment weights matter.  A zero of the
contour kernel can reduce or, for a simple pole, cancel a particular
renormalon contribution.  It does not prove the absence of the associated
operator from full QCD, where anomalous dimensions generally turn poles into
branch points.  The precise weight rule is derived next.

\section{Spectral moments and the \texorpdfstring{$u=2$}{u=2} cancellation}
\label{sec:moments}
The standard total-$\tau$ weight suppresses the leading gluon-condensate
renormalon and is therefore an unusually forgiving extrapolation
target~\cite{Beneke:2008ad,Beneke:2012vb}.  We
first derive the weight-dependent Borel factor so that the subsequent tests
include moments in which the $u=2$ singular behavior is not suppressed.

We write a normalized polynomial Adler weight as
\begin{equation}
 W(x)=\sum_{m=0}^{M}W_m x^m,\qquad W_0=W(0)=1,
 \label{eq:weight-polynomial}
\end{equation}
related to the spectral weight by Eq.~\eqref{eq:spectral-to-adler-weight};
its massless perturbative moment is Eq.~\eqref{eq:general-contour-moment}.

We use the integer-renormalon Borel variable $u=\beta_0t$ and the
convention~\cite{Beneke:1998ui,Beneke:2008ad}
\begin{equation}
 B[\Dhat](u)=\sum_{n=0}^{\infty}
 \frac{c_{n+1,1}}{\beta_0^n n!}\,u^n,
 \qquad
 \Dhat_{\mathcal P}(a)=\frac{1}{\beta_0}
 \int_{\mathcal C_{\mathcal P}}du\,
 e^{-u/(\beta_0a)}B[\Dhat](u),
 \label{eq:borel-normalization}
\end{equation}
where $\mathcal P$ specifies the lateral or principal-value contour when the
positive real axis contains IR singularities.  Expanding the integral term by
term reproduces Eq.~\eqref{eq:adler-series}; the prefactor $1/\beta_0$ is
therefore part of the convention and must not be dropped.

At one-loop running the Borel transform factorizes.  With the branch of
$\ln(-x)$ continuous above and below the timelike axis,
\begin{align}
 B[\delta_W^{(0)}](u)&=K_W(u)B[\Dhat](u),
 \label{eq:borel-factorization}\\
 K_W(u)&=\frac{1}{2\pi i}\oint_{|x|=1}\frac{dx}{x}
 W(x)(-x)^{-u}\nonumber\\
 &=\frac{\sin(\pi u)}{\pi}
 \sum_{m=0}^{M}\frac{W_m}{u-m}.
 \label{eq:Jwu}
\end{align}
Equation~\eqref{eq:Jwu} is derived explicitly in
Appendix~\ref{app:contour}.

Near a positive integer $p$,
\begin{equation}
 K_W(u)=(-1)^pW_p+\cO(u-p)\quad (W_p\neq0),
 \label{eq:integer-limit}
\end{equation}
whereas $K_W(u)=\cO(u-p)$ when $W_p=0$.  Consequently a simple IR pole of
$B[\Dhat](u)$ at $u=p$ survives if $W_p\neq0$ and is cancelled if
$W_p=0$.  Appendix~\ref{sec:order-rule} sharpens this into a statement about the
\emph{order} of the surviving singularity, Eq.~\eqref{eq:order-rule}, which is
what fixes the exponents used by the transformation constructed there.

We compare
\begin{align}
 W_\tau(x)&=(1-x)^3(1+x)=1-2x+2x^3-x^4,
 \label{eq:Wtau}\\
 W_{\rm GC}(x)&=(1-x)^3=1-3x+3x^2-x^3,
 \label{eq:Wgc}\\
 W_{P2}(x)&=(1-x)^2(1+x)=1-x-x^2+x^3.
 \label{eq:Wp2}
\end{align}
Here ``$P2$'' is only a mnemonic for the doubly pinched comparison weight,
whose factor $(1-x)^2$ gives a double zero at the timelike endpoint.  The standard weight has $(W_\tau)_2=0$; the pinched comparison weights have
$(W_{\rm GC})_2=3$ and $(W_{P2})_2=-1$.  Thus both comparison moments retain
the leading $u=2$ singular behavior, with different moment factors.  In the
large-$\beta_0$ limit the Adler singularity is a simple pole, so the zero of
$K_{W_\tau}$ cancels it completely.  In full QCD it is a branch point: a zero
of $K_W$ lowers the branch exponent by one but need not make the Borel
transform analytic.  Thus $W_2=0$ and $W_2\ne0$ distinguish suppressed and
unsuppressed $D=4$ sensitivity; complete cancellation is a special property
of the simple-pole limit~\cite{Beneke:2012vb}.

\subsection{The fixed-order weight map and three facts used throughout}
\label{sec:weight-map}

Three elementary facts about the map from Adler to moment coefficients are
used repeatedly in the rest of the paper.  They are collected here because
none belongs to a particular transformation: they are properties of the weight
map itself, and the strongest results of Parts~III and~IV rest on them.

\paragraph{Triangularity and unit diagonal.}
In FOPT the scale is fixed at $\mu^2=s_0$, and the moment is the power
series~\eqref{eq:fopt-series} in $a_0\equiv a(s_0)$.  On the contour
$Q^2=-s_0x$, so the logarithm of Eq.~\eqref{eq:adler-log-expansion} is
$L=\ln(-x)$, and integrating term by term gives
\begin{equation}
 d_N[W]=\sum_{k=1}^{N}k\,c_{N,k}\,J_{k-1}[W],
 \qquad
 J_l[W]\equiv\frac{1}{2\pi i}\oint_{|x|=1}\frac{dx}{x}\,W(x)\,\ln^l(-x).
 \label{eq:dN-JN}
\end{equation}
The contour runs counterclockwise and $\ln(-x)$ has its cut on the positive
real $x$ axis.  The logarithmic moments $J_l[W]$ are pure numbers fixed by the
weight polynomial; all QCD dynamics resides in the coefficients $c_{N,k}$.

Renormalization-group invariance fixes every $c_{N,k}$ with $k\geq2$ as a
linear combination of lower-order coefficients $c_{m,1}$, $m\leq N-k+1$, with
coefficients that are polynomials in $\beta_0,\ldots,\beta_{N-2}$.
Substituting into Eq.~\eqref{eq:dN-JN} gives a linear map
\begin{equation}
 d_N[W]=\sum_{m=1}^{N}M_{Nm}[W]\,c_{m,1},
 \qquad M_{Nm}[W]=0\quad(m>N).
 \label{eq:linear-weight-map}
\end{equation}
The matrix $M[W]$ is lower triangular.  Its off-diagonal entries are
combinations of $J_1[W],\ldots,J_{N-m}[W]$ weighted by products of $\beta$
coefficients; they describe how the running carries $c_{m,1}$ up to order
$a_0^N$.  Separating the diagonal,
\begin{equation}
 d_N[W]=M_{NN}[W]\,c_{N,1}+R_N[W],
 \qquad
 R_N[W]\equiv\sum_{m=1}^{N-1}M_{Nm}[W]\,c_{m,1},
 \label{eq:triangular-weight-map}
\end{equation}
where the remainder $R_N[W]$ is fixed by $c_{1,1},\ldots,c_{N-1,1}$, by
$\beta_0,\ldots,\beta_{N-2}$, and by $J_1[W],\ldots,J_{N-1}[W]$.

The diagonal comes from the $k=1$ term of Eq.~\eqref{eq:dN-JN}, which has no
logarithm.  With $W(x)=\sum_jW_jx^j$, only the constant term survives the
contour integral, so for a normalized weight
\begin{equation}
 M_{NN}[W]=J_0[W]=\frac{1}{2\pi i}\oint\frac{dx}{x}\,W(x)=W_0=1
 \label{eq:unit-diagonal}
\end{equation}
and $d_N[W]=c_{N,1}+R_N[W]$.  This holds for every normalized polynomial
weight and was verified symbolically for all weights and orders used here.
The map is therefore invertible order by order, $c_{N,1}=d_N[W]-R_N[W]$, and a
shift of $c_{N,1}$ moves the order-$a_0^N$ coefficient of every normalized
moment by the same amount.  At fixed order, all differences between weights
enter through $R_N[W]$.

\paragraph{The transfer identity.}
Let $c_{1,1},\ldots,c_{N,1}$ be known, and let any transformation acting on
the common Adler sequence (a sequence accelerator, a Pad\'e or
conformal-mapping construction, a renormalon model) return
$\widehat c_{N+1,1}$, with error $\Delta c\equiv\widehat c_{N+1,1}-c_{N+1,1}$.
Propagating the forecast through Eq.~\eqref{eq:triangular-weight-map} at order
$N+1$ gives
\begin{equation}
 \widehat d_{N+1}[W]=M_{N+1,N+1}[W]\,\widehat c_{N+1,1}
 +\sum_{m=1}^{N}M_{N+1,m}[W]\,\widehat c_{m,1},
 \label{eq:forecast-dN1}
\end{equation}
where $\widehat c_{m,1}$ are the lower-order values the reconstruction
actually used.  Subtracting the exact $d_{N+1}[W]$, with the same $\beta$
coefficients, gives the exact relation
\begin{equation}
 \widehat d_{N+1}[W]-d_{N+1}[W]
 =M_{N+1,N+1}[W]\,\Delta c
 +\sum_{m=1}^{N}M_{N+1,m}[W]\,\bigl(\widehat c_{m,1}-c_{m,1}\bigr).
 \label{eq:general-error-transfer}
\end{equation}
The first term comes from the forecast itself.  The second comes from any
mismatch in the lower-order inputs and depends on the weight through
$J_1[W],\ldots,J_N[W]$.  If the forecast and the reconstruction use the same
known inputs, the second term vanishes and the unit diagonal gives
\begin{equation}
 \widehat d_{N+1}[W]-d_{N+1}[W]=\widehat c_{N+1,1}-c_{N+1,1}
 \label{eq:arts-error-transfer}
\end{equation}
for every normalized weight (for an unnormalized one the right-hand side is
$W_0\,\Delta c$), and correspondingly
$\widehat\delta^{(0)}_W-\delta^{(0)}_W=a_0^{N+1}\,\Delta c$ at fixed $a_0$.
An error in the last supplied Adler coefficient therefore reaches every
normalized moment with the same absolute size, whatever transformation
produced it: the map does not know how $\widehat c_{N+1,1}$ was obtained.  Two
conditions are essential.  The forecast must be made for the Adler sequence
and then propagated; if each moment series is extrapolated separately,
nothing forces a weight-independent error.  And the lower-order inputs must
agree; any mismatch reintroduces the second term of
Eq.~\eqref{eq:general-error-transfer}.  A corollary is that a combination
$\sum_i\lambda_i d_{N+1}[W_i]$ of normalized moments has forecast error
$(\sum_i\lambda_i)\,\Delta c$, which vanishes when $\sum_i\lambda_i=0$.

Equal absolute errors do not mean equal relative errors,
\begin{equation}
 \frac{\widehat d_{N+1}[W]-d_{N+1}[W]}{d_{N+1}[W]}
 =\frac{\Delta c}{c_{N+1,1}+R_{N+1}[W]} .
 \label{eq:relative-error}
\end{equation}
A moment such as $W_{P2}$ can be small because $c_{N+1,1}$ and $R_{N+1}[W]$
nearly cancel, which amplifies the common error in relative terms.  The
identity is also specific to FOPT: in CIPT the coefficient of $c_{N+1,1}$ is
the weight-dependent contour integral $J^{\rm CI}_{N+1}[W]$ of
Eq.~\eqref{eq:cipt-diagonal} rather than $W_0$.  Section~\ref{sec:tolerance}
turns Eq.~\eqref{eq:arts-error-transfer} into the tolerance that governs every
three-weight test in this paper, and Sec.~\ref{sec:tolerance-cipt} shows what
survives after contour improvement.

\paragraph{The Pad\'e--Shanks defect.}
Let
\begin{equation}
 F(z)=\sum_{n=1}^{\infty}c_nz^n
\end{equation}
be a specified perturbative series, and suppose that
$c_1,\ldots,c_N$ are known, with the relevant denominators assumed to be
nonzero.  After factoring the known leading power,
\begin{equation}
 \frac{F(z)}{z}
 =
 c_1+c_2z+\cdots+c_Nz^{N-1}
 +\mathcal O(z^N),
\end{equation}
the \(N\) supplied coefficients determine a Pad\'e approximant satisfying
\(L+M=N-1\).  The denominator-degree-one choice is therefore the
\([N-2/1]\) approximant of \(F(z)/z\).

Equivalently, the first Shanks, or Aitken-\(\Delta^2\), transformation of the
last three partial sums assumes that the last observed coefficient ratio
persists for one additional order:
\begin{equation}
 \frac{\widehat c_{N+1}^{\,\mathrm{dir}}}{c_N}
 =
 \frac{c_N}{c_{N-1}}.
 \label{eq:local-geometric-ratio}
\end{equation}
It therefore predicts
\begin{equation}
 \widehat c_{N+1}^{\,\mathrm{dir}}
 =
 \frac{c_N^2}{c_{N-1}}.
 \label{eq:direct-pade-forecast-local}
\end{equation}
Thus the first Shanks element and the one-pole \([N-2/1]\) Pad\'e
approximant encode the same local geometric-remainder assumption.

For a known coefficient \(c_n\), the corresponding one-step
Pad\'e--Shanks defect is
\begin{equation}
 \Delta_n
 \equiv
 c_n-\frac{c_{n-1}^2}{c_{n-2}},
 \qquad n\geq2,
 \qquad c_0\equiv1.
 \label{eq:pade-defect}
\end{equation}
Here \(c_0\equiv1\) is only a boundary convention used to define
\(\Delta_2\); it is not counted as an additional supplied perturbative
coefficient when assigning the Pad\'e degrees.

The defect is the signed departure from the direct geometric continuation:
\(\Delta_n>0\) means that \(c_n\) lies above
\(c_{n-1}^2/c_{n-2}\), whereas \(\Delta_n<0\) means that it lies below it.
At the first unknown order, the direct continuation imposes
\begin{equation}
 \widehat c_{N+1}^{\,\mathrm{dir}}
 =
 \frac{c_N^2}{c_{N-1}},
 \qquad
 \widehat\Delta_{N+1}=0.
 \label{eq:zero-forecast-defect}
\end{equation}
The vanishing forecast defect is imposed by the direct first Shanks/Aitken
transformation and its corresponding denominator-degree-one Pad\'e
continuation.  It is not a property of an arbitrary \([L/M]\) approximant
with \(M>1\).

The defect is sequence and normalization dependent.  Applying the same
construction to ordinary coefficients, Borel-normalized coefficients, or
weighted-moment coefficients defines different defects.  Unless explicitly
stated otherwise, Eq.~\eqref{eq:pade-defect} is applied in this paper to the
normalized common-Adler coefficient sequence.  It is the defect variable used
in the cross-weight obstruction of Sec.~\ref{sec:obstruction}; the detailed
Shanks--Pad\'e derivation is given in Appendix~\ref{sec:protocol}.

\parttitle{Part II. Extrapolation algorithms and the validation protocol}

Part~II summarizes the transformation families and, before any of them is
applied, states the accuracy-through-order requirement, the leave-one-order-out
protocol, and the ten-percent acceptance metric used by all later tables, and
shows that this metric is a single tolerance on the Adler coefficient.  It
closes with a brief account of every method that produces a numerical
forecast, Sec.~\ref{sec:methods-brief}.  The complete definitions and
derivations are provided in the appendices.

\section{Forecast families and predeclared validation protocol}
\label{sec:protocol-summary}

The analysis compares three levels of input.  Direct transformations act on
the Adler coefficients or their partial sums; Borel normalization first
removes an assumed factorial factor; and
renormalon-informed transformations additionally receive singularity
positions, exponents, and a Borel prescription.  These are distinct
hypotheses, not interchangeable implementations of one estimator.  Complete
definitions, accuracy-through-order derivations, and recursion formulae are
collected in Appendix~\ref{sec:protocol}.

\paragraph{Accuracy through order.}
\label{par:accuracy-through-order}

Let the known truncation of a formal perturbative series be
\begin{equation}
 F_K(a)=\sum_{n=0}^{K} f_n a^n .
 \label{eq:known-truncated-series}
\end{equation}
A rational approximant or transformed continuation
$\widehat F_K(a)$ is said to have \emph{accuracy through order $K$} if its
Taylor re-expansion reproduces every supplied coefficient:
\begin{equation}
 [a^n]\widehat F_K(a)=f_n,
 \qquad 0\leq n\leq K .
 \label{eq:ato-coefficient-definition}
\end{equation}
Here $[a^n]G(a)$ denotes the coefficient multiplying $a^n$ in the Taylor
expansion of $G(a)$.  Equivalently, the accuracy-through-order condition can
be written as
\begin{equation}
 \widehat F_K(a)-F_K(a)
 =\mathcal O\!\left(a^{K+1}\right).
 \label{eq:ato-remainder-definition}
\end{equation}
Only after this matching condition has been satisfied do we interpret
\begin{equation}
 \widehat f_{K+1}
 \equiv [a^{K+1}]\widehat F_K(a)
 \label{eq:ato-next-coefficient}
\end{equation}
as a one-order-ahead coefficient forecast.

Accuracy through order is an algebraic matching condition.  It is stronger
than numerical stability at one particular value of the coupling.  A
transformation may give a stable estimate of the summed series at a fixed
value of $a$ while changing one or more of the already known coefficients
when it is re-expanded in powers of $a$.  Such a construction remains a
legitimate fixed-coupling extrapolant, but it is not an
accuracy-through-order coefficient forecast.  This distinction is important
below for the canonical $\rho_2$ transformation and for fixed-coupling
implementations of Levin-type transformations.  Detailed method-specific
derivations are given in Appendix~\ref{sec:protocol}.
\paragraph{Balanced Pad\'e family.}
For a series supplied through $N$ coefficients, a Pad\'e approximant
$[L/M]=P_L/Q_M$, with $Q_M(0)=1$, uses all inputs when $L+M=N-1$.
Our baseline is the predeclared near-diagonal family
\begin{equation}
 L+M=N-1,\qquad M\geq1,\qquad |L-M|\leq1,
 \label{eq:balanced-main}
\end{equation}
whose parameter counting, rationale, and validation are given in
Appendices~\ref{sec:protocol} and \ref{app:pade-forecast}.  The first equality
is exact parameter counting; the last inequality is a predeclared
methodological balance condition, not a theorem about the optimal Pad\'e
shape.  It
shares the available degrees as evenly as possible between numerator and
denominator.  Because the degrees are integers, $|L-M|<1$ would select only
the exactly diagonal case $L=M$, whereas Eq.~\eqref{eq:balanced-main} also
retains the adjacent near-diagonal approximants.

The unbalanced extremes clarify the choice.  The approximant $[N-1/0]$ is
only the known Taylor polynomial and predicts every unsupplied coefficient to
be zero; hence $[3/0]$ is not a genuine four-input extrapolation.  The
opposite choice $[0/N-1]$ is mathematically valid but assigns all nontrivial
freedom to the denominator.  Its exclusion from the baseline rests on tested
performance and asymptotic motivation rather than on a theorem that it must
fail at every low order.  Direct coefficient matching with the four known
Adler coefficients gives
$c_{5,1}^{[2/1]}=378.03$, $c_{5,1}^{[1/2]}=445.85$, and
$c_{5,1}^{[0/3]}=379.47$.  A $1\%$ increase of $c_{4,1}$ moves these three
forecasts by $+2.0\%$, $+2.3\%$, and $+2.0\%$, respectively.  The numerical
proximity of $[0/3]$ and $[2/1]$ at this order is therefore an accidental
low-order agreement, not an independent argument for denominator-only
continuation.  The single available lower-order postdiction is likewise not
decisive: at three inputs $[0/2]$ misses the withheld $c_{4,1}$ by $-47.8\%$,
while the balanced $[1/1]$ misses it by $-49.6\%$.

The exactly solvable large-$\beta_0$ benchmark gives the more discriminating
test.  The absolute Adler errors of the denominator-only sequence at forecast
orders $4,8,12,16$ are $558\%$, $353\%$, $0.51\%$, and $5.88\%$, respectively,
showing erratic rather than uniform convergence.  At order twelve its
$W_\tau$, $W_{\rm GC}$, and $W_{P2}$ moment errors are approximately $586\%$,
$277\%$, and $95.8\%$ (Table~\ref{tab:large-beta-errors}).  We therefore retain
$[0/3]$ only as an unbalanced diagnostic and exclude it from the baseline and
from every quoted envelope.  This is a predeclared methodological choice;
the isolated four-loop postdiction alone would not justify it.  Balance does
not exclude every pathology either: the balanced
common-Adler $[1/2]$ approximant contains a nearly cancelling pole--zero pair,
with a pole at $a=0.1072$ and a zero at $a=0.1130$, close to
$a(m_\tau^2)=0.10173$, which is one reason both balanced members are reported
and neither is selected after the target is inspected.
The balanced four-input family is $[2/1]$ and $[1/2]$, while five inputs
select $[2/2]$.

Every validation choice is fixed before the withheld coefficient is exposed.
The forecast must reproduce the supplied series through its last known order;
no approximant is removed after the target is inspected; and errors are
reported by the signed metric of Eq.~\eqref{eq:error}.  The same rules are
applied to the five-loop $\beta$ function, renormalon-sensitive moments, the
pole--$\MSbar$ mass relation, and the exactly solvable large-$\beta_0$ limit.
This separation between definition and test prevents a successful-looking
physical number from retroactively selecting its own algorithm.

\paragraph{Status labels.}\label{sec:status-labels}
Every forecast in the tables carries one of six labels, fixed by the protocol
rather than by the value obtained.  \textsc{base} marks a predeclared
reference continuation before validation and does not by itself imply
acceptance; \textsc{cond} marks a one-order-ahead continuation that has both
accuracy through order and the declared validation record; \textsc{roll}
marks the same declared model rolled forward more than one order past the last
supplied coefficient, and therefore untested by any one-order-ahead
validation; \textsc{indet} marks a rolled value whose envelope contains zero,
so that not even its sign is determined; \textsc{diag} marks a continuation
retained only as a diagnostic, because it fails a declared external test,
depends on an input the method cannot determine, or lacks accuracy through
order; and \textsc{abst} marks a method whose own acceptance certificate
declines to issue a physical prediction.  The epistemic level of
each class of result is stated in Appendix~\ref{sec:claims}.

\section{The three-weight acceptance criterion}
\label{sec:tolerance}
The ten-percent three-weight criterion fixed above is applied to every
construction in Parts~III and~IV and in the appendices.  Because of the unit diagonal of
Sec.~\ref{sec:weight-map}, it is not three conditions but one, and its
resolving power can be stated in advance, independently of any algorithm.

\subsection{The criterion is a tolerance on the Adler coefficient}
\label{sec:tolerance-adler}

The triangular map of Eq.~\eqref{eq:triangular-weight-map} has unit diagonal
for every normalized polynomial weight; this was verified symbolically for all
weights and all orders used here.  Consequently the transfer identity
\eqref{eq:arts-error-transfer} holds, and the declared ten-percent
three-weight criterion is, \emph{for any transformation acting on the common
Adler sequence}, exactly a tolerance on that single coefficient,
\begin{equation}
 \bigl|\widehat c_{N+1,1}-c_{N+1,1}\bigr|
 \;\leq\;\min_{W\in\{W_\tau,W_{\rm GC},W_{P2}\}}
 0.1\,\bigl|d_{N+1}[W]\bigr|.
 \label{eq:adler-tolerance}
\end{equation}
With the exact moments of Table~\ref{tab:qcd-moments},
\begin{equation}
 \bigl|\widehat c_{3,1}-c_{3,1}\bigr|\leq1.4523,
 \qquad
 \bigl|\widehat c_{4,1}-c_{4,1}\bigr|\leq2.1509,
 \label{eq:tolerance-numbers}
\end{equation}
because $W_{P2}$ is the smallest moment at both orders.  Equation
\eqref{eq:tolerance-numbers} is the single most useful reformulation in this
paper.  It says that the withheld-$d_4$ three-weight test, which reads as a
ten-percent criterion, is in fact a $1.69\%$ requirement on
$d_4[W_\tau]=127.0786$, and the withheld-$d_3$ test a $5.51\%$ requirement on
$d_3[W_\tau]=26.3659$.  The
apparent difficulty of $W_{P2}$ is not a property of that moment; it is the
amplification factor $d_4[W_\tau]/d_4[W_{P2}]=5.91$ acting on a
weight-independent error.

\subsection{Resolving power at the order of interest}
\label{sec:resolving-power}

Equation~\eqref{eq:adler-tolerance} can be pushed one order further, and the
result bears directly on how much any moment test can ever settle.  Unit
diagonal makes the order-five moments exactly linear in the unknown
coefficient,
\begin{equation}
 \begin{aligned}
 d_5[W_\tau]&=c_{5,1}+307.783,\\
 d_5[W_{\rm GC}]&=c_{5,1}+1156.925,\\
 d_5[W_{P2}]&=c_{5,1}-512.597,
 \end{aligned}
 \label{eq:d5-linear}
\end{equation}
so $W_{P2}$ remains the smallest moment throughout the range of candidate
values and Eq.~\eqref{eq:adler-tolerance} reads
\begin{equation}
 \bigl|\widehat c_{5,1}-c_{5,1}\bigr|\leq
 0.1\,\bigl|c_{5,1}-512.597\bigr|\simeq24 .
 \label{eq:resolving-five}
\end{equation}
Here $W_{P2}$ remains the most restrictive weight whenever
$|c_{5,1}-512.597|<|c_{5,1}+307.783|$, i.e.\ for $c_{5,1}>102.4$, which covers
the phenomenological range.  The ten-percent three-weight criterion therefore
resolves $c_{5,1}$ only to about $\pm24$ at the representative value
$c_{5,1}=277$: the tolerance ranges from $18$ to $29$ over the Borel--Pad\'e
range $277\pm51$, from $27$ to $21$ over the accuracy-through-order span
$[242,305]$, and shrinks to zero as $c_{5,1}\to512.6$, where $d_5[W_{P2}]$
vanishes and a relative criterion for that moment ceases to be meaningful.  The consequence is worth stating explicitly.  The linear-ratio-fit
value $283$ of Ref.~\cite{Beneke:2008ad} and the Borel--Pad\'e value $277\pm51$ differ by $6$, so no test of
this form could distinguish them at any order of moment data; the full span of
accuracy-through-order continuations collected in
Sec.~\ref{sec:global-comparison}, $241.70\leq c_{5,1}\leq304.71$, is $2.7$
tolerances wide in total.  A criterion that is a genuine discriminator at the
calibration order, where it is $\pm2.15$ on a coefficient of $49.08$, has by
one order later become comparable to the entire disagreement it would be asked
to settle.

\subsection{The same collapse in contour-improved perturbation theory}
\label{sec:tolerance-cipt}

The unit diagonal of Eq.~\eqref{eq:triangular-weight-map} is a property of
fixed-order perturbation theory, so it is fair to ask whether
Eq.~\eqref{eq:adler-tolerance} is an artifact of that choice.  It is not.  The
collapse survives contour improvement for a reason that does not use
triangularity at all, and the resulting test turns out to be very much weaker.

In CIPT the running coupling is kept on the contour and the Adler series is
truncated before integration, Eq.~\eqref{eq:cipt-series}, so that
\begin{equation}
 \delta^{\rm CIPT}_{W,N}=\sum_{n=1}^{N}c_{n,1}\,J^{\rm CI}_n[W],
 \qquad
 J^{\rm CI}_n[W]=\frac{1}{2\pi i}\oint\frac{dx}{x}\,W(x)\,a^n(-s_0x).
 \label{eq:cipt-diagonal}
\end{equation}
There is no triangular map here at all: the map from Adler coefficients to the
moment is \emph{diagonal}, and its diagonal entries $J^{\rm CI}_n[W]$ are
neither unity nor weight independent.  The transfer identity
\eqref{eq:arts-error-transfer} accordingly fails: a forecast error does
\emph{not} produce the same absolute discrepancy in every moment:
\begin{equation}
 \widehat\delta^{\rm CIPT}_{W,N}-\delta^{\rm CIPT}_{W,N}
 =\bigl(\widehat c_{N,1}-c_{N,1}\bigr)\,J^{\rm CI}_N[W].
 \label{eq:cipt-transfer}
\end{equation}
The right-hand side is nevertheless a single number multiplied by a known
weight-dependent factor, and that is all the collapse ever required.  Imposing
the relative criterion on a set $\mathcal W$ of weights gives
\begin{equation}
 \bigl|\widehat c_{N,1}-c_{N,1}\bigr|\;\leq\;
 \epsilon\,\min_{W\in\mathcal W}
 \left|\frac{\delta^{\rm CIPT}_{W,N}}{J^{\rm CI}_N[W]}\right| ,
 \label{eq:cipt-tolerance}
\end{equation}
one tolerance on one coefficient, exactly as in
Eq.~\eqref{eq:adler-tolerance} and with a different amplification factor.
Multi-moment testing is one-dimensional in both schemes.  What distinguishes
them is only how sharp the resulting single constraint is.

We evaluate Eq.~\eqref{eq:cipt-diagonal} by integrating
$da/dt=-i\beta(a)$ along $x=-e^{it}$, $t\in(-\pi,\pi]$, with $a(0)=a(s_0)$.  At the coupling
of Eq.~\eqref{eq:alphas-input} the diagonal entries are those of
Table~\ref{tab:cipt-diagonal}.  Relative to the fixed-order values $a^n$ they
are enhanced by factors $1.06$--$1.56$ for $W_\tau$ and $1.55$--$2.76$ for
$W_{\rm GC}$, whereas for $W_{P2}$ the enhancement at $n=1,2$ ($1.14$, $1.10$)
turns into a suppression at $n\geq3$, down to $0.24$ at $n=5$.

\begin{table}[t]
\centering
\small
\caption{Diagonal entries $J^{\rm CI}_n[W]$ of the CIPT map
\eqref{eq:cipt-diagonal} at $\alpha_s(m_\tau)=0.3196$, $n_f=3$.  Fixed-order
perturbation theory would put $a^n$ in every column; the CIPT map has no unit
diagonal and no weight independence.}
\label{tab:cipt-diagonal}
\input{generated_data/cipt_diagonal_table.tex}
\end{table}

Two independent checks precede any use of these numbers.  Setting
$\beta\equiv0$ freezes the coupling and must return $J^{\rm CI}_n\to a^n$ for
every weight, which it does to machine precision.  More stringently, the
small-$a$ expansion of Eq.~\eqref{eq:cipt-diagonal} must reproduce the columns
of the fixed-order matrix, $J^{\rm CI}_n[W](a)=\sum_{N\geq n}M_{Nn}[W]a^N$;
truncating that sum at $N=4$ leaves a residual which the coefficients
$M_{5n},\ldots,M_{8n}$ predict, and observed and predicted residuals agree to
between $1\times10^{-4}$ and $2\times10^{-3}$ relative for all three weights
and $n=1,2,3$.  This ties the contour integration to the independently computed
fixed-order predictions.  As a physical control, the four-loop
$\delta^{(0)}[W_\tau]$ evaluates to $0.19694$ in FOPT and $0.18198$ in CIPT at
this coupling, reproducing the familiar pattern and separation.

For the two renormalon-sensitive weights, the fixed-order separation is shown
explicitly in Table~\ref{tab:cipt-fopt-separation}.  Here ``separation'' means
the numerical difference between two truncation prescriptions at the same
order and coupling; it is not an experimental contamination and must not be
interpreted as an independently distributed uncertainty.  At four loops it is
only $+1.28\%$ for $W_{\rm GC}$ but $-11.92\%$ for $W_{P2}$; at five loops,
using the illustrative $c_{5,1}=277$, the corresponding differences are
$-1.89\%$ and $-10.06\%$.  The strong weight dependence is precisely why a
single FOPT--CIPT average would hide rather than resolve the asymptotic
prescription issue.

The fifth-order number $277$ in this paragraph and the following tables is a
fixed external reference point: it is the central Borel--Pad\'e estimate of
Ref.~\cite{Boito:2018rwt}.  It permits an identical input to be used when the
two contour prescriptions are compared.  It does not enter the construction
of the BJ--CVRC estimate, does not affect any four-loop entry, and should not
be read as a second preferred value of $c_{5,1}$.

\begin{table}[t]
\centering
\small
\caption{FOPT--CIPT separation for the two $u=2$-sensitive weights at
$\alpha_s(m_\tau)=0.3196$.  The fifth-order rows use $c_{5,1}=277$.}
\label{tab:cipt-fopt-separation}
\input{generated_data/cipt_fopt_separation_table.tex}
\end{table}

\paragraph{The CIPT test is weaker by more than an order of magnitude.}
Evaluating Eq.~\eqref{eq:cipt-tolerance} at $\epsilon=0.1$ against
Eq.~\eqref{eq:adler-tolerance}, with the $N=5$ entries evaluated at the
representative value $c_{5,1}=277$, the central Borel--Pad\'e value of
Ref.~\cite{Boito:2018rwt}, gives
\begin{equation}
 \begin{aligned}
 N=4:&\quad 2.151\ (\text{FOPT})\ \ \text{vs}\ \ 77.7\ (\text{CIPT}),\\
 N=5:&\quad 23.56\ (\text{FOPT})\ \ \text{vs}\ \ 770\ (\text{CIPT}),
 \end{aligned}
 \label{eq:cipt-vs-fopt}
\end{equation}
ratios of $36$ and $33$.  These three-weight figures compare different binding
weights, so Table~\ref{tab:fopt-cipt} also gives the single-weight tolerances
from which they are the minima.

\begin{table}[!tbp]
\centering
\small
\caption{Single-weight coefficient tolerances at $\epsilon=0.1$.  FOPT:
$\epsilon|d_N[W]|$, independent of $\alpha_s$.  CIPT:
$\epsilon|\delta^{\rm CIPT}_{W,N}/J^{\rm CI}_N[W]|$ at the coupling of
Eq.~\eqref{eq:alphas-input}.  The $N=5$ entries use $c_{5,1}=277$.  Bold
entries are the three-weight tolerances. }
\label{tab:fopt-cipt}
\input{generated_data/cipt_weight_table.tex}
\end{table}

The most restrictive weight differs between the schemes, for different
reasons.  In FOPT it is the weight with the smallest $|d_N[W]|$, which is
$W_{P2}$ because its RG-generated part is negative and partly cancels the
Adler coefficient: $d_4[W_{P2}]=c_{4,1}-27.567$ and
$d_5[W_{P2}]=c_{5,1}-512.597$.  A ten-percent moment criterion is therefore an
unusually sharp $4.4\%$ criterion on $c_{4,1}$.  In CIPT the partial sums
differ only moderately between weights, $\delta^{\rm CIPT}_{W,4}=0.182$,
$0.223$, $0.144$ for $W_\tau$, $W_{\rm GC}$, $W_{P2}$, whereas the last-order
contour integrals differ much more,
$J^{\rm CI}_4[W]=(1.48,\,2.87,\,0.63)\times10^{-4}$ and
$J^{\rm CI}_5[W]=(1.16,\,3.01,\,0.26)\times10^{-5}$.  The tightest CIPT
bound therefore comes from $W_{\rm GC}$, which has the largest
$J^{\rm CI}_N$, and $W_{P2}$, the most restrictive weight in FOPT, becomes the
least restrictive in CIPT.  The comparison does not rely on mixing weights:
for $W_{\rm GC}$ alone, FOPT is tighter than CIPT by factors of $3.6$ at $N=4$
and $5.4$ at $N=5$.

CIPT is less selective because the last retained term is only a small fraction
of the contour-improved partial sum: for $W_{\rm GC}$ it is $6.3\%$ of
$\delta^{\rm CIPT}_{W,4}$ and $3.6\%$ of $\delta^{\rm CIPT}_{W,5}$, so a
ten-percent tolerance on the partial sum admits an excursion larger than the
coefficient being tested.  At $N=5$ the permitted excursion, $\pm770$, is $2.8$
times $c_{5,1}$: a ten-percent three-weight test run in CIPT is for practical
purposes vacuous as a constraint on $c_{5,1}$.  Over
$\alpha_s(m_\tau)\in[0.30,0.33]$ the CIPT tolerances change by only $-8\%$ to
$+20\%$ (from $89.8$ at $\alpha_s=0.30$ to $72.5$ at $\alpha_s=0.33$ for $N=4$,
and from $925$ to $705$ for $N=5$), so the conclusion is not an artifact of the
coupling chosen.

The two criteria act on different objects: the FOPT test on the order-$N$
moment coefficient, the CIPT test on the whole partial sum, because CIPT
moments are not power series in one coupling.  Most of the factor of about
thirty-five is this dilution.  On the order-$N$ term alone, CIPT would allow
$\pm\epsilon|c_{N,1}|=\pm4.91$ and $\pm27.7$ for every weight, still looser
than FOPT because FOPT exploits the cancellation in $d_N[W_{P2}]$; on the
partial sum, the FOPT analogue would allow $\pm153$ and $\pm1477$.  For the
coefficient-level criterion used to validate forecasts, the fixed-order test
is therefore the stronger one, and it is still only $\pm2.15$ at the
calibration order and $\pm24$ at the order of interest
(Sec.~\ref{sec:resolving-power}).

\section{The forecast methods in brief}
\label{sec:methods-brief}

Every number quoted for $c_{5,1}$ or $c_{6,1}$ in this paper comes from one of
the methods summarized in this section.  For each method we state what it acts
on, what it is told beyond the four known coefficients, the rule that produces
the next coefficient, and how it fares on the single yardstick of
Sec.~\ref{sec:tolerance}.  The physical-order numbers are reported in Appendix~\ref{sec:qcd-forecast}, and collected in the global comparison of
Sec.~\ref{sec:global-comparison}.  Complete definitions,
derivations, and benchmark tables are in the appendices listed in
Table~\ref{tab:methods-brief}.

Two distinctions organize the list.  The first is the level of input
introduced in Sec.~\ref{sec:protocol-summary}.  Data-driven transformations
infer their remainder model from the coefficients themselves; Borel
normalization adds an assumed factorial growth; and renormalon-informed
constructions are also told singularity positions and, in some cases,
exponents.  The second distinction is the object acted on.  A method applied
once to the common Adler sequence and then propagated through
Eq.~\eqref{eq:triangular-weight-map} obeys the transfer
identity~\eqref{eq:arts-error-transfer}, so its three-weight test is the single
tolerance $|\widehat c_{4,1}-c_{4,1}|\leq2.1509$.  A method applied to each
moment series separately does not obey it, and must be tested weight by
weight.

\begin{table*}[!tbp]
\centering
\caption{The forecasting methods of this paper.  ``Blind test'' is the
three-input prediction of the withheld fourth order: for methods acting on the
common Adler sequence it is $\widehat c_{4,1}$ and its distance from
$c_{4,1}=49.0757$ in units of the tolerance $2.1509$; for moment-wise methods
it is the signed error of the withheld $d_4[W_{P2}]$, the binding weight.  The
status refers to the physical $c_{5,1}$ and uses the labels of
Sec.~\ref{sec:status-labels}.}
\label{tab:methods-brief}
\scriptsize
\setlength{\tabcolsep}{3pt}
\begin{tabularx}{\textwidth}{@{}>{\raggedright\arraybackslash}p{0.20\textwidth}
>{\raggedright\arraybackslash}p{0.12\textwidth}
>{\raggedright\arraybackslash}X
>{\raggedright\arraybackslash}p{0.16\textwidth}
c
>{\raggedright\arraybackslash}p{0.13\textwidth}@{}}
\toprule
method & acts on & supplied structure & blind test & status & details\\
\midrule
direct Pad\'e / Shanks / Wynn & each moment & none & $W_{P2}$: $+129.9\%$
 & \textsc{diag} & Sec.~\ref{sec:brief-pade}; Apps.~\ref{sec:pade-all-weight-predictions}, \ref{sec:protocol}\\
Borel-normalized Pad\'e & each moment & factorial growth & $W_{P2}$: $+244.9\%$
 & \textsc{diag} & Sec.~\ref{sec:brief-pade}; Apps.~\ref{sec:other-physical}, \ref{sec:protocol}\\
common-Adler Pad\'e / Borel--Pad\'e & Adler & none / factorial growth
 & $24.75$ / $37.13$; $11.3$ / $5.6$ tol & \textsc{diag}
 & Sec.~\ref{sec:brief-pade}; Apps.~\ref{sec:pade-all-weight-predictions}, \ref{sec:other-physical}\\
D-log Pad\'e (direct, Borel; moment or Adler-projected) & moment or Adler
 & one effective branch point; Borel version also factorial growth
 & $W_{P2}$: $\geq+42.6\%$; Adler: $39.06$, $4.7$ tol & \textsc{diag}
 & Sec.~\ref{sec:brief-dlog}; Apps.~\ref{sec:other-physical}, \ref{sec:dlog-pade}\\
$\eta_n$- and $\alpha$-modified Wynn & each moment & predeclared parameter grid
 & $W_{P2}$: $+119$ to $+130\%$ & \textsc{diag}
 & Sec.~\ref{sec:brief-sequence}; Apps.~\ref{sec:sequence-physical-envelope}, \ref{sec:regularization}\\
Brezinski $\theta_2$ & each moment & inverse-power remainder
 & $W_{P2}$: $+71.9\%$ & \textsc{diag}
 & Sec.~\ref{sec:brief-sequence}; Apps.~\ref{sec:sequence-physical-envelope}, \ref{sec:remainder-models}\\
Levin-$u$, canonical $\rho_2$ & each moment & remainder estimate; fixed coupling
 & --- & \textsc{diag}
 & Sec.~\ref{sec:brief-sequence}; Apps.~\ref{sec:other-physical}, \ref{sec:remainder-models}\\
renormalon annihilator ($m=3$) & each moment
 & positions, exponents, scanned $\gamma_{-1}$ & $W_{P2}$: $+271$ to $+320\%$
 & \textsc{diag} & Sec.~\ref{sec:brief-annihilator};
 Apps.~\ref{sec:qcd-annihilator}, \ref{sec:annihilator}--\ref{sec:threshold}\\
one-mode $u=2$ continuation & Adler & $u=2$, $\gamma_2=128/81$
 & $32.83$; $7.55$ tol & \textsc{diag}
 & Sec.~\ref{sec:brief-annihilator}; Apps.~\ref{sec:qcd-u2-only-forecast}, \ref{sec:u2-only}\\
ARTS/PORT & Adler (defects) & position $u_\star=2$ only
 & $50.08$; $0.47$ tol (passes) & \textsc{abst}
 & Sec.~\ref{sec:brief-arts}; Apps.~\ref{sec:arts-physical}, \ref{sec:defect-transport}\\
linear ratio fit, Ref.~\cite{Beneke:2008ad} & $W_\tau$ term ratios & none
 & $\simeq52$; fails & --- & Sec.~\ref{sec:brief-closure}; App.~\ref{sec:cvrc-bj}\\
BJ--CVRC & Adler & BJ-type singularities and exponents; one free amplitude
 & calibrated on $c_{4,1}$ & \textsc{cond}$^{\dagger}$
 & Sec.~\ref{sec:brief-closure}; Secs.~\ref{sec:obstruction}--\ref{sec:cvrc}\\
\bottomrule
\end{tabularx}
\end{table*}

\subsection{Rational approximants}
\label{sec:brief-pade}

A Pad\'e approximant $[L/M]$ to $F(a)/a$ with $L+M=N-1$ reproduces the $N$
supplied coefficients and predicts the next one by re-expansion.  The balanced
rule~\eqref{eq:balanced-main} selects $[1/2]$ and $[2/1]$ at four inputs.  With
three inputs the only balanced member is $[1/1]$, and its forecast is the
geometric continuation of the last ratio,
\begin{equation}
 \widehat d_{N+1}^{\,[1/1]}=\frac{d_N^2}{d_{N-1}},
 \qquad
 \widehat d_{N+1}^{\,\rm Borel}=\frac{N}{N-1}\,\frac{d_N^2}{d_{N-1}} .
 \label{eq:brief-pade11}
\end{equation}
The second form is the same approximant applied to the Borel coefficients
$d_n/(n-1)!$, which presumes factorial growth $d_n\sim A^n\Gamma(n)$.  The first
form is also the first Shanks transform and the first even element of Wynn's
$\varepsilon$ algorithm, so these are one construction rather than three.

Each version can be applied in two ways: to each moment series separately, or
once to the common Adler sequence followed by
Eq.~\eqref{eq:triangular-weight-map}.  On the moments, the direct form passes
$W_\tau$ and $W_{\rm GC}$ at the five-percent level but misses $W_{P2}$ by
$+129.9\%$ (Table~\ref{tab:pade-all-weight}), and Borel normalization makes
every weight worse (Table~\ref{tab:qcd-loo}).  On the Adler sequence both
forms fail the single tolerance: $\widehat c_{4,1}=24.75$ (direct) and $37.13$
(Borel) lie $11.3$ and $5.6$ tolerances from the known value.  The two
balanced direct moment-space approximants are retained only as the simplest
accuracy-through-order reference continuation.  Because the $W_{P2}$ miss
violates the same predeclared criterion used to exclude the Borel-normalized
form, both are classified uniformly as diagnostics (\textsc{diag}); neither
contributes to an accepted forecast envelope.
A $\Gamma$-normalized variant, which also inserts a known large-order exponent,
Eq.~\eqref{eq:gamma-forecast}, is used only for the pole--$\MSbar$ mass
benchmark, the one series whose exponent is established independently.

\subsection{D-log Pad\'e approximants}
\label{sec:brief-dlog}

A branch point $F(z)\sim(1-z/z_0)^{-1-\gamma}$ becomes the simple pole
$(1+\gamma)/(z_0-z)$ of the logarithmic derivative $G=F'/F$.  The D-log method
approximates $G$ by a Pad\'e approximant and recovers the next coefficient of
$F$ from the exact recursion $(n+1)f_{n+1}=\sum_{k=0}^{n}g_kf_{n-k}$,
Eq.~\eqref{eq:dlog-recursion}.  It can therefore encode one singularity
position \emph{and} exponent from the data.  The Borel version applies the same
steps to the Borel-transformed series.

At three inputs the logarithmic derivative has only two known terms and can
represent only one effective singularity.  The resulting Borel D-log $[0/1]$
forecast coincides exactly with the $\theta_2$ forecast of
Sec.~\ref{sec:brief-sequence},
\begin{equation}
 \widehat d_4^{\,\mathrm{B\,D\text{-}log}\,[0/1]}=\widehat d_4^{\,\theta_2}
 =\frac{2d_3^2}{d_2}-\frac{d_2d_3}{d_1},
 \label{eq:brief-dlog-theta}
\end{equation}
as follows from Eqs.~\eqref{eq:dlog-g012}--\eqref{eq:dlog-d4} applied to
$d_n/(n-1)!$, so the two methods share one blind record.  No three-input D-log
realization passes $W_{P2}$: the smallest error is $+42.6\%$
(Table~\ref{tab:dlog-weight-test}).  Projected onto the Adler sequence, Borel
D-log gives $\widehat c_{4,1}=39.06$, $4.7$ tolerances low.  All four D-log
rows are therefore diagnostics.  The balanced Borel D-log sequence does reach
persistent ten-percent accuracy in the large-$\beta_0$ limit, but only from
order $4$, $6$, $11$, or $12$, depending on the observable
(Table~\ref{tab:dlog-thresholds}).

\subsection{Shanks/Wynn regularizations and remainder-model transformations}
\label{sec:brief-sequence}

These transformations act on the partial sums of a moment series and differ
in the remainder they are built to cancel.  Two predeclared one-parameter
deformations of the denominator of the shifted first Wynn element are tested,
$\eta_n=k\,\Delta^2S_n$ and $\alpha\,\Delta S_{n+1}$.  Their forecasts have
the closed forms
\begin{equation}
 \widehat d_{N+j}^{\,(\eta)}=\frac{1}{1+k}\,\frac{d_N^2}{d_{N-1}}
 \left(\frac{d_N}{d_{N-1}}\right)^{\!j-1},
 \qquad
 \widehat d_{N+j}^{\,(\alpha)}=\frac{d_N^2}{d_{N-1}}
 \left[(1+\alpha)\frac{d_N}{d_{N-1}}\right]^{j-1},
 \label{eq:brief-wynn-reg}
\end{equation}
on the grids $k,\alpha\in\{0.01,\ldots,0.05\}$.  The $\eta_n$ deformation
rescales the geometric forecast by $1/(1+k)$.  The $\alpha$ deformation
leaves the first unseen coefficient unchanged and first enters the second.

Brezinski's $\theta_2$ is built for logarithmic and inverse-power remainders
rather than exponential ones.  With $A,B,C=d_{N-2},d_{N-1},d_N$ its
accuracy-through-order continuation predicts
\begin{equation}
 \widehat d_{N+1}^{\,(\theta)}=\frac{2C^2}{B}-\frac{BC}{A}.
 \label{eq:brief-theta}
\end{equation}
Wynn's canonical $\rho_2$ is a fixed-coupling accelerator whose
re-expansion does not reproduce the last supplied coefficient, so it is not an
admissible coefficient forecast.  The accuracy-through-order Levin-$u$
transformation, with the term-based remainder estimate of
Eq.~\eqref{eq:levin-u-formal-remainder}, reduces at four inputs exactly to
Eq.~\eqref{eq:brief-theta}, Eq.~\eqref{eq:beta-levin-theta-identity}, and is
therefore not counted separately.  Fixed-coupling evaluations of Levin-$u$
and canonical $\rho_2$ yield only anchor-dependent effective coefficients
(App.~\ref{sec:other-physical}); they are diagnostics and enter neither the
global comparison of Sec.~\ref{sec:global-comparison} nor any envelope.

Apart from the two fixed-coupling rows, all of these methods retain accuracy
through order, but none passes the weight test: on the withheld $d_4[W_{P2}]$
the shifted-Wynn, $\eta_n$, and $\alpha$ rows err by $+119$ to $+130\%$ and
$\theta_2$ by $+71.9\%$ (Table~\ref{tab:regularized-qcd-loo}).  All of them
therefore carry \textsc{diag} status at physical order.

\subsection{Renormalon-annihilating sequence transformations}
\label{sec:brief-annihilator}

Pad\'e and sequence transformations must infer their effective modes from the
same short list they extrapolate.  The annihilator is told them instead.  It
models the prescription-matched partial sums as
$S_N=S+\sum_{j=1}^{m}A_j\,\Omega_N^{(u_j)}$, with one declared remainder mode
for each Borel singularity at $u_j$, and forms
\begin{equation}
 \mathcal R_{m,N}=\sum_{k=0}^{m}\lambda_kS_{N+k},
 \qquad
 \sum_{k=0}^{m}\lambda_k=1,
 \qquad
 \sum_{k=0}^{m}\lambda_k\,\Omega_{N+k}^{(u_j)}=0\quad(j=1,\ldots,m).
 \label{eq:brief-annih-sum}
\end{equation}
The weights $\lambda_k$ preserve the Borel sum $S$, remove the $m$ declared
modes, and are fixed before any coefficient is used; they have the closed form
of Eq.~\eqref{eq:lambda-closed}.  For coefficient forecasts the
coupling-independent dual is used,
\begin{equation}
 d_n=\sum_{j=1}^{m}A_j\,\Gamma(n+\gamma_j)\left(\frac{\beta_0}{u_j}\right)^{\!n-1}:
 \label{eq:brief-annih-coeff}
\end{equation}
$m$ consecutive coefficients fix the amplitudes $A_j$, and the next coefficient
follows.  The positions and confluent multiplicities are fixed by the
order rule of Eq.~\eqref{eq:order-rule}; in full QCD the infrared exponents
are $\gamma_p=p\beta_1/\beta_0^2$, Eq.~\eqref{eq:qcd-exponent}, while the
ultraviolet exponent $\gamma_{-1}$ is not fixed and is scanned.

In the exactly solvable large-$\beta_0$ limit, with the exact
prescription-matched tail of Eq.~\eqref{eq:exact-mode}, the annihilator
converges and avoids the finite-order spikes of the balanced Pad\'e table.
Even there, persistent percent-level accuracy needs eleven to thirteen
coefficients (Table~\ref{tab:threshold}).  At physical order the $m=3$ model
misses the withheld $d_4[W_{P2}]$ by $+271$ to $+320\%$ across the
$\gamma_{-1}$ scan (Table~\ref{tab:qcd-annihilator}), and its $c_{5,1}$ moves
from $306$ to $374$ with an input it cannot determine.  Its rows are
diagnostics.

The shortest mode list, $m=1$ with only the gluon-condensate singularity, is
the \emph{one-mode $u=2$ continuation}.  Applied to the Adler sequence,
Eq.~\eqref{eq:brief-annih-coeff} reduces to the single-renormalon ratio
\begin{equation}
 c_{n+1,1}=\bigl(n+\gamma_2\bigr)\frac{\beta_0}{2}\,c_{n,1},
 \qquad \gamma_2=\frac{2\beta_1}{\beta_0^2}=\frac{128}{81}.
 \label{eq:brief-u2}
\end{equation}
It postdicts $\widehat c_{4,1}=32.83$, $7.55$ tolerances low, and is a
diagnostic (App.~\ref{sec:qcd-u2-only-forecast}).

\subsection{ARTS/PORT: renormalon transport of the Pad\'e defect}
\label{sec:brief-arts}

The Adler-projected renormalon-transported Shanks transformation (ARTS) keeps
the geometric Shanks forecast as a baseline and models only its defect
$\Delta_n=c_{n,1}-c_{n-1,1}^2/c_{n-2,1}$, Eq.~\eqref{eq:pade-defect}.  For a
single factorial mode $c_{n,1}=K\,\Gamma(n+\gamma)(\beta_0/u)^{n-1}$, the defect
ratio $q_n=\Delta_n/\Delta_{n-1}$ obeys $q_{n+1}=q_n+\beta_0/u$, with the
exponent $\gamma$ cancelling.  Transporting the last defect with a declared
nearest position $u_\star$, and nothing else, gives the position-only
ratio-transport (PORT) forecast
\begin{equation}
 \widehat c_{N+1,1}^{\,\rm ARTS}=\frac{c_{N,1}^2}{c_{N-1,1}}
 +\Delta_N\left(\frac{\Delta_N}{\Delta_{N-1}}+\frac{\beta_0}{u_\star}\right),
 \qquad c_{0,1}\equiv1 ,
 \label{eq:brief-arts}
\end{equation}
which is exact for a single mode and not for a sum of modes.  ``Adler-projected''
means that the formula acts once on the common Adler sequence and every
moment is then reconstructed through Eq.~\eqref{eq:triangular-weight-map}, so
the transfer identity applies.  Two known defects are required, so ARTS is
undefined on the withheld-$c_{3,1}$ fold.

With $u_\star=2$, ARTS postdicts $\widehat c_{4,1}=50.08$, $0.47$ tolerances
from the known value.  Apart from the calibrated closure below, it is the only
construction in Table~\ref{tab:partIII-summary} that passes, and the same
formula reproduces the withheld pole--$\MSbar$ coefficient to better than four
percent.  Its order-adaptive multi-mode extension issues a forecast only after
a rolling certificate on earlier defects has passed.  For $m$ fitted defect
modes the certificate requires at least $m+3$ known defects.  The physical
three-mode version therefore needs six defects, $\Delta_2,\ldots,\Delta_7$,
and hence seven nontrivial Adler coefficients; only four are known.  Even its
one-mode version needs four defects while only three exist.  OAMDT therefore
cannot certify $c_{5,1}$ and abstains
(\textsc{abst}).  Appendix~\ref{sec:arts-physical} reports the resulting
diagnostic value and explains why it is not promoted.

\subsection{The literature ratio estimate and the conditional BJ--CVRC closure}
\label{sec:brief-closure}

Two renormalon-motivated constructions require separate discussion because
they use related physical ideas but have different logical status.

\paragraph{The Beneke--Jamin linear-ratio estimate.}
The estimate quoted in Ref.~\cite{Beneke:2008ad} is obtained by fitting a
straight line to the successive ratios of the fixed-order $W_\tau$
perturbative terms and extrapolating that line by one order.  Applied at the
physical four-coefficient order, this prescription gives
\begin{equation}
 \widehat c_{5,1}^{\,\mathrm{ratio}}
 =282.8231\simeq283 .
 \label{eq:brief-bj-ratio-c5}
\end{equation}
The value $283$ is therefore a linear term-ratio extrapolation.  It is used as
an input to the principal five-parameter Beneke--Jamin model; it is not the
output of that model.  It must also be distinguished from the separate
four-parameter predictive variant of Ref.~\cite{Beneke:2008ad}, which gives
$c_{5,1}=280.1247\simeq280.12$.

The same ratio prescription can be tested one order lower.  With only the
first three Adler coefficients supplied, the line through the two available
term ratios reduces algebraically to the $\theta_2$ continuation of
Eq.~\eqref{eq:brief-theta} and gives
\begin{equation}
 \widehat c_{4,1}^{\,\mathrm{ratio}}
 =52.0838 .
 \label{eq:brief-bj-ratio-c4}
\end{equation}
Its absolute error is
\begin{equation}
 \left|
 \widehat c_{4,1}^{\,\mathrm{ratio}}-c_{4,1}
 \right|
 =3.0081,
 \label{eq:brief-bj-ratio-error}
\end{equation}
which exceeds the three-weight tolerance $2.1509$ by a factor of $1.399$.
The largest corresponding moment error is $13.986\%$, for $W_{P2}$.
Consequently, the linear-ratio estimate fails the predeclared acceptance
criterion, even though its numerical value is close to the range usually
associated with renormalon models.  The detailed comparison is given in
App.~\ref{sec:cvrc-bj}.

\paragraph{The BJ--CVRC construction.}
The \emph{cross-validated renormalon closure} (CVRC) of Part~IV is not a new
renormalon-model class.  It is a calibration-and-falsification protocol
applied to a specified pure-power truncation of the Beneke--Jamin Borel
architecture.  The ansatz retains the declared leading renormalon positions
and their renormalization-group exponents, including the alternating
ultraviolet singularity at $u=-1$, but replaces the full branch-cut structure
by its leading powers and omits the associated subleading cut coefficients.
We therefore use ``BJ-type'' for the Borel ansatz, ``CVRC'' for the protocol,
and ``BJ--CVRC'' only when both are meant.

At order $N$, the declared basis contains $N$ constrained columns, consisting
of infrared-renormalon terms and low-degree Borel-plane monomials selected by
the nesting rule of Eq.~\eqref{eq:cvrc-nesting}.  One additional column is
included for the leading ultraviolet singularity at $u=-1$, whose amplitude
is denoted by $\rho$.  Thus there are $N+1$ amplitudes subject to $N$
coefficient-matching conditions.

For any fixed $\rho$, the remaining $N$ amplitudes are determined by a linear
system so that the model reproduces
$c_{1,1},\ldots,c_{N,1}$ exactly.  Every member of the resulting
one-dimensional family therefore has accuracy through the supplied order.
Because both the matching equations and coefficient extraction are linear in
the Borel amplitudes, every one-order-ahead forecast is affine in $\rho$:
\begin{equation}
 \widehat c_{N+1,1}(\rho)
 =
 \kappa_{0,N}+\kappa_{1,N}\rho .
 \label{eq:brief-cvrc}
\end{equation}
Here $\kappa_{0,N}$ is the forecast of the closure-free member
$\rho=0$ and is a linear functional of the supplied coefficients.  By
contrast, $\kappa_{1,N}$ is the structural response to the declared $u=-1$
ultraviolet contribution: it is fixed by the nesting rule, singularity
positions, exponents, and $\beta_0$, and is independent of the supplied
$c_{n,1}$.  Neither quantity is fitted to the unknown coefficient
$c_{N+1,1}$.  Equation~\eqref{eq:brief-cvrc} is derived in
Sec.~\ref{sec:cvrc-structure} and written explicitly as
Eq.~\eqref{eq:cvrc-affine}.

\paragraph{Meaning and role of the ordered fold.}
Perturbative coefficients form an ordered sequence, so the procedure does not
remove an arbitrary coefficient from the middle of the series.  At a known
order $N$, an ordered leave-one-order-out fold supplies the consecutive
lower-order set
\begin{equation}
 \mathcal D_{N-1}
 =
 \{c_{1,1},c_{2,1},\ldots,c_{N-1,1}\}
 \label{eq:brief-cvrc-fold-input}
\end{equation}
and withholds the next known coefficient $c_{N,1}$.  The lower-order affine
family predicts
\begin{equation}
 \widehat c_{N,1}(\rho)
 =
 \kappa_{0,N-1}+\kappa_{1,N-1}\rho .
 \label{eq:brief-cvrc-fold-forecast}
\end{equation}
The highest-order physical fold fixes the closure amplitude through
\begin{equation}
 \widehat c_{N,1}(\rho_\star)=c_{N,1},
 \qquad
 \rho_\star
 =
 \frac{c_{N,1}-\kappa_{0,N-1}}
      {\kappa_{1,N-1}} .
 \label{eq:brief-cvrc-calibration}
\end{equation}
This is a one-step rolling-origin calibration, rather than ordinary
leave-one-out cross-validation.

The calibration coefficient is reproduced exactly after $\rho_\star$ has
been fixed and therefore does not provide independent validation.  The
nontrivial question is whether one common closure amplitude remains compatible
with the lower-order fold, all three moment weights, and the external
benchmarks.  Equivalently, the admissible $\rho$ intervals obtained at
different known orders must have a nonempty intersection.  This condition is
not automatic: the misspecification scan of
Sec.~\ref{sec:cvrc-misspec} rejects several altered singularity positions and
exponents because no common $\rho$ satisfies the declared tolerance.

The terminology ``cross-validated'' is therefore used here in the restricted
sense of a predeclared cross-order consistency and falsification protocol.  It
does not mean that the calibration fold is a statistically independent blind
test.  The strictly recursive test, in which $\rho$ is fixed at one known
order and the next coefficient is predicted without using it, is reported
separately in Sec.~\ref{sec:cvrc-drift}; the pole-mass and large-$\beta_0$
benchmarks provide the genuinely blind external tests.

Finally, the alternating $u=-1$ contribution allows the BJ--CVRC family to
reverse the observed Pad\'e-defect sign.  It is therefore not
defect-sign-preserving and can evade the obstruction of
Sec.~\ref{sec:obstruction}, unlike the common-Adler rational continuations
considered there.  After calibration, tolerance propagation, and the declared
structural variations, the resulting physical-order estimate is the
conditional forecast
\begin{equation}
 c_{5,1}^{\mathrm{BJ\text{-}CVRC}}
 =248\pm80 ,
 \label{eq:brief-cvrc-result}
\end{equation}
as detailed in Eq.~\eqref{eq:cvrc-final}.  The quoted uncertainty is a
conditional model-and-method interval, not a statistical measurement and not
a substitute for a direct six-loop calculation.

\parttitle{Part III. External tests at known orders and to all orders}

\section{Summary of external validation}
\label{sec:external-summary}
Part~III summarizes the outcome of applying the protocol.  The four external tests are ordered by
increasing structural specificity: the scheme-dependent but exactly known
five-loop $\beta$ function, two renormalon-sensitive full-QCD moments at the
known order, the nearby $u=1/2$ pole--$\MSbar$ mass renormalon, and the exactly
solvable large-$\beta_0$ limit in which coefficients and
prescription-dependent Borel sums are available to arbitrary order.

A reader who wants the outcome before the detail can take it from
Table~\ref{tab:partIII-summary}, whose constructions are defined in
Sec.~\ref{sec:methods-brief}.  For every construction that acts first on
the common Adler sequence, the transfer identity
\eqref{eq:arts-error-transfer} supplies the same one-dimensional yardstick, the withheld-$c_{4,1}$ tolerance
$|\widehat c_{4,1}-c_{4,1}|\leq2.1509$ of Eq.~\eqref{eq:tolerance-numbers}, independently of the moment used for the test.  Moment-by-moment transformations
do not obey this collapse and are tested separately in
App.~\ref{sec:pade-weight-certificate}.  The pattern in the common-Adler rows
is the subject of the Proposition of
Sec.~\ref{sec:obstruction}, which explains it without reference to any
implementation detail: the constructions that reproduce the observed low-order
acceleration overshoot $c_{5,1}$ badly, and those that impose renormalon
asymptotics undershoot the known $c_{4,1}$.  Part~III and Appendix~\ref{sec:qcd-forecast} may be read as the evidence;
Sec.~\ref{sec:obstruction} explains why the two requirements conflict
inside the defect-sign-preserving class.

\begin{table}[!ht]
\centering
\small
\caption{Representative constructions, each expressed as a forecast of the
Adler coefficient and placed on the single yardstick supplied by the transfer
identity: the forecast of the withheld $c_{4,1}=49.0757$ from
$c_{1,1},c_{2,1},c_{3,1}$, against the three-weight tolerance
$2.1509$.  ``$\times$ tol'' is $|\widehat c_{4,1}-c_{4,1}|$ in units of that
tolerance, so a value $\leq1$ passes the declared criterion for all three
weights simultaneously.  The linear-ratio-fit row acts on the $W_\tau$ moment
series and is converted to $\widehat c_{4,1}$ afterwards; it is the estimator
behind the literature value $c_{5,1}\approx283$ (App.~\ref{sec:cvrc-bj}).  The
CVRC row is exact by construction, since its closure is calibrated on
$c_{4,1}$.}
\label{tab:partIII-summary}
\begin{tabular}{lrrc}
\toprule
construction & $\widehat c_{4,1}$ & $\times$ tol & verdict\\
\midrule
direct Shanks / Pad\'e (Adler)      & $24.75$ & $11.31$ & fails\\
one-mode $u=2$ continuation         & $32.83$ & $7.55$  & fails\\
Borel--Pad\'e (Adler)               & $37.13$ & $5.55$  & fails\\
linear ratio fit, Ref.~\cite{Beneke:2008ad} fn.~2
                                    & $52.0838$ & $1.399$  & fails\\
\addlinespace
ARTS/PORT, single mode $u=2$        & $50.08$ & $0.47$  & passes\\
CVRC (calibration fold)             & $49.08$ & $0$     & by construction\\
\bottomrule
\end{tabular}
\end{table}

The outcomes of the four external tests can each be stated in one sentence.
(i)~None of the tested short-series transformations predicts the known
five-loop $n_f=3$ coefficient $\beta_4$ to ten percent from the four lower
coefficients (Appendix~\ref{sec:beta-benchmark}).  (ii)~The two
$u=2$-sensitive moments $W_{\rm GC}$ and $W_{P2}$ give incompatible moment-wise
continuations, so a transformation applied separately to each moment does not
define a single Adler continuation (Appendix~\ref{sec:qcd-moment-test}; see also
App.~\ref{sec:pade-weight-certificate}).  (iii)~The pole--$\MSbar$ mass
relation is forecast more accurately only when its earlier factorial onset and
leading renormalon information are supplied
(Appendix~\ref{sec:pole-mass}).  (iv)~In the exactly solvable large-$\beta_0$
limit, Pad\'e, D-log, and renormalon-informed constructions do become
accurate, but only after an observable-dependent onset; with exact renormalon
input, persistent percent-level control requires eleven to thirteen
coefficients (Appendices~\ref{sec:largebeta} and~\ref{sec:threshold}).

The derivations and order-by-order tables behind these statements are given in
Appendices~\ref{sec:beta-benchmark}--\ref{sec:threshold}.  This placement keeps
the acceptance logic and its outcome in the main text while separating
the evidence audit from the physical forecast.

\parttitle{Part IV. The closure obstruction and the cross-validated
renormalon closure}

Part~III exposes a common tension: continuations that reproduce the observed
low-order acceleration tend to overshoot $c_{5,1}$, whereas imposed renormalon
asymptotics tend to undershoot the known $c_{4,1}$.  Part~IV first proves the
corresponding defect-sign obstruction and then tests the class that evades it.
That class already includes the Beneke--Jamin model
\cite{Beneke:2008ad}; our contribution is a cross-validation criterion for a
specified pure-power BJ-type ansatz, not a new renormalon-model class.

\section{A no-go for defect-sign-preserving transformations}
\label{sec:obstruction}
Before asking which constructions are admissible it is worth stating precisely
what any of them has to do.  The single Adler tolerance of
Sec.~\ref{sec:tolerance} and one exact statement about the Pad\'e defect
together fix the target so narrowly that most of the algorithm space is
excluded on arithmetic grounds alone.

\subsection{The defect sequence and its required sign reversal}

For the Adler sequence the one-step Pad\'e/Shanks defect of
Eq.~\eqref{eq:pade-defect} reads
\begin{equation}
 \Delta_n=c_{n,1}-\frac{c_{n-1,1}^2}{c_{n-2,1}},
 \qquad c_{0,1}\equiv1,
 \label{eq:defect-def}
\end{equation}
and we define
$q_n=\Delta_n/\Delta_{n-1}$.  From the four known coefficients,
\begin{equation}
 \begin{aligned}
 (\Delta_2,\Delta_3,\Delta_4)&=(0.63982,\;3.68200,\;24.32311),\\
 (q_3,q_4)&=(5.75473,\;6.60595).
 \end{aligned}
 \label{eq:defects-known}
\end{equation}
The defects are positive and grow by a factor of six per order.  Two
consequences follow immediately.

\paragraph{(i) Passing the withheld-$c_{4,1}$ test requires reproducing that
growth.}  Combining Eqs.~\eqref{eq:tolerance-numbers} and
\eqref{eq:defect-def} with the known $c_{2,1},c_{3,1}$,
\begin{equation}
 \widehat c_{4,1}\in[46.925,\,51.227]
 \iff
 \widehat\Delta_4\in[22.172,\,26.474]
 \iff
 \frac{\widehat\Delta_4}{\Delta_3}\in[6.022,\,7.190].
 \label{eq:d4-defect-window}
\end{equation}

\paragraph{(ii) Any literature-range $c_{5,1}$ requires a sign reversal.}
The geometric baseline at the next order is
$c_{4,1}^2/c_{3,1}=378.028$, so
\begin{equation}
 \widehat\Delta_5=\widehat c_{5,1}-378.028 .
\end{equation}
Table~\ref{tab:required-delta5} evaluates this for the values quoted in the
literature and in Sec.~\ref{sec:global-comparison}.  Every one of them
requires $\widehat\Delta_5<0$, whereas $\Delta_2,\Delta_3,\Delta_4>0$.

\begin{table}[t]
\centering
\caption{Defect $\widehat\Delta_5=\widehat c_{5,1}-c_{4,1}^2/c_{3,1}$ implied
by candidate values of $c_{5,1}$.  The three known defects are
$+0.640$, $+3.682$, $+24.323$.}
\label{tab:required-delta5}
\begin{tabular}{lrr}
\toprule
source & $c_{5,1}$ & $\widehat\Delta_5$\\
\midrule
lower edge of this work's CVRC window & $229.8$ & $-148.2$\\
CVRC central value, Eq.~\eqref{eq:cvrc-final} & $248.0$ & $-130.0$\\
Ref.~\cite{Beneke:2008ad}, Eq.~(3.10) & $283$ & $-95.0$\\
Ref.~\cite{Beneke:2008ad}, model & $280$ & $-98.0$\\
direct Pad\'e $[2/1]$, Sec.~\ref{sec:global-table} & $304.7$ & $-73.3$\\
\addlinespace
ARTS/PORT, Sec.~\ref{sec:global-table} & $566.1$ & $+188.0$\\
\bottomrule
\end{tabular}
\end{table}

\subsection{The obstruction}

Call a transformation of the \emph{common Adler sequence}
\emph{defect-sign-preserving} at order $N$ if, whenever
$\Delta_2,\ldots,\Delta_N$ are positive and increasing, its forecast satisfies
$\widehat\Delta_{N+1}\geq0$.  Ordinary Shanks/Pad\'e saturates the bound with
$\widehat\Delta_{N+1}=0$; ARTS/PORT with a positive transported defect, and a
single factorial mode $K\Gamma(n+\gamma)(\beta_0/u_\star)^{n-1}$ with $K>0$ and
$u_\star>0$, are strictly inside it.

\begin{quote}
\textbf{Proposition.}  Let $\mathcal T$ act on the common Adler sequence and be
defect-sign-preserving.  Then
\begin{enumerate}[leftmargin=1.6em,label=(\roman*)]
 \item at $N=4$, because
 $\Delta_2,\Delta_3,\Delta_4=+0.640,+3.682,+24.323$ are positive and
 increasing, $\mathcal T$ returns
 $\widehat c_{5,1}\geq c_{4,1}^2/c_{3,1}=378.028$;
 \item the withheld-$c_{4,1}$ three-weight test does \emph{not} exclude
 $\mathcal T$, since by Eq.~\eqref{eq:d4-defect-window} the test window
 $\widehat\Delta_4\in[22.172,26.474]$ lies entirely in
 $\widehat\Delta_4>0$.
\end{enumerate}
Consequently a defect-sign-preserving Adler-space transformation may pass the
withheld-$d_4$ three-weight test, but it can never return a $c_{5,1}$ below
$378.028$, and in particular never a value in the range
$[229,305]$ spanned by the literature and by
Sec.~\ref{sec:global-comparison}.  The proposition does not apply to BJ--CVRC:
its explicit alternating $u=-1$ contribution allows
$\widehat\Delta_{N+1}$ to change sign, so BJ--CVRC is not
defect-sign-preserving.  Nor does it apply to transformations performed on an
individual moment rather than on the common Adler sequence.
\end{quote}

Part (i) is the definition applied to Eq.~\eqref{eq:defects-known}; part (ii)
is Eq.~\eqref{eq:d4-defect-window}.  The content is in the conjunction:
passing the weight test and returning a literature-range fifth coefficient are
not merely difficult together, they are incompatible inside this class.

The Proposition is testable against the global comparison of
Sec.~\ref{sec:global-comparison}, and it passes.  Every continuation in that
comparison that acts on the common Adler
sequence \emph{and is defect-sign-preserving} returns $c_{5,1}\geq378.03$:
\begin{equation}
 \begin{aligned}
 \text{common-Adler Pad\'e }[1/2],[2/1]&:\;445.85,\;378.03,\\
 \text{common-Adler Borel--Pad\'e }[1/2],[2/1]&:\;665.81,\;504.04,\\
 \text{Adler-projected direct, Borel D-log}&:\;445.31,\;568.70,\\
 \text{ARTS/PORT}&:\;566.07 .
 \end{aligned}
 \label{eq:adler-space-rows}
\end{equation}
The $[2/1]$ entry saturates the bound exactly, as it must: it is the Shanks
baseline.  Every other row of that comparison lies
\emph{below} $378.03$ for one of exactly two reasons.  Either it acts on a
moment series or on the annihilator's own sum space rather than on the common
Adler sequence, the direct and Borel-normalized moment Pad\'es,
the direct and Borel D-log Pad\'es, the shifted-Wynn grid, $\theta_2$,
Levin-$u$, $\rho_2$, and the annihilator, in which case the Proposition
simply does not apply; the mixed signs of their three weight errors (for
instance $+5.15\%$, $-3.67\%$, $+129.9\%$ for direct Shanks) confirm directly
that they are not common-Adler maps, since a common-Adler map has a
weight-independent absolute error.  Or it is a common-Adler map that is
\emph{not} defect-sign-preserving, and the only such rows are the three
cross-validated-closure rows of Sec.~\ref{sec:cvrc}, which is what places them
in the escaping class.

\begin{figure}[t]
\centering
\includegraphics[width=0.57\textwidth]{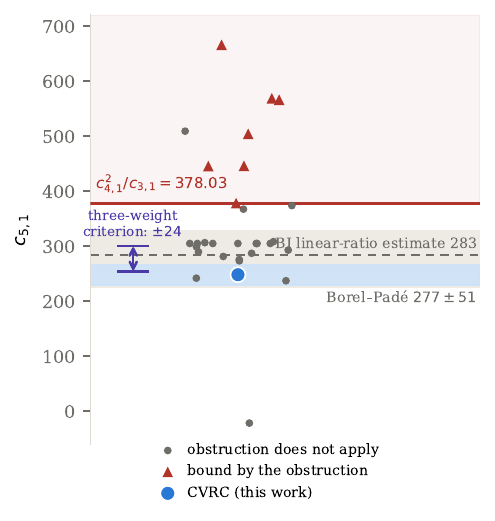}
\caption{Classification of the fifth-order continuations by the hypotheses of
the defect-sign obstruction.  Red triangles are common-Adler maps that are
defect-sign-preserving and therefore lie at or above
$c_{4,1}^2/c_{3,1}=378.03$.  Gray circles denote rows for which the
Proposition does not apply because they act in moment or transformed-sum
space, average several approximants, or introduce a sign-reversing
renormalon structure.  The blue point is the BJ--CVRC central value.  The
gray band is the independent Borel--Pad\'e range $277\pm51$; the light-blue
band is only the CVRC structural closure window $[229.8,266.2]$, not the
adopted total interval $248\pm80$.  The arrow shows the approximate
$\pm24$ resolving power of the order-five three-weight criterion near the
literature range, and the dashed line marks the separate BJ linear-ratio
estimate $283$.}
\label{fig:defect-partition}
\end{figure}

Figure~\ref{fig:defect-partition} displays this logical partition without
assigning false independence to the individual rows.  It is intentionally
different from the detailed two-panel comparison of
Fig.~\ref{fig:global-comparison}: the present figure asks only whether the
Proposition applies, whereas the later figure identifies every method and also
shows its rolled sixth-order output.

The Proposition explains the pattern of Part~III and
Appendix~\ref{sec:qcd-forecast} without reference to any
implementation detail.  ARTS satisfies Eq.~\eqref{eq:d4-defect-window}
exactly, $\widehat\Delta_4=25.331$, and is therefore bound by (i) to
$c_{5,1}\geq378.03$; its own formula gives $566.07$.  Conversely, the
constructions that impose a renormalon sign structure are bound to miss
$c_{4,1}$: the one-mode $u=2$ continuation of
App.~\ref{sec:qcd-u2-only-forecast} is low by $33.11\%$, and the annihilator
of App.~\ref{sec:qcd-annihilator} misses the known $d_4$ over a range that
runs from $-7\%$ for $W_\tau$ to nearly a factor of four for $W_{P2}$,
depending on weight and on the ultraviolet exponent it cannot fix.  Neither is
a failure of implementation.

The escape route is now visible.  A transformation that can do both must
(a) not be defect-sign-preserving, and (b) generate the sign reversal from
declared structure rather than from the data, because the data contain no
sign reversal.  In the Borel plane the structures that do this are the
singularities on the negative-$u$ axis, whose contributions alternate; the
leading one is the ultraviolet renormalon at $u=-1$, and its amplitude is the
direction that four coefficients constrain least.  This is not a hypothetical
escape: renormalon models of exactly this kind have been built before, in
particular that of Ref.~\cite{Beneke:2008ad}, and the Proposition is best
read as an explanation of why such models succeed where the rational and
sequence-transformation families surveyed in Part~III and
Appendix~\ref{sec:qcd-forecast} cannot.  What
Sec.~\ref{sec:cvrc} adds is to make that amplitude an explicit undetermined
parameter rather than a fitted one, so that the criterion of
Sec.~\ref{sec:tolerance} can be applied to the resulting family and can reject
it.

\section{Cross-validation of a Beneke--Jamin-type ansatz}
\label{sec:cvrc}
The obstruction of Sec.~\ref{sec:obstruction} says that a viable continuation
must generate a defect sign reversal from declared structure.  The leading
renormalon skeleton required for that purpose is not new: it is the one used
by Beneke and Jamin~\cite{Beneke:2008ad}, including the alternating $u=-1$
ultraviolet contribution.  Our Borel ansatz retains the same leading
singularity positions and renormalization-group exponents, but replaces the
full cuts by pure powers and omits their subleading coefficients.  It is
therefore a \emph{Beneke--Jamin-type ansatz}, not the full Beneke--Jamin model.

The new element is the cross-validated renormalon-closure criterion, abbreviated
CVRC.  One singularity amplitude $\rho$ is left free, which makes every
forecast affine in $\rho$.  The three-weight criterion then selects an
admissible interval or rejects the declared structure; six of the nine
alternatives scanned in Sec.~\ref{sec:cvrc-misspec} are rejected.  Thus ``CVRC'' below
labels the calibration and falsification procedure, while ``BJ-type'' labels
the Borel ansatz.  We use ``BJ--CVRC'' only when both are meant.

\subsection{The ultraviolet exponent used by the renormalon-informed tests}
\label{sec:uv-exponent-fix}

The declared structure needs one input that the physical-order annihilator
of App.~\ref{sec:qcd-annihilator} left open: the ultraviolet exponent.  There
it was scanned over the predeclared grid $\gamma_{-1}\in\{-0.90,-0.79,0\}$
(Sec.~\ref{sec:global-table}), and App.~\ref{sec:pole-annihilator} uses
$\gamma_{-1}=b+1$.  In the mode convention of this paper,
$B\supset K/(u_\star-u)^{1+\gamma}$, the value $\gamma_{-1}=-2b$ describes a
\emph{simple} pole at $u=-1$ dressed by the running coupling, whereas
Table~\ref{tab:order-rule} and the large-$\beta_0$ transform of
App.~\ref{sec:largebeta} have a \emph{double} pole at $u=-1$.  Expanding Eq.~\eqref{eq:large-beta-borel} about $u=-1$ using
only the $k=2$ term of the sum, which alone is singular there, gives
\begin{equation}
 B[\Dhat](u)=\pi B_{\rm BJ}[\Dhat](u)\;\xrightarrow[u\to-1]{}\;
 \frac{4}{9}\,e^{-5/3}\,\frac{1}{(1+u)^2}
 =\frac{0.083945}{(1+u)^2},
 \label{eq:uv-residue}
\end{equation}
which the numerical Cauchy projection reproduces.  Dressing the double pole
with the same renormalization-group factor that produced
Eq.~\eqref{eq:qcd-exponent} shifts the exponent by $-2b$ and gives
\begin{equation}
 \alpha_{-1}\equiv1+\gamma_{-1}=2-\frac{\beta_1}{\beta_0^2}
 =1.209877,\qquad \gamma_{-1}=+0.209877
 \label{eq:uv-exponent}
\end{equation}
for $n_f=3$, which is the value $\tilde\gamma_1=1.21$ used in the
renormalon model of Ref.~\cite{Beneke:2008ad}.  The predeclared grid
$\{-0.90,-0.79,0\}$ therefore brackets a region that the large-$\beta_0$
structure excludes, and the annihilator rows labeled \textsc{diag} in the global comparison of
Sec.~\ref{sec:global-table} are diagnostics of a mis-specified
ultraviolet exponent as well as of an underdetermined amplitude.

Two clarifications prevent this choice from being over-read.
First, adopting Eq.~\eqref{eq:uv-exponent} does \emph{not} rescue the
annihilator: Table~\ref{tab:qcd-annihilator} shows its withheld-$d_4$ error
growing monotonically as $\gamma_{-1}$ increases through zero, so the
large-$\beta_0$-motivated exponent makes that particular construction worse,
not better.  The
negative conclusion of App.~\ref{sec:qcd-annihilator} therefore stands and is
if anything strengthened.  Second, what Eq.~\eqref{eq:uv-exponent} fixes is the
\emph{declared structure available to the closure construction}; everything
below uses it, and Sec.~\ref{sec:cvrc-misspec} shows that this value is not
merely preferable but necessary, with $\gamma_{-1}=-0.79$ the admissible set of
the construction of Sec.~\ref{sec:cvrc} is empty.

\subsection{Declared structure and the model at order \texorpdfstring{$N$}{N}}
\label{sec:cvrc-structure}

Using the Borel convention of Eq.~\eqref{eq:borel-normalization}, a singularity
$N_p(p-u)^{-\alpha_p}$ contributes
\begin{equation}
 c_{n,1}\big|_p
 =N_p\,p^{-\alpha_p-n+1}\,
 \frac{\Gamma(\alpha_p+n-1)}{\Gamma(\alpha_p)}\,\beta_0^{\,n-1},
 \label{eq:cvrc-mode}
\end{equation}
and a Borel-plane monomial $u^j$ contributes $\beta_0^j\,j!$ to $c_{j+1,1}$
and nothing to any other order.  The two families are complementary: the
singularities carry the large-$n$ behavior, the monomials carry the analytic
(non-renormalon) part confined to low orders.  This is the leading
singularity-plus-polynomial architecture of Ref.~\cite{Beneke:2008ad}; the
pure-power truncation used here is specified explicitly so that it is not
mistaken for that reference's full cut model.

The declared exponents are those of Eq.~\eqref{eq:qcd-exponent} and
Eq.~\eqref{eq:uv-exponent},
\begin{equation}
 \alpha_p=1+\frac{p\beta_1}{\beta_0^2}\;(p>0),
 \qquad
 \alpha_{-k}=2-\frac{k\beta_1}{\beta_0^2}\;(k>0),
 \label{eq:cvrc-exponents}
\end{equation}
so that for $n_f=3$
\begin{equation}
 \alpha_2=2.580247,\quad
 \alpha_3=3.370370,\quad
 \alpha_4=4.160494,\quad
 \alpha_{-1}=1.209877 .
\end{equation}
No exponent is fitted.

\paragraph{Nesting rule.}  At order $N$ the model uses $N$ \emph{constrained}
columns, seeded by the leading infrared singularity $u=2$ and the monomial
$u^0$, and then alternating between the next singularity and the next
monomial:
\begin{equation}
 \{u{=}2,\;u^0\},\;
 \{u{=}3\},\;\{u^1\},\;\{u{=}4\},\;\{u^2\},\ldots
 \label{eq:cvrc-nesting}
\end{equation}
To these is added one further column, the \emph{closure}: the leading
ultraviolet singularity $u=-1$ with exponent \eqref{eq:uv-exponent}, whose
amplitude is called $\rho$.  The model therefore has $N+1$ parameters for $N$
constraints.

Two distinct singularity lists are used, and their roles must not be
conflated.  In \emph{BJ--CVRC-minimal} (abbreviated CVRC-minimal), the
constrained singular columns contain only the infrared tower
$u=2,3,4,\ldots$; the leading ultraviolet mode $u=-1$ is still present, but
only as the free closure direction.  In \emph{BJ--CVRC-complete}
(CVRC-complete), the constrained list additionally interleaves the
subleading ultraviolet tower,
\begin{equation}
 u=2,-2,3,-3,4,-4,\ldots,
 \label{eq:cvrc-complete-list}
\end{equation}
while $u=-1$ remains the same free closure mode.  In both variants the
singularity columns are alternated with the polynomial columns
$u^0,u^1,u^2,\ldots$ according to Eq.~\eqref{eq:cvrc-nesting}; hence each
order-$N$ construction still contains exactly $N$ constrained columns plus
one free amplitude $\rho$.

The names describe model content, not an accuracy ranking.  CVRC-complete is
closer to the full singularity tower in the exactly solvable
large-$\beta_0$ laboratory and becomes superior once sufficiently many
coefficients are available.  At low order it is over-parameterized and can be
substantially worse.  The physical four-coefficient result therefore uses
CVRC-minimal.  CVRC-complete is used only for the high-order
large-$\beta_0$ validation of Sec.~\ref{sec:cvrc-largebeta}; it is not an
alternative physical estimate of $c_{5,1}$.

\subsection{The closure family is exactly one-dimensional and affine}
\label{sec:cvrc-affine-family}

Let $f_j(n)$, $j=1,\ldots,N$, denote the $N$ constrained columns selected by
the nesting rule, and let $g(n)$ be the $u=-1$ closure column.  The model
coefficient at order $n$ is
\begin{equation}
 c^{\rm mod}_{n,1}(\rho)
 =\sum_{j=1}^{N}a_j(\rho)f_j(n)+\rho\,g(n).
 \label{eq:cvrc-coefficient-model}
\end{equation}
For the supplied orders define
\begin{equation}
 (X_N)_{nj}=f_j(n),\qquad
 (\bm v_N)_n=g(n),\qquad
 \bm c_N=(c_{1,1},\ldots,c_{N,1})^{\mathsf T}.
 \label{eq:cvrc-matrix-definitions}
\end{equation}
The matching equations are therefore
\begin{equation}
 X_N\bm a(\rho)+\rho\bm v_N=\bm c_N.
 \label{eq:cvrc-matrix-system}
\end{equation}
The rank guard described below ensures that $X_N$ is nonsingular for every
physical-order construction used here.  Hence the full $N\times(N+1)$ matrix
$[X_N\;\bm v_N]$ has rank $N$ and nullity one.  Its solution set is the affine
line
\begin{equation}
 \begin{pmatrix}\bm a(\rho)\\ \rho\end{pmatrix}
 =
 \begin{pmatrix}X_N^{-1}\bm c_N\\0\end{pmatrix}
 +\rho
 \begin{pmatrix}-X_N^{-1}\bm v_N\\1\end{pmatrix},
 \qquad
 \bm a(\rho)=X_N^{-1}\bm c_N-\rho X_N^{-1}\bm v_N.
 \label{eq:cvrc-parameter-line}
\end{equation}
Substitution back into Eq.~\eqref{eq:cvrc-matrix-system} gives
\begin{equation}
 X_N\bm a(\rho)+\rho\bm v_N
 =\bm c_N-\rho\bm v_N+\rho\bm v_N=\bm c_N.
 \label{eq:cvrc-ato-proof}
\end{equation}
Thus every value of $\rho$ reproduces all supplied coefficients exactly:
accuracy through order holds on the entire affine line, not only after the
closure has been calibrated.

At a future order $m>N$, let
$\bm x_m^{\mathsf T}=(f_1(m),\ldots,f_N(m))$ and $v_m=g(m)$.  Equations
\eqref{eq:cvrc-coefficient-model} and \eqref{eq:cvrc-parameter-line} give
\begin{equation}
 \widehat c_{m,1}(\rho)
 =\underbrace{\bm x_m^{\mathsf T}X_N^{-1}\bm c_N}
 _{\displaystyle\kappa_{0,N}^{(m)}}
 +\rho\underbrace{\left(v_m-\bm x_m^{\mathsf T}X_N^{-1}\bm v_N\right)}
 _{\displaystyle\kappa_{1,N}^{(m)}}.
 \label{eq:cvrc-general-affine}
\end{equation}
For the one-step target $m=N+1$ the superscript is suppressed, so
\begin{equation}
 \widehat c_{N+1,1}(\rho)=\kappa_{0,N}+\kappa_{1,N}\,\rho
 \label{eq:cvrc-affine}
\end{equation}
with
\begin{equation}
 \kappa_{0,N}=\bm x_{N+1}^{\mathsf T}X_N^{-1}\bm c_N,
 \qquad
 \kappa_{1,N}=v_{N+1}-\bm x_{N+1}^{\mathsf T}X_N^{-1}\bm v_N.
 \label{eq:cvrc-kappas}
\end{equation}
The intercept is linear in the supplied coefficient vector.  The slope is
strictly structural: it depends on the declared columns, their exponents and
$\beta_0$, but not on any supplied $c_{n,1}$.

For the physical $n_f=3$ Adler series, the $N=2$ and $N=3$ one-step families
and the first two future rows of the $N=4$ family are
\begin{equation}
 \begin{aligned}
 \widehat c_{3,1}(\rho)&=6.60484+24.50000\,\rho,\\
 \widehat c_{4,1}(\rho)&=31.67880-287.96007\,\rho,\\
 \widehat c_{5,1}(\rho)&=395.34025+2439.20878\,\rho,\\
 \widehat c_{6,1}^{(N=4\ {\rm roll})}(\rho)
 &=3447.71581+5720.03859\,\rho .
 \end{aligned}
 \label{eq:cvrc-lines}
\end{equation}
The fourth line is a two-order roll from the fixed $N=4$ design; its slope is
$\kappa_{1,4}^{(6)}=5720.03859$.  It is not the fresh $N=5$ one-step forecast.
Rebuilding the nested design at $N=5$ adds the $u=4$ singular column and gives
$\kappa_{1,5}=-40991.93887$.  The two slopes therefore come from different
design matrices and need not agree.

Rebuilding the declared design independently at every $N$ gives the fresh
one-step structural slopes
\begin{equation}
 \begin{aligned}
 \kappa_{1,2},\;\kappa_{1,3},\;\kappa_{1,4}&=+24.50,\;-287.96,\;+2\,439.21,\\
 \kappa_{1,5},\;\kappa_{1,6},\;\kappa_{1,7}&=-4.0992\times10^{4},\;+5.2424\times10^{5},
 \;-1.1320\times10^{7}.
 \end{aligned}
 \label{eq:closure-slopes}
\end{equation}
The raw closure mode satisfies
\begin{equation}
 \frac{g(n+1)}{g(n)}
 =-\beta_0\bigl(\alpha_{-1}+n-1\bigr),
 \label{eq:cvrc-uv-ratio}
\end{equation}
so it alternates and grows factorially.  Equation~\eqref{eq:cvrc-kappas}
shows that $\kappa_{1,N}$ is the residual closure contribution after the first
$N$ closure coefficients have been interpolated by the constrained basis.
For the declared nesting that residual retains the alternating sign and rapid
growth shown in Eq.~\eqref{eq:closure-slopes}.  This is a verified property of
the declared structure, not a theorem for every possible singularity list or
column ordering.

The alternating ultraviolet mode does exactly the job identified at the end
of Sec.~\ref{sec:obstruction}: one negative $\rho$ lowers
$\widehat c_{3,1}$, raises $\widehat c_{4,1}$ and lowers
$\widehat c_{5,1}$ relative to the pure infrared model.  A
defect-sign-preserving transformation has no such degree of freedom.  Finally,
$\rho=0$ recovers the pure infrared model, which returns
$\widehat c_{4,1}=31.68$ and therefore fails
Eq.~\eqref{eq:d4-defect-window}, in agreement with the Proposition.

\subsection{Fixing the closure: one-fold leave-one-order-out}
\label{sec:cvrc-closure}

Within the order-$N$ accuracy-through-order family, matching the four known
coefficients leaves $\rho$ undetermined because every member reproduces those
inputs.  The protocol fixes $\rho$ from the preceding order: the model built
from $c_{1,1},\ldots,c_{N-1,1}$ is required to reproduce the next known
coefficient $c_{N,1}$.  This is the single available calibration fold, not an
independent validation test.  By Eq.~\eqref{eq:cvrc-affine} it is one linear
equation with a unique solution,
\begin{equation}
 \rho_\star=\frac{c_{N,1}-\kappa_{0,N-1}}{\kappa_{1,N-1}}
 \label{eq:cvrc-rho}
\end{equation}
so that the resulting CVRC transformation from
$(c_{1,1},\ldots,c_{N,1})$ to
$\widehat c_{N+1,1}$ is the closed-form linear transformation
\begin{equation}
 \widehat c^{\rm BJ\text{-}CVRC}_{N+1,1}
 =\kappa_{0,N}+\kappa_{1,N}\,\frac{c_{N,1}-\kappa_{0,N-1}}{\kappa_{1,N-1}}
 \label{eq:cvrc-master}
\end{equation}
For a fixed declared structure, the slopes $\kappa_{1,N-1}$ and
$\kappa_{1,N}$ are structural constants, while each intercept $\kappa_0$ is a
linear functional of its supplied coefficient vector.  Consequently
Eq.~\eqref{eq:cvrc-master} is linear in the full set of supplied coefficients
(or affine in the remaining coefficients if the normalization $c_{1,1}=1$ is
held fixed).  It would become nonlinear only if singularity positions,
exponents or the nesting rule were retuned from the same data; no such retuning
is performed here.  The target coefficient $c_{N+1,1}$ does not enter the
construction.

At physical order, $N=4$ and Eqs.~\eqref{eq:cvrc-lines}
give
\begin{equation}
 \rho_\star=\frac{49.07570-31.67880}{-287.96007}=-0.060414 .
 \label{eq:rho-star}
\end{equation}
The negative sign is a statement about the physical series: the ultraviolet
mode must enter with the sign that suppresses, not enhances, the fifth-order
coefficient relative to the pure infrared model.

\subsection{The three-weight test, and exactly what it certifies}
\label{sec:cvrc-weight-test}

Table~\ref{tab:cvrc-moment} applies the calibrated model to both
leave-one-order-out folds of Table~\ref{tab:qcd-loo}.  All six entries lie
inside the declared ten-percent band.  No other construction examined in this
paper achieves that: direct Shanks fails two of six, Borel Shanks fails all
six, the D-log family fails $W_{P2}$, and ARTS, which needs two Pad\'e
defects and therefore three prior coefficients, is structurally undefined on
the three $d_3$ entries and is tested only on the other three.

\begin{table*}[!tbp]
\centering
\caption{CVRC on the two leave-one-order-out folds of
Table~\ref{tab:qcd-loo}, with $\rho_\star=-0.060414$ fixed by
Eq.~\eqref{eq:cvrc-rho}.  The $d_4$ fold is the calibration fold and is exact
by construction.  For comparison the last two columns repeat the
direct-Shanks and ARTS entries, with ``n/a'' where the method is not defined.}
\label{tab:cvrc-moment}
\small
\setlength{\tabcolsep}{5pt}
\begin{tabular}{llrrrrr}
\toprule
target & weight & exact & CVRC & CVRC error & direct Shanks & ARTS\\
\midrule
$d_3$ & $W_\tau$      & $26.36593$  & $25.11960$  & $-4.727\%$ & $+2.65\%$ & n/a\\
$d_3$ & $W_{\rm GC}$  & $34.48416$  & $33.23784$  & $-3.614\%$ & $-3.63\%$ & n/a\\
$d_3$ & $W_{P2}$      & $14.52303$  & $13.27671$  & $-8.582\%$ & $+25.3\%$ & n/a\\
\addlinespace
$d_4$ & $W_\tau$      & $127.07862$ & $127.07862$ & $0$ (fold) & $+5.15\%$ & $+0.79\%$\\
$d_4$ & $W_{\rm GC}$  & $214.13978$ & $214.13978$ & $0$ (fold) & $-3.67\%$ & $+0.47\%$\\
$d_4$ & $W_{P2}$      & $21.50873$  & $21.50873$  & $0$ (fold) & $+129.9\%$& $+4.69\%$\\
\bottomrule
\end{tabular}
\end{table*}

It is essential to say precisely what Table~\ref{tab:cvrc-moment} is and is
not, because the comparison in its last two columns is not like for like.

\paragraph{It is not six blind forecasts.}  The $d_4$ row is exact because
that fold \emph{defines} $\rho_\star$.  Nor is the $d_3$ row blind: by
Eq.~\eqref{eq:cvrc-rho}, $\rho_\star$ uses the data-dependent intercept
$\kappa_{0,3}$, which is built from
$c_{1,1},c_{2,1},c_{3,1}$, together with the structural slope
$\kappa_{1,3}$.  Thus $c_{3,1}$ feeds back into its own reconstruction.
Perturbing $c_{3,1}$ by $\pm5\%$ moves $\rho_\star$
from $-0.0604$ to $-0.0493$ and $-0.0716$, and moves
$\widehat c_{3,1}(\rho_\star)$ from $5.125$ to $5.398$ and $4.852$.  The
direct-Shanks and Borel-Shanks columns of Table~\ref{tab:qcd-loo}, by
contrast, \emph{are} blind on all six entries.  Ranking CVRC above them on a
count of entries passed would therefore be a category error, and we do not
make that ranking.

\paragraph{It is a consistency certificate for the declared structure.}  What
Table~\ref{tab:cvrc-moment} does establish is that a \emph{single} closure
value reproduces both withheld orders, for all three weights, inside the
declared band.  Equivalently, the admissible set
\begin{equation}
 \mathcal R=\Bigl\{\rho:\;
 \max_{W,\,N\in\{2,3\}}
 \bigl|\text{rel.\ error of }\widehat d_{N+1}[W](\rho)\bigr|\leq10\%\Bigr\}
 \label{eq:admissible-set}
\end{equation}
is nonempty.  That is not automatic: $\mathcal R$ is the intersection of two
intervals cut out of the same one-dimensional family by two different orders,
and a mis-specified singularity list makes it empty.
Section~\ref{sec:cvrc-misspec} shows that it does.

\paragraph{The genuinely blind statements are elsewhere.}  They are the
pole-mass test of Sec.~\ref{sec:cvrc-pole}, in which $\rho$ is fixed from
$r_3$ and $r_4$ is then forecast with no knowledge of $r_4$; the
large-$\beta_0$ table of Sec.~\ref{sec:cvrc-largebeta}, in which the whole
protocol is run recursively against exactly known targets; and the negative
result of Sec.~\ref{sec:cvrc-drift}.

\subsection{Misspecification scan: the admissible set is discriminating}
\label{sec:cvrc-misspec}

Because $\mathcal R$ is nonempty only when the declared structure is
compatible with two known orders at once, its emptiness is a usable test of
that structure.  Table~\ref{tab:cvrc-misspec} scans the declared inputs one at
a time, in the same supervised spirit as
App.~\ref{sec:supervised-scan}, and reports whether $\mathcal R$ survives.

\begin{table}[t]
\centering
\caption{Misspecification scan of the declared structure.  ``empty'' means
that no single closure value satisfies the ten-percent three-weight criterion
on both folds, so the declared structure is rejected outright.  Only the
leading infrared position and the ultraviolet exponent are varied; all other
inputs are held at their declared values and the remaining infrared
singularities are retained as later columns.  With four coefficients only the
first two singularity columns are ever reached, so the $u=3$ row is a
re-ordering of $\{u{=}2,u{=}3\}$ rather than a displacement, and its verdict
comes from the two-coefficient fold alone.  }
\label{tab:cvrc-misspec}
\small
\begin{tabular}{llr}
\toprule
declared input varied & admissible set $\mathcal R$ & $c_{5,1}$\\
\midrule
\emph{baseline}: $u_{\rm IR}=2$, $\alpha_{-1}=1.20988$
 & $[-0.0679,-0.0529]$ & $247.98$\\
\addlinespace
$\alpha_{-1}=1-\beta_1/\beta_0^2=0.20988$ (renormalon-informed test) & empty & ---\\
$\alpha_{-1}=1$ (simple UV pole)             & $[-0.0879,-0.0710]$ & $258.39$\\
$\alpha_{-1}=2$ (undressed double pole)      & $[-0.0292,-0.0228]$ & $210.52$\\
\addlinespace
UV singularity moved to $u=-1.5$             & empty & ---\\
UV singularity moved to $u=-2$               & empty & ---\\
\addlinespace
leading IR singularity at $u=1$              & empty & ---\\
leading IR singularity at $u=3/2$            & $[-0.0585,-0.0460]$ & $290.40$\\
leading IR singularity at $u=5/2$            & empty & ---\\
leading IR singularity at $u=3$              & empty & ---\\
\bottomrule
\end{tabular}
\end{table}

The scan rejects the ultraviolet exponent used in the renormalon-informed
tests, both displaced ultraviolet
positions, and three of the four displaced infrared positions.  It does not resolve $\alpha_{-1}$ finely: the
declared value $1.20988$, the simple pole $\alpha_{-1}=1$ and the undressed
double pole $\alpha_{-1}=2$ all survive, with central values spanning
$210.5\leq c_{5,1}\leq258.4$.  This is a real but limited discriminating
power, and it is the honest measure of how much structural information four
coefficients carry.  It is quoted as a systematic in
Sec.~\ref{sec:cvrc-forecast}.

\subsection{The closure drift, and an honest negative}
\label{sec:cvrc-drift}

The calibration fold is exact by construction, so it is not itself a test.
The corresponding test is whether $\rho$ is \emph{stable} between folds.
Solving Eq.~\eqref{eq:cvrc-rho} one order lower, from
$\widehat c_{3,1}(\rho)=c_{3,1}$, gives
\begin{equation}
 \rho_\star^{(c_3\ \rm fold)}=-0.009544,
 \qquad
 \rho_\star^{(c_4\ \rm fold)}=-0.060414,
 \label{eq:closure-drift}
\end{equation}
a drift of $0.0509$.  Running the protocol strictly recursively, fixing
$\rho$ on the $c_{3,1}$ fold and then forecasting $c_{4,1}$ blind, gives
$\widehat c_{4,1}=34.427$, low by $29.9\%$, with three-weight errors
$-11.5\%$, $-6.8\%$ and $-68.1\%$.  We record this explicitly.  It means that
the CVRC forecast of $c_{5,1}$ is conditional on the closure being stable
between the fourth and fifth orders, and the drift
\eqref{eq:closure-drift} is the quantitative measure of how far the series is
from the regime in which that assumption is safe.

Section~\ref{sec:cvrc-largebeta} shows that this is the expected
pre-asymptotic behavior rather than a defect of the construction: in the
large-$\beta_0$ laboratory the same strictly recursive protocol is not even
determined at $N=4$, where its design is rank deficient and it abstains, is off
by $+24\%$ at $N=5$, and meets ten percent persistently from $N=8$.  The physical Adler series sits at $N=4$, inside the first of those
regimes and one order below the first at which the tower is even defined.

\subsection{Blind pole--\texorpdfstring{$\MSbar$}{MSbar} mass test}
\label{sec:cvrc-pole}

The pole-mass series~\cite{Marquard:2015qpa,Marquard2016} provides an
external, genuinely recursive test: $\rho$ is
fixed from the withheld-$r_3$ fold and $r_4$ is then forecast blind.  The
declared structure is Eq.~\eqref{eq:cvrc-exponents} with the infrared tower
$u=\tfrac12,1,\tfrac32,2,\ldots$ and the same $u=-1$ closure, and
$\beta_0=(11-2n_\ell/3)/4$.

\begin{table}[t]
\centering
\caption{Blind four-loop pole-mass coefficient from CVRC.  Nothing but
$r_1,r_2,r_3$, the declared exponents, and $\beta_0,\beta_1$ enters.}
\label{tab:cvrc-pole}
\begin{tabular}{crrrr}
\toprule
$n_\ell$ & exact $r_4$ & CVRC & error & $\rho_\star$\\
\midrule
$3$ & $1702.71$ & $1742.38$ & $+2.33\%$ & $+0.1233$\\
$4$ & $1235.63$ & $1306.70$ & $+5.75\%$ & $+0.0735$\\
$5$ & $839.18$  & $939.00$  & $+11.90\%$& $+0.0030$\\
\bottomrule
\end{tabular}
\end{table}

Table~\ref{tab:cvrc-pole} improves on the direct ($-22\%$), Borel
($+16$--$18\%$) and $\Gamma$-normalized ($+9.5$--$12.1\%$) forecasts of
Table~\ref{tab:pole-test} for $n_\ell=3,4$, and is comparable to the last of
them for $n_\ell=5$.  It does not improve on ARTS
(Table~\ref{tab:port-pole}), which reaches $1.60$--$3.84\%$ on this
benchmark.  We state this asymmetry plainly: ARTS is the better transformation
for the nearby $u=1/2$ renormalon, CVRC is the better transformation for the
three-weight $\tau$ moments, and the two facts are consistent with the
Proposition of Sec.~\ref{sec:obstruction}, since the pole-mass defects do not
require a sign reversal at the tested order.

\subsection{All-order large-\texorpdfstring{$\beta_0$}{beta0} validation and the
required order}
\label{sec:cvrc-largebeta}

\begin{figure*}[!tbp]
 \centering
 \includegraphics[width=0.7\textwidth]{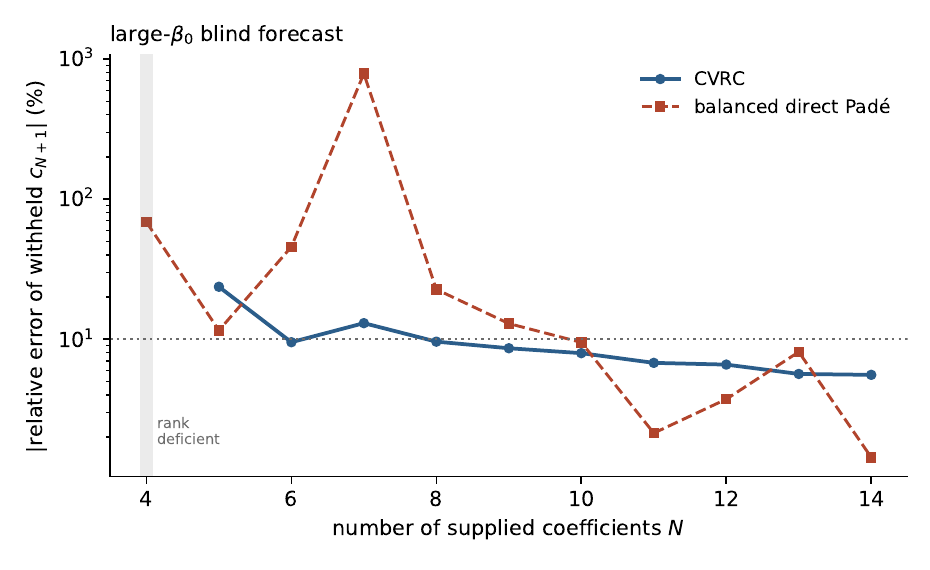}
 \caption{Strictly recursive blind forecast error of the withheld
 large-$\beta_0$ Adler coefficient, CVRC-minimal against the balanced direct
 Pad\'e on identical input.  The dotted line is the declared ten-percent
 criterion.  The CVRC curve begins at $N=5$ because the order-three
 CVRC-minimal design is rank deficient and the method abstains at $N=4$; that
 abscissa is shaded.  The corresponding CVRC-complete curve is tabulated in
 Table~\ref{tab:cvrc-largebeta}; it lies below $1.5\%$ at every $N\geq8$ and
 below one percent from $N=12$.}
 \label{fig:cvrc-largebeta}
\end{figure*}

In the large-$\beta_0$ limit the declared structure is not a model: by
Table~\ref{tab:order-rule} the exact singularity content of
Eq.~\eqref{eq:large-beta-borel} is a simple pole at $u=2$ and double poles at
every other nonzero integer except $u=1$, where the transform is regular, so
$\alpha_2=1$ and $\alpha_p=2$ elsewhere, and the closure is again $u=-1$.
Table~\ref{tab:cvrc-largebeta} runs the strictly recursive protocol of
Sec.~\ref{sec:cvrc-drift} at each order, and Fig.~\ref{fig:cvrc-largebeta}
compares the resulting CVRC-minimal errors with those of the balanced direct
Pad\'e approximant on identical input.

\begin{table*}[!tbp]
\centering
\caption{Strictly recursive blind forecast of the withheld large-$\beta_0$
Adler coefficient $c_{N+1}$ from the first $N$.  CVRC-minimal constrains the
infrared tower only; CVRC-complete interleaves the subleading ultraviolet
tower $u=-2,-3,\ldots$.  The last column is the balanced direct Pad\'e on the
same input.  The entry marked \emph{singular} is an abstention: the order-three
CVRC-minimal design matrix is exactly rank deficient, so the closure is not
determined at that order and no forecast is issued (see the text).}
\label{tab:cvrc-largebeta}
\small
\setlength{\tabcolsep}{4pt}
\begin{tabular}{rrrrrrr}
\toprule
$N$ & exact $c_{N+1}$ & CVRC-minimal & error & CVRC-complete & error
& balanced Pad\'e\\
\midrule
 $4$ & $787.827$              & \multicolumn{2}{c}{singular}                         & $-65.7317$             & $-108.34\%$ & $-68.67\%$\\
 $5$ & $-1991.38$             & $-2463.11$              & $+23.69\%$ & $-7249.98$             & $+264.07\%$ & $-11.56\%$\\
 $6$ & $9.85728\times10^{4}$  & $1.07961\times10^{5}$   & $+9.52\%$  & $1.03053\times10^{5}$  & $+4.55\%$   & $-45.70\%$\\
 $7$ & $-1.07762\times10^{6}$ & $-1.21799\times10^{6}$  & $+13.03\%$ & $-1.11470\times10^{6}$ & $+3.44\%$   & $+787.01\%$\\
 $8$ & $2.77521\times10^{7}$  & $3.04189\times10^{7}$   & $+9.61\%$  & $2.79793\times10^{7}$  & $+0.82\%$   & $-22.68\%$\\
 $9$ & $-5.38781\times10^{8}$ & $-5.85233\times10^{8}$  & $+8.62\%$  & $-5.46064\times10^{8}$ & $+1.35\%$   & $-12.94\%$\\
$10$ & $1.39591\times10^{10}$ & $1.50692\times10^{10}$  & $+7.95\%$  & $1.41226\times10^{10}$ & $+1.17\%$   & $-9.49\%$\\
$11$ & $-3.59839\times10^{11}$& $-3.84235\times10^{11}$ & $+6.78\%$  & $-3.63705\times10^{11}$& $+1.07\%$   & $+2.14\%$\\
$12$ & $1.05429\times10^{13}$ & $1.12376\times10^{13}$  & $+6.59\%$  & $1.06118\times10^{13}$ & $+0.65\%$   & $-3.73\%$\\
$13$ & $-3.25921\times10^{14}$& $-3.44335\times10^{14}$ & $+5.65\%$  & $-3.27830\times10^{14}$& $+0.59\%$   & $-8.12\%$\\
$14$ & $1.09265\times10^{16}$ & $1.15347\times10^{16}$  & $+5.57\%$  & $1.09962\times10^{16}$ & $+0.64\%$   & $-1.43\%$\\
\bottomrule
\end{tabular}
\end{table*}

Strict large-$\beta_0$ contains poles rather than branch cuts, because the
running beyond $\beta_0$ is switched off; a ``subleading cut coefficient'' is
therefore not defined in that limit.  The controlled analogue is to restore
the exact subleading ultraviolet pole tower omitted by CVRC-minimal.  The
comparison in Table~\ref{tab:cvrc-largebeta} is deliberately that test.  At
$N=8$ the absolute error drops from $9.61\%$ to $0.82\%$, and over
$N=8,\ldots,14$ it drops from $5.57$--$9.61\%$ to $0.59$--$1.35\%$.  Thus
subleading singularity content is numerically important once it is resolved.
The same comparison also prevents an overclaim: at the physical input order
$N=4$, CVRC-minimal is singular and CVRC-complete fails by $108\%$, so the
large-$\beta_0$ limit cannot certify a $\pm30$ physical structural band.  That
band is instead the observed full-QCD shift under the declared exponent,
closure-window, and Beneke--Jamin cut variations; the independent $\pm74$
method component is retained precisely because the sparse-order laboratory
does not validate the structural band by itself.

Three statements follow, and they sharpen the coefficient-count conclusion of
App.~\ref{sec:threshold}.  All three are phrased as \emph{persistence}
statements, i.e.\ they quote the first order beyond which the criterion is met
at every tabulated order, which is the same convention used for the
annihilator thresholds of Table~\ref{tab:threshold}.
\begin{enumerate}[leftmargin=2em]
 \item CVRC-complete meets the ten-percent criterion at every $N\geq6$;
 CVRC-minimal first meets it at $N=6$ but relapses to $13.03\%$ at $N=7$ and
 meets it persistently only from $N=8$.  The balanced direct Pad\'e on the same
 input meets it persistently only from $N=10$, and with much larger excursions
 below that ($+787\%$ at $N=7$).
 \item CVRC-complete is already at the percent level from $N=8$: its errors are
 $0.82$, $1.35$, $1.17$, $1.07$, $0.65$, $0.59$ and $0.64$ percent at
 $N=8,\ldots,14$, so it is
 within $1.5\%$ at every $N\geq8$ and persistently within $1\%$ from $N=12$.
 The renormalon annihilator requires eleven to thirteen
 coefficients for persistent percent-level control on the same laboratory.
 The fair comparison is therefore: at $N=8$--$11$, where the annihilator is not
 yet at percent level, CVRC-complete already is; at $N\geq12$ the two are
 comparable.  CVRC-minimal never reaches one percent in the tabulated range,
 flattening at $5$--$8\%$ because it omits the subleading ultraviolet tower
 that dominates the large-$\beta_0$ asymptotics.
 \item Below $N=6$ the method is either undetermined or badly wrong, and
 CVRC-complete is the worse of the two because it is over-parameterized there.
 At $N=4$ CVRC-minimal does not fail, it \emph{abstains}: the order-three
 design is exactly rank deficient.  Its three columns are the $u=2$ simple
 pole, the monomial $u^0$ and the $u=3$ double pole, sampled at $n=0,1,2$.  The
 monomial touches $n=0$ only, so the rank is decided by the $2\times2$ block at
 $n=1,2$, where the ratio of the two singular columns is
 \begin{equation}
  \frac{3^{-2-n}\,(n+1)!}{2^{-1-n}\,n!}=\frac{2(n+1)}{9}\Big(\frac23\Big)^{n}
  =\frac{8}{27}
  \qquad\text{at both }n=1\text{ and }n=2 ,
  \label{eq:rank-defect}
 \end{equation}
 so the two rows are proportional and the block is singular.  The determinant
 vanishes identically, not to within round-off, and a least-squares solver
 would return a minimum-norm answer that looks like a forecast but is fixed by
 the solver's regularization rather than by the data.  Both implementations
 accompanying this paper test the design for rank deficiency before solving and
 abstain when it is deficient.  The test is the ratio of smallest to largest
 pivot in a partial-pivoted Gaussian elimination on the column-normalized
 design, compared against $10^{-12}$; this is used in preference to the
 determinant, which is a product of $N$ pivots and therefore cannot distinguish
 a merely ill-conditioned high-order design from a genuinely rank-deficient
 low-order one.  The degenerate case returns exactly zero, while across every
 other design used in this paper the statistic stays above $2.6\times10^{-6}$,
 six orders of magnitude clear of the threshold, so nothing else is affected.  Read together with the two preceding items this
 \emph{strengthens} the coefficient-count conclusion: below $N=6$ the procedure
 is not merely inaccurate, at $N=4$ it is not defined.  The defect is specific
 to the large-$\beta_0$ exponents: Eq.~\eqref{eq:rank-defect} is an identity
 only because $\alpha_2=1$ and $\alpha_3=2$ there, and the physical
 renormalization-group values $\alpha_2=2.580$, $\alpha_3=3.370$ of
 Eq.~\eqref{eq:cvrc-exponents} break the proportionality, so the physical
 order-three design used to fix $\rho_\star$ is well conditioned.  Finally, the
 physical forecast below uses CVRC-minimal because declaring the subleading
 ultraviolet tower on four coefficients is not conservative but harmful.
\end{enumerate}

\subsection{The calibrated Borel model in closed form}
\label{sec:cvrc-model}

The closure rule has a simple structural reading that makes the calibrated
model explicit.  At order $N$ the model contains all the columns of the order
$N-1$ model plus one new one.  The closure $\rho_\star$ of
Eq.~\eqref{eq:cvrc-rho} is defined by demanding that the order $N-1$ model
already reproduce $c_{N,1}$; that solution therefore reproduces
$c_{1,1},\ldots,c_{N,1}$, and since it is a member of the order-$N$ column
span, the order-$N$ solve returns it \emph{with the new column at zero
amplitude}.  Hence
\begin{quote}
\textbf{Closure $=$ null newest column.}  $\rho_\star$ is the unique value for
which the amplitude of the column added at order $N$ vanishes; equivalently,
the order-$N$ and order-$(N-1)$ CVRC-calibrated BJ-type ans\"atze coincide.
\end{quote}

The identity has a deflationary corollary that we state rather than leave for
the reader to discover.  Because the closure kills the newest column, the
calibrated model at $\rho_\star$ is the plain four-parameter exact fit of the
three singularities and $u^0$ to the four known coefficients: at physical order
the leave-one-order-out rule selects the same model that ordinary exact fitting
selects, and in that sense it supplies no constraint that fitting had not
already supplied.  It is also, term for term, the variant that
Ref.~\cite{Beneke:2008ad} obtains by setting $d_1^{\rm PO}=0$
(App.~\ref{sec:cvrc-bj}).  The closure rule does real work at the lower folds,
$N=2$ and $N=3$, where it is what generates the admissible closure interval of
Sec.~\ref{sec:cvrc-forecast} below and the misspecification rejections of
Table~\ref{tab:cvrc-misspec}; it does none at $N=4$.  Readers should therefore
not read the phrase ``fixed by cross-validation'' as implying that the physical
central value carries information beyond an exact fit.  It does not.  The
calibration-and-consistency protocol supplies the admissible interval and the
misspecification test; it does not supply independent information for the
physical central value.

The consequence is that the physical CVRC forecast is not an
$N$-dependent recipe but a single Borel function.  With the declared exponents
\eqref{eq:cvrc-exponents} and the four known coefficients,
\begin{equation}
 \begin{aligned}
 B[\Dhat](u)={}&\frac{6.463246}{(2-u)^{2.580247}}
 -\frac{26.662923}{(3-u)^{3.370370}}\\
 &-\frac{0.060414}{(1+u)^{1.209877}}+0.637084 ,
 \end{aligned}
 \label{eq:cvrc-borel-model}
\end{equation}
in the normalization \eqref{eq:borel-normalization}.  All four amplitudes are
fixed exactly by $c_{1,1},\ldots,c_{4,1}$; the $u=-1$ amplitude \emph{is} the
closure $\rho_\star=-0.060414$ of Eq.~\eqref{eq:rho-star}, which by the
null-newest-column identity is what the exact four-parameter fit returns
anyway.  Equation \eqref{eq:cvrc-borel-model} is the pure-power BJ-type ansatz
defined above.  It shares the leading singularity positions, exponents, and
polynomial organization of Ref.~\cite{Beneke:2008ad}, but it is not identical
to that model because its subleading cut structure is absent.  The positions
and exponents come from the renormalization group and the amplitudes from the
four known coefficients.  At this order the leave-one-order-out condition adds
no information beyond the exact fit.

Every Adler coefficient now follows from Eq.~\eqref{eq:cvrc-mode} without
further input,
\begin{equation}
 c_{n,1}=\sum_{p}N_p\,p^{-\alpha_p-n+1}
 \frac{\Gamma(\alpha_p+n-1)}{\Gamma(\alpha_p)}\,\beta_0^{\,n-1}
 +\delta_{n1}\,p_0 ,
 \label{eq:cvrc-any-order}
\end{equation}
with $p\in\{2,3,-1\}$ and the convention $p^{-\alpha-n+1}\to
(-1)^{n-1}|p|^{-\alpha-n+1}$ for $p<0$.  Rolling the closure rule forward
instead of using Eq.~\eqref{eq:cvrc-any-order} gives the same numbers: at every
step the recomputed closure returns $\rho_\star$ to $2\times10^{-17}$, so the
CVRC map is a fixed point of its own recursion and the rolled forecast is
unambiguous.

\subsection{Physical forecast, closure window, and structural systematics}
\label{sec:cvrc-forecast}

All results in this subsection use CVRC-minimal.  No physical-order number is
obtained from CVRC-complete.  With $\rho_\star=-0.060414$ from
Eq.~\eqref{eq:rho-star},
Eq.~\eqref{eq:cvrc-lines} gives
\begin{equation}
 \widehat c_{5,1}^{\rm BJ\text{-}CVRC}=247.98,
 \qquad
 \widehat c_{6,1}^{\rm BJ\text{-}CVRC}=3102.1,
 \label{eq:cvrc-central}
\end{equation}
and, through the inverse map \eqref{eq:global-inverse-map},
$d_5[W_\tau]=555.76$ and $d_6[W_\tau]=1872.62$.

Three \emph{structural} uncertainties are quoted below, and none of them is
statistical.  They are not the whole error: the method uncertainty of the
transformation itself is separated out and quantified in
Sec.~\ref{sec:cvrc-method-error}, and it is the larger contribution.

\paragraph{Closure window.}  Instead of the single point $\rho_\star$, retain
the whole admissible set \eqref{eq:admissible-set}.  From
Eq.~\eqref{eq:cvrc-lines} this is the interval
\begin{equation}
 \rho\in[-0.06788,\,-0.05294]
 \;\Longrightarrow\;
 \begin{cases}
 c_{5,1}\in[229.8,\,266.2],\\
 c_{6,1}\in[3059,\,3145],
 \end{cases}
 \label{eq:cvrc-window}
\end{equation}
symmetric about the central values because the family is affine.

\paragraph{The $u=3$ exponent.}  The $D=6$ operators mixing into the $u=3$
singularity carry an anomalous dimension that
Eq.~\eqref{eq:cvrc-exponents} omits.  Varying $\alpha_3$ from $3.00$ (no
renormalization-group dressing at all) to $3.58$ (an illustrative dressing;
\emph{not} the value of Ref.~\cite{Beneke:2008ad}, whose $u=3$ exponent is
the pure renormalization-group $3.370370$ that we adopt, see
App.~\ref{sec:cvrc-bj}) moves the central value over
$c_{5,1}\in[235.3,257.8]$ and $c_{6,1}\in[3021,3170]$, i.e.\
$^{+9.8}_{-12.7}$ and $^{+68}_{-81}$.

\paragraph{The ultraviolet exponent.}  Table~\ref{tab:cvrc-misspec} shows that
the admissible set does not resolve $\alpha_{-1}$ finely.  The declared value
$2-\beta_1/\beta_0^2=1.20988$ follows from the same argument that fixes
$\alpha_2$, and is the value used in Ref.~\cite{Beneke:2008ad}; a reader who
declines that step and admits the whole surviving range
$\alpha_{-1}\in[1,2]$ obtains central values $c_{5,1}\in[210.5,258.4]$ at the
declared $\alpha_3$, or $[198.2,268.2]$ over the full
$\alpha_3\times\alpha_{-1}$ grid, and, taking the union of the corresponding
closure windows, the much wider envelope
$c_{5,1}\in[174.8,284.6]$, $c_{6,1}\in[2955,3456]$.  We do not adopt that
envelope, but we record it, because it is the correct statement of what four
coefficients determine if the renormalization-group dressing of the
ultraviolet double pole is not assumed.

Combining the first two envelopes linearly, and keeping them asymmetric,
gives the \emph{structural} (parametric) uncertainty of the declared model,
\begin{equation}
 c_{5,1}^{\rm BJ\text{-}CVRC}=248^{+28}_{-31}\Big|_{\rm struct},
 \qquad
 c_{6,1}^{\rm BJ\text{-}CVRC}=3102^{+111}_{-124}\Big|_{\rm struct},
 \label{eq:cvrc-struct}
\end{equation}
with the wider $\alpha_{-1}$-scanned alternative recorded above.

\subsection{The method uncertainty, and why it dominates}
\label{sec:cvrc-method-error}

Equation~\eqref{eq:cvrc-struct} propagates the ambiguity of the declared
structure.  It does \emph{not} contain the error of the transformation itself,
and at $N=4$ that error is the larger of the two.  We take the unusual step of
quantifying it, because everything needed to do so is already in this paper.

Three measurements bound the one-order-ahead accuracy of CVRC-minimal at the
order at which it is being used here.
\begin{enumerate}[leftmargin=2em]
 \item \textbf{The large-$\beta_0$ laboratory offers no support at $N=4$.}
 Table~\ref{tab:cvrc-largebeta} is unambiguous: at $N=4$ the minimal design is
 rank deficient and the method abstains, CVRC-complete returns $-108.34\%$,
 and the balanced direct Pad\'e returns $-68.67\%$.  The first order at which
 the minimal tower is defined at all is $N=5$, where its error is $+23.69\%$;
 it does not meet ten percent persistently until $N=8$.  The physical Adler
 series sits at $N=4$, one order \emph{below} the first laboratory point.
 \item \textbf{The one blind test at physical order misses by
 $\bm{-29.9\%}$.}  This is the honest negative of
 Sec.~\ref{sec:cvrc-drift}: run strictly recursively, with $\rho$ fixed one
 order lower still, the same map returns $\widehat c_{4,1}=34.43$ against
 $c_{4,1}=49.0757$.
 \item \textbf{The pole-mass test is better but does not transfer.}
 Table~\ref{tab:cvrc-pole} gives $2.3$--$11.9\%$ from only three input
 coefficients.  That series is dominated by the \emph{nearby} $u=1/2$
 renormalon, whose asymptotics set in far earlier than those of a
 $u=2$-dominated Adler function; the same asymmetry is what makes ARTS/PORT
 succeed there and fail here (Sec.~\ref{sec:cvrc-pole}).  It is therefore an
 optimistic bound, not a representative one.
\end{enumerate}

The demonstrated accuracy at this order therefore lies between roughly $12\%$
on an easier problem and $30\%$ on the actual one.  We adopt
\begin{equation}
\sigma_{\rm method}=30\%\times c_{5,1}^{\rm BJ\text{-}CVRC}=74 ,
 \label{eq:cvrc-method-error}
\end{equation}
chosen as the larger of the $+23.69\%$ error at the first large-$\beta_0$
order where the tower used here is defined and the $29.9\%$ magnitude of the
strictly recursive physical miss one order lower.  This choice directly
brackets the only same-observable blind failure rather than assuming that its
relative size improves at the next order.  Two further facts fix the \emph{shape} of the
band.  First, the large-$\beta_0$ errors of CVRC-minimal are uniformly
positive ($+23.69,+9.52,+13.03,+9.61,\ldots$), whereas the physical recursive
test \emph{undershoots}; the laboratory does not predict even the sign of the
physical error, so no sign-informed correction is legitimate and the band must
be symmetric.  Second, once $\sigma_{\rm method}$ is included the asymmetry of
Eq.~\eqref{eq:cvrc-struct} is numerically irrelevant, so we symmetrize the
structural half-width to $30$.  Adding the two in quadrature,
\begin{equation}
 c_{5,1}^{\rm BJ\text{-}CVRC}=248\pm30\,(\text{struct})\pm74\,(\text{method})
 =248\pm80 .
 \label{eq:cvrc-final}
\end{equation}

\paragraph{$c_{6,1}$ is not advanced as a prediction.}  Every validation in
this work (moments, pole mass, and the all-order large-$\beta_0$ protocol)
tests \emph{one-order-ahead} forecasting.  The $c_{6,1}$ entry of
Eq.~\eqref{eq:cvrc-struct} is the second component of the $N=4$ closure
family, i.e.\ two orders past the last supplied coefficient, and by the
criterion just applied to $c_{5,1}$ it is unvalidated.  We therefore demote it
to the status the rest of the rolled sequence carries:
\begin{equation}
 c_{6,1}^{\rm BJ\text{-}CVRC}=3102^{+111}_{-124}\Big|_{\rm struct},
 \qquad \textsc{roll},
 \label{eq:cvrc-c6}
\end{equation}
reported in Table~\ref{tab:cvrc-high-order} alongside
$c_{7,1}$--$c_{12,1}$ rather than quoted as a result.  Applying the method
uncertainty of Eq.~\eqref{eq:cvrc-method-error} to it, together with the
measured cost of a one-order roll, would give something of order
$3100\pm900$, which is not worth quoting.  The third component
$\widehat c_{7,1}=1.30\times10^4$ is likewise not advanced as a forecast.

\begin{figure*}[!tbp]
 \centering
 \includegraphics[width=0.95\textwidth]{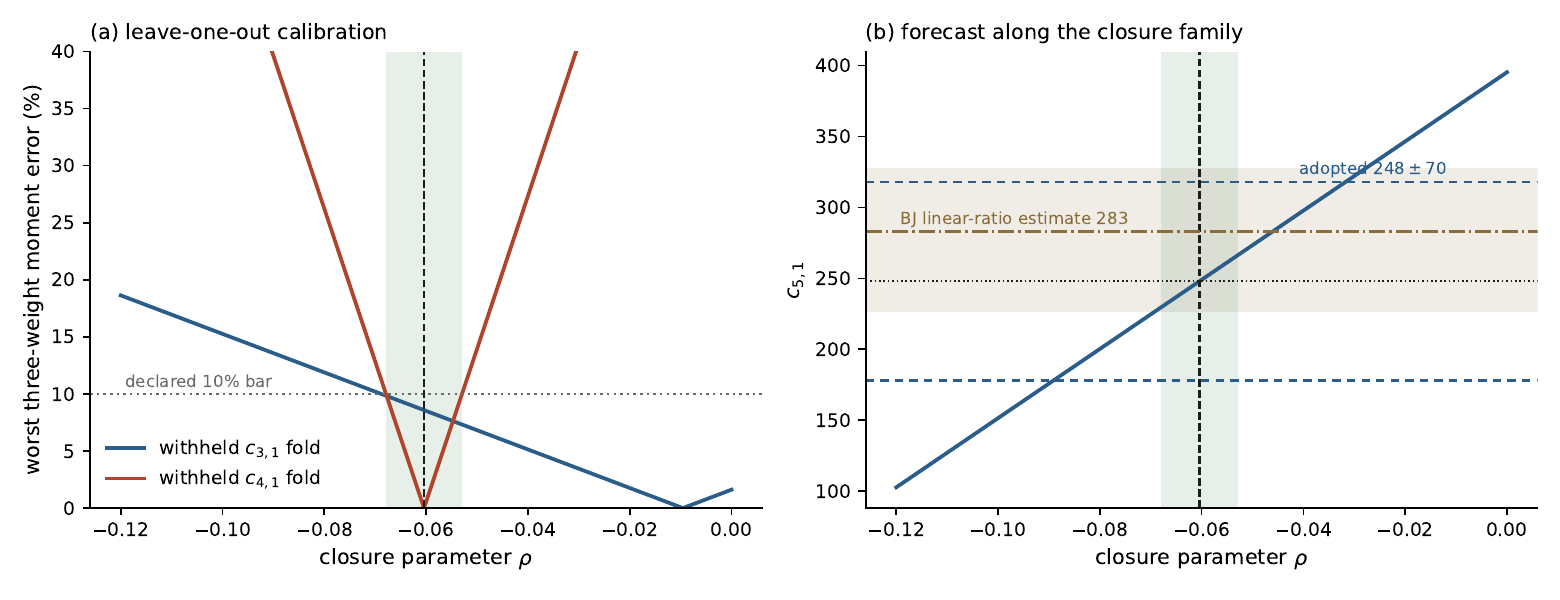}
 \caption{(a)~Worst three-weight relative error on each leave-one-order-out
 fold as a function of the closure parameter $\rho$.  The dashed vertical line
 is $\rho_\star$ of Eq.~\eqref{eq:rho-star}; the shaded band is the admissible
 window \eqref{eq:cvrc-window}.  (b)~The forecast
 $\widehat c_{5,1}(\rho)$ along the same family, with the shaded literature
 band $277\pm51$ of Ref.~\cite{Boito:2018rwt} and the
 value $283$ commonly attributed to Ref.~\cite{Beneke:2008ad} (in fact its
Eq.~(3.10) input; see App.~\ref{sec:cvrc-bj}).  The dashed blue
 pair is the adopted result \eqref{eq:cvrc-final}, $248\pm80$: the vertical
 admissible window is only the closure part of the \emph{structural} band
 \eqref{eq:cvrc-struct} and is much narrower than the total, which is
 dominated by the method uncertainty of Sec.~\ref{sec:cvrc-method-error}.}
 \label{fig:cvrc-closure}
\end{figure*}

\paragraph{What this costs, and why we pay it.}  Equation
\eqref{eq:cvrc-final} is a weaker statement than
Eq.~\eqref{eq:cvrc-struct} would have been, and weaker than the
Borel--Pad\'e determination $277\pm51$ it agrees with.  We regard that as the
correct outcome rather than a disappointment: it is the uncertainty implied by
the tests this paper itself declared and ran, and quoting a $\pm12\%$ band
alongside a $-29.9\%$ blind miss at the same order would not survive contact
with a careful reader.  The contribution of the construction is not a tighter
number; it is that a value in this range is returned by the only continuation
examined here that passes the declared acceptance criterion, from a structure
whose singularities and exponents are declared in advance.
Figure~\ref{fig:cvrc-closure} displays the calibration and the closure family.

\parttitle{Part V. Global comparison of the continuations}

Part~V places the single advanced result beside every other continuation
constructed in this paper, generated from the same inputs and mapped into one
convention, and states which comparisons and combinations are legitimate.  It
introduces no new method and changes no status label.

\section{Global comparison of \texorpdfstring{$c_{5,1}$}{c51} and
\texorpdfstring{$c_{6,1}$}{c61} across all approximants}
\label{sec:global-comparison}
Sections~\ref{sec:methods-brief}--\ref{sec:cvrc} and
Appendix~\ref{sec:qcd-forecast} introduced each approximant together with its
own validation record, and each therefore reported its physical continuation
in a different place and in a different normalization.  A reader who wants to know
what the four known coefficients actually permit for $c_{5,1}$ and $c_{6,1}$
needs those numbers side by side, generated from one set of inputs and one
inverse renormalization-group map.  This section supplies that comparison.  It
introduces no new method and no new assumption: every row is a continuation
already defined in Sec.~\ref{sec:methods-brief} and
Appendix~\ref{sec:qcd-forecast} and is placed in a common convention using the
same inverse map.  The status labels are those of
Sec.~\ref{sec:status-labels}, with the epistemic levels of
App.~\ref{sec:claims}, and the section ends by stating explicitly which
combinations of rows are legitimate and which are not.

\subsection{Common inputs and the single inverse map}
\label{sec:global-inputs}

Every entry uses the same two input sequences.  The Adler series is
Eq.~\eqref{eq:known-adler},
\begin{equation}
 \begin{aligned}
 c_{1,1}&=1, &c_{2,1}&=1.639821204,\\
 c_{3,1}&=6.371014, &c_{4,1}&=49.07570 .
 \end{aligned}
 \label{eq:global-adler-input}
\end{equation}
and the standard $\tau$ moment series generated from it by the exact contour
integration of Sec.~\ref{sec:weight-map} is
\begin{equation}
 \begin{aligned}
 d_1&=1, &d_2&=5.202321204,\\
 d_3&=26.365928485, &d_4&=127.078615760 .
 \end{aligned}
 \label{eq:global-moment-input}
\end{equation}
No entry uses any other input, and in particular no entry uses a fitted
normalization, a tuned parameter chosen after inspecting an unknown-order
forecast, or an external estimate of $c_{5,1}$.

Methods that forecast moment coefficients return $d_5$ and $d_6$; methods that
forecast Adler coefficients return $c_{5,1}$ and $c_{6,1}$ directly.  The two
are related by the sixth-order triangular FOPT map of
Eq.~\eqref{eq:d5d6-map}, used here in both directions,
\begin{align}
 c_{5,1}&=d_5+391.546-14.25\,c_{4,1},\nonumber\\
 c_{6,1}&=d_6+7860.51-45.1119\,c_{4,1}-17.8125\,c_{5,1},
 \label{eq:global-inverse-map}
\end{align}
with $c_{4,1}$ fixed at its known value.  Because
Eq.~\eqref{eq:global-inverse-map} is exact and linear, the comparison is not
affected by where in the chain a given method chooses to act; what differs
between rows is only the continuation model, which is the object of interest.
The map is triangular, so an error $\delta c_{5,1}$ propagates into
$c_{6,1}$ with the fixed multiplier $-17.8125$.  A method that overestimates
$c_{5,1}$ will therefore tend to underestimate $c_{6,1}$ at fixed $d_6$, and
the anticorrelation visible in Table~\ref{tab:global-comparison} between the
two columns is partly this algebraic effect rather than independent
information.

Two operational conventions are stated once here and applied uniformly.
First, an approximant that is defined only for one further coefficient is
rolled forward once: the predicted $d_5$ (or $c_{5,1}$) is appended to the
input sequence and the same construction, with rebalanced degrees where
applicable, produces the second coefficient.  This is the same rolling rule
used in the large-$\beta_0$ tests of
Apps.~\ref{sec:largebeta} and \ref{sec:annihilator-validation}, so the rows
inherit the convergence behavior documented there.  Second, the parameterized
families are propagated over their complete predeclared grids,
$k,\alpha\in\{0.01,\dots,0.05\}$ and
$\gamma_{-1}\in\{-0.90,-0.79,0\}$, and no grid point is removed after the
fact.

\subsection{The comparison table}
\label{sec:global-table}

Table~\ref{tab:global-comparison} collects every continuation.  The status
column uses the labels defined in Sec.~\ref{sec:status-labels}.

\begin{table*}[!tbp]
\centering
\caption{Coefficient continuations obtained from the four known inputs in
Eqs.~\eqref{eq:global-adler-input} and
\eqref{eq:global-moment-input}.  Moment-space rows quote the predicted
$d_5$ and $d_6$ for the standard $\tau$ weight; Adler-space rows quote the
moment coefficients implied by Eq.~\eqref{eq:global-inverse-map}.
Status: \textsc{base}, prevalidation reference; \textsc{cond},
validated conditional forecast; \textsc{diag}, diagnostic only;
and \textsc{abst}, the method's own certificate abstains.  Formal Levin-$u$
is represented by the algebraically identical $\theta_2$ row and is therefore
not duplicated.  Canonical $\rho_2$ does not reproduce the last supplied
coefficient and therefore has no admissible coefficient forecast.
Anchor-dependent
``effective coefficients'' extracted from fixed-coupling transformed sums are
not included in this table or in any forecast envelope.  The dagger on the
CVRC rows applies to $c_{5,1}$ only; their $c_{6,1}$ entries are rolled
two-order-ahead values.}
\label{tab:global-comparison}
\scriptsize
\setlength{\tabcolsep}{3pt}

\begin{tabular}{@{}llrrrrc@{}}
\toprule
method
& specification
& $d_5$
& $d_6$
& $c_{5,1}$
& $c_{6,1}$
& status\\
\midrule

\multicolumn{7}{l}{\itshape
 Rational approximants to the $W_\tau$ moment series}\\

\quad direct Pad\'e
& $[1/2]$
& $549.48$
& $1\,738.24$
& $241.70$
& $3\,079.56$
& \textsc{diag}\\

\quad direct Pad\'e
& $[2/1]$
& $612.49$
& $2\,952.10$
& $304.71$
& $3\,171.04$
& \textsc{diag}\\

\quad Borel-normalized Pad\'e
& $[1/2]$
& $286.05$
& $-6\,761.97$
& $-21.73$
& $-728.21$
& \textsc{diag}\\

\quad Borel-normalized Pad\'e
& $[2/1]$
& $816.66$
& $6\,560.23$
& $508.88$
& $3\,142.48$
& \textsc{diag}\\

\addlinespace
\multicolumn{7}{l}{\itshape
 Rational approximants to the common Adler series}\\

\quad common-Adler Pad\'e
& $[1/2]$
& $753.63$
& $6\,440.82$
& $445.85$
& $4\,145.81$
& \textsc{diag}\\

\quad common-Adler Pad\'e
& $[2/1]$
& $685.81$
& $3\,998.96$
& $378.03$
& $2\,911.94$
& \textsc{diag}\\

\quad common-Adler Borel--Pad\'e
& $[1/2]$
& $973.60$
& $19\,576.36$
& $665.81$
& $13\,363.16$
& \textsc{diag}\\

\quad common-Adler Borel--Pad\'e
& $[2/1]$
& $811.82$
& $9\,802.54$
& $504.04$
& $6\,470.98$
& \textsc{diag}\\

\addlinespace
\multicolumn{7}{l}{\itshape
 Logarithmic-derivative (D-log) Pad\'e approximants}\\

\quad direct D-log Pad\'e
& balanced
& $597.24$
& $2\,762.02$
& $289.46$
& $3\,252.61$
& \textsc{diag}\\

\quad Borel D-log Pad\'e
& balanced
& $544.76$
& $1\,539.08$
& $236.98$
& $2\,964.53$
& \textsc{diag}\\

\quad Adler-projected direct D-log
& balanced
& $753.09$
& $6\,600.42$
& $445.31$
& $4\,314.98$
& \textsc{diag}\\

\quad Adler-projected Borel D-log
& balanced
& $876.48$
& $13\,180.44$
& $568.70$
& $8\,697.05$
& \textsc{diag}\\

\addlinespace
\multicolumn{7}{l}{\itshape
 Accuracy-through-order nonlinear sequence transformations}\\

\quad shifted Wynn/Shanks
& --
& $612.49$
& $2\,952.10$
& $304.71$
& $3\,171.04$
& \textsc{diag}\\

\quad $\eta_n$-modified Wynn
& $k=0.01$
& $606.43$
& $2\,922.87$
& $298.65$
& $3\,249.83$
& \textsc{diag}\\

\quad $\eta_n$-modified Wynn
& $k=0.02$
& $600.48$
& $2\,894.22$
& $292.70$
& $3\,327.08$
& \textsc{diag}\\

\quad $\eta_n$-modified Wynn
& $k=0.03$
& $594.65$
& $2\,866.12$
& $286.87$
& $3\,402.83$
& \textsc{diag}\\

\quad $\eta_n$-modified Wynn
& $k=0.04$
& $588.94$
& $2\,838.56$
& $281.15$
& $3\,477.12$
& \textsc{diag}\\

\quad $\eta_n$-modified Wynn
& $k=0.05$
& $583.33$
& $2\,811.53$
& $275.54$
& $3\,549.99$
& \textsc{diag}\\

\quad $\alpha$-modified Wynn
& $\alpha=0.01$
& $612.49$
& $2\,981.62$
& $304.71$
& $3\,200.56$
& \textsc{diag}\\

\quad $\alpha$-modified Wynn
& $\alpha=0.02$
& $612.49$
& $3\,011.14$
& $304.71$
& $3\,230.08$
& \textsc{diag}\\

\quad $\alpha$-modified Wynn
& $\alpha=0.03$
& $612.49$
& $3\,040.66$
& $304.71$
& $3\,259.61$
& \textsc{diag}\\

\quad $\alpha$-modified Wynn
& $\alpha=0.04$
& $612.49$
& $3\,070.19$
& $304.71$
& $3\,289.13$
& \textsc{diag}\\

\quad $\alpha$-modified Wynn
& $\alpha=0.05$
& $612.49$
& $3\,099.71$
& $304.71$
& $3\,318.65$
& \textsc{diag}\\

\quad Brezinski $\theta_2$ / formal Levin-$u$
& identical ATO forecasts
& $580.94$
& $2\,495.85$
& $273.16$
& $3\,276.85$
& \textsc{diag}\\

\addlinespace
\multicolumn{7}{l}{\itshape
 Renormalon-transported defect forecasts}\\

\quad ARTS/PORT, single mode
& $u=2$
& $873.85$
& $12\,631.16$
& $566.07$
& $8\,194.67$
& \textsc{abst}\\

\addlinespace
\multicolumn{7}{l}{\itshape
 BJ--CVRC conditional closure
 (Sec.~\ref{sec:cvrc})}\\

\quad BJ--CVRC
& $\rho_\star=-0.06041$
& $555.76$
& $1\,872.62$
& $247.98$
& $3\,102.14$
& \textsc{cond}$^{\dagger}$\\

\quad BJ--CVRC
& window, lower edge
& $537.54$
& $1\,505.40$
& $229.76$
& $3\,059.42$
& \textsc{cond}$^{\dagger}$\\

\quad BJ--CVRC
& window, upper edge
& $573.98$
& $2\,239.87$
& $266.20$
& $3\,144.87$
& \textsc{cond}$^{\dagger}$\\

\addlinespace
\multicolumn{7}{l}{\itshape
 Renormalon-annihilating transformations}\\

\quad annihilator, $m=1$
& $u=2$, $\gamma_2=128/81$
& $615.87$
& $2\,121.87$
& $308.09$
& $2\,280.69$
& \textsc{diag}\\

\quad annihilator, $m=3$
& $\gamma_{-1}=-0.79$
& $674.90$
& $4\,535.72$
& $367.12$
& $3\,643.06$
& \textsc{diag}\\

\quad annihilator, $m=3$
& $\gamma_{-1}=-0.90$
& $681.37$
& $4\,494.39$
& $373.59$
& $3\,486.47$
& \textsc{diag}\\

\quad annihilator, $m=3$
& $\gamma_{-1}=0$
& $613.99$
& $5\,114.27$
& $306.21$
& $5\,306.60$
& \textsc{diag}\\

\bottomrule
\end{tabular}

\vspace{2pt}
\begin{minipage}{0.98\textwidth}
\footnotesize
$^{\dagger}$ The BJ--CVRC status applies to the $c_{5,1}$ entries.  The
$c_{6,1}$ entries are rolled two-order-ahead values and have
\textsc{roll} status.
\end{minipage}
\end{table*}

\subsection{Graphical comparison across all approximants}
\label{sec:global-figure}

\begin{figure*}[!tbp]
 \centering
 \includegraphics[width=0.90\textwidth,keepaspectratio]{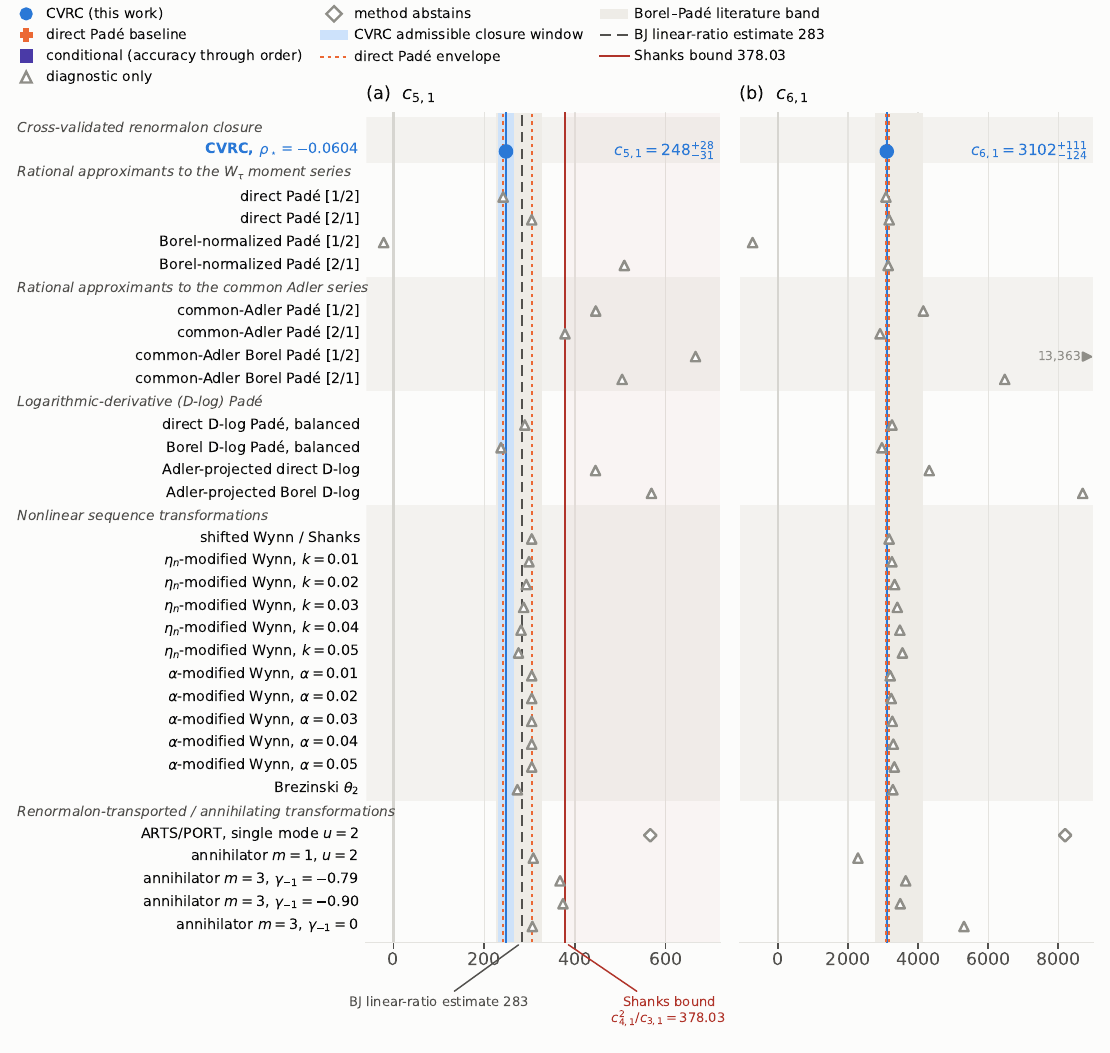}
 \caption{Graphical comparison of every $c_{5,1}$ (panel a) and $c_{6,1}$
 (panel b) continuation obtained in this work.    Rows are grouped by method
 family.  Symbol shape and color together encode the status label of
 Table~\ref{tab:global-comparison}, with the cross-validated-closure rows split
 off into their own class, so no reading depends on color alone.  The
 cross-validated renormalon closure is the top row, with its admissible closure
 window \eqref{eq:cvrc-window} shown as the blue band; the dashed orange lines
 are the direct-Pad\'e envelope and the grey band is the Borel--Pad\'e
 literature range ($277\pm51$ and $3460\pm690$).  In panel~(a) the solid red
 line is the Shanks bound $c_{4,1}^2/c_{3,1}=378.03$ of the Proposition of
 Sec.~\ref{sec:obstruction}; the faint shading marks the half-line
 $\widehat c_{5,1}\geq378.03$ to which every defect-sign-preserving
 common-Adler transformation is confined, and every such row indeed lies inside
 it, while everything to the left is a moment-space or sum-space continuation,
 to which the Proposition does not apply, or CVRC, which is constructed to
violate its hypothesis.  Formal Levin-$u$ is represented by the identical
$\theta_2$ point, and canonical $\rho_2$ is not shown because it has no
accuracy-through-order coefficient forecast.  One point falls outside the
plotted range and is drawn as an edge arrow carrying its value.  The
clustering of the accuracy-through-order rows against the scatter of the
diagnostic rows is the quantitative statement that four coefficients
underdetermine the continuation model.}
 \label{fig:global-comparison}
\end{figure*}

Unlike the logical partition in Fig.~\ref{fig:defect-partition},
Figure~\ref{fig:global-comparison} identifies the complete comparison
method by method and displays both forecast orders.  Every continuation of
Table~\ref{tab:global-comparison} appears,
including the one-mode $u=2$ annihilator row and the cross-validated
renormalon closure of Sec.~\ref{sec:cvrc}.  Rows are grouped by method family; the status label is carried
by symbol shape as well as by color, a filled circle for CVRC, a filled
cross for the direct-Pad\'e baseline, a filled square for the
accuracy-through-order rows, an open triangle for diagnostics and an open
diamond for abstentions, so the figure survives monochrome printing and
color-vision deficiency.  Formal Levin-$u$ is represented by the identical
$\theta_2$ point and is not duplicated.  Canonical $\rho_2$ is absent because
it fails accuracy through order and therefore supplies no coefficient
forecast.  The only off-scale point is the $c_{6,1}$ entry of the common-Adler
Borel--Pad\'e $[1/2]$ continuation; it is drawn as an edge arrow carrying its
value rather than rescaled away.

Three reference structures are overlaid.  The blue band is the CVRC admissible
closure window \eqref{eq:cvrc-window}; the dashed orange pair is the
direct-Pad\'e envelope of Eq.~\eqref{eq:global-pade-box}; the grey band is the
Borel--Pad\'e literature range of Ref.~\cite{Boito:2018rwt}.  The most
informative addition is the solid red line in panel~(a) at
$c_{4,1}^2/c_{3,1}=378.03$.  By the Proposition of
Sec.~\ref{sec:obstruction} no defect-sign-preserving common-Adler
transformation can return a value to the left of it, and the figure confirms
this without exception: every common-Adler row sits on or beyond the line, the common-Adler Pad\'e $[2/1]$ sits exactly \emph{on} it, being the Shanks
baseline itself, while every row to its left either acts on a moment
series or on the annihilator's own sum space, to which the Proposition does not
apply, or is CVRC, which is constructed to violate the hypothesis.

\subsection{What the table shows}
\label{sec:global-reading}

\paragraph{The accuracy-through-order rows cluster, but fail the declared weight test.}
Throughout this subsection the three cross-validated-closure rows are counted
separately and excluded from the spans, for the reason given in
Sec.~\ref{sec:global-envelopes}: they are three points of one one-parameter
family, not three continuations, and they are the only rows that pass every
declared weight test on both leave-one-order-out folds
(Table~\ref{tab:cvrc-moment}).  The ARTS/PORT row passes the withheld-$d_4$
test but is undefined on the withheld-$d_3$ fold, and its own certificate
abstains (Sec.~\ref{sec:brief-arts}).  Their union,
$c_{5,1}\in[229.8,266.2]$, is the closure-window part of the structural band
\eqref{eq:cvrc-struct}; the result actually advanced,
Eq.~\eqref{eq:cvrc-final}, is wider because it adds the method uncertainty of
Sec.~\ref{sec:cvrc-method-error}.
With that convention the fourteen accuracy-through-order Pad\'e/Wynn rows
span
$241.70\le c_{5,1}\le304.71$ and $3079.56\le c_{6,1}\le3549.99$, a relative
spread of $23\%$ and $14\%$ about their midpoints.  Admitting the diagnostic
rows widens the $c_{5,1}$ range to $[-21.73,665.81]$ and the $c_{6,1}$ range
to $[-728.21,13\,363.16]$; both then include zero and both change sign.  This
contrast is the single most informative feature of the comparison.  It is not
an argument that the diagnostic methods are intrinsically bad --- several of
them converge in the large-$\beta_0$ tests of
Apps.~\ref{sec:largebeta}, \ref{sec:annihilator-validation}, and
\ref{sec:dlog-largebeta} once enough coefficients are supplied --- but a
demonstration that at four coefficients the choice of continuation model, not
the data, controls the answer.

\paragraph{Normalization dominates the Pad\'e rows.}
The direct and Borel-normalized rational approximants are built from
identical input and differ only by the weight applied before the rational
fit.  Their $c_{5,1}$ predictions are $241.70$ and $304.71$, and $-21.73$ and
$508.88$, respectively.  The Borel-normalized pair straddles zero, and its
$[1/2]$ member returns a negative $c_{6,1}$.  A factorial weight is the correct asymptotic
normalization for a renormalon-dominated series, so the instability of these
rows at four coefficients is precisely the statement that the series has not
yet entered its asymptotic regime.  The same conclusion was reached
independently in App.~\ref{sec:largebeta}, where the balanced Borel--Wynn
table shows large low-order excursions before converging.

\paragraph{Projection order matters more than the approximant.}
The rows built on the common Adler series and then propagated through the
weight maps sit systematically higher than the corresponding moment-space
rows: $445.85$ and $378.03$ against $241.70$ and $304.71$ for the direct
approximants, and $665.81$ and $504.04$ against $-21.73$ and $508.88$ for the
Borel-normalized ones.  The same pattern appears in the D-log family, where
Adler projection moves $c_{5,1}$ from $289.46$ to $445.31$.  This is the
numerical counterpart of the operation-order discussion of
App.~\ref{sec:arts-order}: forecasting the common Adler coefficient first and
reconstructing the moments afterwards is a genuinely different transformation
from forecasting a moment directly, and at this order the difference is larger
than the spread within either family.

\paragraph{The regularized Wynn families move $c_{6,1}$, not $c_{5,1}$.}
Across its full grid the $\alpha$ family leaves $c_{5,1}$ at $304.71$ exactly
and moves $c_{6,1}$ from $3200.56$ to $3318.65$.  This is the algebraic
statement proved in App.~\ref{sec:alpha}: the parameter cancels from the
one-step forecast and first appears in the second.  A flat $c_{5,1}$ column is
therefore an identity and must not be read as evidence of stability.  The
$\eta_n$ family does move both, monotonically and in opposite directions, so
that its $c_{5,1}$ decreases while its $c_{6,1}$ increases; through
Eq.~\eqref{eq:global-inverse-map} the second effect is partly the propagated
image of the first.

\paragraph{The renormalon-informed rows are the widest spread of all.}
The single-mode $u=2$ annihilator gives
$(c_{5,1},c_{6,1})=(308.09,2280.69)$, but its blind $c_{4,1}$ error is
$33.11\%$ and it fails the all-weight certificate.  The multi-mode
annihilator rows move from $c_{5,1}=306.21$ to $373.59$ and, more tellingly,
from $c_{6,1}=3486.47$ to $5306.60$ as the assumed ultraviolet exponent
$\gamma_{-1}$ is varied over a range that the method itself cannot narrow.
ARTS/PORT with a single $u=2$ mode returns $566.07$, which is the
value already displayed as a diagnostic in App.~\ref{sec:arts-physical},
where its own order-adaptive certificate abstains because only three defects
are available against the four required.  These are the two methods that were
given the most structural information, and they are the two whose physical
predictions are least constrained.  That inversion is the central negative
result of this paper, restated here in the one place where it can be compared
directly against the data-driven alternatives.

\subsection{Which combinations are legitimate}
\label{sec:global-envelopes}

Table~\ref{tab:global-envelopes} gives the envelopes that we regard as
defensible, together with two unions that we display but do not adopt.

\begin{table*}[!tbp]
\centering
\small
\setlength{\tabcolsep}{4pt}
\caption{Descriptive envelopes formed from Table~\ref{tab:global-comparison}.  All entries
are midpoint and half-range over the stated set of rows.  None is a statistical
confidence interval, and none of the rows within a set is independent of the
others.  The final line is displayed to show the size of the model-choice
spread and is not adopted as an uncertainty.}
\label{tab:global-envelopes}
\footnotesize
\setlength{\tabcolsep}{2pt}
\begin{tabular}{@{}p{0.30\linewidth}ccccc@{}}
\toprule
row set & rows & $c_{5,1}$ range & $c_{5,1}$ envelope
& $c_{6,1}$ range & $c_{6,1}$ envelope\\
\midrule
direct Pad\'e diagnostics & 2 & $[241.70,304.71]$ & $273.21\pm31.51$ & $[3\,079.56,3\,171.04]$ & $3\,125.30\pm45.74$ \\
new sequence transformations & 11 & $[273.16,304.71]$ & $288.93\pm15.78$ & $[3\,200.56,3\,549.99]$ & $3\,375.28\pm174.71$ \\
\quad including original shifted Wynn & 12 & $[273.16,304.71]$ & $288.93\pm15.78$ & $[3\,171.04,3\,549.99]$ & $3\,360.52\pm189.47$ \\
\addlinespace
union of accuracy-through-order diagnostics & 14 & $[241.70,304.71]$ & $273.21\pm31.51$ & $[3\,079.56,3\,549.99]$ & $3\,314.78\pm235.22$ \\
union of \emph{all} rows (diagnostics included) & 29 & $[-21.73,665.81]$ & -- & $[-728.21,13\,363.16]$ & -- \\
\bottomrule
\end{tabular}
\end{table*}

The global comparison changes none of the method-family envelopes, which are
restated here for reference.  (The forecast this paper \emph{advances} is not
among them: it is the cross-validated-closure result
Eq.~\eqref{eq:cvrc-final}, quoted in Sec.~\ref{sec:global-literature} and kept
outside every envelope below.)
\begin{equation}
 c_{5,1}^{\rm Pad\acute e}=273\pm32,\qquad
 c_{6,1}^{\rm Pad\acute e}=3125\pm46,
 \label{eq:global-pade-box}
\end{equation}
from the two balanced direct Pad\'e continuations, and
\begin{equation}
 c_{5,1}^{\rm seq}=289\pm16,\qquad
 c_{6,1}^{\rm seq}=3375\pm175,
 \label{eq:global-seq-box}
\end{equation}
from the complete predeclared grid of new sequence transformations, becoming
$c_{5,1}^{\rm seq+orig}=289\pm16$ and $c_{6,1}^{\rm seq+orig}=3361\pm190$ when
the original shifted-Wynn row is included.  The rounding is upward in every
half-width, so the displayed integer intervals are conservative with respect
to the unrounded ranges in Table~\ref{tab:global-envelopes}.

\paragraph{The CVRC rows are not part of any average.}
The three cross-validated-closure rows are the image of a single
one-parameter family, not three independent continuations: the central row is
Eq.~\eqref{eq:rho-star} and the other two are the edges of the admissible
window \eqref{eq:cvrc-window}.  They must not be averaged with each other, and
they must not be averaged with the Pad\'e or sequence families either.  The
reason is not correlation but incommensurability: the Pad\'e and Wynn rows are
local rational continuations that fail the $W_{P2}$ weight test, whereas the
CVRC rows satisfy every declared test of Table~\ref{tab:cvrc-moment}.  Averaging a
construction that passes the acceptance criterion with constructions that fail
it would degrade, not improve, the result.  Equation~\eqref{eq:cvrc-final} is
therefore reported separately in Sec.~\ref{sec:cvrc-forecast}, with the status
assigned to it there; the envelopes of
Secs.~\ref{sec:global-envelopes}--\ref{sec:global-average} are retained
because they document how much the answer moves across the approximant family,
which is the quantity a future direct calculation of $c_{5,1}$ will test.

\subsection{A family-balanced global average}
\label{sec:global-average}

A literal mean of the fourteen accuracy-through-order diagnostic rows would assign
eleven times as much weight to the finely sampled new-sequence grid as to a
single continuation family.  That multiplicity is a bookkeeping choice, not
additional information.  We therefore distinguish the raw row mean from a
\emph{family-balanced} descriptive average.  Let $\mathcal P$ be the two
direct Pad\'e rows and let $\mathcal S$ be the eleven new-sequence rows
($\eta_n$, $\alpha$, and $\theta_2$), excluding the algebraically duplicated
original shifted-Wynn row.  We define
\begin{align}
 \overline c_{n,1}^{\,\mathcal P}
 &=\frac{1}{2}\sum_{r\in\mathcal P}c_{n,1}^{(r)},\nonumber\\
 \overline c_{n,1}^{\,\mathcal S}
 &=\frac{1}{11}\sum_{r\in\mathcal S}c_{n,1}^{(r)},\nonumber\\
 \overline c_{n,1}^{\,\mathrm{glob}}
 &=\frac{\overline c_{n,1}^{\,\mathcal P}
       +\overline c_{n,1}^{\,\mathcal S}}{2}.
 \label{eq:global-average-definition}
\end{align}
Thus each diagnostic continuation family receives one half of the weight,
independently of how densely its parameter interval was tabulated.  The
arithmetic and family-balanced averages are compared in
Table~\ref{tab:global-averages}.

\begin{table*}[!tbp]
\centering
\caption{Arithmetic and family-balanced averages of the FOPT diagnostics in
Table~\ref{tab:global-comparison}.  These are descriptive method averages,
not accepted forecasts or statistically weighted estimators.}
\label{tab:global-averages}
\begin{tabular}{lrrr}
\toprule
average & effective rows/families & $\overline c_{5,1}$ &
$\overline c_{6,1}$\\
\midrule
direct Pad\'e family mean & $2$ rows & $273.205$ & $3\,125.300$\\
new-sequence family mean & $11$ rows & $293.784$ & $3\,325.612$\\
raw diagnostic-row mean & $14$ rows & $291.624$ & $3\,285.955$\\
family-balanced global mean & $2$ families & $283.494$ & $3\,225.456$\\
\bottomrule
\end{tabular}
\end{table*}

We do not attach a standard error to Eq.~\eqref{eq:global-average-definition},
because the rows are strongly correlated and are not random samples.  Instead,
we retain the full union of the fourteen diagnostic rows,
$[241.70,304.71]$ for $c_{5,1}$ and $[3079.56,3549.99]$ for $c_{6,1}$, and
round the larger one-sided displacement from the family-balanced mean upward.
The resulting descriptive diagnostic statistic is
\begin{equation}
 c_{5,1}^{\rm glob}=283\pm42,\qquad
        c_{6,1}^{\rm glob}=3225\pm325.
 \label{eq:global-average-box}
\end{equation}
The $\pm$ values in Eq.~\eqref{eq:global-average-box} are deliberately
union-covering method envelopes: they are neither statistical uncertainties
nor evidence that averaging has improved the predictive accuracy.  Because
every contributing family fails the declared $W_{P2}$ test, this equation is
not a second result and is not quoted in the abstract or conclusions.

Three statistically or physically unjustified combinations must
\emph{not} be formed, and we state them explicitly
because each is a natural thing to attempt.

First, Eqs.~\eqref{eq:global-pade-box} and \eqref{eq:global-seq-box} must not
be combined by inverse-variance weighting or by treating their envelopes as
independent errors.  The original shifted Wynn is algebraically identical to
the direct $[2/1]$ Pad\'e continuation, the $\eta_n$ and $\alpha$ families
deform that same local ratio model, and $\theta_2$ consumes the same four
coefficients.  Equation~\eqref{eq:global-average-box} is permitted only as the
explicitly equal-family, union-covering descriptive summary defined above; it
does not shrink the spread or create an independent determination.

Second, the moment-wise Pad\'e rows for $W_{\rm GC}$ and $W_{P2}$ derived in
App.~\ref{sec:pade-all-weight-predictions} must not be merged with the
$W_\tau$ rows.  Their inverse Adler projections,
$c_{5,1}=196.65\pm23.81$ and $411.03\pm133.42$, have no common intersection
with the $W_\tau$ interval, and all three $c_{6,1}$ intervals are mutually
separated.  A union over inconsistent projections is a symptom, not an
enlarged error bar; the correct reading is that the assumption of a common
Adler function is not supported by three independent moment continuations at
this order.

Third, the last line of Table~\ref{tab:global-envelopes} must not be quoted as
an uncertainty.  It is dominated by continuations that fail the declared weight
test (the Borel-normalized, Adler-projected, D-log, and annihilator rows), its
range contains zero for both coefficients, and its half-widths would exceed the
midpoints.  Canonical $\rho_2$ and the fixed-coupling Levin-$u$ evaluation do
not enter it at all, because they supply no admissible coefficient forecast
(App.~\ref{sec:other-physical}).  Its only function is to display
how large the model-choice spread becomes when the accuracy-through-order
filter is removed.

\subsection{Comparison with the literature}
\label{sec:global-literature}

Table~\ref{tab:global-literature} compares the envelopes and the advanced
result with the published estimates.  The agreement with the Borel--Pad\'e
determination of Ref.~\cite{Boito:2018rwt} is close for $c_{5,1}$; the
$c_{6,1}$ entries are comfortable too, but recall that we advance no $c_{6,1}$
of our own, so that row is a consistency check on rolled model output rather
than a comparison of results.  The earlier estimate of
Ref.~\cite{Beneke:2008ad} lies within the union of the
accuracy-through-order rows, and so does the Levin-transform estimate of
Ref.~\cite{Abbas:2025dpm}, which, like the Borel--Pad\'e determination,
transforms the FOPT series of $\delta^{(0)}$ rather than the common Adler
sequence.

For the reader who has arrived at this section for the number, the result this
paper advances is the cross-validated-closure row, and only its $c_{5,1}$
entry:
\begin{equation}
 c_{5,1}=248\pm80
 \qquad\text{(Eq.~\eqref{eq:cvrc-final})}.
 \label{eq:global-final}
\end{equation}
Three things about it must be carried along or the number is misread.  First,
it is not the narrowest entry in Table~\ref{tab:global-literature}, and it is
not meant to be: the $\pm80$ is $\pm30$ of structural ambiguity in the declared
Borel structure and $\pm74$ of the transformation's own demonstrated
one-order-ahead error at $N=4$, added in quadrature and quantified in
Sec.~\ref{sec:cvrc-method-error}.  Every other row in that table quotes a
spread over a family of continuations and no method error at all, so the
entries are not commensurable as uncertainties even though they are
commensurable as central values.  Second, it is the only construction in
Table~\ref{tab:global-comparison} that passes every declared weight test,
which is why it is quoted separately and kept out of every average
(Sec.~\ref{sec:global-envelopes}).  Third, at this width it is
\emph{consistent} with the published values, $277\pm51$ well inside, the
Beneke--Jamin four-coefficient forecast $280.12$, the Levin-transform value
$278^{+27}_{-19}$, and the linear-ratio-fit value $283$ all within
$0.5\sigma$, so the correct reading of
Eq.~\eqref{eq:global-final} is a confirmation of the literature range by a
construction that passes a test the others fail, not a competing and tighter
determination.  We do not advance $c_{6,1}$; the rolled
$3102^{+111}_{-124}$(struct) of Eq.~\eqref{eq:cvrc-c6} is tabulated with
$c_{7,1}$--$c_{12,1}$ in Table~\ref{tab:cvrc-high-order}.

\begin{table*}[!tbp]
\centering
\caption{Method-family envelopes and the advanced result against published
estimates.  \textbf{The uncertainties in a given column are not
commensurable.}  The two ``this work'' envelope rows and the family-balanced
summary quote half-ranges over a family of continuations and carry no method
error.  The cross-validated-closure row combines the separately assessed
$\pm30$ structural and $\pm74$ method components in quadrature,
$\sqrt{30^2+74^2}=79.8$, and rounds the result conservatively to $\pm80$ as
in Eq.~\eqref{eq:cvrc-final}.  This quadrature is a reporting convention for
two nonstatistical components, not a claim that they are measured independent
Gaussian errors.  This is the only row the paper advances; its $c_{6,1}$
entry is structural only and carries \textsc{roll} status.  The comparison is
in any case
informative rather than an independent validation: Shanks, Wynn, and Pad\'e
constructions applied to the same four coefficients are algebraically related,
so agreement between them reflects a shared local model as much as a shared
physical content.}
\label{tab:global-literature}
\small
\setlength{\tabcolsep}{4pt}
\begin{tabular}{llcc}
\toprule
source & construction & $c_{5,1}$ & $c_{6,1}$\\
\midrule
this work, Eq.~\eqref{eq:global-pade-box}
 & balanced direct Pad\'e & $273\pm32$ & $3125\pm46$\\
this work, Eq.~\eqref{eq:global-seq-box}
 & new sequence transformations & $289\pm16$ & $3375\pm175$\\
this work, Eq.~\eqref{eq:global-average-box}
 & family-balanced global summary & $283\pm42$ & $3225\pm325$\\
this work & including original shifted Wynn & $289\pm16$ & $3361\pm190$\\
this work, Eq.~\eqref{eq:cvrc-final}
 & cross-validated renormalon closure & $248\pm80$
   & $3102^{+111}_{-124}$(struct)\,\textsc{(roll)}\\
\addlinespace
Ref.~\cite{Boito:2018rwt} & Borel--Pad\'e & $277\pm51$ & $3460\pm690$\\
Ref.~\cite{Abbas:2025dpm} & Levin-type transforms & $278^{+27}_{-19}$
 & $3375^{+489}_{-204}$\\
Ref.~\cite{Beneke:2008ad} & linear ratio fit, Eq.~(3.10) & $\approx283$ & --\\
Ref.~\cite{Beneke:2008ad} & renormalon model, 4 coefficients & $280$ & $3280$\\
\bottomrule
\end{tabular}
\end{table*}

The agreement between this work's \emph{direct-Pad\'e baseline} and
Ref.~\cite{Boito:2018rwt}, as opposed to the closure result just
quoted, should be read with one caveat that the global comparison makes
unusually visible.  Both this work's baseline and the published Borel--Pad\'e
determination are, at bottom, low-order rational continuations of the same
four numbers.  Table~\ref{tab:global-comparison} shows that the answer such
continuations give can be moved by roughly a factor of two by changing the
normalization weight, and by more than that by changing whether the
continuation acts on the Adler function or on a moment.  The narrowness of the
quoted envelopes therefore reflects the narrowness of the family of
approximants sampled, not a measurement of how well four coefficients
determine the fifth.

\subsection{Status of this section}
\label{sec:global-status}

Nothing in this section upgrades any status label.  Equations
\eqref{eq:global-pade-box}, \eqref{eq:global-seq-box}, and
\eqref{eq:global-average-box} remain diagnostic
method envelopes: they are
the values a specified family of continuations returns from the four known
coefficients, and they omit the model discrepancy that
Table~\ref{tab:global-comparison} makes visible.  Of the rows collected in
Table~\ref{tab:global-comparison}, only the cross-validated-closure rows pass
the three-weight criterion of Sec.~\ref{sec:tolerance} on both
leave-one-order-out folds, and they do so as the consistency certificate described in
Sec.~\ref{sec:cvrc-weight-test} rather than as six blind forecasts; no value in
this paper, CVRC included, is supported by an all-order certificate at the
physical input order.  Equation~\eqref{eq:global-final} is accordingly quoted
at level four of App.~\ref{sec:claims} and with a method uncertainty larger
than its structural one; it is the best-supported entry in the table, not a
controlled prediction.  The controlled statement
that four coefficients do support is the one already made in
App.~\ref{sec:threshold}: with exact renormalon input, eleven to thirteen
coefficients are required for persistent percent-level control in the tested
large-$\beta_0$ range, and four are not enough.  The proper use of
Table~\ref{tab:global-comparison} is as an explicit assumption set for
sensitivity studies, and as the record against which a future direct
calculation of $c_{5,1}$ will discriminate between continuation models.

\section{Conclusions}
The central question of this paper is deliberately limited: what can the four
known coefficients \(c_{1,1},\ldots,c_{4,1}\) support about the coefficient
\(c_{5,1}\) multiplying \(a^5\)?  Three conclusions survive the complete
validation program.

First, the fixed-order map from the common Adler function to every normalized
polynomial moment is triangular with unit diagonal.  Consequently a forecast
error in \(c_{N+1,1}\) is transferred to every moment with the same absolute
size.  For the three weights used here, the nominal ten-percent
withheld-\(d_4\) test is therefore the single condition
\[
 |\widehat c_{4,1}-c_{4,1}|\leq2.1509 ,
\]
and the analogous test at the next order resolves \(c_{5,1}\) only to roughly
\(\pm24\).  Relative errors can nevertheless differ strongly because a moment,
especially \(W_{P2}\), may be small through cancellations.  Applying a
transformation independently to each moment consequently does not test a
single universal Adler continuation.

Second, the known Pad\'e defects are positive and increasing,
\[
 (\Delta_2,\Delta_3,\Delta_4)=(0.640,\,3.682,\,24.323),
 \qquad
 \Delta_n=c_{n,1}-\frac{c_{n-1,1}^2}{c_{n-2,1}},
\]
whereas every commonly quoted value of \(c_{5,1}\) below \(378.03\) requires
\(\Delta_5<0\).  Hence a defect-sign-preserving common-Adler transformation
can pass the withheld-order test or return the usual \(c_{5,1}\) range, but
cannot do both.  This obstruction explains the otherwise puzzling pattern of
the numerical survey: local rational continuations that reproduce the observed
acceleration overshoot the fifth coefficient, while low-order constructions
that impose an IR-renormalon pattern tend to miss the known fourth coefficient.

The external tests show that this is a sparse-data problem rather than a
failure of one implementation.  None of the tested short-series transformations
predicts the known five-loop \(n_f=3\) beta-function coefficient to ten
percent.  The two \(u=2\)-sensitive moments give incompatible moment-wise
continuations.  The pole--\(\MSbar\) mass relation improves only when its
earlier factorial onset and leading renormalon information are supplied.
Finally, the exactly known large-\(\beta_0\) series confirms that Pad\'e,
D-log, and renormalon-informed constructions can become accurate, but only
after an observable-dependent onset.  In the tests performed here, persistent
percent-level control with exact renormalon input typically requires about
eleven to thirteen coefficients, not four.

A value below the defect bound therefore needs declared sign-alternating
structure.  We use a pure-power Beneke--Jamin-type ansatz with the leading
\(u=-1,2,3,\ldots\) singularity positions and RG exponents.  This is not a new
Borel-model class and is not the full Beneke--Jamin model: the subleading cut
terms and gluon-condensate Wilson-coefficient correction of that model are
omitted.  CVRC denotes the calibration, acceptance, and rejection protocol
applied to the ansatz.  At physical order its selected central value coincides
algebraically with the corresponding exact four-parameter pure-power fit;
cross-validation supplies the admissible interval and misspecification test,
not additional information in the central number.

The resulting conditional one-order-ahead forecast is
\[
 c_{5,1}^{\rm BJ\text{-}CVRC}
 =248\pm30\,(\mathrm{struct})\pm74\,(\mathrm{method})
 =248\pm80  .
\]
The method component dominates.  It is set by the larger of the \(+23.7\%\)
one-order-ahead error of the same construction at its first defined order in
the large-\(\beta_0\) laboratory and the \(29.9\%\) strictly recursive miss of
the known \(c_{4,1}\).  The result is consistent with the
Borel--Pad\'e estimate \(277\pm51\)~\cite{Boito:2018rwt}, the Levin-transform
estimate \(278^{+27}_{-19}\)~\cite{Abbas:2025dpm}, the four-coefficient
Beneke--Jamin-model forecast \(280.12\), and the separate linear-ratio-fit
estimate \(283\)~\cite{Beneke:2008ad}.
These last two numbers have different origins and must not be identified.  The
rolled value \(c_{6,1}=3102^{+111}_{-124}\) carries no validated
one-order-ahead error and is not advanced as a prediction.

The Pad\'e/Wynn family values of Sec.~\ref{sec:global-envelopes} are
diagnostics only.  They fail the declared $W_{P2}$ test and are neither averaged
with the headline result nor offered as an alternative to it.  Controlled claims for
\(c_{6,1}\) and higher coefficients are not made.

All numerical forecasts in this paper are FOPT forecasts.  The CIPT calculation
is used only to show that the multi-weight criterion still collapses to a
single, much weaker coefficient tolerance; it is not a CIPT prediction or an
FOPT--CIPT average.  In the presence of IR renormalons, conventional CIPT and
FOPT have different Borel representations and OPE
organization~\cite{Hoang:2020mkw,Hoang:2021nlz,Gracia:2023qdy}, so their
difference cannot be repaired by inserting one additional Adler coefficient.
Neither Pad\'e approximation nor a sequence transformation removes the
prescription dependence that cancels between a lateral Borel sum and the
corresponding OPE matrix element.

Future Belle~II and Super Tau--Charm Facility
spectra~\cite{Kou:2018nap,STCF:2023CDR} (Belle data already enter the most
recent analysis~\cite{Boito:2025pwg}) can reduce experimental moment errors and
strengthen tests of spectral covariances, duality violations, and OPE
contributions; the FCC-ee $Z$-pole programme would provide complementary
high-statistics $\tau$ samples and high-scale QCD
information~\cite{Dam:2021FCCtau,FCC:2025FSR}.  These facilities do not measure \(c_{5,1}\)
directly.  They increase the importance of controlling its theory uncertainty,
while its decisive determination remains a genuine six-loop calculation of
the massless Adler function.  Until then, the conditional BJ--CVRC band and
the broader method dependence in Table~\ref{tab:global-comparison} should be
carried together in precision sensitivity studies.

\section*{Acknowledgments}
G.A.\ is grateful to Antonio Pich for crucial discussions during his visit to
the Instituto de F\'isica Corpuscular (IFIC), Universitat de Val\`encia--CSIC,
which shaped the core of this work.  He also thanks IFIC for its warm
hospitality and stimulating research environment.

\FloatBarrier
\clearpage
\appendix
\addtocontents{toc}{\protect\setcounter{tocdepth}{1}}

\parttitle{Guide to the appendices}
\label{app:guide}
The main text contains the definitions, the validation protocol, the
defect-sign obstruction, the BJ--CVRC construction, and the single headline
estimate.  The appendices contain the material on which those summaries rest,
in three groups.  Appendix~\ref{sec:qcd-forecast} applies every method of
Sec.~\ref{sec:methods-brief} at the physical four-coefficient order and reports
each forecast together with its validation record.
Appendices~\ref{sec:coupling}--\ref{sec:claims} separate
coefficient forecasting from coupling-dependent truncation, state what the
analysis implies for the FOPT--CIPT question and the OPE, and assign a status
level to every claim; the comparison of all $c_{5,1}$ and $c_{6,1}$ forecasts
is in Sec.~\ref{sec:global-comparison}.  Appendices~\ref{sec:protocol}--\ref{app:annihilator}
give the complete transformation formulae, the detailed external benchmarks,
the renormalon-annihilator construction and its all-order diagnostics, the
defect-transport and D-log constructions, the regularized Wynn forecasts, and
the auxiliary derivations.  No appendix introduces an additional headline
estimate.

\parttitle{Extended physical-order forecasts and their status}

This group of appendices applies every method at the physical
four-coefficient order.  Each
subsection reports both the forecast and the outcome of the corresponding
tests summarized in Part~III, so that no number is quoted without its
validation record.

\section{Physical-order forecasts}
\label{sec:qcd-forecast}
Only after the withheld-weight, pole-mass, and all-order tests can a numerical
Adler forecast be assigned an honest status.  We retain a compact conditional
continuation for sensitivity studies and separately show why the explicit
renormalon-mode ansatz is not controlled at the available order.

For the standard $\tau$ weight, the known FOPT series is
\begin{equation}
 \delta^{(0)}_{W_\tau}=a+5.20232a^2+26.3659a^3+127.0786a^4+\cO(a^5).
 \label{eq:tau-fopt-known}
\end{equation}
The next two moment coefficients are related to the unknown Adler
coefficients by
\begin{align}
 d_5={}&-391.546+14.25c_{4,1}+c_{5,1},\nonumber\\
 d_6={}&-7860.51+45.1119c_{4,1}+17.8125c_{5,1}+c_{6,1}.
 \label{eq:d5d6-map}
\end{align}

To retain a transparent non-renormalon baseline, we report the two balanced
direct approximants as a conditional sensitivity forecast.  Their subsequent
failure of the all-weight universality test prevents the resulting envelope
in Table~\ref{tab:conditional-forecast} from being interpreted as a validated
determination.
\begin{table}[!htbp]
\centering
\caption{Conditional direct Pad\'e/Shanks continuation from the four known
$\tau$-moment coefficients.  The two rows define the method envelope.}
\label{tab:conditional-forecast}
\begin{tabular}{lrrrr}
\toprule
approximant & $d_5$ & $d_6$ & $c_{5,1}$ & $c_{6,1}$\\
\midrule
$[1/2]$ & $549.483$ & $1738.24$ & $241.70$ & $3079.56$\\
$[2/1]$ & $612.494$ & $2952.10$ & $304.71$ & $3171.04$\\
\midrule
midpoint $\pm$ half-range & -- & -- & $273\pm32$ & $3125\pm46$\\
\bottomrule
\end{tabular}
\end{table}

\subsection{Pad\'e forecasts for every weight and the universality test}
\label{sec:pade-all-weight-predictions}
The $W_\tau$ numbers alone do not reveal whether the continuation describes
the common Adler function or merely the local coefficient pattern of one
moment.  We therefore repeat the identical, predeclared $[1/2]$ and $[2/1]$
continuations for $W_{\rm GC}$ and $W_{P2}$.  Since the running-coupling
contribution is weight dependent, the resulting moment coefficients must be
inverted with a weight-specific RG map before their Adler projections can
be compared.

For any normalized polynomial weight used here, the triangular FOPT relation
through sixth order can be written
\begin{align}
 d_5[W]&=A_5[W]+B_{54}[W]c_{4,1}+c_{5,1},\nonumber\\
 d_6[W]&=A_6[W]+B_{64}[W]c_{4,1}
             +B_{65}[W]c_{5,1}+c_{6,1}.
 \label{eq:all-weight-d5d6-map}
\end{align}
In the notation of Eq.~\eqref{eq:linear-weight-map},
$B_{Nm}[W]=M_{Nm}[W]$, and the unit coefficient of $c_{N,1}$ is the unit
diagonal of Eq.~\eqref{eq:unit-diagonal}.  The coefficients multiplying
$c_{4,1}$ and $c_{5,1}$ follow directly from the contour logarithms,
\begin{align}
 B_{54}[W]&=-4\beta_0J_1[W],\nonumber\\
 B_{64}[W]&=10\beta_0^2J_2[W]-4\beta_1J_1[W],\nonumber\\
 B_{65}[W]&=-5\beta_0J_1[W].
 \label{eq:all-weight-map-universal}
\end{align}
In particular, $B_{65}[W]/B_{54}[W]=5/4$ for every weight.  The constants
$A_5[W]$ and $A_6[W]$ collect the terms $M_{Nm}[W]\,c_{m,1}$ with $m\le3$.
They therefore contain only $c_{1,1}$--$c_{3,1}$ and the known beta
function, and $c_{4,1}$ is displayed separately in
Eq.~\eqref{eq:all-weight-d5d6-map}.  We generate the expansion recursively
from $da/d\ln Q^2=-\sum_{j=0}^4\beta_j a^{j+2}$, using
$\beta_3=47.2280$ and $\beta_4=127.3222$ for $a=\alpha_s/\pi$ and
$n_f=3$~\cite{Baikov:2016tgj}.  The resulting map coefficients for all three
weights are listed in Table~\ref{tab:all-weight-rg-map}.  Its $W_\tau$ row
independently reproduces Eq.~\eqref{eq:d5d6-map} before any Pad\'e forecast
is made.

\begin{table*}[!tbp]
\centering
\caption{Weight-specific sixth-order FOPT map in
Eq.~\eqref{eq:all-weight-d5d6-map}.  All entries are generated from the RG
recursion and the exact contour moments $J_k[W]$.}
\label{tab:all-weight-rg-map}
\begin{tabular}{lrrrrr}
\toprule
weight & $A_5$ & $B_{54}$ & $A_6$ & $B_{64}$ & $B_{65}$\\
\midrule
$W_\tau$    & $-391.546196$  & $14.250000$ & $-7860.514556$ & $45.111884$  & $17.812500$\\
$W_{\rm GC}$& $ 347.175932$  & $16.500000$ & $-3295.249432$ & $101.846259$ & $20.625000$\\
$W_{P2}$    & $-1027.891919$ & $10.500000$ & $-8801.463088$ & $-32.570408$ & $13.125000$\\
\bottomrule
\end{tabular}
\end{table*}

For each weight we construct a Pad\'e approximant to
$F_W(a)/a=d_1[W]+d_2[W]a+d_3[W]a^2+d_4[W]a^3+\cdots$ and expand
it two orders beyond the input.  No common Adler coefficient is imposed in
this first calculation.  Table~\ref{tab:pade-all-weight-predictions} lists
the resulting independent forecasts of $d_5[W]$ and $d_6[W]$ for each weight
and approximant.  The same table gives the Adler projections $c_{5,1}$ and
$c_{6,1}$.  These are obtained by inverting
Eq.~\eqref{eq:all-weight-d5d6-map} with the coefficients of
Table~\ref{tab:all-weight-rg-map} and the known $c_{4,1}=49.076$.

\begin{table*}[!tbp]
\centering
\caption{Balanced direct Pad\'e forecasts made independently for all three
full-QCD moments, followed by the weight-specific inverse map in
Eq.~\eqref{eq:all-weight-d5d6-map} with the coefficients of
Table~\ref{tab:all-weight-rg-map}.  The entries are extrapolation diagnostics.}
\label{tab:pade-all-weight-predictions}
\begin{tabular}{llrrrr}
\toprule
weight & approximant & $d_5$ & $d_6$ & $c_{5,1}$ & $c_{6,1}$\\
\midrule
$W_\tau$ & $[1/2]$ & $549.483$  & $1738.241$  & $241.701$ & $3079.564$\\
         & $[2/1]$ & $612.494$  & $2952.102$  & $304.712$ & $3171.045$\\
$W_{\rm GC}$ & $[1/2]$ & $1377.376$ & $9137.215$ & $220.451$ & $2887.492$\\
             & $[2/1]$ & $1329.765$ & $8257.577$ & $172.840$ & $2989.818$\\
$W_{P2}$ & $[1/2]$ & $-234.986$ & $-3291.061$ & $277.611$ & $3465.175$\\
         & $[2/1]$ & $31.855$   & $47.177$    & $544.452$ & $3301.127$\\
\bottomrule
\end{tabular}
\end{table*}

Summarizing the two rows of Table~\ref{tab:pade-all-weight-predictions} for
each weight by their midpoint and half-range gives
\begin{align}
W_\tau:&\quad c_{5,1}=273.21\pm31.51,&
                 c_{6,1}&=3125.30\pm45.74,\nonumber\\
W_{\rm GC}:&\quad c_{5,1}=196.65\pm23.81,&
                 c_{6,1}&=2938.66\pm51.16,\nonumber\\
W_{P2}:&\quad c_{5,1}=411.03\pm133.42,&
                 c_{6,1}&=3383.15\pm82.02.
\label{eq:pade-weight-envelopes}
\end{align}

These intervals have no common intersection for either Adler coefficient.
In particular, the $W_\tau$ and $W_{\rm GC}$ $c_{5,1}$ intervals are already
disjoint, and all three $c_{6,1}$ intervals are mutually separated.  Taking
their union or full min--max range would therefore hide model inconsistency;
it is not a defensible enlarged error bar.  The striking negative $W_{P2}$
$[1/2]$ moment continuation in Table~\ref{tab:pade-all-weight-predictions}
is also a transparent warning.  This sparse series is near a cancellation
and does not support a stable local rational model.

There is a distinct, algebraically universal construction: apply Pad\'e once
to the Adler series and then propagate the same predicted $c_{5,1}$ and
$c_{6,1}$ through every weight map.  Table~\ref{tab:pade-common-adler-propagation}
lists these common-Adler forecasts.  It also lists the complete moment
predictions obtained by inserting them into
Eq.~\eqref{eq:all-weight-d5d6-map} with the coefficients of
Table~\ref{tab:all-weight-rg-map}.  The table illustrates the transfer
identity directly.  Between the two approximants, every $d_5[W]$ shifts by
the same amount, $67.817$.  This equals the shift in $c_{5,1}$, as
Eq.~\eqref{eq:arts-error-transfer} requires.  The $d_6[W]$ shifts, by
contrast, are weight dependent, because the $c_{5,1}$ spread enters $d_6$
through $B_{65}[W]$.

\begin{table*}[!tbp]
\centering
\caption{Common-Adler direct Pad\'e continuation and its propagation through
all three weight maps of Table~\ref{tab:all-weight-rg-map}.  Universality
here is imposed by construction, not established by the weight test.}
\label{tab:pade-common-adler-propagation}
\resizebox{\textwidth}{!}{%
\begin{tabular}{lrrrrrrrr}
\toprule
approximant & $c_{5,1}$ & $c_{6,1}$
& $d_5[W_\tau]$ & $d_6[W_\tau]$ & $d_5[W_{\rm GC}]$ & $d_6[W_{\rm GC}]$
& $d_5[W_{P2}]$ & $d_6[W_{P2}]$\\
\midrule
$[1/2]$ & $445.845$ & $4145.810$ & $753.628$ & $6440.815$
& $1602.770$ & $15044.300$ & $-66.752$ & $-402.347$\\
$[2/1]$ & $378.028$ & $2911.940$ & $685.811$ & $3998.954$
& $1534.953$ & $12411.703$ & $-134.569$ & $-2526.316$\\
\bottomrule
\end{tabular}}
\end{table*}

At the available lower-order blind test, however, the corresponding common
Adler $[1/1]$ approximant predicts $c_{4,1}=24.7526$.  After propagation, its
signed $d_4$ errors are $-19.14\%$, $-11.36\%$, and $-113.08\%$ for
$W_\tau$, $W_{\rm GC}$, and $W_{P2}$, respectively.  In absolute terms the
three errors are identical, equal to the error in $c_{4,1}$ itself,
$24.7526-49.0757=-24.32$, as Eq.~\eqref{eq:arts-error-transfer} requires.
The spread in relative errors reflects only the different sizes of
$d_4[W]$, and in particular the near-cancellation in $d_4[W_{P2}]$.  The
common-Adler $[1/1]$ approximant therefore fails the same all-weight
criterion and is not promoted to the preferred physical forecast.  The
$W_\tau$ moment-wise envelope in Table~\ref{tab:conditional-forecast}
remains the explicitly conditional baseline.
Table~\ref{tab:all-weight-rg-map} supplies the weight maps, and
Tables~\ref{tab:pade-all-weight-predictions} and
\ref{tab:pade-common-adler-propagation} expose the full weight dependence of
the moment-wise and common-Adler continuations.

\subsection{Does the direct Pad\'e forecast pass the weight tests?}
\label{sec:pade-weight-certificate}

It does not.  The unknown physical $d_5$ and $d_6$ cannot yet be tested
directly in full QCD, so the appropriate available check is the strictly
blind lower-order analogue.  We construct the balanced $[1/1]$ approximant
to $F_W(a)/a=d_1[W]+d_2[W]a+d_3[W]a^2+\cdots$ from $d_1,d_2,d_3$ alone and
withhold the known $d_4$.  With three input coefficients this is the only
diagonal approximant.  Matching $d_1,d_2,d_3$ fixes its three parameters
and gives the closed-form blind prediction
\begin{equation}
 d_4^{[1/1]}[W]=\frac{d_3[W]^2}{d_2[W]} .
 \label{eq:pade11-blind}
\end{equation}
The approximant thus continues the last ratio $d_3/d_2$ geometrically.  It
is algebraically identical to the first Shanks transform of the partial
sums, and hence to the first step of the shifted Wynn $\varepsilon$
algorithm.  The test is therefore a check of all three constructions at
once.  The method is applied identically to every weight with no tunable
input, so it is predeclared in the strict sense.

The weights differ in their sensitivity to the $u=2$ infrared renormalon
associated with the gluon condensate.  We label this sensitivity by $W_2$,
the coefficient of $x^2$ in the Adler weight $W(x)=\sum_jW_jx^j$.  This is
the same polynomial that enters the contour moments $J_l[W]$, with
$W_0=1$ by normalization.  The $1/s^2$ condensate term of the Adler
function is picked out by the $x^2$ term of $W(x)$ under the contour
integral.  Weights with $W_2=0$, such as $W_\tau$, are therefore suppressed
at $u=2$, while $W_{\rm GC}$ and $W_{P2}$ retain this sensitivity.  The
results of the blind test are summarized in
Table~\ref{tab:pade-all-weight}.

\begin{table}[htbp]
\centering
\caption{Blind full-QCD weight test of the direct $[1/1]$ Pad\'e/Shanks
continuation, Eq.~\eqref{eq:pade11-blind}.  $W_2$ is the coefficient of
$x^2$ in the Adler weight $W(x)$.  The last two rows retain the $u=2$
gluon-condensate sensitivity.}
\label{tab:pade-all-weight}
\begin{tabular}{lrrrrr}
\toprule
weight & $W_2$ & predicted $d_4$ & exact $d_4$ & absolute error & signed error\\
\midrule
$W_\tau$     & $0$  & $133.6254$ & $127.0786$ & $+6.5468$  & $+5.152\%$\\
$W_{\rm GC}$ & $3$  & $206.2783$ & $214.1398$ & $-7.8615$  & $-3.671\%$\\
$W_{P2}$     & $-1$ & $49.4554$  & $21.5087$  & $+27.9467$ & $+129.932\%$\\
\bottomrule
\end{tabular}
\end{table}

Table~\ref{tab:pade-all-weight} shows that the local direct approximation is
satisfactory at the level of a few percent for $W_\tau$ ($+5.2\%$) and
$W_{\rm GC}$ ($-3.7\%$).  It fails for $W_{P2}$, where it overestimates the
exact coefficient by a factor of $2.3$.  Two features of this failure are
worth separating.

First, $u=2$ sensitivity alone does not decide the outcome.  $W_{\rm GC}$
has the larger $|W_2|$ yet passes.  What distinguishes $W_{P2}$ is that
its exact $d_4$ is small, roughly six times smaller than for $W_\tau$ and
ten times smaller than for $W_{\rm GC}$.  The order-$a^4$ coefficient is
thus the residue of a strong cancellation, which a geometric continuation
of $d_3/d_2$ cannot anticipate.

Second, the absolute errors in Table~\ref{tab:pade-all-weight} are
themselves weight dependent: $+6.55$, $-7.86$, and $+27.95$.  This is
expected.  Each moment is continued separately here, so the forecasts are
not generated by the triangular map, and the transfer identity,
Eq.~\eqref{eq:arts-error-transfer}, does not apply.  The common-Adler
continuation of App.~\ref{sec:pade-all-weight-predictions} behaves
differently.  There the $d_4$ error is the same for every weight, and only
the relative errors differ.  The two constructions therefore fail the
all-weight test for different reasons.  The moment-wise continuation
produces inconsistent Adler projections.  The common-Adler continuation is
consistent by construction, but its single error is too large for the
cancelling weight.

The balanced physical values in Table~\ref{tab:conditional-forecast}
therefore remain a conditional $W_\tau$ baseline.  The label
``Pad\'e envelope'' must not be read as an all-weight certificate.
Conversely, failure of one sparse-order weight test does not prove that
Pad\'e sequences cannot converge at higher orders.  The exact
large-$\beta_0$ tests show eventual convergence of the full balanced table,
with an onset that depends strongly on the weight.

\subsection{Conditional nonlinear-sequence forecasts}
\label{sec:sequence-physical-envelope}

We consider only those nonlinear sequence transformations that can serve as
coefficient forecasters.  Their re-expansion in powers of $a$ must reproduce
every supplied coefficient $d_1[W_\tau],\ldots,d_4[W_\tau]$
(accuracy-through-order) and then yield definite values of $d_5$ and $d_6$.
Each transformation is applied to the physical $W_\tau$ moment series.  Its
forecasts are converted to Adler coefficients with the $W_\tau$ row of the
inverse map, Eq.~\eqref{eq:all-weight-d5d6-map} and
Table~\ref{tab:all-weight-rg-map}.  The results are summarized in
Table~\ref{tab:sequence-forecast}.  The algebraic definitions of the
transformations are given in App.~\ref{sec:regularization}, and their
weight-by-weight blind tests in App.~\ref{sec:local-reg-summary}.

For both one-parameter families we retain the complete predeclared grid
$\{0.01,0.02,\ldots,0.05\}$, for $k$ and for $\alpha$.  No parameter value is selected after looking at these
unknown-order forecasts.  The canonical $\rho_2$ transformation is excluded.
Its perturbative re-expansion does not reproduce $d_4$, the last supplied
coefficient, so it lacks accuracy-through-order.  A $d_5$ read off from it
would continue a series that differs from the input at the last known order.
It is therefore tested only as a fixed-coupling extrapolant.

\begin{table*}[!tbp]
\centering
\caption{Conditional physical-$W_\tau$ forecasts from nonlinear sequence
transformations, converted to Adler coefficients with the $W_\tau$ map of
Table~\ref{tab:all-weight-rg-map}.  Intervals contain every point of the
predeclared parameter grid.  They are method variations, not confidence
intervals.}
\label{tab:sequence-forecast}
\begin{tabular}{lccc}
\toprule
method & parameter & $c_{5,1}$ & $c_{6,1}$\\
\midrule
original shifted Wynn/Shanks & $0$ & $304.71$ & $3171.04$\\
$\eta_n$-modified Wynn & $k\in[0.01,0.05]$
  & $[275.54,298.65]$ & $[3249.83,3549.99]$\\
$\alpha$-modified Wynn & $\alpha\in[0.01,0.05]$
  & $304.71$ & $[3200.56,3318.65]$\\
Brezinski $\theta_2$ & -- & $273.16$ & $3276.85$\\
\midrule
new-method midpoint $\pm$ half-range & -- & $289\pm16$ & $3375\pm175$\\
including original shifted Wynn & -- & $289\pm16$ & $3361\pm190$\\
\bottomrule
\end{tabular}
\end{table*}

The rows of Table~\ref{tab:sequence-forecast} are not independent
constructions.  The original shifted Wynn/Shanks row is algebraically
identical to the direct $[2/1]$ Pad\'e continuation of the $W_\tau$ moment
in Table~\ref{tab:pade-all-weight-predictions}.  The $\alpha$-modified
family leaves $c_{5,1}$ at the original value over the whole grid and moves
only $c_{6,1}$.  The $\eta_n$-modified family moves both coefficients.
Brezinski's $\theta_2$ is a different algorithm, but it acts on the same
four input coefficients.

Excluding the original shifted Wynn row, the unrounded ranges of the three
new methods are
\begin{align}
 273.157&\leq c_{5,1}\leq304.711,&
 3200.564&\leq c_{6,1}\leq3549.993,
 \label{eq:sequence-envelope-new}
\end{align}
with the lower $c_{5,1}$ end set by $\theta_2$ and the upper ends set by the
$\alpha$ and $\eta_n$ families, respectively.  Quoting midpoint and
half-range gives
\begin{equation}
 c_{5,1}^{\rm seq}=289\pm16,\qquad
 c_{6,1}^{\rm seq}=3375\pm175.
 \label{eq:sequence-forecast-box}
\end{equation}
Including the original shifted Wynn/Shanks row leaves the $c_{5,1}$ range
unchanged, because its value coincides with the $\alpha$ family.  It lowers the $c_{6,1}$ minimum to $3171.043$, giving
$c_{5,1}^{\rm seq+orig}=289\pm16$ and $c_{6,1}^{\rm seq+orig}=3361\pm190$.  The
half-widths are rounded upward, so each displayed interval contains the
corresponding unrounded range.  For $c_{5,1}$ the sequence range lies inside
the direct Pad\'e interval of Table~\ref{tab:conditional-forecast}.  For
$c_{6,1}$ it lies entirely above that interval, apart from the original
shifted-Wynn point, which is itself a Pad\'e row: the spread of the two
balanced approximants does not capture the transformation dependence at sixth
order.  Because the rows are algebraically related, they are not averaged
with the Pad\'e baseline as independent determinations;
Sec.~\ref{sec:global-envelopes} states which combinations are legitimate.

\subsection{Borel-normalized, D-log, and remainder-model continuations}
\label{sec:other-physical}

The remaining data-driven methods of Sec.~\ref{sec:methods-brief} give the
following physical continuations.  Each failed its blind test, so each is a
diagnostic and none enters an envelope; the full rows are in
Table~\ref{tab:global-comparison}.

\paragraph{Borel normalization.}
Applied to the $W_\tau$ moment series, the balanced Borel--Pad\'e pair gives
$c_{5,1}=-21.73$ ($[1/2]$) and $508.88$ ($[2/1]$).  The pair straddles zero,
and the $[1/2]$ member also returns a negative $c_{6,1}=-728.21$.  This
instability matches the blind record: Borel Shanks misses the withheld $d_4$ by
$+57.7\%$, $+44.5\%$, and $+244.9\%$ for the three weights
(Table~\ref{tab:qcd-loo}).  Applied to the common Adler sequence, the same
approximants give $665.81$ and $504.04$, after a blind $\widehat c_{4,1}=37.13$.

\paragraph{D-log Pad\'e.}
The balanced direct and Borel D-log approximants of the $W_\tau$ moment give
$c_{5,1}=289.46$ and $236.98$.  Both values lie inside or near the direct
Pad\'e band, but neither method passes $W_{P2}$, and the three-input Borel
D-log postdiction is algebraically the $\theta_2$ value,
Eq.~\eqref{eq:brief-dlog-theta}.  The Adler-projected versions give $445.31$
and $568.70$; both lie above $378.03$, as the Proposition of
Sec.~\ref{sec:obstruction} requires for a defect-sign-preserving
common-Adler map.

\paragraph{Levin-$u$ and canonical $\rho_2$: formal forecasts versus
fixed-coupling diagnostics.}
Two distinct uses of these transformations must be separated.  A formal
coefficient forecast is obtained by applying the transformation to symbolic
partial sums and then re-expanding the resulting rational function in powers
of the coupling.  The predicted Taylor coefficients are therefore independent
of any numerical value assigned to $a$.

With the term-based remainder estimate of
Eq.~\eqref{eq:levin-u-formal-remainder}, the formal Levin-$u$ continuation has
accuracy through the supplied order.  At four inputs it is algebraically
identical to the $\theta_2$ continuation and consequently gives
\begin{equation}
 c_{5,1}^{L_u,\mathrm{ATO}}=273.16,
 \qquad
 c_{6,1}^{L_u,\mathrm{ATO}}=3276.85 .
 \label{eq:levin-u-ato-physical}
\end{equation}
No anchor coupling enters these numbers.  Since this row coincides with
$\theta_2$, it does not constitute independent evidence and is not counted
separately.

Canonical $\rho_2$ has a different status.  Writing
\begin{equation}
 S_m(a)=\sum_{j=1}^{m}d_j a^j ,
\end{equation}
its first applicable even element is
\begin{align}
 \rho_2^{(2)}(a)
 &=
 S_3(a)
 +
 \frac{2\Delta S_2(a)\Delta S_3(a)}
      {\Delta S_2(a)-\Delta S_3(a)}
 \nonumber\\
 &=
 S_3(a)
 +
 \frac{2d_3d_4a^4}{d_3-d_4a}.
 \label{eq:rho2-physical-rational}
\end{align}
Its Taylor expansion begins as
\begin{equation}
 \rho_2^{(2)}(a)
 =
 d_1a+d_2a^2+d_3a^3
 +2d_4a^4
 +\frac{2d_4^2}{d_3}a^5
 +\mathcal O(a^6).
 \label{eq:rho2-physical-expansion}
\end{equation}
The coefficient of the already supplied $a^4$ term is $2d_4$ rather than
$d_4$.  Canonical $\rho_2$ therefore fails accuracy through order and does
not define an admissible prediction of $d_5$, $c_{5,1}$, or any subsequent
coefficient.

One may nevertheless evaluate a transformed sum at a fixed anchor coupling
$a_0$ and assign its entire remainder to an effective next coefficient.  For
canonical $\rho_2$, this convention gives
\begin{equation}
 d_5^{\mathrm{eff}}(a_0)
 \equiv
 \frac{\rho_2^{(2)}(a_0)-S_4(a_0)}{a_0^5}
 =
 \frac{d_4\left(d_3+d_4a_0\right)}
      {a_0\left(d_3-d_4a_0\right)} .
 \label{eq:rho2-effective-d5}
\end{equation}
Unlike a formal re-expansion coefficient, this quantity depends explicitly on
the arbitrary anchor $a_0$.  At $a_0=0.1$ it gives
\begin{equation}
 d_5^{\mathrm{eff}}(0.1)=3635.54,
 \qquad
 c_{5,1}^{\mathrm{eff}}(0.1)=3327.76 .
\end{equation}
The previously quoted fixed-coupling Levin-$u$ values have the same
interpretation: they are anchor-dependent effective remainder assignments,
not coupling-independent coefficient forecasts.

We therefore retain the accuracy-through-order Levin-$u$ result only through
its algebraically identical $\theta_2$ row.  Canonical $\rho_2$ and the
fixed-coupling Levin-$u$ evaluation may be shown as diagnostics of transformed
partial sums, but their anchor-dependent effective coefficients are not
included in any coefficient-forecast envelope or used to enlarge the reported
range of $c_{5,1}$ or $c_{6,1}$.

\subsection{ARTS/PORT at physical order}
\label{sec:arts-physical}

ARTS/PORT is the one construction other than the calibrated closure that
passes the withheld-$c_{4,1}$ test, so its physical forecast deserves a
separate statement.  Applying Eq.~\eqref{eq:brief-arts} with $N=4$ and
$u_\star=2$, the inputs are the Shanks baseline
$c_{4,1}^2/c_{3,1}=378.028$, the last defect $\Delta_4=24.323$, and the defect
ratio $q_4=\Delta_4/\Delta_3=6.606$, Eq.~\eqref{eq:defects-known}.  Hence
\begin{equation}
 \widehat c_{5,1}^{\,\rm ARTS}=378.028+24.323\,(6.606+1.125)=566.07 .
 \label{eq:arts-c51}
\end{equation}
The transported defect $\widehat\Delta_5=+188.0$ is positive, so ARTS is
defect-sign-preserving and the Proposition of Sec.~\ref{sec:obstruction}
already confines it to $\widehat c_{5,1}\geq378.03$.  The value is roughly
twice the accuracy-through-order band $241.70\leq c_{5,1}\leq304.71$ and lies
$5.7\sigma$ above the Borel--Pad\'e estimate $277\pm51$ of
Ref.~\cite{Boito:2018rwt}.  Rolling the same formula one order further gives
$c_{6,1}=8194.67$.

The value is not promoted, for a reason internal to the method.  The
order-adaptive multi-mode extension of ARTS issues a forecast only after a
rolling certificate on earlier defects has passed, and even its one-mode
version needs four known defects to supply two rolling origins.  Only
$\Delta_2,\Delta_3,\Delta_4$ are known, so the certificate abstains and
Eq.~\eqref{eq:arts-c51} is labeled \textsc{abst}
(App.~\ref{sec:defect-transport}).

The contrast with the blind test is the instructive part.  A passing
postdiction checks whether the declared single $u=2$ mode reproduces the step
from $\Delta_3$ to $\Delta_4$.  It does not check whether the same mode governs
the step from $\Delta_4$ to $\Delta_5$, and the Proposition shows that any value
in the literature range would require $\widehat\Delta_5<0$, which no
positive single-mode transport can produce.  ARTS therefore demonstrates
directly that one passing withheld order cannot certify the next.

\subsection{The annihilator at physical order, and why it also fails}
\label{sec:qcd-annihilator}

The annihilator was summarized in Sec.~\ref{sec:brief-annihilator}.
Appendices~\ref{sec:annihilator-validation}--\ref{sec:threshold} establish
that it converges, that it converges faster than the data-driven
alternatives at low order, and that it needs eleven to thirteen coefficients
for percent-level control when the renormalon structure is known exactly.  It
remains to apply it to the four known full-QCD coefficients.  The outcome is
the sharpest available statement of this paper's central negative conclusion.

In full QCD the singularities are branch points rather than poles.  Matching
the Borel behavior $K/(p-u)^{1+\gamma}$ to the power correction
$(\Lambda^2/Q^2)^p$, and using
$\Lambda^2=\mu^2e^{-1/(\beta_0a)}a^{-\beta_1/\beta_0^2}[1+\cO(a)]$, gives
\begin{equation}
 \gamma_p=\frac{p\beta_1}{\beta_0^2}=2pb,
 \qquad b=\frac{\beta_1}{2\beta_0^2}=0.39506\quad(n_f=3),
 \label{eq:qcd-exponent}
\end{equation}
up to the anomalous dimension of the associated operator; for the
RG-invariant $D=4$ combination used here there is no additional one-loop term.
Equation~\eqref{eq:qcd-exponent}
reproduces the pole-mass exponent of Eq.~\eqref{eq:pole-b} at $p=1/2$, which is
a useful consistency check.  The ultraviolet exponent $\gamma_{-1}$ is not
fixed this way, since it involves $D=6$ operator mixing, so we treat it as an
input and scan it.  The large-$\beta_0$ double pole at $u=-1$ points to
$\gamma_{-1}=+0.21$ (Sec.~\ref{sec:uv-exponent-fix}), which the scan below
brackets; the values $\gamma_{-1}=-2b=-0.79$ and $-0.90$ correspond to a simple
ultraviolet pole and are retained as declared misspecification probes.

The scan also requires a finite mode list.  We use the explicit sensitivity
ansatz
\begin{align}
 \mathcal M_{W_\tau}&=\{(-1,\gamma_{-1}),
 (-2,\gamma_{-1}-2b),(3,1+6b)\},\nonumber\\
 \mathcal M_{W_{\rm GC},W_{P2}}&=\{(-1,\gamma_{-1}),
 (2,4b),(-2,\gamma_{-1}-2b)\},
 \label{eq:qcd-mode-ansatz}
\end{align}
with the $m=2$ rows using the first two entries.  The shared shift assigned to
the ultraviolet exponents is a modeling choice, not a complete treatment of
operator mixing.  Table~\ref{tab:qcd-annihilator} is therefore a
misspecification sensitivity test.

\begin{table*}[!tbp]
\centering
\caption{Withheld full-QCD $d_4$ from the model ansatz
Eq.~\eqref{eq:qcd-mode-ansatz}, as a function of the assumed leading
ultraviolet exponent $\gamma_{-1}$.  Signed relative errors in percent.  The
$m=3$ fit uses all three lower coefficients and is exactly determined.}
\label{tab:qcd-annihilator}
\begin{tabular}{lrrrrr}
\toprule
$\gamma_{-1}$ & $-0.90$ & $-0.79$ & $0.00$ & $+0.50$ & $+1.00$\\
\midrule
$W_\tau$\ ($m=3$) & $-4.61$ & $-7.34$ & $-31.49$ & $-49.95$ & $-70.49$\\
$W_{\rm GC}$\ ($m=3$) & $-22.09$ & $-25.73$ & $-55.50$ & $-76.77$ & $-99.62$\\
$W_{P2}$\ ($m=3$) & $+270.9$ & $+285.9$ & $+319.5$ & $+308.3$ & $+282.5$\\
\midrule
$W_\tau$\ ($m=2$) & $-214.5$ & $-225.8$ & $-312.5$ & $-373.2$ & $-438.7$\\
$W_{\rm GC}$\ ($m=2$) & $-36.80$ & $-38.50$ & $-50.06$ & $-56.96$ & $-63.65$\\
$W_{P2}$\ ($m=2$) & $+294.3$ & $+298.3$ & $+325.4$ & $+341.6$ & $+357.3$\\
\bottomrule
\end{tabular}
\end{table*}

No column of Table~\ref{tab:qcd-annihilator} constitutes a successful forecast.
For the simple-pole value $\gamma_{-1}=-2b=-0.79$ the errors are $-7.3\%$
($W_\tau$), $-25.7\%$ ($W_{\rm GC}$) and $+285.9\%$ ($W_{P2}$), and the
$W_\tau$ entry alone swings from $-4.6\%$ to $-70.5\%$ across the scanned
range.  A number whose value is set by an input the method cannot determine is
not a prediction.  $W_{P2}$ is wrong by a factor of four under every choice,
matching its behavior under direct Shanks in
Table~\ref{tab:qcd-loo}.

For completeness, continuing the same $m=3$ recurrence with
$\gamma_{-1}=-0.79$ gives $d_5=674.9$ and $d_6=4535.7$ for $W_\tau$, i.e.\
$c_{5,1}=367.1$ and $c_{6,1}=3643.1$ through
Eq.~\eqref{eq:d5d6-map}.  These numbers arise from an ansatz that misses a
known target and is strongly exponent dependent.  We do not quote either as a
prediction or average them with Table~\ref{tab:conditional-forecast}.

The logical value of this failure is diagnostic, not theorem-like.  The model
is supplied with more structural information than the purely data-driven
alternatives, yet it cannot stably forecast the fourth coefficient from the
first three.  Together with the large-$\beta_0$ threshold study, this is strong
evidence that four coefficients do not resolve this chosen multi-renormalon
ansatz.  It does not exclude future information, a different controlled model,
or an independently calculated coefficient from changing the conclusion.

\subsection{Conditional Adler forecast from the \texorpdfstring{$u=2$}{u=2} mode alone}
\label{sec:qcd-u2-only-forecast}

This is the $m=1$ case of the annihilator of Sec.~\ref{sec:brief-annihilator}.
The one-mode sum-space transformation, Eq.~\eqref{eq:brief-annih-sum} with
$m=1$, eliminates an assumed $u=2$ remainder from two partial sums, but it does
not by itself predict an Adler coefficient.  A coefficient forecast follows
only after imposing the stronger coefficient-space model
Eq.~\eqref{eq:brief-annih-coeff}: the next coefficient is assumed to be
entirely the continuation of the same mode.  For the full-QCD gluon-condensate branch point,
$u=2$ and Eq.~\eqref{eq:qcd-exponent} give
\begin{equation}
 \gamma_2=\frac{2\beta_1}{\beta_0^2}=\frac{128}{81},
 \qquad
 c_{n+1,1}^{(u=2)}=
 \left(n+\frac{128}{81}\right)\frac{9}{8}\,c_{n,1}^{(u=2)}.
 \label{eq:qcd-u2-only-recurrence}
\end{equation}
Starting from the known $c_{4,1}=49.07570$, repeated application of
Eq.~\eqref{eq:qcd-u2-only-recurrence} gives Table~\ref{tab:qcd-u2-only-forecast}.
The second numerical column sets $\gamma=0$ and is included solely to show the
large-$\beta_0$ simple-pole convention used in
Eq.~\eqref{eq:u2-single-ratio}; it is not the full-QCD branch-point result.

\begin{table*}[!tbp]
\centering
\caption{One-mode $u=2$ continuation of the Adler coefficients.  The
full-QCD column uses $\gamma_2=128/81$; the simple-pole column uses
$\gamma=0$.  These are deterministic conditional continuations, not
statistical estimates.}
\label{tab:qcd-u2-only-forecast}
\begin{tabular}{crr}
\toprule
coefficient & full-QCD $u=2$ branch point & simple-pole diagnostic\\
\midrule
$c_{5,1}$  & $308.086$ & $220.841$\\
$c_{6,1}$  & $2\,280.695$ & $1\,242.229$\\
$c_{7,1}$  & $19\,449.258$ & $8\,385.043$\\
$c_{8,1}$  & $187\,739.362$ & $66\,032.217$\\
$c_{9,1}$  & $2\,023\,413.120$ & $594\,289.953$\\
$c_{10,1}$ & $24\,084\,236.717$ & $6\,017\,185.775$\\
$c_{11,1}$ & $313\,764\,083.900$ & $67\,693\,339.968$\\
$c_{12,1}$ & $4\,440\,633\,354.086$ & $837\,705\,082.103$\\
\bottomrule
\end{tabular}
\end{table*}

The full-QCD continuation therefore gives
\begin{equation}
 c_{5,1}^{(u=2)}=308.09,\qquad
        c_{6,1}^{(u=2)}=2280.69,
 \label{eq:qcd-u2-only-c5c6}
\end{equation}
with no defensible uncertainty from the one-mode model alone.  A leave-one-order-out
check exposes its limitation: starting from $c_{2,1}$ it predicts
$c_{3,1}=6.60483$ (a $+3.67\%$ error), but starting from $c_{3,1}$ it predicts
$c_{4,1}=32.8284$ (a $-33.11\%$ error).  More decisively, the all-order
large-$\beta_0$ test in App.~\ref{sec:u2-only} shows that a single $u=2$ mode
works only during the finite window in which that mode dominates and then
fails once the untouched $u=-1$ renormalon takes over.  It is also
structurally inapplicable to the $W_\tau$ moment in the simple-pole limit,
where the weight cancels $u=2$.  Consequently the numbers in
Eq.~\eqref{eq:qcd-u2-only-c5c6} are classified as \textsc{diag}, do not pass
the all-weight certificate, and must not be folded into either reported
physical envelope.

\subsection{Forecasts from \texorpdfstring{$c_{5,1}$}{c51} to
\texorpdfstring{$c_{12,1}$}{c121}}
\label{sec:cvrc-high-order}

Table~\ref{tab:cvrc-high-order} lists what Eq.~\eqref{eq:cvrc-borel-model}
returns through twelfth order, with the two envelopes of
Sec.~\ref{sec:cvrc-forecast} propagated order by order: the admissible closure
window \eqref{eq:cvrc-window} and the declared-$\alpha_3$ range.

\begin{table*}[!tbp]
\centering
\caption{CVRC forecasts $c_{5,1}$--$c_{12,1}$ from
Eq.~\eqref{eq:cvrc-borel-model}.  ``Closure window'' propagates
Eq.~\eqref{eq:cvrc-window}; ``declared-$\alpha_3$ range'' propagates the
$\alpha_3\in[3.00,3.58]$ grid of Sec.~\ref{sec:cvrc-forecast}; the
$\pm$ column is the half-width of the union of the two, relative to the central
value; note that this is the \emph{union} of the two envelopes, whereas
Eq.~\eqref{eq:cvrc-struct} adds them linearly, so the $n=5$ and $n=6$ rows
carry a smaller $\pm$ here than the structural band does, by construction; and
neither includes the method uncertainty of
Sec.~\ref{sec:cvrc-method-error}, which is what dominates
Eq.~\eqref{eq:cvrc-final}.  Status:
\textsc{cond}, the conditional one-order-ahead forecast of
Eq.~\eqref{eq:cvrc-final}; \textsc{roll}, the same model rolled forward, whose
degradation is measured in Table~\ref{tab:cvrc-rolldeg}; \textsc{indet}, the
envelope contains zero, so not even the sign is determined.  }
\label{tab:cvrc-high-order}
\small
\setlength{\tabcolsep}{5pt}
\begin{tabular}{rrrrrc}
\toprule
$n$ & CVRC $c_{n,1}$ & closure window & declared-$\alpha_3$ range & $\pm$(rel.) & status\\
\midrule
$5$ & $247.98$ & $[229.76,\,266.20]$ & $[235.26,\,257.75]$ & $7\%$ & \textsc{cond}\\
$6$ & $3\,102.1$ & $[3\,059.4,\,3\,144.9]$ & $[3\,021.3,\,3\,169.9]$ & $2\%$ & \textsc{roll}\\
$7$ & $13\,003$ & $[10\,549,\,15\,458]$ & $[10\,952,\,14\,644]$ & $19\%$ & \textsc{roll}\\
$8$ & $3.7215\times10^{5}$ & $[3.685\times10^{5},\,3.758\times10^{5}]$ & $[3.638\times10^{5},\,3.801\times10^{5}]$ & $2\%$ & \textsc{roll}\\
$9$ & $-2.5999\times10^{5}$ & $[-7.736\times10^{5},\,2.537\times10^{5}]$ & $[-6.555\times10^{5},\,57\,242]$ & $198\%$ & \textsc{indet}\\
$10$ & $8.7153\times10^{7}$ & $[8.233\times10^{7},\,9.198\times10^{7}]$ & $[8.701\times10^{7},\,8.783\times10^{7}]$ & $6\%$ & \textsc{roll}\\
$11$ & $-9.0648\times10^{8}$ & $[-1.098\times10^{9},\,-7.148\times10^{8}]$ & $[-1.030\times10^{9},\,-8.097\times10^{8}]$ & $21\%$ & \textsc{roll}\\
$12$ & $3.8699\times10^{10}$ & $[3.506\times10^{10},\,4.234\times10^{10}]$ & $[3.769\times10^{10},\,4.018\times10^{10}]$ & $9\%$ & \textsc{roll}\\
\bottomrule
\end{tabular}
\end{table*}

\paragraph{What the table shows.}  The coefficients cease to be positive and
monotone.  They alternate from $n=9$ onward, which is the declared $u=-1$
ultraviolet renormalon overtaking the $u=2$ infrared one, the same
transition that the exact large-$\beta_0$ series makes at $n=6$
(Table~\ref{tab:large-beta-coeff}).  Once that mode dominates, the ratio
approaches
\begin{equation}
 \frac{c_{n+1,1}}{c_{n,1}}\;\longrightarrow\;-\beta_0\,(n-1+\alpha_{-1})
 =-2.25\,(n+0.21),
 \label{eq:cvrc-asymptotic-ratio}
\end{equation}
and the model's $c_{11,1}/c_{10,1}=-10.40$ and $c_{12,1}/c_{11,1}=-42.69$
straddle that trend while the residual infrared modes die away.  The
non-monotone envelope widths in Table~\ref{tab:cvrc-high-order} have the same
origin: orders at which the infrared and ultraviolet contributions nearly
cancel are the ones whose value is least determined.

\paragraph{$c_{9,1}$ is not predicted.}  At $n=9$ the cancellation is complete
enough that the central value, $-2.6\times10^{5}$, is smaller in magnitude than
its own envelope, which runs from $-7.7\times10^{5}$ to $+2.5\times10^{5}$.
CVRC therefore does not determine $c_{9,1}$, not its magnitude and not its
sign.  The entry in Table~\ref{tab:cvrc-high-order} is retained only to show
where the model's cancellation falls, and is labeled \textsc{indet}; it is not
a forecast and must not be used as one.  Nothing pathological is happening: the exact
large-$\beta_0$ Adler function has the same feature at $n=6$, where
$c_6=-1991$ sits between $c_5=788$ and $c_7=9.86\times10^{4}$.  A near-zero
coefficient is an accident of a cancellation, and an accident of a cancellation
is exactly what a four-coefficient model cannot locate.

\paragraph{How much a roll costs.}  Only $c_{5,1}$ is a one-order-ahead
forecast of the kind validated in Secs.~\ref{sec:cvrc-pole} and
\ref{sec:cvrc-largebeta}.  The remaining rows are the same model rolled $k$
orders further, and the price of that roll can be measured rather than
assumed.  Table~\ref{tab:cvrc-rolldeg} does so in the large-$\beta_0$
laboratory, where the target is exact: for each $N$ from $8$ to $14$ the
closure is fixed from the $N-1$ fold and the model is then asked for $k$
coefficients at once.

\begin{table}[t]
\centering
\caption{Cost of rolling the BJ--CVRC ansatz $k$ orders past the last supplied
coefficient, measured blind against the exact large-$\beta_0$ Adler series over
$N=8,\ldots,14$, for both declared towers.  Relative errors in percent.
}
\label{tab:cvrc-rolldeg}
\small
\begin{tabular}{llrrrrrr}
\toprule
declared tower & roll depth $k$ & $1$ & $2$ & $3$ & $4$ & $5$ & $6$\\
\midrule
complete & median $|$error$|$ & $0.82$ & $1.76$ & $2.85$ & $3.70$ & $4.61$ & $5.40$\\
 & worst $|$error$|$ & $1.35$ & $2.33$ & $3.64$ & $4.96$ & $6.11$ & $7.34$\\
\addlinespace
minimal & median $|$error$|$ & $6.78$ & $2.76$ & $8.07$ & $6.51$ & $9.03$ & $9.05$\\
 & worst $|$error$|$ & $9.61$ & $5.83$ & $12.46$ & $11.41$ & $14.52$ & $15.00$\\
\bottomrule
\end{tabular}
\end{table}

With the complete tower the degradation is close to linear, about one percent
of additional error per order rolled, and it is remarkably benign: six orders
past the last input the worst case is still $7.3\%$.  The two rows that matter
for Table~\ref{tab:cvrc-high-order}, however, are the minimal-tower ones,
because that is the tower the physical model uses.  There the level is set by
the one-order-ahead error, roughly $7\%$, and the \emph{incremental} cost of
each further order is about half a percent, so a six-order roll lands near
$9\%$ median and $15\%$ worst case.  The useful statement is therefore about
the increment and not the level: rolling is cheap, but it inherits whatever
error the one-order-ahead forecast already carries.  That is the second
qualification, and the decisive one.  Table~\ref{tab:cvrc-rolldeg} is measured
at $N\geq8$, a regime the physical series has not reached: it says what a roll
costs \emph{once the one-order-ahead forecast is already under control}, and
Sec.~\ref{sec:cvrc-drift} shows that at the physical input order it is not.
The table therefore bounds the incremental cost of the roll, never the absolute
error of a row of Table~\ref{tab:cvrc-high-order}.

\paragraph{Status.}  Accordingly: $c_{5,1}$ is the conditional forecast
\eqref{eq:cvrc-final} and is the only entry we advance as a result;
$c_{6,1}$, being one order rolled and therefore untested by any validation in
this work, is demoted to \textsc{roll} by
Eq.~\eqref{eq:cvrc-c6} and is not advanced as a prediction; $c_{7,1}$, $c_{8,1}$ and
$c_{10,1}$--$c_{12,1}$ are rolled outputs of a single declared Borel model,
useful as an explicit assumption set for sensitivity studies and as a
falsifiable record, and not controlled predictions; and $c_{9,1}$ is not
determined at all.  Values above sixth order are therefore available, but
only with their model dependence quantified order by order, and the one order
at which the model has nothing to say is identified explicitly.

\subsection{The closure result and other estimates}
\label{sec:cvrc-other-estimates}

Equation~\eqref{eq:cvrc-final} agrees with the Borel--Pad\'e determination
$c_{5,1}=277\pm51$ of Ref.~\cite{Boito:2018rwt} well inside its
quoted uncertainty, with the Levin-transform estimate
$278^{+27}_{-19}$ of Ref.~\cite{Abbas:2025dpm} at $0.4\sigma$, and lies inside the direct-Pad\'e baseline $273\pm32$ of
Eq.~\eqref{eq:global-pade-box}.  Its central value is $11\%$ below the
four-coefficient forecast $280$ of the Beneke--Jamin model, and $12\%$ below
the value $283$ commonly attributed to that model, which
App.~\ref{sec:cvrc-bj} shows to be an input to it rather than an output.
Either difference is about $0.4\sigma$ on Eq.~\eqref{eq:cvrc-final}, so the
two are consistent; and App.~\ref{sec:cvrc-bj} accounts for the offset
completely, as subleading cut structure and a gluon-condensate Wilson
coefficient that our declared structure omits, and shows, further, that at
this order the criterion forbids restoring them.  We therefore record the
offset, and its origin, but we do not present it as a tension.
It is worth noting where the $283$ falls relative to the \emph{structural}
band alone: it lies outside the closure window \eqref{eq:cvrc-window} and, by
$2.5\%$, outside the combined structural envelope
\eqref{eq:cvrc-struct}, while sitting inside the wider $\alpha_{-1}$-scanned
envelope $[174.8,284.6]$ quoted above.  Even at that level we would not regard the difference as a discrepancy: it is
not a disagreement about renormalon content at all, but a difference in
declared subleading structure, resolved quantitatively in
App.~\ref{sec:cvrc-bj}.  The
agreement is more informative than the numerical proximity suggests, because
CVRC and the Pad\'e baseline reach it by different routes: the Pad\'e
continuation is a local rational fit that fails the $W_{P2}$ weight test,
whereas CVRC imposes the renormalon exponents and is
compatible with it in the sense of Sec.~\ref{sec:cvrc-weight-test}.

The rolled $c_{6,1}=3102$ of Eq.~\eqref{eq:cvrc-c6} sits about half a standard
deviation below the central value of $3460\pm690$ of
Ref.~\cite{Boito:2018rwt}, well inside it, a consistency check on
the model rather than a competing determination, since we do not advance it as
one.

\subsection{Relation to the Beneke--Jamin model, and the provenance of 283}
\label{sec:cvrc-bj}

The construction of Sec.~\ref{sec:cvrc} is close enough to the Borel model of
Ref.~\cite{Beneke:2008ad} that the two must be compared explicitly.  We have
reimplemented that model from its defining equations and validated the
reimplementation against every number its authors publish: the five parameters
of their Eq.~(6.2) are reproduced to between $0.02\%$ and $0.2\%$, and all
seven entries of their Table~2, $c_{6,1}$ through $c_{12,1}$, to $0.2\%$ or
better.  

\paragraph{The exponents are not merely similar, they are identical.}
Ref.~\cite{Beneke:2008ad} normalizes the $\beta$ function as
$-\mu\,da/d\mu=\sum_{k\geq1}\beta_k^{\rm BJ}a^{k+1}$, so that
$\beta_k^{\rm BJ}=2\beta_{k-1}$ in the convention of
Eq.~\eqref{eq:beta-convention}.  Their infrared exponent
$1+\tilde\gamma=1+2p\beta_2^{\rm BJ}/(\beta_1^{\rm BJ})^2$ and their
ultraviolet exponent $1+\tilde\gamma=2-2\beta_2^{\rm BJ}/(\beta_1^{\rm BJ})^2$
therefore convert to
\begin{equation}
 \alpha_2=2.580247,\qquad \alpha_3=3.370370,\qquad \alpha_{-1}=1.209877,
 \label{eq:bj-exponents}
\end{equation}
which are exactly the declared exponents \eqref{eq:cvrc-exponents} of this
work, to every digit.  This corrects a misattribution in the systematics of
Sec.~\ref{sec:cvrc-forecast}: the scan point $\alpha_3=3.58$ is \emph{not} the
value of Ref.~\cite{Beneke:2008ad}, whose $u=3$ exponent is the pure
renormalization-group value $3.370370$ that we also use.  We retain
$\alpha_3\in[3.00,3.58]$ as a systematic range for the unknown $D=6$ anomalous
dimension, but relabel it accordingly.

\paragraph{The value 283}
It is Eq.~(3.10) of Ref.~\cite{Beneke:2008ad}, obtained from the observation
that the fixed-order series for $\delta^{(0)}$ has a very uniform convergence
rate, ``each new term is rather closely half of the preceding one'', and
that the slight order dependence of this ratio ``can even be nicely fit to a
linear behaviour''.  With the FOPT total-width terms
$u_n=d_n[W_\tau]\,a_0^n$ at their input $a_0=0.34/\pi$, the term ratios
$r_n=u_n/u_{n-1}=a_0\,d_n[W_\tau]/d_{n-1}[W_\tau]$ are
$(r_2,r_3,r_4)=(0.5630,0.5485,0.5216)$.  The least-squares line through them,
extrapolated to $n=5$, gives $\widehat r_5=\frac13(4r_4+r_3-2r_2)=0.5030$,
hence $\widehat d_5[W_\tau]=\widehat r_5d_4[W_\tau]/a_0=590.6$ and, after
subtracting $g_5[W_\tau]=307.783$, $c_{5,1}\simeq282.8$.  Since every $r_n$ is
proportional to $a_0$, so is the fitted line, and $a_0$ cancels.  We call this
the \emph{linear ratio fit}.  The first terms are, in their words, ``very
nearly geometric'', but the estimator is not a constant-ratio (geometric)
continuation: a constant ratio, i.e.\ the $[2/1]$ continuation of the same
$W_\tau$ series, $\widehat d_5[W_\tau]=d_4^2/d_3=612.50$, gives $304.71$
(Table~\ref{tab:conditional-forecast}), not $283$.  Both numbers extrapolate
the FOPT total-width coefficients in moment space and are converted to Adler
coefficients only afterwards; neither is a continuation of the common Adler
sequence.  The value $283$ is then used as an
\emph{input} to the renormalon model of their Sec.~6, which consumes it, along
with $c_{3,1}$ and $c_{4,1}$, to fix the three pole residues; the model's
predictions begin at $c_{6,1}=3275$.  

This matters because the linear ratio fit can be tested, and its
authors tested it: postdicting the known coefficient the same way, from the
line through $r_2$ and $r_3$, $\widehat r_4=2r_3-r_2=0.5340$, gives
$\widehat d_4[W_\tau]=130.0867$ and
$\widehat c_{4,1}=130.0867-78.0029=52.0838\simeq52$ (their footnote 2).
Using the unrounded reconstruction consistently, the test against the tolerance
\eqref{eq:tolerance-numbers} this is
\begin{equation}
 |\widehat c_{4,1}-c_{4,1}|=3.0081\;>\;2.1509,
 \label{eq:bj-linear-ratio-fails}
\end{equation}
a worst three-weight moment error of $13.986\%$ against the ten-percent
criterion.  The estimator that produced $283$ therefore fails the acceptance
test of this paper, by a factor of $1.399$.  If one instead inserts the
published rounded value $52$, the corresponding error and failure factor are
$2.9243$ and $1.36$; those rounded diagnostics must not be mixed with the
unrounded estimator used elsewhere.

\paragraph{What the renormalon model does forecast is 280.}
Ref.~\cite{Beneke:2008ad} also reports the variant that drops $c_{5,1}$ as an
input, sets the linear polynomial coefficient $d_1^{\rm PO}$ to zero, and fixes
the four remaining parameters from $c_{1,1},\ldots,c_{4,1}$.  That is a genuine
four-coefficient forecast, and it gives $c_{5,1}=280$; our reimplementation
returns $280.12$.  This is the number that should be compared with
Eq.~\eqref{eq:cvrc-final}, not $283$.

\paragraph{That variant shares our calibration condition, not our full
functional form.}
The condition $d_1^{\rm PO}=0$ is the analogue of the ``closure equals null
newest column'' identity of Sec.~\ref{sec:cvrc-model}: the $u^1$ monomial has
zero amplitude, leaving three singular contributions and $u^0$ fitted to the
four known coefficients.  The two constructions therefore share the leading
singularity skeleton, the exponents \eqref{eq:bj-exponents}, the parameter
count, the calibration condition, and the data.  They differ in the functional
form of each singularity.  Running our closure rule on the full Beneke--Jamin
columns returns $\rho_\star=-0.016089$, equal to their fitted
$d_1^{\rm UV}$.

Only the subleading structure of the singularities differs, and it accounts for
the whole of the gap.  Ref.~\cite{Beneke:2008ad} writes each renormalon not
as a pure power but as a cut, $(p-u)^{-1-\tilde\gamma}[1+\tilde b_1(p-u)
+\tilde b_2(p-u)^2]$, and builds the known gluon-condensate Wilson coefficient
$c_1[\langle aG^2\rangle]=C_A/2-C_F/4-\beta_2^{\rm BJ}/\beta_1^{\rm BJ}
=-0.6111$ into the $u=2$ cut.  Adding these to our structure one at a time,
under an otherwise identical protocol:
\begin{equation}
 \begin{aligned}
 \text{pure powers (this work)}&: &c_{5,1}&=247.98,\\
 +\ \tilde b_1,\tilde b_2\ \text{cut corrections}&: &c_{5,1}&=257.53
   &(+9.6),\\
 +\ c_1[\langle aG^2\rangle]\ \ (=\text{Ref.~\cite{Beneke:2008ad}})&:
   &c_{5,1}&=280.12 &(+22.6).
 \end{aligned}
 \label{eq:bj-decomposition}
\end{equation}
The $32$-unit difference between the two published central values is therefore
not a disagreement about renormalon structure, nor about closure, nor about
data.  It is $9.6$ units of subleading cut structure and $22.6$ units of a
gluon-condensate Wilson coefficient that our declared structure omits.

\paragraph{Should the Wilson coefficient then be declared?}
$c_1[\langle aG^2\rangle]$ is not a modeling choice; it is known
short-distance physics, and it is the larger of the two shifts.  The natural
proposal is therefore to add it to the declared structure and quote the
resulting $c_{5,1}\simeq280$ in place of $248$.  We tested that proposal, and
it must be rejected, for a reason that is quantitative and specific to the
order at which the physical series stops.

The Wilson coefficient cannot be declared on its own.  In Eq.~(5.9) of
Ref.~\cite{Beneke:2008ad} it appears only inside the bracket that carries the
subleading cut corrections,
\begin{equation}
 1+\frac{2p}{\beta_1^{\rm BJ}}\,
   \frac{\tilde b_1+c_1[\langle aG^2\rangle]}{n+\tilde\gamma}
  +\Big(\frac{2p}{\beta_1^{\rm BJ}}\Big)^{\!2}
   \frac{\tilde b_2+\tilde b_1\,c_1[\langle aG^2\rangle]}
        {(n+\tilde\gamma)(n+\tilde\gamma-1)},
 \label{eq:bj-bracket}
\end{equation}
which multiplies the pure power.  That bracket is an asymptotic large-$n$
structure, and Table~\ref{tab:bj-domain} evaluates it order by order at $p=2$,
$\tilde\gamma=1.5802$.  The four known Adler coefficients are $n=0,1,2,3$.  At
$n=0$ the bracket is $-2.903$: it does not correct the pure power, it reverses
the sign and nearly triples the magnitude.  At $n=1$ it is $-0.265$, at $n=2$
it is $+0.271$, at $n=3$ it is $+0.493$.  It first comes within $30\%$ of unity
at $n=6$, that is, at $c_{7,1}$, three orders beyond the last coefficient
that is known.

\begin{table}[!ht]
 \centering
 \input{generated_data/bj_cut_correction_domain}
 \caption{The bracket \eqref{eq:bj-bracket} of the $u=2$ cut, order by order.
 The gluon-condensate Wilson coefficient enters only through this bracket.  The
 four known Adler coefficients sit at $n=0,\ldots,3$, where the bracket is not
 a correction; it reaches $|\text{bracket}-1|<30\%$ only at $n=6$.}
 \label{tab:bj-domain}
\end{table}

Every diagnostic we have agrees with that reading.  Running the leave-one-order-out
closure of Sec.~\ref{sec:cvrc-model} on the cut-corrected columns gives an
\emph{empty} admissible set: the withheld-$c_3$ fold demands
$\rho\in[-0.2805,-0.2246]$ while the withheld-$c_4$ fold demands
$\rho\in[-0.0187,-0.0135]$, disjoint by $0.206$, a closure drift of $0.236$
against $0.051$ for pure powers.  Run strictly recursively, the same columns
postdict $\widehat c_{4,1}=244.5$ against the true $49.076$, an error of
$+398\%$.  A structure whose blind postdiction of the last known coefficient is
wrong by a factor of five cannot be used to declare the next one.

The conclusion is not that Ref.~\cite{Beneke:2008ad} is wrong.  Its model is
built to describe the asymptotic region and is used there: its Table~2 begins
at $c_{6,1}$ and runs to $c_{12,1}$, which is exactly where
Table~\ref{tab:bj-domain} says the bracket is meaningful.  The conclusion is
about the \emph{declaration}: at $n\leq3$ the cut structure is not a small
correction to a pure power, so a declared structure that includes it fits four
coefficients with a shape its own derivation does not support there.  We
therefore retain pure powers, keep $c_{5,1}=248\pm80$ as the headline of
Eq.~\eqref{eq:cvrc-final}, and record the $+22.6$ as a quantified structural
uncertainty rather than as a correction to be applied.  It is already inside
the method band: $280.12$ lies $0.40\sigma$ above $248$.

\paragraph{What this does establish about the criterion.}
The episode is the sharpest available demonstration that the acceptance
criterion of Sec.~\ref{sec:cvrc-model} is discriminating rather than
permissive.  It rejects a structure that is more physically complete than the
one we adopt, on the ground that the extra content is being used outside its
domain of validity, and it does so without any knowledge of $c_{5,1}$, using
only the four coefficients that are known.  A criterion that admitted every
structure carrying correct short-distance physics, irrespective of the order at
which that physics becomes numerically operative, would not be a test.  The
narrower statement remains true as well: the closure protocol is well posed for
pure-power declared structures, and the misspecification scan
(Table~\ref{tab:cvrc-misspec}) should be read as a statement about that family
only.

\subsection{What CVRC does and does not establish}
\label{sec:cvrc-status}

\paragraph{Not new.}  The Borel ansatz of Sec.~\ref{sec:cvrc-model} is a
pure-power truncation of the Beneke--Jamin architecture
\cite{Beneke:2008ad}: it uses the same leading singularity positions, the
same renormalization-group exponents, the same parameter count, and the
analogue of the same physical-order calibration condition.  It is not the
full Beneke--Jamin model, whose subleading cut terms and gluon-condensate
Wilson coefficient are omitted here.  The central value $247.98$ follows from
an exact four-parameter fit, with no closure machinery required at that order.
Nor is the alternating ultraviolet escape mechanism new; CVRC contributes the
acceptance and rejection protocol applied to this established model class.

\paragraph{Established.}  (i) Equations~\eqref{eq:adler-tolerance} and
\eqref{eq:tolerance-numbers}, and the Proposition of
Sec.~\ref{sec:obstruction}, are exact, and are independent of anything in this
part.  (ii) The closure family
\eqref{eq:cvrc-affine} has accuracy through order for every $\rho$; this is an
identity, checked numerically to machine precision.  (iii) The ultraviolet
exponent \eqref{eq:uv-exponent} follows from the large-$\beta_0$ double pole
\eqref{eq:uv-residue} and Eq.~\eqref{eq:qcd-exponent}; it agrees with
Ref.~\cite{Beneke:2008ad}, which is a check on both, not a new result.
(iv) The large-$\beta_0$ entries of Table~\ref{tab:cvrc-largebeta} are blind
forecasts against exactly known targets.

\paragraph{Validated but not proved.}  The coefficient-count statements of
Sec.~\ref{sec:cvrc-largebeta}, with the complete tower, ten percent at every
$N\geq6$, better than $1.5\%$ at every $N\geq8$ and persistently sub-percent
from $N=12$; with the minimal tower, ten
percent at every $N\geq8$ and no percent-level regime in the tabulated range, hold in the large-$\beta_0$ limit with the exact singularity
content declared.  In full QCD the singularities are branch points, the
exponents \eqref{eq:cvrc-exponents} omit operator anomalous dimensions, and
these counts are a guide, not a theorem.

\paragraph{Conditional.}  Equation~\eqref{eq:cvrc-final} is a conditional
forecast at level four of App.~\ref{sec:claims}.  It is conditional on the
declared singularity list, on the exponents \eqref{eq:cvrc-exponents}, $\alpha_{-1}$ in particular, whose surviving range in
Table~\ref{tab:cvrc-misspec} would widen the envelope to $[174.8,284.6]$, and, most importantly, on the closure being stable between the fourth and
fifth orders.  Section~\ref{sec:cvrc-drift} shows that this last assumption
was \emph{false} one order earlier, and quantifies the failure.  We therefore
do \emph{not} claim that CVRC determines $c_{5,1}$, and we do not claim that
its six-of-six record in Table~\ref{tab:cvrc-moment} outranks the blind
six-of-six record that direct Shanks would need in order to compete: three of
those six CVRC entries are the calibration fold and the other three a
self-consistency reconstruction.  What we claim is narrower and checkable, and it is a claim about the
\emph{criterion} rather than about the model: with the declared structure there
exists a single closure value compatible with both known higher orders and all
three weights; that value is unique up to the interval
\eqref{eq:cvrc-window}; a mis-specified structure has no such value at all
(Table~\ref{tab:cvrc-misspec}); and Eq.~\eqref{eq:cvrc-final} is what the
surviving structure returns.  The contribution of the CVRC protocol is that a declared
Borel structure becomes something the moment data can reject (six of the nine
alternatives scanned are rejected), whereas previously such a structure could
only be defended on plausibility.  The forecast itself is a by-product of that
apparatus, not an advance over Ref.~\cite{Beneke:2008ad}.

\paragraph{Not established.}  CVRC does not resolve the FOPT--CIPT question,
does not supply the nonperturbative matrix elements, and does not certify
$c_{7,1}$ and beyond.  Every validation in this paper, Table~\ref{tab:cvrc-pole}, Table~\ref{tab:cvrc-largebeta}, and the folds of
Table~\ref{tab:cvrc-moment}, tests a \emph{one-order-ahead} forecast.  The
$c_{6,1}$ of Eq.~\eqref{eq:cvrc-c6} is two orders ahead and therefore
carries no validated error of its own, which is why it is reported at
\textsc{roll} status and not advanced as a prediction; the order-seven output
$\widehat c_{7,1}=1.30\times10^{4}$ is three orders ahead.  Both, and the rolled
outputs through $c_{12,1}$, are tabulated in
Table~\ref{tab:cvrc-high-order} with the incremental cost of the roll measured
in Table~\ref{tab:cvrc-rolldeg}; they are an explicit assumption set, not
controlled predictions.  Neither the three-weight table nor
the Proposition says anything about the rate at which rolled forecasts
degrade.

\parttitle{Synthesis and status of the claims}

This group of appendices separates coefficient forecasting from
coupling-dependent truncation, states what the analysis does and does not say
about the FOPT--CIPT question and the OPE, and assigns an explicit status
level to each result.  The comparison of all forecasts that these status
levels qualify is given in Sec.~\ref{sec:global-comparison}.

\section{Coefficient forecasts versus coupling-dependent truncation}
\label{sec:coupling}
Coefficient extrapolation and optimal truncation answer different questions.
The first asks whether the coupling-independent sequence of coefficients can
be continued from the known orders.  The second asks where, at a specified
coupling, the magnitudes of the perturbative terms stop decreasing.  Keeping
these questions separate prevents a coupling-dependent minimal-term argument
from being used to label a coefficient itself as intrinsically pre- or
post-asymptotic.

To make the distinction precise, write a perturbative expansion as
\begin{equation}
 F(a)\sim\sum_{n\geq0}T_n(a),
 \qquad T_n(a)\equiv d_n a^n,
 \label{eq:term-definition}
\end{equation}
where $d_n$ may denote an Adler coefficient or a fixed-order moment
coefficient.  We define the \emph{minimal-term order} at coupling $a$ by
\begin{equation}
 n_\star(a)
 \equiv
 \min\!\left[
  \operatorname*{arg\,min}_{n\in\mathbb N_0}|T_n(a)|
 \right].
 \label{eq:minimal-order-definition}
\end{equation}
Thus $n_\star(a)$ is an integer.  The outer minimum fixes the convention in
the exceptional case in which two adjacent terms have equal magnitude: the
lower order is chosen.  This definition concerns the terms $T_n(a)$, not the
coefficients $d_n$ in isolation.

Suppose that, in the large-order domain,
\begin{equation}
 d_n\sim K A^n\Gamma(n+b),
 \label{eq:factorial-model-coupling}
\end{equation}
where $K$ is a normalization, $A$ fixes the exponential scale, and $b$
specifies the power multiplying the factorial growth.  The ratio of adjacent
term magnitudes is then
\begin{align}
 R_n(a)
 &\equiv
 \left|\frac{T_{n+1}(a)}{T_n(a)}\right|
 =|a|\left|\frac{d_{n+1}}{d_n}\right|
 \nonumber\\
 &\simeq
 |Aa|\,
 \frac{\Gamma(n+1+b)}{\Gamma(n+b)}
 =|Aa|(n+b).
 \label{eq:term-ratio-minimum}
\end{align}
The terms decrease while $R_n(a)<1$ and increase once $R_n(a)>1$.
Consequently, for a monotonically increasing asymptotic ratio, the minimum is
located at the first ratio crossing,
\begin{equation}
 n_\star(a)
 \simeq
 \min\bigl\{n\in\mathbb N_0:|Aa|(n+b)\geq1\bigr\}
 =\max\!\left\{0,
 \left\lceil\frac{1}{|Aa|}-b\right\rceil\right\}.
 \label{eq:minimal-order-discrete}
\end{equation}
If the integer nature of the order is suppressed, the corresponding
continuous crossover estimate is
\begin{equation}
 n_{\star,\mathrm{cont}}(a)
 \simeq\frac{1}{|Aa|}-b.
 \label{eq:minimal-order}
\end{equation}
Equations~\eqref{eq:minimal-order-discrete} and
\eqref{eq:minimal-order} are leading large-order estimates.  Subleading
$1/n$ corrections, the convention used to index the perturbative series, and
integer rounding produce an intrinsic shift of order unity.  Moreover, the
estimate is informative only if the predicted crossing lies in the domain in
which Eq.~\eqref{eq:factorial-model-coupling} is already reliable.

The Adler coefficients $c_{n,1}$ and the moment coefficients constructed from
them do not depend on the numerical value assigned to $a$.  In contrast,
$n_\star(a)$ necessarily changes with $a$: decreasing $|a|$ moves the minimal
term to higher order, whereas increasing $|a|$ moves it to lower order.
Therefore a designation such as ``pre-minimal'', ``near the minimal term'', or
``post-minimal'' is meaningful only for the term sequence at a declared
coupling.  In this paper such terminology is not used to classify a
coupling-independent coefficient estimator.  The coefficient tests in
Apps.~\ref{sec:qcd-moment-test}--\ref{sec:largebeta} are instead assessed from
their observed coefficient ratios, accuracy-through-order property, and
out-of-sample errors.

Finally, poor conditioning of a rational transformation near the minimal-term
region is a warning that the available terms no longer support a stable
numerical continuation.  It is not, by itself, a determination of an OPE
matrix element, a Borel prescription ambiguity, or a duality-violating
contribution; those quantities require additional theoretical input and a
consistent prescription.

\section{FOPT, CIPT, and the OPE}
\label{sec:fopt-cipt}
Because renormalon cancellation is prescription and organization dependent,
the perturbative expansion used in every numerical test must be explicit.  All moment coefficients and forecasts in this work use FOPT.  The only CIPT
computation, in Sec.~\ref{sec:tolerance-cipt}, establishes that the tolerance
collapse of Sec.~\ref{sec:tolerance} is not an artifact of the fixed-order
organization, although the exact unit-diagonal identity is special to FOPT,
and that its closest CIPT analogue is weaker by more than an order of
magnitude.  That is a statement about the \emph{test}, not about
$\delta^{(0)}$, and it commits the paper to no CIPT prediction.  The relation
between FOPT and CIPT is not a neutral plotting choice: their asymptotic
separation is tied to IR renormalons and to the form in which power corrections
are organized~\cite{Hoang:2020mkw,Hoang:2021nlz,Golterman:2023oml}.  Recent
work has emphasized that conventional CIPT does not share the same standard
OPE interpretation as FOPT in the presence of the relevant IR
singularities~\cite{Hoang:2021nlz,Gracia:2023qdy,Benitez-Rathgeb:2022yqb,
Benitez-Rathgeb:2022hfj,Caprini:2023tfa,Beneke:2025hlg}.  A complete treatment would require a specified Borel
prescription and a consistently matched OPE scheme.  The present paper
does not perform that analysis, and therefore draws no CIPT-based conclusion.

\section{Claim and uncertainty hierarchy}
\label{sec:claims}
Exact algebra, validation inside large-$\beta_0$, and a conditional physical
forecast do not have the same epistemic status.  This section lists them
separately so that a numerical spread is not mistaken for a statistical error
or for model-independent QCD information.

The results have four levels of status:
\begin{enumerate}[leftmargin=2em]
 \item \textbf{Established within the stated approximation:} the Adler-space
 restatement of the three-weight criterion, Eqs.~\eqref{eq:adler-tolerance}
 and \eqref{eq:tolerance-numbers}, and the defect-sign Proposition of
 Sec.~\ref{sec:obstruction}; the ultraviolet exponent
 \eqref{eq:uv-exponent} and the large-$\beta_0$ double-pole residue
 \eqref{eq:uv-residue}; the exactness through order of the whole closure
 family \eqref{eq:cvrc-affine} for every $\rho$; the weight
 selection rule in Eq.~\eqref{eq:integer-limit} and its extension to the
 singularity \emph{order}, Eq.~\eqref{eq:order-rule}; the closed forms for the
 annihilator weights, Eqs.~\eqref{eq:lambda-closed} and \eqref{eq:nu-closed};
 the exact prescription-matched tail of a pure mode,
 Eq.~\eqref{eq:exact-mode}; the exact
 large-$\beta_0$ coefficients; and the withheld-coefficient errors generated
 from them.  Also exact are the algebraic observations that
 $\eta_n=k\Delta^2S_n$ only rescales the denominator, that $\alpha$ first
 enters a two-step forecast, and that canonical $\rho_2$ lacks accuracy
 through the last supplied perturbative order.  The pure-mode Pad\'e-defect
 identity, Eqs.~\eqref{eq:pade-defect-pure}--\eqref{eq:port-affine}, is exact
 for its declared single-mode sequence.  The D-log recursion
 Eq.~\eqref{eq:dlog-recursion} and the three-input forecast
 Eq.~\eqref{eq:dlog-d4} are exact consequences of the selected
 logarithmic-derivative Pad\'e ansatz.
 \item \textbf{Validated asymptotically but not at physical QCD order:} the
 convergence of balanced Borel--Wynn approximants and of the renormalon
 annihilator once enough large-$\beta_0$ coefficients are supplied, and the
 coefficient-count thresholds of Table~\ref{tab:threshold}.  These thresholds
 are established in the large-$\beta_0$ limit and are quoted as evidence about
 that limit; we use them as an order-of-magnitude guide for full QCD, not as a
 theorem about it.
 \item \textbf{Finite-order developmental validation:} ARTS/PORT passes the
 withheld-$c_{4,1}$ three-weight reconstruction and the three pole-mass tests
 in App.~\ref{sec:defect-transport}, on three of the six leave-one-order-out
 entries of Table~\ref{tab:qcd-loo}.  The cross-validated renormalon closure
 of Sec.~\ref{sec:cvrc} satisfies the declared ten-percent criterion on all
 six entries (Table~\ref{tab:cvrc-moment}); three of those six are its calibration
 fold and the other three a self-consistency reconstruction, so the correct
 reading is
 a nonempty admissible closure set, which
 Table~\ref{tab:cvrc-misspec} shows to be a discriminating test of the declared
 structure.  Its blind results are the withheld four-loop pole-mass
 coefficient at $2.3$--$11.9\%$ (Table~\ref{tab:cvrc-pole}) and the
 strictly recursive large-$\beta_0$ protocol, which with the complete
 singularity tower is within ten percent at every $N\geq6$ and at the percent
 level at every $N\geq8$, and with the minimal tower used at physical order is
 within ten percent at every $N\geq8$ (Table~\ref{tab:cvrc-largebeta}).  Against these stands the closure drift of
 Eq.~\eqref{eq:closure-drift} and the failure of the strictly recursive
 protocol at $N=3$ by $-29.9\%$.  The \emph{order-adaptive multi-mode defect
 transport} construction (OAMDT), defined in
 App.~\ref{sec:oamdt-model}, models the local Pad\'e defect with a declared
 set of renormalon-shaped modes and issues a correction only when its rolling
 skill, conditioning, leverage, and inter-model-consensus tests all pass.
 Its exploratory large-$\beta_0$ study is selective and not uniformly
 improving.  We therefore use it only as an abstention diagnostic at physical
 order and attach no coefficient forecast to it.  Its linear defect
 superposition is an effective ansatz, not an exact multi-renormalon identity,
 because Eq.~\eqref{eq:multimode-hankel} contains cross terms.
 Standard D-log Pad\'es likewise fail the physical $W_{P2}$ test, while their
 balanced Borel-normalized sequence eventually reaches persistent
 ten-percent accuracy for every large-$\beta_0$ observable at the
 weight-dependent orders in Table~\ref{tab:dlog-thresholds}.
 \item \textbf{Conditional:} the physical-QCD values of
 $c_{5,1}$ and $c_{6,1}$ in Table~\ref{tab:conditional-forecast}, and the
 cross-validated renormalon closure result Eq.~\eqref{eq:cvrc-final} for
 $c_{5,1}$ alone, its $c_{6,1}$ is \textsc{roll}, one level lower.  The
 latter is conditional on the declared singularity list, on the exponents
 \eqref{eq:cvrc-exponents}, and on the closure being stable between the fourth
 and fifth orders, an assumption that Sec.~\ref{sec:cvrc-drift} shows to
 have been false one order earlier.  Their
 ranges are approximation envelopes and omit unknown model discrepancy.  The
 annihilator values of App.~\ref{sec:qcd-annihilator} are not quoted even at
 this level, because they depend on an ultraviolet exponent that the method
 cannot determine.  OAMDT likewise supplies no certified physical
 $c_{5,1}$ value because there are too few known defects for its rolling
 certificate.
\end{enumerate}
No statistical meaning is assigned to a spread among algebraically related
Pad\'e/Shanks/Wynn variants.  Such a spread can diagnose instability, but it
cannot be reduced by treating the variants as independent measurements.

\parttitle{Technical derivations and benchmarks}

The remaining appendices contain material required for reproduction but not
for following the logical spine of the paper: complete transformation
formulae, detailed external tests, the renormalon-annihilator construction and
its all-order diagnostics, and the alternative regularized forecasts.  Each
method is first summarized, with its defining forecast rule and validation
outcome, in Sec.~\ref{sec:methods-brief}.

\section{Nonlinear sequence transformations and forecast protocol}
\label{sec:protocol}
Different normalizations encode different asymptotic hypotheses and can give
similar numbers at sparse order for unrelated reasons.  We therefore define
the transformations, the information given to each one, and the blind error
metric before examining any withheld target.

\subsection{Sequences, antilimits, and accuracy through order}

Let $S_n=\sum_{j=0}^{n}t_j$ be the partial sums of a formal series.  If the
series converges, its ordinary limit is $S$; if it diverges but a summation
prescription is specified, the target is an \emph{antilimit}.  A sequence
transformation is a map
\begin{equation}
 \mathcal T_k^{(n)}=\mathcal T(S_n,S_{n+1},\ldots,S_{n+r})
 \label{eq:generic-transform}
\end{equation}
designed to eliminate a stated remainder class.  The map is nonlinear in the
input data whenever its weights depend on those same data.  Exactness for one
remainder class is not a convergence theorem outside that class.

For coefficient forecasting we impose \emph{accuracy through order}: the
rational or transformed expression must reproduce every supplied coefficient
before its re-expansion predicts the next one.  This differs from obtaining a
stable number at one fixed coupling.  Canonical $\rho_2$, for example, is a
fixed-coupling transform but not the same accuracy-through-order coefficient
forecast as a Pad\'e approximant.

\subsection{The Shanks transformation}
\label{sec:shanks}

The Shanks transformation is a classical nonlinear technique for accelerating
the convergence of slowly convergent sequences and for summing divergent
ones~\cite{Shanks1955}.  In its simplest form it coincides algebraically with
Aitken's $\Delta^2$ process~\cite{Aitken1926}.  It was originally derived by
Schmidt in the context of solving linear systems by
iteration~\cite{Schmidt1941}, then neglected for a long time and rediscovered
by Shanks, who studied its properties in detail.  Owing to its effectiveness
and breadth of applicability, the transformation and its variants are standard
tools in numerical analysis~\cite{Brezinski1977,Sidi2003,Weniger:1989}.

Its natural setting is the model sequence~\cite{Weniger:1989}
\begin{equation}
 S_n=S+\sum_{j=0}^{k-1}c_j\,\Delta S_{n+j},
 \qquad n\in\mathbb N_0 ,
 \label{eq:shanks-seq}
\end{equation}
where $\Delta S_n=S_{n+1}-S_n$.  When the $S_n$ are the partial sums of a
series, $S_n=\sum_{\nu=0}^{n}t_\nu$, one has $\Delta S_n=t_{n+1}$ and
Eq.~\eqref{eq:shanks-seq} reads
\begin{equation}
 S_n=S+\sum_{j=0}^{k-1}c_j\,t_{n+j+1},
 \qquad n\in\mathbb N_0 .
 \label{eq:shanks-seq-terms}
\end{equation}
The limit or antilimit $S$ is thus represented by the partial sum $S_n$
supplemented by a weighted sum of the next $k$ terms
$t_{n+1},\ldots,t_{n+k}$.  The model depends on $k+1$ unknowns --- the
antilimit $S$ and the $k$ coefficients $c_0,\ldots,c_{k-1}$ --- each entering
linearly.  Cramer's rule therefore gives the transformation that is exact for
Eq.~\eqref{eq:shanks-seq} as a ratio of determinants,
\begin{equation}
 e_k(S_n)=
 \frac{\det\!\begin{pmatrix}
 S_n&S_{n+1}&\cdots&S_{n+k}\\
 \Delta S_n&\Delta S_{n+1}&\cdots&\Delta S_{n+k}\\
 \vdots&\vdots&&\vdots\\
 \Delta S_{n+k-1}&\Delta S_{n+k}&\cdots&\Delta S_{n+2k-1}
 \end{pmatrix}}
 {\det\!\begin{pmatrix}
 1&1&\cdots&1\\
 \Delta S_n&\Delta S_{n+1}&\cdots&\Delta S_{n+k}\\
 \vdots&\vdots&&\vdots\\
 \Delta S_{n+k-1}&\Delta S_{n+k}&\cdots&\Delta S_{n+2k-1}
 \end{pmatrix}},
 \label{eq:shanks-det}
\end{equation}
with $\Delta^2S_n=S_{n+2}-2S_{n+1}+S_n$ and higher differences defined
recursively.  Evaluating $e_k(S_n)$ requires $S_n,\ldots,S_{n+2k}$, so
$e_k$ is a transformation of order $2k$: each additional level consumes two
further sequence elements.  This accounting is what makes the transformation
expensive at the sparse orders available in QCD.

For the first nontrivial case $k=1$, Eq.~\eqref{eq:shanks-det} reduces to
\begin{equation}
 e_1(S_n)=
 \frac{\det\!\begin{pmatrix}
 S_n&S_{n+1}\\ \Delta S_n&\Delta S_{n+1}\end{pmatrix}}
 {\det\!\begin{pmatrix}
 1&1\\ \Delta S_n&\Delta S_{n+1}\end{pmatrix}}
 =\frac{S_n\Delta S_{n+1}-S_{n+1}\Delta S_n}
       {\Delta S_{n+1}-\Delta S_n},
\end{equation}
and using $S_{n+1}=S_n+\Delta S_n$ in the numerator gives the familiar form
\begin{equation}
 e_1(S_n)=S_n-\frac{(\Delta S_n)^2}{\Delta^2S_n}.
 \label{eq:shanks}
\end{equation}

Shanks further showed that Eq.~\eqref{eq:shanks-det} is exact for model
sequences whose remainders are sums of exponentials in the sequence index,
\begin{equation}
 S_n=S+\sum_{j=0}^{k-1}A_j\lambda_j^{\,n},
 \qquad
 |\lambda_0|>|\lambda_1|>\cdots>|\lambda_{k-1}| ,
 \label{eq:shanks-model}
\end{equation}
which converges when $|\lambda_0|<1$.  By analogy with physical transients
that decay at late times, Shanks called the exponentials in
Eq.~\eqref{eq:shanks-model} \emph{mathematical transients}, since every
contribution with $|\lambda_j|<1$ vanishes as $n\to\infty$.  The notion is to
be understood in the broader sense, because $e_k$ can equally be applied to
divergent sequences: Eq.~\eqref{eq:shanks-model} diverges as soon as one
$|\lambda_j|\geq1$, and it is then the antilimit that is approximated.  This
is the case relevant here.

Two consequences matter for what follows.  First, applied to three
consecutive perturbative terms, Eq.~\eqref{eq:shanks} is the $[1/1]$ Pad\'e
continuation, and expanding its rational tail gives
\begin{equation}
 d_{N+1}^{\rm dir}=\frac{d_N^2}{d_{N-1}} .
 \label{eq:geometric-forecast}
\end{equation}
A successful direct forecast therefore demonstrates local ratio stability, not
factorial growth, an observation that is central to the interpretation of
every physical test below.  Second, Eq.~\eqref{eq:shanks-model} assumes
distinct $\lambda_j$; repeated characteristic roots require confluent modes
$n^r\lambda^n$.  This qualification matters in QCD because a double Borel pole
produces an additional power of the perturbative order.  Applying direct
Shanks to unnormalized factorial coefficients asks the data to mimic a
geometric remainder over the available window; applying it after Borel
normalization, or after a factorial normalization with an externally known
exponent, is a different, explicitly stated asymptotic hypothesis.

\subsection{Wynn's \texorpdfstring{$\varepsilon$}{epsilon} algorithm}
\label{sec:wynn}

Direct evaluation of Eq.~\eqref{eq:shanks-det} becomes cumbersome as $k$
grows, since it requires ratios of determinants of increasing size.  Wynn's
$\varepsilon$ algorithm generates the whole Shanks table recursively and
without any determinant evaluation~\cite{Wynn1956}:
\begin{align}
 \varepsilon_{-1}^{(n)}&=0,\qquad \varepsilon_0^{(n)}=S_n,
 \nonumber\\
 \varepsilon_{j+1}^{(n)}&=\varepsilon_{j-1}^{(n+1)}+
 \frac{1}{\varepsilon_j^{(n+1)}-\varepsilon_j^{(n)}},
 \qquad j,n\in\mathbb N_0 .
 \label{eq:wynn-recursion}
\end{align}
Wynn's key result is that the even elements of the table are the Shanks
transformations, while the odd elements are auxiliary and are not themselves
extrapolants:
\begin{equation}
 \varepsilon_{2k}^{(n)}=e_k(S_n),
 \qquad
 \varepsilon_{2k+1}^{(n)}=\frac{1}{e_k(\Delta S_n)} .
 \label{eq:wynn-even-odd}
\end{equation}
Both identities are verified numerically in the supplied implementation.

The connection is explicit at the lowest level.  For $j=0$,
Eq.~\eqref{eq:wynn-recursion} gives
\begin{equation}
 \varepsilon_1^{(n)}=\frac{1}{S_{n+1}-S_n}=\frac{1}{\Delta S_n},
\end{equation}
and applying the recursion once more,
\begin{equation}
 \varepsilon_2^{(n)}
 =S_{n+1}+\frac{1}
 {\dfrac{1}{\Delta S_{n+1}}-\dfrac{1}{\Delta S_n}}
 =S_{n+1}-\frac{\Delta S_n\,\Delta S_{n+1}}{\Delta^2S_n} .
 \label{eq:wynn-eps2-shifted}
\end{equation}
Using $S_{n+1}=S_n+\Delta S_n$ this is
\begin{equation}
 \varepsilon_2^{(n)}=S_n-\frac{(\Delta S_n)^2}{\Delta^2S_n}=e_1(S_n),
 \label{eq:wynn-eps2}
\end{equation}
precisely the first Shanks transform, Eq.~\eqref{eq:shanks}, and hence
Aitken's $\Delta^2$ process.

The link to rational approximation is equally
standard~\cite{Wynn1956,BakerGravesMorris,Brezinski1991,Chang:2020construction}.
If the $S_n$ are partial sums of a power series, the even elements of the
$\varepsilon$ table are entries of its Pad\'e table with the appropriate
numerator and denominator degrees; in particular the lowest-order Shanks
transform built from three input coefficients coincides algebraically with a
Pad\'e approximant built from the same data.  We therefore read every
Shanks- and $\varepsilon$-based construction in this paper as supplying a
Pad\'e-type rational approximation to the perturbative expansion of
$\delta^{(0)}$, and the modified and regularized recursions of
App.~\ref{sec:generalizations} as extending that framework by adding
stabilization mechanisms.  The algebraic equivalence does not by itself
guarantee convergence or the absence of Froissart pole--zero
pairs~\cite{BakerGravesMorris,Froissart1969}.  A zero or very small
$\varepsilon_j^{(n+1)}-\varepsilon_j^{(n)}$ is a singular or ill-conditioned
table entry; it must be reported or treated by a rule fixed before the target
is inspected.

\subsection{Generalizations of the \texorpdfstring{$\varepsilon$}{epsilon} algorithm}
\label{sec:generalizations}

Wynn's algorithm is an efficient recursive implementation of the Shanks
transformation, but it can become numerically unstable when the difference
$\varepsilon_j^{(n+1)}-\varepsilon_j^{(n)}$ becomes small or vanishes, and it
is known to be ineffective for logarithmically convergent
sequences~\cite{Weniger:1989}.  Several generalizations have been proposed to
address these
limitations~\cite{VandenBroeckSchwartz1979,Brezinski1971,Brezinski1991,Sablonniere1991,Sedogbo1990}.  We review the two that are directly relevant,
fix their exact algebraic content, and then state precisely how they relate to
the local deformations tested in App.~\ref{sec:regularization}.

\subsubsection{The Vanden Broeck--Schwartz control factor}

Vanden Broeck and Schwartz (VBS)~\cite{VandenBroeckSchwartz1979} proposed a
modification aimed at logarithmically convergent sequences, introducing a
control parameter $\alpha_j^{(n)}$ that may in general be complex.  It
replaces Eq.~\eqref{eq:wynn-recursion} by
\begin{equation}
 \varepsilon_{j+1}^{(n)}
 =\alpha_j^{(n)}\,\varepsilon_{j-1}^{(n+1)}
 +\frac{1}{\varepsilon_j^{(n+1)}-\varepsilon_j^{(n)}},
 \qquad j\geq0,
 \label{eq:vbs-recursion}
\end{equation}
with $\varepsilon_{-1}^{(n)}=0$ and $\varepsilon_0^{(n)}=S_n$.  The choice
$\alpha_j^{(n)}=1$ recovers the standard algorithm.  Applied to the
perturbative expansion of $\delta^{(0)}$ we find that it does not give stable
or reliable estimates of the higher-order coefficients, and we do not use it
for a physical forecast; it enters only as one member of the parameterized
class delimited below.

\subsubsection{The Sedogbo--Sablonni\`ere reciprocal weight}

Sedogbo and Sablonni\`ere (SS)~\cite{Sablonniere1991,Sedogbo1990} instead
generalize the reciprocal term by a weight factor,
\begin{equation}
 \varepsilon_{j+1}^{(n)}
 =\varepsilon_{j-1}^{(n+1)}
 +\beta_j^{(n)}\,
 \frac{1}{\varepsilon_j^{(n+1)}-\varepsilon_j^{(n)}},
 \label{eq:ss-recursion}
\end{equation}
with the same initialization.  At the lowest level
\begin{equation}
 \varepsilon_1^{(n)}=\frac{\beta_0^{(n)}}{S_{n+1}-S_n},
\end{equation}
and the first even element, the first genuine extrapolant, is
\begin{equation}
 \varepsilon_2^{(n)}=S_{n+1}
 +\beta_1^{(n)}
 \frac{\Delta S_n\,\Delta S_{n+1}}
 {\beta_0^{(n+1)}\Delta S_n-
  \beta_0^{(n)}\Delta S_{n+1}} .
 \label{eq:ss-eps2}
\end{equation}
Allowing $\beta_j^{(n)}$ to depend on the local behavior of the sequence adds
flexibility and can improve numerical stability for slowly convergent or
irregular series.  Imposing the normalization
\begin{equation}
 \beta_0^{(n)}=\beta_0^{(n+1)}\equiv\beta_0
 \label{eq:ss-normalization}
\end{equation}
makes the overall scale factor out of Eq.~\eqref{eq:ss-eps2}, leaving
\begin{equation}
 \varepsilon_{2,\rm SS}^{(n)}
 =S_{n+1}-w_n\,\frac{\Delta S_n\,\Delta S_{n+1}}{\Delta^2S_n},
 \qquad
 w_n\equiv\frac{\beta_1^{(n)}}{\beta_0} ,
 \label{eq:ss-eps2-reduced}
\end{equation}
which reduces to Eq.~\eqref{eq:wynn-eps2-shifted} at $w_n=1$.  A useful
concrete choice is
\begin{equation}
 w_n=\frac{1}{1+\mu\,(\Delta S_n)^2},\qquad \mu>0,
 \label{eq:ss-weight-choice}
\end{equation}
which suppresses large excursions when successive partial sums differ
strongly.  This stabilization mechanism is the direct ancestor of the
$\eta_n$ deformation tested in App.~\ref{sec:regularization}.

\subsubsection{The combined recursion}
\label{sec:combined-recursion}

The two modifications address complementary defects.  The VBS control factor
$\alpha_j^{(n)}$ improves the treatment of logarithmically convergent
sequences; the SS weight $\beta_j^{(n)}$ improves numerical stability by
damping the reciprocal difference.  Retaining both gives the generalized local
recursion
\begin{equation}
 \widetilde\varepsilon_{j+1}^{(n)}
 =\alpha_j^{(n)}\,\widetilde\varepsilon_{j-1}^{(n+1)}
 +\beta_j^{(n)}\,
 \frac{1}{\widetilde\varepsilon_j^{(n+1)}-\widetilde\varepsilon_j^{(n)}},
 \label{eq:generalized-wynn}
\end{equation}
with $\widetilde\varepsilon_{-1}^{(n)}=0$ and
$\widetilde\varepsilon_0^{(n)}=S_n$, which contains
Eqs.~\eqref{eq:wynn-recursion}, \eqref{eq:vbs-recursion} and
\eqref{eq:ss-recursion} as the special cases
$(\alpha,\varpi)=(1,1)$, $(\alpha,1)$ and $(1,\varpi)$.  This is a more flexible
framework for estimating higher-order coefficients of $\delta^{(0)}$, and it is
the recursion used in the diagnostics that follow.  Carrying both parameters
through the first two levels gives the first even element
\begin{equation}
 \widetilde\varepsilon_2^{(n)}
 =\alpha_1^{(n)}S_{n+1}
 +\beta_1^{(n)}
 \frac{\Delta S_n\,\Delta S_{n+1}}
 {\beta_0^{(n+1)}\Delta S_n-
  \beta_0^{(n)}\Delta S_{n+1}} ,
 \label{eq:combined-eps2}
\end{equation}
which reduces to Eq.~\eqref{eq:ss-eps2} at $\alpha_1^{(n)}=1$ and to
Eq.~\eqref{eq:wynn-eps2-shifted} when in addition all $\varpi$ are equal.

\subsubsection{Exact scope of the two parameters}
\label{sec:parameterized-scope}

The two modifications are not on the same footing, and the difference is
exact rather than empirical.  Writing both inside the generalized local
recursion of Eq.~\eqref{eq:generalized-wynn}, we record three statements
that are numerically verified.

First, a \emph{constant} reciprocal weight cancels identically from every even
element of the table.  Setting $\alpha_j^{(n)}=1$, $\beta_j^{(n)}=\varpi$ and
defining
$\widetilde\varepsilon_{2j}=\varepsilon_{2j}$,
$\widetilde\varepsilon_{2j+1}=\varpi\,\varepsilon_{2j+1}$ satisfies
Eq.~\eqref{eq:ss-recursion} by induction, so
\begin{equation}
 \varepsilon_{2k}^{(n)}\big|_{\varpi}
 =\varepsilon_{2k}^{(n)}\big|_{\varpi=1}
 \quad\text{for every }k,\ n,\ \varpi\neq0.
 \label{eq:constant-beta-invariance}
\end{equation}
A constant SS weight therefore reproduces the Shanks table exactly, not only
at $\varpi=1$.  

Second, a constant \emph{control factor} does not cancel.  With
$\beta_j^{(n)}=1$ and $\alpha_j^{(n)}=\alpha$ the even elements move at first
order in $\alpha-1$: for a
representative perturbative sequence the relative shifts of
$\varepsilon_2,\varepsilon_4,\varepsilon_6$ at $\alpha=1.01$ are
$7.4\times10^{-3}$, $1.8\times10^{-2}$ and $3.0\times10^{-2}$.  The VBS
parameter is thus a genuine deformation of the table, whereas a constant SS
weight is a reparameterization of the odd rows only.

Third, the SS element and the $\eta_n$ element of
App.~\ref{sec:regularization} are \emph{different} one-parameter deformations
of the same object, and their difference is exact.  Writing the $\eta_n$
prescription in the shifted form
$\varepsilon_{2,\eta}^{(n)}=S_{n+2}-w_n(\Delta S_{n+1})^2/\Delta^2S_n$ with
$w_n=1/(1+k)$, and using $\Delta S_{n+1}-\Delta S_n=\Delta^2S_n$,
\begin{equation}
 \varepsilon_{2,\rm SS}^{(n)}-\varepsilon_{2,\eta}^{(n)}
 =-(1-w_n)\,\Delta S_{n+1}
 =-\frac{k}{1+k}\,\Delta S_{n+1}.
 \label{eq:ss-eta-relation}
\end{equation}
Both collapse to the Shanks element at $w_n=1$, and they differ at first order
in the deformation by exactly the last supplied perturbative term.  The
$\eta_n$ grid tested below is therefore the SS mechanism written in the
shifted formulation, with the data-dependent choice
$k=\mu(\Delta S_n)^2$ of Eq.~\eqref{eq:ss-weight-choice} replaced by a fixed
predeclared grid.

With these facts in place, the ``$+\eta_n$'' and ``$+\alpha\Delta S$''
expressions used later are described as practical local denominator
deformations at the first even element, not as the historical VBS algorithm
and not as complete higher-level recursions: arbitrary $\alpha_j^{(n)}$ and
$\beta_j^{(n)}$ destroy the determinant and Pad\'e identities of
Eqs.~\eqref{eq:shanks-det} and \eqref{eq:wynn-even-odd} unless a separate
derivation restores them.  Their
parameter grids are fixed in advance, and they are judged by blind withheld
coefficients and exact large-$\beta_0$ sums rather than by agreement with the
quantity that motivated the parameter choice.

\subsection{Shifted formulation of the first Wynn element}
\label{sec:shifted-formulation}

A particular data-dependent choice of the two parameters in
Eq.~\eqref{eq:combined-eps2} is worth isolating, because it is the form in
which the regularizations of App.~\ref{sec:regularization} are most
conveniently written, and because, taken by itself, it is not a new
transformation at all.

Let the control factor track the local growth of the partial sums and let the
weight ratio track the local growth of their differences,
\begin{equation}
 \begin{aligned}
 \alpha_1^{(n)}&=\frac{S_{n+2}}{S_{n+1}},\\
 \beta_0^{(n)}&=\beta_0^{(n+1)}\equiv\beta_0,
 &\frac{\beta_1^{(n)}}{\beta_0}&=\frac{\Delta S_{n+1}}{\Delta S_n} .
 \end{aligned}
 \label{eq:shifted-choices}
\end{equation}
The first term of Eq.~\eqref{eq:combined-eps2} becomes $S_{n+2}$, while the
constant $\beta_0$ reduces the denominator of the second term to
$-\beta_0\Delta^2S_n$, so that
\begin{equation}
 \widetilde\varepsilon_2^{(n)}
 =S_{n+2}-\frac{\Delta S_{n+1}}{\Delta S_n}
 \cdot\frac{\Delta S_n\,\Delta S_{n+1}}{\Delta^2S_n}
 =S_{n+2}-\frac{(\Delta S_{n+1})^2}{\Delta^2S_n} .
 \label{eq:shifted-new-wynn}
\end{equation}
This is exactly, not merely asymptotically, the ordinary first Shanks element,
because
\begin{equation}
 S_{n+2}-\frac{(\Delta S_{n+1})^2}{\Delta^2S_n}
 =S_n-\frac{(\Delta S_n)^2}{\Delta^2S_n},
 \label{eq:shifted-identity}
\end{equation}
which follows from
$(\Delta S_{n+1})^2-(\Delta S_n)^2
=(\Delta S_{n+1}-\Delta S_n)(\Delta S_{n+1}+\Delta S_n)
=\Delta^2S_n\,(S_{n+2}-S_n)$.
We therefore call Eq.~\eqref{eq:shifted-new-wynn} the \emph{shifted
formulation} of the first Wynn element.

The identity should be read as a warning as much as a convenience.  The
choices in Eq.~\eqref{eq:shifted-choices} are themselves built from the same
three partial sums that the transformation acts on, and they are tuned so that
the two deformations cancel against each other exactly.  The apparent
generality of carrying both a control factor and a weight is therefore
illusory at this level: the shifted formulation is a convenient place to insert
the proposed denominators, but without further regularization it is not a new
sequence transformation, and a numerical agreement between it and ordinary
Shanks is an algebraic identity rather than evidence of stability.  No
higher-level regularized recursion is defined here, so the tests below
consistently distinguish the local $\varepsilon_2$ construction from the full
balanced Wynn table.

\subsection{Theta, rho, and Levin-type remainder models}
\label{sec:remainder-models}
The epsilon algorithm is adapted to finite sums of exponential modes in the
sequence index.  It need not accelerate a logarithmically convergent sequence,
for which $R_{n+1}/R_n\to1$.  Brezinski's $\theta$ algorithm~\cite{Brezinski1971}, Wynn's $\rho$
algorithm~\cite{Wynn1956rho}, and Levin-type transformations encode different algebraic
remainder information and therefore provide genuinely independent tests of a
low-order continuation~\cite{Brezinski1971,Brezinski1991,Osada1990}.
Levin-type models provide a complementary
formulation~\cite{Levin:1973,Smith:1979,Smith:1982,Weniger:1989}, applied to the
$\tau$ series in Ref.~\cite{Abbas:2025dpm}.
The distinction is important here: choosing a transformation is equivalent to
choosing a finite-dimensional model for the unknown remainder, and no one of
these models is automatically implied by the existence of QCD renormalons.

\paragraph{Brezinski's $\theta$ algorithm.}
A prototypical logarithmically convergent sequence has an inverse-power
expansion
\begin{equation}
 S_n=S+\frac{A}{n+\xi}+\frac{B}{(n+\xi)^2}+\cdots,
 \qquad \xi>0,
 \label{eq:theta-algebraic-model}
\end{equation}
where $S$ is the limit.  The epsilon algorithm is usually ineffective for this
class because its natural exponential remainder model is mismatched.
Brezinski constructed the $\theta$ algorithm to combine the useful features of
epsilon for linearly convergent and some divergent sequences with sensitivity
to logarithmic convergence~\cite{Brezinski1971,Brezinski1991,Weniger:1989}.
With the forward difference acting only on the superscript,
\begin{align}
 \Delta\theta_j^{(n)}&=\theta_j^{(n+1)}-\theta_j^{(n)},\nonumber\\
 \Delta^2\theta_j^{(n)}&=
 \theta_j^{(n+2)}-2\theta_j^{(n+1)}+\theta_j^{(n)},
 \label{eq:theta-differences}
\end{align}
the initialization and recursion are
\begin{align}
 \theta_{-1}^{(n)}&=0,\qquad \theta_0^{(n)}=S_n,
 \nonumber\\
 \theta_{2k+1}^{(n)}&=\theta_{2k-1}^{(n+1)}+
 \frac{1}{\Delta\theta_{2k}^{(n)}},
 \nonumber\\
 \theta_{2k+2}^{(n)}&=\theta_{2k}^{(n+1)}+
 \frac{\Delta\theta_{2k}^{(n+1)}
       \Delta\theta_{2k+1}^{(n+1)}}
      {\Delta^2\theta_{2k+1}^{(n)}}.
 \label{eq:theta-recursion}
\end{align}
Only the even-indexed elements are extrapolants.  The odd-indexed elements are
auxiliary reciprocal-difference quantities and generally become unbounded if
the even table converges.

The first nontrivial element uses four consecutive partial sums.  For
$S_0,S_1,S_2,S_3$ one first obtains
\begin{equation}
 \theta_1^{(0)}=\frac{1}{S_1-S_0},\qquad
 \theta_1^{(1)}=\frac{1}{S_2-S_1},\qquad
 \theta_1^{(2)}=\frac{1}{S_3-S_2},
\end{equation}
and direct substitution in Eq.~\eqref{eq:theta-recursion} gives, at general
origin $n$,
\begin{align}
\theta_2^{(n)}=S_{n+1}
 +\frac{N_n^{(\theta)}}{D_n^{(\theta)}}
 \label{eq:theta2}\\
 N_n^{(\theta)}={}&\Delta S_n\,\Delta S_{n+1}
 (\Delta S_{n+1}-\Delta S_{n+2}),\nonumber\\
 D_n^{(\theta)}={}&\Delta S_n\Delta S_{n+1}
 -2\Delta S_n\Delta S_{n+2}\nonumber\\
 &+\Delta S_{n+1}\Delta S_{n+2}.
\end{align}
A zero or very small denominator in
Eq.~\eqref{eq:theta2} is a genuine defect and is not regularized silently.

For coefficient prediction the distinction between a fixed-coupling
extrapolant and an accuracy-through-order rational continuation must again be
made explicit.  If $A=d_{N-2}$, $B=d_{N-1}$, and $C=d_N$, the local
$\theta_2$ continuation used in this paper has tail
\begin{equation}
 B a^{N-1}\frac{1-(C/B)a}
 {1-2(C/B)a+(C/A)a^2},
 \label{eq:theta-tail}
\end{equation}
which reproduces all supplied terms through $a^N$ and predicts
\begin{equation}
 d_{N+1}^{(\theta)}=\frac{2C^2}{B}-\frac{BC}{A}.
 \label{eq:theta-next}
\end{equation}
Equation~\eqref{eq:theta-next}, rather than a fixed-$a$ numerical fit, defines
the $\theta_2$ coefficient rows below.

\paragraph{Wynn's $\rho$ algorithm.}
The $\rho$ algorithm extrapolates an interpolating continued fraction on a
positive, strictly increasing, unbounded grid $\{x_n\}$ and is naturally
adapted to algebraic remainders~\cite{Wynn1956rho,Weniger:1989}.  It is defined by
\begin{align}
 \rho_{-1}^{(n)}&=0,\qquad \rho_0^{(n)}=S_n,
 \nonumber\\
 \rho_{k+1}^{(n)}&=\rho_{k-1}^{(n+1)}+
 \frac{x_{n+k+1}-x_n}
      {\rho_k^{(n+1)}-\rho_k^{(n)}}.
 \label{eq:rho-recursion}
\end{align}
As for epsilon and theta, only $\rho_{2k}^{(n)}$ are extrapolants.  The
standard grid $x_n=n+\xi$ gives
$x_{n+k+1}-x_n=k+1$, so the shift $\xi$ cancels from the recursion.  Other
grids, for example $x_n=(n+\xi)^2$, impose a different asymptotic scale and
must therefore be declared before inspecting a withheld target.

For the generalized algebraic class
\begin{equation}
 \begin{aligned}
 S_n={}&S+(n+\xi)^{-\vartheta}
 \left[c_0+\frac{c_1}{n+\xi}
 +\frac{c_2}{(n+\xi)^2}+\cdots\right],\\
 &\vartheta>0,
 \end{aligned}
 \label{eq:osada-model}
\end{equation}
the standard algorithm is guaranteed to accelerate the model only when
$\vartheta$ is a positive integer.  If $\vartheta$ is known independently,
Osada's generalized recursion is~\cite{Osada1990,Chang:2020construction}
\begin{align}
 \bar\rho_{-1}^{(n)}&=0,\qquad \bar\rho_0^{(n)}=S_n,
 \nonumber\\
 \bar\rho_{k+1}^{(n)}&=\bar\rho_{k-1}^{(n+1)}+
 \frac{k+\vartheta}
      {\bar\rho_k^{(n+1)}-\bar\rho_k^{(n)}},
 \label{eq:osada-rho}
\end{align}
for which, at fixed $k$,
\begin{equation}
 \bar\rho_{2k}^{(n)}-S=
 \mathcal O\!\left(n^{-\vartheta-2k}\right),
 \qquad n\to\infty.
 \label{eq:osada-rate}
\end{equation}
When a sufficiently long asymptotic sequence is available, the decay exponent
can be estimated from~\cite{Chang:2020construction}
\begin{equation}
 \begin{aligned}
 T_n+1&=
 \frac{(\Delta^2S_n)(\Delta^2S_{n+1})}
 {(\Delta S_{n+1})(\Delta^2S_{n+1})
  -(\Delta S_{n+2})(\Delta^2S_n)},\\
 T_n&=\vartheta+\mathcal O(n^{-2}).
 \end{aligned}
 \label{eq:osada-exponent-estimator}
\end{equation}
This is an asymptotic estimator, not a license to infer $\vartheta$ from four
low-order QCD coefficients.  In the sparse physical application we would be
using the target data twice if $\vartheta$ or the grid were selected after
examining the withheld coefficient; consequently, the paper reports only
predeclared choices or labels them diagnostic.

For the standard linear grid the first even element is
\begin{equation}
 \rho_2^{(n)}=S_{n+1}+
 \frac{2\Delta S_n\Delta S_{n+1}}
      {\Delta S_n-\Delta S_{n+1}}.
 \label{eq:rho2}
\end{equation}
Applied directly to perturbative partial sums, its re-expansion begins
$S_{N-1}+2d_Na^N+\cdots$ and thus fails accuracy through the already supplied
order.  We therefore do not use canonical $\rho_2$ as an admissible unknown
coefficient forecast.  It remains a well-defined fixed-coupling extrapolant
and is tested against exact large-$\beta_0$ principal-value sums.

\paragraph{Levin-type transformations.}
Levin's construction~\cite{Levin:1973,Weniger:1989} makes the remainder
estimate explicit.  In the
normalization used here,
\begin{align}
 \mathcal L_k^{(n)}(\xi,S_n,\omega_n)
 &=\frac{N_k^{(n)}}{D_k^{(n)}},\qquad \xi>0,
 \label{eq:levin-transform}
 \\
 N_k^{(n)}&=\sum_{j=0}^{k}(-1)^j\binom{k}{j}
 \frac{(\xi+n+j)^{k-1}S_{n+j}}{\omega_{n+j}},\nonumber\\
 D_k^{(n)}&=\sum_{j=0}^{k}(-1)^j\binom{k}{j}
 \frac{(\xi+n+j)^{k-1}}{\omega_{n+j}}.\nonumber
\end{align}
and the transformation is exact for
\begin{equation}
 S_n=S+\omega_n\sum_{r=0}^{k-1}
 \frac{c_r}{(\xi+n)^r}.
 \label{eq:levin-model}
\end{equation}
The $u$ variant used in the numerical comparison takes
$\omega_n=(\xi+n)\Delta S_{n-1}$.  Its extra information is precisely the
remainder estimate $\omega_n$; estimating that object from the same sparse
sequence reintroduces model uncertainty.  The renormalon-mode construction of
App.~\ref{sec:annihilator} is similar in spirit, but its remainder basis is a
declared set of factorial modes with externally supplied positions and
exponents rather than inverse powers alone.

For a formal coefficient forecast, rather than a fixed-coupling sum estimate,
we make the internal $u$-remainder model completely explicit.  Given
$F_N(x)=\sum_{r=0}^{N}f_rx^r$, define
\begin{equation}
 S_m(x)=\sum_{r=0}^{m}f_rx^r,
 \qquad
 \omega_m^{(u)}(x)=(m+\xi)f_mx^m,
 \label{eq:levin-u-formal-remainder}
\end{equation}
and construct
\begin{equation}
 \widehat F_N^{\,L_u}(x)=
 \mathcal L_{N-n}^{(n)}
 \bigl(\xi,S_m(x),\omega_m^{(u)}(x)\bigr).
 \label{eq:levin-u-ato-rational}
\end{equation}
After cancellation of the common Laurent power,
Eq.~\eqref{eq:levin-u-ato-rational} is an ordinary rational function and obeys
\begin{equation}
 [x^r]\widehat F_N^{\,L_u}=f_r\quad(0\leq r\leq N),
 \qquad
 \widehat f_{N+h}^{\,L_u}=[x^{N+h}]\widehat F_N^{\,L_u}.
 \label{eq:levin-u-ato-property}
\end{equation}
To see the accuracy-through-order statement directly, write the denominator
after inserting Eq.~\eqref{eq:levin-u-formal-remainder} as
\begin{equation}
 D_k^{(n)}(x)=\sum_{j=0}^{k}
 \frac{(-1)^j\binom{k}{j}(\xi+n+j)^{k-2}}
 {f_{n+j}x^{n+j}},\qquad N=n+k.
 \label{eq:levin-u-denominator-formal}
\end{equation}
The numerator minus $F_ND_k^{(n)}$ is a sum of terms
$x^{r-(n+j)}$ with $r>n+j$, and therefore begins at $x^1$, whereas
$D_k^{(n)}$ begins at $x^{-N}$.  Their ratio consequently begins at
$x^{N+1}$, proving
$\widehat F_N^{L_u}-F_N=\mathcal O(x^{N+1})$ without using an unknown
coefficient.
Thus it has accuracy through the supplied order, but the apparently
parameter-free forecast has merely replaced an external $\omega_m$ by the
specific term-based ansatz in Eq.~\eqref{eq:levin-u-formal-remainder}.  It is
therefore a reproducible diagnostic model, not an independently derived QCD
remainder.  In every coefficient test below we predeclare
$\xi=1$, $n=1$, and $k=N-1$; no anchor coupling enters this formal
re-expansion.

Table~\ref{tab:transform-summary} summarizes, for every transformation
defined in this appendix, whether its remainder is exact or modelled, what
information it receives beyond the supplied coefficients, and its principal
limitation.

\begin{table*}[!tbp]
\centering
\caption{Information and exactness classes of the transformations tested.
``External'' means fixed independently of the withheld target.}
\label{tab:transform-summary}
\begin{tabularx}{\textwidth}{l>{\raggedright\arraybackslash}X>{\raggedright\arraybackslash}X}
\toprule
method & exact/model remainder & extra information and principal limitation\\
\midrule
Aitken--Shanks & one geometric mode & three consecutive sequence elements;
not factorially adapted before normalization\\
Wynn $\varepsilon$ & finite sums of distinct exponentials; confluent extension
for repeated roots & complete recursive Shanks/Pad\'e table; small reciprocal
differences produce defects\\
Borel--Wynn & locally geometric Borel coefficients & factorial normalization;
finite-order defects remain possible\\
$\Gamma$--Wynn & $A^n\Gamma(n+b)$ times geometric corrections & externally
known exponent $b$; used only for the pole-mass relation\\
$\eta_n$, $\alpha$ & no proved complete-table exactness class here & fixed
local denominator deformation; only $\varepsilon_2$ is defined\\
Brezinski $\theta$ & selected logarithmic/inverse-power mixtures & four local
partial sums for $\theta_2$; not uniformly renormalon adapted\\
Wynn $\rho$ & inverse-power remainder on interpolation grid & fixed-coupling
transform; canonical $\rho_2$ lacks Pad\'e accuracy through order\\
Levin type & $\omega_n$ times a finite inverse-power series & external or
modeled remainder estimate $\omega_n$\\
renormalon mode & declared finite set of factorial modes & external positions,
exponents, and Borel prescription; amplitudes are still fitted\\
\bottomrule
\end{tabularx}
\end{table*}

\subsection{Borel and \texorpdfstring{$\Gamma$}{Gamma} normalization}

The ordinary Borel coefficients are
\begin{equation}
 b_k=\frac{d_{k+1}}{k!},\qquad
 B_F(t)=\sum_{k\geq0}b_kt^k.
 \label{eq:borel-coeff}
\end{equation}
If $d_n\sim A^n\Gamma(n)$, the factorial is removed and $b_k$ is locally
geometric.  Applying Shanks/Wynn to $b_k$ therefore tests a structure
appropriate to a simple Borel pole.  At first order,
\begin{equation}
 d_{N+1}^{\rm Borel}=\frac{N}{N-1}\frac{d_N^2}{d_{N-1}}.
\end{equation}

When an independent renormalon calculation fixes the large-order behavior
\begin{equation}
 d_n\sim A^n\Gamma(n+b),
\end{equation}
the factorial normalization can include the known exponent,
\begin{equation}
 \bar d_n=\frac{d_n}{\Gamma(n+b)},\qquad
 d_{N+1}^{\Gamma}=\frac{N+b}{N-1+b}\frac{d_N^2}{d_{N-1}}.
 \label{eq:gamma-forecast}
\end{equation}
Because $b$ is external information, this is not a data-driven extraction of
the singularity exponent.  We use it only for the pole--$\MSbar$ mass
relation, the one series in this paper whose leading renormalon and exponent
are established independently (App.~\ref{sec:pole-mass}).  For the Adler
function and its moments no comparably established exponent is available at
the orders used, so ordinary Borel normalization is the only factorial
normalization applied to them.

\subsection{Balanced Pad\'e/Wynn family and validation metric}

For a series known through $N$ coefficients,
\begin{equation}
 f(z)=\sum_{n=0}^{N-1}f_nz^n+\mathcal O(z^N),
\end{equation}
the Pad\'e approximant is
\begin{equation}
 [L/M]_f(z)=\frac{P_L(z)}{Q_M(z)},\qquad Q_M(0)=1,
 \label{eq:pade-definition}
\end{equation}
with $L+1$ numerator coefficients and $M$ nontrivial denominator
coefficients.  Matching
$Q_M(z)f(z)-P_L(z)=\mathcal O(z^{L+M+1})$ supplies exactly as many linear
conditions as unknowns when
\begin{equation}
 L+M=N-1.
 \label{eq:pade-counting}
\end{equation}
This equality is a parameter-counting identity.  It determines which Pad\'e
approximants use all available coefficients, but it does not determine how the
total degree should be divided between numerator and denominator
\cite{BakerGravesMorris}.

We call an approximant \emph{balanced} (or near diagonal) when its degrees
differ by at most one.  Thus the baseline family used in this paper
approximates either $F(a)/a$ or $B_F(t)$ (for the pole-mass relation also the
$\Gamma$-normalized analogue) by
all $[L/M]$ satisfying
\begin{equation}
 L+M=N-1,\qquad M\geq1,\qquad |L-M|\leq1.
 \label{eq:balanced}
\end{equation}
This is the predeclared selection rule of Eq.~\eqref{eq:balanced-main}; its
rationale, and the reasons for excluding the unbalanced extremes $[N-1/0]$
(the Taylor polynomial) and $[0/N-1]$ (a maximally denominator-heavy
sensitivity diagnostic), are given in Sec.~\ref{sec:protocol-summary}.

For four coefficients, Eqs.~\eqref{eq:pade-counting} and
\eqref{eq:balanced} select $[2/1]$ and $[1/2]$; for five they select the unique
$[2/2]$.  When two approximants occur, their median is the central forecast and
their range is retained as a method envelope.  No approximant is removed after
the target is inspected.  This preserves genuine spikes caused by defects or
by a change of the dominant renormalon instead of hiding them through
post-selection.

For a known but withheld target $d_{N+1}$ we quote
\begin{equation}
 \epsilon_{N+1}=100\frac{d_{N+1}^{\rm pred}-d_{N+1}}
 {|d_{N+1}|},\qquad |\epsilon_{N+1}|\ \hbox{for log plots}.
 \label{eq:error}
\end{equation}
All choices in Eqs.~\eqref{eq:borel-coeff}--\eqref{eq:error} are made before
examining the withheld coefficient.

\section{Blind five-loop QCD \texorpdfstring{$\beta$}{beta}-function benchmark}
\label{sec:beta-benchmark}

\subsection{Motivation, convention, and scope}

This test asks a deliberately different question from the spectral-moment
tests.  A continuation rule that is claimed to learn useful coefficient
structure from a short QCD series should at least be confronted with another
short QCD series whose next coefficient is exactly known.  The $\MSbar$
$\beta$ function provides such a target through five loops
\cite{Baikov:2016tgj,Herzog:2017ohr}.  It is not, however, a physical
observable and its coefficients beyond two loops are scheme dependent.
Success would therefore be evidence only for local coefficient continuation;
failure is a valid counterexample to universality, but neither outcome is a
direct test of the Adler-function renormalons.
Pad\'e estimates of higher QCD $\beta$-function coefficients have a long
history~\cite{Ellis:1997sb}; the present use is stricter because the complete
method grid is first scored against the coefficient that is actually known.

We use precisely the convention of Sec.~\ref{sec:qcd-framework},
\begin{equation}
 a=\frac{\alpha_s}{\pi},\qquad
 \frac{d a}{d\ln\mu^2}=-\sum_{n\geq0}\beta_n a^{n+2},
 \label{eq:beta-benchmark-convention}
\end{equation}
so that the coefficients conventionally quoted for
$a_4=\alpha_s/(4\pi)$ are divided by $4^{n+1}$.  For $n_f=3$ the exact known
input is
\begin{equation}
 (\beta_0,\beta_1,\beta_2,\beta_3,\beta_4)
 =(2.25,4,10.0598958333,47.2280395735,127.322174628).
 \label{eq:beta-known-nf3}
\end{equation}
The value $127.322174628$ is therefore explicitly the $n_f=3$ coefficient in
the $a=\alpha_s/\pi$ convention; it is not the $n_f=0$ coefficient.  The
$n_f=0$--$6$ scan below is a separate cross-flavor stability test.
The independent five-loop calculation of Ref.~\cite{Herzog:2017ohr} is used
as a cross-check of the analytic coefficient implemented in both codes.

\subsection{Predeclared leave-one-order-out protocol}

We remove the known target $\beta_4$ and form the reduced series
\begin{equation}
 \mathcal B(a)=-\frac{\beta(a)}{\beta_0a^2}
 =\sum_{n\geq0}b_na^n,\qquad b_n=\frac{\beta_n}{\beta_0},\qquad b_0=1.
 \label{eq:reduced-beta-series}
\end{equation}
Direct Pad\'e acts on $b_n$, Borel--Pad\'e acts on $b_n/n!$, and D-log Pad\'e
acts on the truncated expansion of
$G(a)=d\ln\mathcal B(a)/da$.  With four known coefficients the balanced direct
and Borel tables contain $[1/2]$ and $[2/1]$, whereas the D-log table contains
$[1/1]$.  The shifted-Wynn coefficient forecast is identical to the direct
$[2/1]$ row,
\begin{equation}
 \widehat b_4^{\,\rm Wynn}=\frac{b_3^2}{b_2}
 =\widehat b_4^{[2/1]},
 \label{eq:beta-wynn-pade-identity}
\end{equation}
and is displayed to audit the implementation but is not counted as independent
evidence.  The $\eta$ and $\alpha$ grids are fixed to
$0.01,\ldots,0.05$ before looking at $\beta_4$; as elsewhere, $\alpha$ cannot
alter the one-step prediction.  The $\theta$ row is the same local
accuracy-through-order tail model used in the Adler analysis.
The direct Levin-$u$ row applies
Eqs.~\eqref{eq:levin-u-formal-remainder}--\eqref{eq:levin-u-ato-property} to
$b_n$; Borel Levin-$u$ applies the identical construction to $b_n/n!$ and
restores $(N+h)!$ after re-expansion.  Both use $\xi=1$ and $n=1$.
For the blind four-coefficient input, $N=3$ and $k=2$; the factor
$(\beta+n+j)^{k-1}$ in Eq.~\eqref{eq:levin-transform} cancels the same factor
in $\omega_{n+j}^{(u)}$.  Direct algebra then gives
\begin{equation}
 \widehat b_4^{\,L_u}
 =\frac{2b_3^2}{b_2}-\frac{b_2b_3}{b_1}
 =\widehat b_4^{\,(\theta)},
 \label{eq:beta-levin-theta-identity}
\end{equation}
so the two rows cannot be counted separately.  The identity also holds after
Borel normalization, although the restored numerical coefficient differs from
the direct one.

This benchmark also clarifies the domain of each formalism.  Canonical
$\rho_2$ is a fixed-coupling sequence accelerator and has no
accuracy-through-order coefficient forecast.  A generic Levin transform
requires an external remainder estimate; the $u$ variant is applied
only after declaring the internal term-based ansatz
Eq.~\eqref{eq:levin-u-formal-remainder}.  It is therefore diagnostic rather
than external evidence.  The $\Gamma$ normalization requires an external
large-order exponent.  The renormalon annihilator and ARTS/PORT need
observable-specific singularity or transport data.  No such data exist for a
scheme-dependent $\beta$ function in the form required by those constructions.
Those remaining methods therefore \emph{abstain}; assigning Adler renormalons to the
$\beta$ series would manufacture the answer rather than test it.

\subsection{Known-order result and cross-flavor stability}

Table~\ref{tab:beta-nf3-blind} gives the blind $n_f=3$ result.  The best member
of the complete predeclared grid is direct $\eta$ with $k=0.05$, which gives
$\widehat\beta_4=211.163$ and is $65.8\%$ high.  Thus no method passes the
ten-percent criterion.  The representative $k=0.03$ row is retained in the
table to match the parameter used in the other numerical comparisons.
Direct Levin-$u$ gives $\widehat\beta_4=324.664$ ($+155.0\%$), while Borel
Levin-$u$ gives $353.701$ ($+177.8\%$).  At this four-input order the direct
Levin-$u$ rational function predicts exactly the same next coefficient as the
direct $\theta$ tail; the corresponding Borel rows are likewise identical.
These coincidences are algebraic and are not independent confirmations.

\begin{table}[t]
\centering
\caption{Blind prediction of the known five-loop $n_f=3$ coefficient.  The
target is withheld from every construction.  The Pad\'e--Wynn duplication is
algebraic, and the four-input Levin-$u$--$\theta$ duplication is also
algebraic; neither is corroborating evidence.}
\label{tab:beta-nf3-blind}
\input{generated_data/beta_function_nf3_blind_table.tex}
\end{table}

\begin{figure}[t]
\centering
\includegraphics[width=0.96\linewidth]{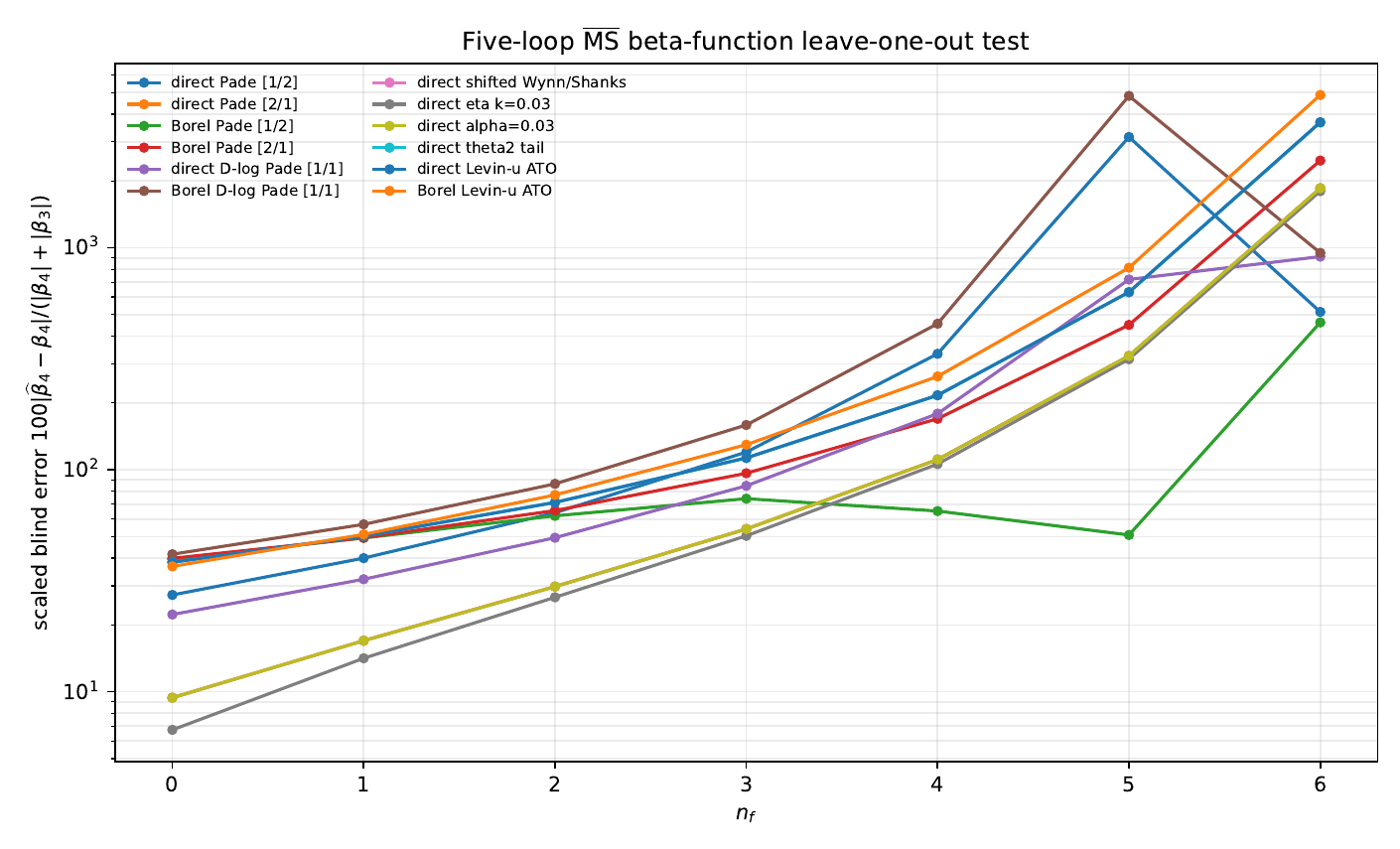}
\caption{Leave-one-order-out prediction of the known five-loop $\MSbar$
$\beta$-function coefficient.  Lines show predeclared representatives; the
vertical scale is logarithmic.  The use of Eq.~\eqref{eq:beta-scaled-error}
keeps the near-zero target at $n_f=6$ visible without treating it as an
infinite relative error.  The direct Levin-$u$ curve lies exactly on the
direct $\theta$ curve at this input order by
Eq.~\eqref{eq:beta-levin-theta-identity}.}
\label{fig:beta-blind}
\end{figure}

The calculation is repeated without retuning for $n_f=0,\ldots,6$.  Ordinary
relative errors become misleading when the exact target is small, so
Fig.~\ref{fig:beta-blind} plots the scaled diagnostic
\begin{equation}
 E_4(n_f)=100\frac{|\widehat\beta_4-\beta_4|}
 {|\beta_4|+|\beta_3|}.
 \label{eq:beta-scaled-error}
\end{equation}
Across $n_f=0,\ldots,5$, the median absolute relative error of the best family,
the direct $\eta$ grid, is $51.7\%$; its largest flavor-wise grid median is
$708\%$.
For direct Levin-$u$ the corresponding cross-flavor median and maximum are
$123.5\%$ and $1418.5\%$; Borel Levin-$u$ gives $138.6\%$ and $1827.4\%$.
The failure is therefore not an isolated $n_f=3$ accident.

There is also a direct physical reason not to expect the Adler extrapolants to
transfer.  Beyond two loops the $\beta$-function coefficients are
scheme-dependent, and the five-loop coefficient results from large
color/flavor structures with cancellations that are unrelated to the
$u=-1,2,3$ singularity pattern of a physical Adler function.  Four positive
low-order coefficients therefore do not encode a stable geometric or
factorially normalized remainder for this auxiliary series.  The uniformly
poor blind record is evidence against universality of the extrapolants, not
evidence for a new renormalon of the $\beta$ function.

\subsection{Conditional higher coefficients and the required negative conclusion}

For completeness, the same rules are applied to all five known coefficients.
Table~\ref{tab:beta-nf3-forecast} lists the resulting $\beta_5$ and rolled
$\beta_6$ outputs.  Intervals are min--max method-grid envelopes, not
uncertainties.  The direct and Borel--Pad\'e rows are the unique balanced
$[2/2]$ continuations; the two D-log entries arise from $[1/2]$ and $[2/1]$.
The direct Levin-$u$ continuation gives
$(\widehat\beta_5,\widehat\beta_6)=(-124.05,-3101.88)$, whereas Borel
Levin-$u$ gives $(-443.26,-8546.81)$.  Their sign reversal relative to the
known positive coefficients is consistent with their failed blind test and
not evidence for a physical sign change.

\begin{table*}[!tbp]
\centering
\caption{Conditional higher-order $n_f=3$ $\beta$-function outputs in the
convention of Eq.~\eqref{eq:beta-benchmark-convention}.  No entry has passed
the blind five-loop test, and $\beta_6$ compounds a model-generated $\beta_5$.
The values must not be interpreted as QCD determinations.}
\label{tab:beta-nf3-forecast}
\input{generated_data/beta_function_nf3_forecast_table.tex}
\end{table*}

The complete applicable-method span is
\begin{equation}
 -443.3\leq\widehat\beta_5\leq3495.9,
 \qquad
 -4.0523\times10^4\leq\widehat\beta_6\leq2384.6.
 \label{eq:beta-forecast-span}
\end{equation}
The sign changes and multi-order spread are larger than any defensible method
envelope.  Consequently this work quotes \emph{no} global average or preferred
prediction for $\beta_5$ or $\beta_6$.  Equation~\eqref{eq:beta-forecast-span}
is a falsifiable record of conditional extrapolations.  This negative result
is also a control on the Adler analysis: agreement among a small subset of
nonlinear continuations cannot by itself be promoted to a universal QCD
forecast.

\section{Full-QCD test with a nonvanishing \texorpdfstring{$x^2$}{x2} weight}
\label{sec:qcd-moment-test}
This leave-one-order-out test asks whether accuracy for $W_\tau$ transfers to
two weights that retain the $u=2$ sensitivity, without using a renormalon model
or the withheld coefficient in the construction.  The logarithmic contour
moments $J_k[W]$ of Eq.~\eqref{eq:dN-JN} can be evaluated in closed
form~\cite{Beneke:2008ad,Penarrocha:2001ig}.  Expanding the running coupling about $a\equiv a(s_0)$ gives
\begin{align}
 a(-s_0x)={}&a-\beta_0La^2+(\beta_0^2L^2-\beta_1L)a^3
 \nonumber\\
 &+\left(-\beta_0^3L^3+\frac52\beta_0\beta_1L^2-\beta_2L\right)a^4
 +\cO(a^5),
 \label{eq:running-expand}
\end{align}
where $L=\ln(-x)$ and the $\beta_j$ are those of
Eq.~\eqref{eq:beta-convention}.  Insertion of Eq.~\eqref{eq:running-expand} into Eq.~\eqref{eq:general-contour-moment}
yields
\begin{align}
 d_1={}&c_{1,1}J_0,\nonumber\\
 d_2={}&c_{2,1}J_0-c_{1,1}\beta_0J_1,\nonumber\\
 d_3={}&c_{3,1}J_0-2c_{2,1}\beta_0J_1
       +c_{1,1}(\beta_0^2J_2-\beta_1J_1),\nonumber\\
 d_4={}&c_{4,1}J_0-3c_{3,1}\beta_0J_1
 +c_{2,1}(3\beta_0^2J_2-2\beta_1J_1)\nonumber\\
 &+c_{1,1}\left(-\beta_0^3J_3+\frac52\beta_0\beta_1J_2-\beta_2J_1\right).
 \label{eq:moment-coeffs}
\end{align}
The contour moments and resulting FOPT coefficients are given in Table~\ref{tab:qcd-moments}.
\begin{table*}[!tbp]
\centering
\caption{Known full-QCD moment coefficients for $n_f=3$.}
\label{tab:qcd-moments}
\begin{tabular}{lrrrrrrr}
\toprule
weight & $J_1$ & $J_2$ & $J_3$ & $d_1$ & $d_2$ & $d_3$ & $d_4$\\
\midrule
$W_\tau$ & $-1.58333$ & $0.390687$ & $3.97757$
& $1$ & $5.20232$ & $26.3659$ & $127.079$\\
$W_{\rm GC}$ & $-1.83333$ & $1.43235$ & $2.12205$
& $1$ & $5.76482$ & $34.4842$ & $214.140$\\
$W_{P2}$ & $-1.16667$ & $-1.01209$ & $4.98676$
& $1$ & $4.26482$ & $14.5230$ & $21.5087$\\
\bottomrule
\end{tabular}
\end{table*}

The ratios for $W_\tau$ are $5.202$, $5.068$, and $4.820$, whereas those
for $W_{\rm GC}$ are $5.765$, $5.982$, and $6.210$.  The latter rise, as
expected when the unsuppressed IR contribution becomes more important, but
four orders are still too few to display factorial scaling.
For $W_{P2}$ the ratios are $4.265$, $3.405$, and $1.481$; competition among
the moment factors is already severe and defeats a local geometric forecast.

\begin{table*}[!tbp]
\centering
\caption{Leave-one-order-out forecasts in full QCD.  Signed errors are in
percent.  The target coefficient was not used in constructing the forecast.}
\label{tab:qcd-loo}
\begin{tabular}{llrrr}
\toprule
weight & target & exact & direct Shanks & Borel Shanks\\
\midrule
$W_\tau$ & $d_3$ & $26.3659$ & $27.0641\;( +2.65\%)$
& $54.1283\;( +105.3\%)$\\
$W_\tau$ & $d_4$ & $127.079$ & $133.625\;( +5.15\%)$
& $200.438\;( +57.7\%)$\\
$W_{\rm GC}$ & $d_3$ & $34.4842$ & $33.2332\;( -3.63\%)$
& $66.4663\;( +92.7\%)$\\
$W_{\rm GC}$ & $d_4$ & $214.140$ & $206.278\;( -3.67\%)$
& $309.417\;( +44.5\%)$\\
$W_{P2}$ & $d_3$ & $14.5230$ & $18.189\;( +25.3\%)$
& $36.378\;( +150.5\%)$\\
$W_{P2}$ & $d_4$ & $21.5087$ & $49.455\;( +129.9\%)$
& $74.183\;( +244.9\%)$\\
\bottomrule
\end{tabular}
\end{table*}

Because the triangular map of Eq.~\eqref{eq:triangular-weight-map} has unit
diagonal, the three relative errors in each row of Table~\ref{tab:qcd-loo} are
not independent once a transformation acts on the common Adler sequence.
Section~\ref{sec:tolerance} converts the ten-percent criterion of this table
into the equivalent Adler-space tolerances
$|\widehat c_{3,1}-c_{3,1}|\leq1.4523$ and
$|\widehat c_{4,1}-c_{4,1}|\leq2.1509$, which is the form used in the
BJ--CVRC analysis of Part~IV.

The nonzero-$W_2$ application is therefore successful as an interpolation,
but it gives a negative answer to a stronger question: the currently known
orders do not yet select the factorially normalized extrapolation.  We do not
use Table~\ref{tab:qcd-loo} as proof that subleading renormalons have been
resolved.

\section{Nearby-renormalon test: pole--\texorpdfstring{$\MSbar$}{MSbar} mass relation}
\label{sec:pole-mass}
Moment coefficients through fourth order may still be pre-asymptotic.  The
pole--$\MSbar$ relation supplies a complementary stress test: the heavy-quark
pole mass has a leading IR renormalon at the nearby position $u=1/2$, whose
exponent is known independently, and its factorial growth is visible much
earlier than in the total-$\tau$ moment~\cite{Ball:1995ni,Beneke:1998ui}.  At
$\mu=\overline m(\overline m)$,
\begin{equation}
 \frac{M_{\rm pole}}{\overline m(\overline m)}
 =1+\sum_{n=1}^{4}r_na^n+\cO(a^5),
 \label{eq:pole-series}
\end{equation}
with
\begin{align}
 r_1={}&\frac43,\nonumber\\
 r_2={}&13.4434-1.04137n_\ell,\nonumber\\
 r_3={}&190.595-26.655n_\ell+0.652694n_\ell^2.
 \label{eq:pole-lower}
\end{align}
We use the updated four-loop results
\begin{equation}
 r_4=1702.71,\ 1235.63,\ 839.18
 \quad\hbox{for}\quad n_\ell=3,4,5,
 \label{eq:pole-r4}
\end{equation}
obtained from Refs.~\cite{Marquard:2015qpa,Marquard2016}.  The
leading-renormalon exponent~\cite{Beneke:1998ui} is
\begin{equation}
 b=\frac{\beta_1}{2\beta_0^2}
 =0.39506,\ 0.36960,\ 0.32892
 \quad\hbox{for}\quad n_\ell=3,4,5.
 \label{eq:pole-b}
\end{equation}

\begin{table*}[!tbp]
\centering
\caption{Prediction of the withheld four-loop pole-mass coefficient.  The
last column is the factorial normalization with the externally known exponent
of Eq.~\eqref{eq:pole-b}, Eq.~\eqref{eq:gamma-forecast}.}
\label{tab:pole-test}
\begin{tabular}{crrrr}
\toprule
$n_\ell$ & exact $r_4$ & direct & Borel normalized & $\Gamma(n+b)$ normalized\\
\midrule
3 & $1702.71$ & $1315.33\;(-22.75\%)$ & $1972.99\;(+15.87\%)$
& $1864.51\;(+9.50\%)$\\
4 & $1235.63$ & $960.86\;(-22.24\%)$ & $1441.29\;(+16.64\%)$
& $1366.35\;(+10.58\%)$\\
5 & $839.18$ & $658.34\;(-21.55\%)$ & $987.51\;(+17.68\%)$
& $941.02\;(+12.14\%)$\\
\bottomrule
\end{tabular}
\end{table*}

The same test at three loops, using only $r_1$ and $r_2$, gives absolute
errors of $30.9$--$31.6\%$ (direct), $36.8$--$38.2\%$ (Borel), and
$17.7$--$21.1\%$ (known exponent).  Including the independently known
exponent gives the smallest error at both orders, whereas plain Borel
normalization is better than the direct forecast only at four loops.  Even
the best of the three remains at the $10$--$20\%$ level, which is not precise
enough to justify percent-level high-order predictions from three or four
coefficients.

This section deliberately freezes the pole-mass series as an external
benchmark before the renormalon-annihilating transformation is introduced.
Appendix~\ref{sec:pole-annihilator} returns to exactly the same withheld target
after deriving the coefficient-space annihilator; no mode is selected from the
known value of $r_4$.

\section{All-order \texorpdfstring{large-$\beta_0$}{large-beta0} validation}
\label{sec:largebeta}
A few withheld physical coefficients cannot demonstrate asymptotic
convergence.  The large-$\beta_0$ limit provides a controlled laboratory with
arbitrarily many coefficients, a known Borel transform, and a declared
principal-value target, so finite-order successes and failures can be placed
on an actual trajectory.

\subsection{Exact Borel transforms}

The large-$\beta_0$ limit provides coefficients to arbitrary order and is
therefore an external validation laboratory~\cite{Beneke:1998ui,
LovettTurner:1994hx,Ball:1995ni,Neubert:1995gd,Boito:2018rwt}.  We quote the
exact transform in the Borel normalization of
Refs.~\cite{Beneke:2008ad,Boito:2018rwt}, with Borel variable $t$ conjugate to
$\alpha_s$ and $u=\beta_1^{\rm BJ}t/(2\pi)$, where
$\beta_1^{\rm BJ}=2\beta_0=9/2$ in our convention~\eqref{eq:beta-convention}.
This normalization differs from Eq.~\eqref{eq:borel-normalization} by a
factor $\pi$, $B_{\rm BJ}[\,\cdot\,](u)=B[\,\cdot\,](u)/\pi$, and in the
$\MSbar$ scheme
\begin{equation}
 B_{\rm BJ}[\Dhat](u)=\frac{32}{3\pi}\frac{e^{5u/3}}{2-u}
 \sum_{k=2}^{\infty}\frac{(-1)^kk}
 {[k^2-(1-u)^2]^2}.
 \label{eq:large-beta-borel}
\end{equation}
It has a simple IR pole at $u=2$ and double poles at the remaining integer
renormalons.  The moment transforms follow without another diagrammatic
calculation:
\begin{equation}
 B_{\rm BJ}[\delta_W^{(0)}](u)=K_W(u)B_{\rm BJ}[\Dhat](u).
 \label{eq:large-beta-moment}
\end{equation}
For $W_\tau$, Eq.~\eqref{eq:large-beta-moment} reduces to
\begin{equation}
 B_{\rm BJ}[\delta^{(0)}_{W_\tau}](u)=
 \frac{12}{(1-u)(3-u)(4-u)}\frac{\sin(\pi u)}{\pi u}
 B_{\rm BJ}[\Dhat](u),
 \label{eq:large-beta-tau}
\end{equation}
which explicitly cancels $u=2$.

If
\begin{equation}
 B_{\rm BJ}[\delta](u)=\sum_{n\geq0}b_nu^n,
\end{equation}
the perturbative coefficient of $a^{n+1}$ is
\begin{equation}
 d_{n+1}=\pi b_n n!\beta_0^n,
 \qquad \beta_0=\frac94.
 \label{eq:coeff-from-borel}
\end{equation}
We extract $b_n$ through order 20 by a Cauchy projection,
\begin{equation}
 b_n=\frac{1}{2\pi r^n}\int_0^{2\pi}d\theta\,
 B_{\rm BJ}(re^{i\theta})e^{-in\theta},
 \label{eq:cauchy}
\end{equation}
using 1024 Fourier points and the closed trigamma representation
Eq.~\eqref{eq:trigamma-closed}, rather than a truncated $k$ sum.  Independent
projections at $r=0.50$ and $0.62$ agree to better than
$7\times10^{-66}$ in the supplied 70-digit build; the analysis aborts if the
relative difference exceeds $2\times10^{-40}$.  Published low-order
coefficients are also enforced as regression tests.

\begin{table*}[!tbp]
\centering
\caption{First ten exact large-$\beta_0$ coefficients.  Scientific notation
is used where useful.}
\label{tab:large-beta-coeff}
\begin{tabular}{rrrrr}
\toprule
$n$ & Adler $c_n$ & $W_\tau$: $d_n$ & $W_{\rm GC}$: $d_n$ & $W_{P2}$: $d_n$\\
\midrule
1 & $1.00000$ & $1.00000$ & $1.00000$ & $1.00000$\\
2 & $1.55649$ & $5.11899$ & $5.68149$ & $4.18149$\\
3 & $15.7112$ & $28.7790$ & $35.8035$ & $18.7591$\\
4 & $24.8320$ & $156.674$ & $228.946$ & $67.8305$\\
5 & $787.827$ & $900.780$ & $1578.52$ & $204.844$\\
6 & $-1991.38$ & $4867.31$ & $11213.1$ & $-625.861$\\
7 & $9.85728\times10^4$ & $3.22843\times10^4$ & $9.15452\times10^4$ & $-9.37228\times10^3$\\
8 & $-1.07762\times10^6$ & $1.53373\times10^5$ & $7.47172\times10^5$ & $-2.14053\times10^5$\\
9 & $2.77521\times10^7$ & $1.99419\times10^6$ & $8.02855\times10^6$ & $-9.37506\times10^5$\\
10 & $-5.38781\times10^8$ & $-3.29562\times10^6$ & $6.78979\times10^7$ & $-4.04875\times10^7$\\
\bottomrule
\end{tabular}
\end{table*}

The Adler coefficients alternate from order six because the double UV pole at
$u=-1$ becomes dominant.  Moment integration reduces this UV pole to a simple
one and delays sign alternation.  For $W_{\rm GC}$, the surviving $u=2$ pole
further extends the same-sign regime.  These transitions make the test more
demanding than a single-pole toy model.
The opposite $W_2$ residue in $W_{P2}$ produces cancellations and an earlier
sign change, giving a deliberately different noncancelling stress test.

\subsection{Withheld-coefficient forecasts}

\begin{figure*}[!tbp]
 \centering
 \includegraphics[width=\linewidth]{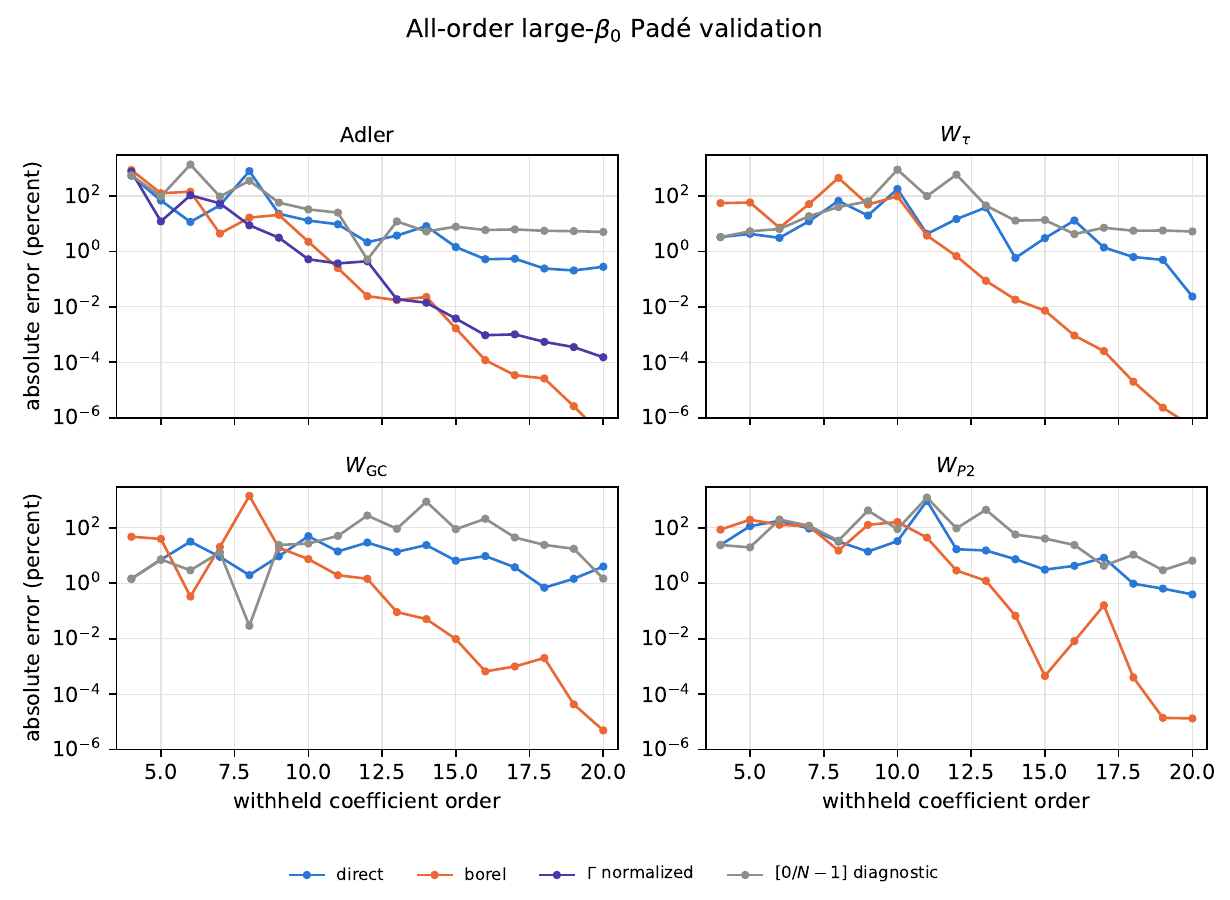}
 \caption{All-order validation of balanced Pad\'e/Shanks/Wynn forecasts.  The
 standard $\tau$ moment and the $u=2$-sensitive moment show large finite-order
 spikes as the dominant renormalon changes.  Borel-normalized approximants
 converge rapidly once the relevant structure is resolved.}
 \label{fig:large-beta-errors}
\end{figure*}

Table~\ref{tab:large-beta-errors} and Fig.~\ref{fig:large-beta-errors} show
the absolute relative error of the first coefficient not used in each
approximant.

\begin{table*}[!tbp]
\centering
\caption{Selected absolute withheld-coefficient errors (percent) in the
large-$\beta_0$ limit.  Large isolated errors are retained; they are not
removed as outliers.  The rows marked ``diag.'' use the unbalanced
denominator-only approximant $[0/N-1]$, which is excluded from the baseline;
they show that it does not exhibit stable convergence across the four
observables as coefficients are added.}
\label{tab:large-beta-errors}
\setlength{\tabcolsep}{9pt}
\input{generated_data/large_beta0_pade_errors.tex}
\end{table*}

Three conclusions follow.
\begin{enumerate}[leftmargin=2em]
 \item Direct Shanks can be accurate at low order for a moment even when it is
 not on a convergent trajectory.  The small order-four errors for
 $W_\tau$ and $W_{\rm GC}$ are examples.
 \item Borel normalization is asymptotically appropriate but not uniformly
 superior at sparse order.  Its order-eight errors for the moments include
 genuine failures caused by finite-order rational defects and mixed
 singularities.
 \item Once sufficient coefficients are supplied, the Borel--Wynn sequence
 passes a stringent certificate.  At order 12 its errors are $0.024\%$,
 $0.674\%$, $1.455\%$, and $2.898\%$ for the Adler, $W_\tau$,
 $W_{\rm GC}$, and $W_{P2}$ series.  At order 16 the first three are below
 $10^{-3}\%$, while $W_{P2}$ is at $0.00821\%$.
\end{enumerate}
The certificate validates the construction as an asymptotic method.  It does
not imply that the four known full-QCD coefficients lie in its convergence
regime.

\section{Renormalon-annihilating sequence transformations}
\label{sec:annihilator}
Shanks and Pad\'e approximants must infer their effective modes from the same
short coefficient list they extrapolate.  To isolate the value of external
renormalon information, we construct a transformation whose \emph{sum-space
weights} are fixed in advance by a specified mode model.  Its coefficient-space
dual still fits amplitudes to data, and every conclusion is conditional on the
chosen positions, exponents, prescription, and mode truncation.

\subsection{Construction}

Suppose, as a model, that the prescription-matched partial sums have the form
\begin{equation}
 S_N=S+\sum_{j=1}^{m}A_j\Omega_N^{(u_j)}+\cdots,
 \label{eq:model-multi-ren}
\end{equation}
where $S$ is the chosen Borel sum, the $A_j$ are unknown amplitudes, and
$\Omega_N^{(u_j)}$ are specified remainder modes attached to singularities at
$u=u_j$.  The ellipsis contains omitted singularities, analytic background,
and subleading corrections; it is the source of model error.  Define
\begin{equation}
 \mathcal R_{m,N}^{(u_1,\ldots,u_m)}=\sum_{k=0}^{m}\lambda_kS_{N+k}.
 \label{eq:R-m-def}
\end{equation}
Inserting Eq.~\eqref{eq:model-multi-ren} gives
\begin{equation}
 \mathcal R_{m,N}
 =S\sum_{k=0}^{m}\lambda_k
 +\sum_{j=1}^{m}A_j\sum_{k=0}^{m}\lambda_k\Omega_{N+k}^{(u_j)}+\cdots,
\end{equation}
so that preserving $S$ and removing the selected modes requires
\begin{equation}
 \sum_{k=0}^{m}\lambda_k=1,
 \qquad
 \sum_{k=0}^{m}\lambda_k\Omega_{N+k}^{(u_j)}=0,
 \quad j=1,\ldots,m.
 \label{eq:lambda-conditions}
\end{equation}
These are $m+1$ linear equations for $m+1$ unknowns,
\begin{equation}
 M\lambda=(1,0,\ldots,0)^{T},\qquad
 M_{0k}=1,\quad M_{jk}=\Omega_{N+k}^{(u_j)},
 \label{eq:lambda-system-general}
\end{equation}
whence $\lambda_k=(M^{-1})_{k0}$.  Because only ratios
\begin{equation}
 r_k^{(u)}=\frac{\Omega_{N+k}^{(u)}}{\Omega_N^{(u)}},\qquad r_0^{(u)}=1,
 \label{eq:ratio-def}
\end{equation}
enter, the overall normalization of each mode is irrelevant and the rows of
$M$ may be rescaled at will.  For $m=3$ and modes $u=2,3,-1$, writing
$\sigma_k=r_k^{(2)}$, $\tau_k=r_k^{(3)}$, $\chi_k=r_k^{(-1)}$ and
\begin{equation}
 \begin{aligned}
 \mathcal D_k&=\det\begin{pmatrix}
 \sigma_{\ell_1}&\sigma_{\ell_2}&\sigma_{\ell_3}\\
 \tau_{\ell_1}&\tau_{\ell_2}&\tau_{\ell_3}\\
 \chi_{\ell_1}&\chi_{\ell_2}&\chi_{\ell_3}\end{pmatrix},
 \\
 \{\ell_1,\ell_2,\ell_3\}&=\{0,1,2,3\}\setminus\{k\}.
 \end{aligned}
\end{equation}
Cramer's rule gives
$\lambda_k=(-1)^k\mathcal D_k/(\mathcal D_0-\mathcal D_1+\mathcal D_2-\mathcal D_3)$ and hence
\begin{equation}
 \mathcal R_{3,N}^{(2,3,-1)}
 =\frac{\mathcal D_0S_N-\mathcal D_1S_{N+1}+\mathcal D_2S_{N+2}-\mathcal D_3S_{N+3}}
 {\mathcal D_0-\mathcal D_1+\mathcal D_2-\mathcal D_3}.
 \label{eq:R-three-mode-final}
\end{equation}
The $\lambda_k$ are fixed functions of $N$, $a$ and the chosen modes.  No
coefficient value enters them, but the transformed estimate
$\mathcal R_{m,N}$ of course depends on the input partial sums.

\subsection{Closed form for the weights}
\label{sec:closed-form}

Determinants of this size are unnecessary.  Suppose all modes share a common
exponent $\gamma$ and take the leading remainder estimate
$\Omega_N^{(u)}=\Gamma(N+1+\gamma)(\beta_0/u)^Na^{N+1}$, so that
\begin{equation}
 r_k^{(u)}=\varphi_k\,z_u^{\,k},
 \qquad
 \varphi_k=\prod_{p=1}^{k}(N+p+\gamma),
 \qquad
 z_u=\frac{\beta_0a}{u},
 \label{eq:ratio-product}
\end{equation}
with $\varphi_0=1$.  Set $\mu_k=\lambda_k\varphi_k$.  The annihilation conditions in
Eq.~\eqref{eq:lambda-conditions} then read $\sum_k\mu_kz_j^k=0$ for
$j=1,\ldots,m$, i.e.\ the polynomial $P(z)=\sum_{k=0}^{m}\mu_kz^k$ of degree
$m$ has the $m$ prescribed roots $z_1,\ldots,z_m$.  Hence
$P(z)=\mu_m\prod_j(z-z_j)$ and, with $e_q$ the elementary symmetric
polynomials, $\mu_k=\mu_m(-1)^{m-k}e_{m-k}(z_1,\ldots,z_m)$.  Imposing
$\sum_k\lambda_k=1$ fixes $\mu_m$.  With
$E_r=e_r(z_1,\ldots,z_m)$ and
$D=\sum_{l=0}^{m}(-1)^{m-l}E_{m-l}/\varphi_l$, the result is
\begin{equation}
 \lambda_k=\frac{(-1)^{m-k}E_{m-k}}{\varphi_kD},
 \qquad k=0,\ldots,m.
 \label{eq:lambda-closed}
\end{equation}
The transformation is therefore the $\varphi_k$-weighted coefficient filter of the
polynomial $\prod_j(z-z_j)$.  The weights are uniquely determined when the
generalized Vandermonde system has full rank and the normalization denominator
in Eq.~\eqref{eq:lambda-closed} is nonzero.  Distinct $z_j$ make the
equal-exponent Vandermonde part full rank, but do not by themselves guarantee
good numerical conditioning.  Conditioning degrades as two positions approach
one another and can also degrade at special $(N,a)$ values.  Equation~\eqref{eq:lambda-closed} agrees with the
determinant solution of Eq.~\eqref{eq:lambda-system-general} to $10^{-45}$ in
our implementation.  For unequal $\gamma_j$ no product formula exists and we
solve Eq.~\eqref{eq:lambda-system-general} directly and monitor its condition
number; no blanket nonsingularity claim is made for arbitrary parameters.

\subsection{The exact remainder mode}
\label{sec:exact-mode}

The mode function used above is the one suggested by large-order behavior,
but it is \emph{not} the remainder.  This distinction is not academic: it is
the difference between a transformation that converges and one that does not.

Consider an isolated Borel singularity
$B_p(u)\simeq K_p(1-u/p)^{-1-\gamma}$.  This normalization keeps a real mode
real also for $p<0$ and noninteger $\gamma$.  Its
Taylor coefficients give, with $d_{n+1}=\pi b_nn!\beta_0^n$,
\begin{equation}
 \begin{aligned}
 T_n&\equiv d_{n+1}a^{n+1}
 =\widehat C\,\Gamma(n+1+\gamma)\,z^n,\\
 z&=\frac{\beta_0a}{p},
 &\widehat C&=\frac{\pi K_pa}{\Gamma(1+\gamma)}.
 \end{aligned}
 \label{eq:single-mode-term}
\end{equation}
The remainder after $N$ terms is the principal value of $\sum_{n\geq N}T_n$,
which is \emph{not} $T_N$: the tail is itself factorially divergent.  Using
$\Gamma(n+1+\gamma)z^n=z^n\int_0^\infty dt\,t^{n+\gamma}e^{-t}$ and summing the
geometric series inside the integral,
\begin{equation}
 \Omega_N^{(u,\gamma)}(a)
 =z^N\,\mathrm{PV}\!\int_0^\infty dt\,
 \frac{t^{N+\gamma}e^{-t}}{1-zt},
 \qquad z=\frac{\beta_0a}{u}.
 \label{eq:exact-mode}
\end{equation}
This is the exact prescription-matched tail for the pure coefficient mode
generated by an isolated singularity of order $1+\gamma$, and reduces to
$\Gamma(N+1+\gamma)$ as $z\to0$, which is the mode function of
App.~\ref{sec:closed-form}.  The correction is an $\mathcal O(1)$ factor that
depends strongly on $N$: for $z=0.1125$ (the $u=2$ mode at $a=0.1$) the ratio
$\Omega_N/[\Gamma(N+1)z^N]$ equals $1.70$ at $N=2$, $2.18$ at $N=6$ and
$-1.28$ at $N=10$, the sign change occurring as $N$ passes the minimal term.
A transformation built on the $z\to0$ mode therefore annihilates the wrong
function once $Nz\sim1$, and we find that it does not converge; the same
transformation built on Eq.~\eqref{eq:exact-mode} does.  This is quantified in
App.~\ref{sec:annihilator-validation}.

For $u<0$ the integral in Eq.~\eqref{eq:exact-mode} is unambiguous.  For
$u>0$ we evaluate it on a ray $t\in e^{i\theta}\mathbb R_+$ and take the real
part, using the same prescription as for the Borel sum.  A finite list then
reproduces only the corresponding finite pure-mode contribution; omitted
singularities and analytic background remain.

\subsection{Coefficient-space dual}
\label{sec:coefficient-dual}

Equation~\eqref{eq:R-m-def} estimates $S$ at a given coupling.  The
principal validation metric of this paper is instead a withheld coefficient, and
for that we use the dual construction, which is coupling independent.  Model
the coefficients themselves,
\begin{equation}
 d_n=\sum_{j=1}^{m}A_j\,\omega_n^{(u_j)},
 \qquad
 \omega_n^{(u,\gamma)}=\Gamma(n+\gamma)\left(\frac{\beta_0}{u}\right)^{\!n-1},
 \label{eq:coeff-model}
\end{equation}
which follows from Eq.~\eqref{eq:single-mode-term}; shifting the displayed
power by one merely rescales each fitted amplitude.  The $m$ consecutive coefficients $d_{n},\ldots,d_{n+m-1}$
determine the amplitudes, and then
\begin{equation}
 d_{n+m}^{\rm pred}=\sum_{j}A_j\omega_{n+m}^{(u_j)}.
 \label{eq:coeff-forecast}
\end{equation}
Equivalently, the sequence obeys an $(m{+}1)$-term linear recurrence
$\sum_k\nu_kd_{n+k}=0$ with known, $n$-dependent coefficients.  Repeating the
argument of App.~\ref{sec:closed-form} with $y_j=\beta_0/u_j$ and
$C_k=\prod_{p=0}^{k-1}(n+\gamma+p)$ gives, for equal exponents,
writing $E_r\equiv e_r(y_1,\ldots,y_m)$,
\begin{align}
 \nu_k&=\frac{(-1)^{m-k}E_{m-k}}{C_k}
 \label{eq:nu-closed}\\
 d_{n+m}^{\rm pred}
 &=C_m\sum_{k=0}^{m-1}
 \frac{(-1)^{m-k+1}E_{m-k}}{C_k}\,d_{n+k}.
\end{align}
For $m=1$ this collapses to $d_{n+1}=(n+\gamma)(\beta_0/u)d_n$, the familiar
single-renormalon ratio.  The dual is exact only when the finite-mode model
Eq.~\eqref{eq:coeff-model} holds.  Perturbative coefficients have no
principal-value ambiguity, but the mode truncation remains a model choice.

\subsection{External \texorpdfstring{$u=1/2$}{u=1/2} benchmark: the
pole--\texorpdfstring{$\MSbar$}{MSbar} mass relation}
\label{sec:pole-annihilator}

Appendix~\ref{sec:pole-mass} defined and froze the pole-mass benchmark before
the annihilator was introduced.  We now apply the coefficient-space dual to
that unchanged withheld target.  The pole--$\MSbar$ relation is the physical
series in this paper whose leading renormalon is both close and unambiguous.
Its established mode is $u=1/2$ with exponent
$\gamma=b=\beta_1/(2\beta_0^2)$, the same external input used in
Eq.~\eqref{eq:gamma-forecast}.  The difference is that the annihilator also
supplies the geometric ratio $\beta_0/u=2\beta_0$ and fits only its amplitude,
whereas the $\Gamma$-normalized forecast leaves that ratio to the data.

For $m=1$, Eq.~\eqref{eq:nu-closed} becomes
$r_{n+1}=(n+b)(2\beta_0)r_n$, so the withheld four-loop coefficient is
predicted from $r_3$ alone.  This is the only column whose mode content is
fixed solely by the established leading pole-mass renormalon.  For diagnostic
purposes, $m=2$ adds a candidate ultraviolet mode at $u=-1$, and $m=3$ adds a
candidate infrared mode at $u=3/2$.  The implementation takes
$\gamma_{-1}=b+1$ and $\gamma_{3/2}=3b$; these are model assumptions, so the
last two columns are sensitivity studies rather than parameter-free
predictions.

The leading-only $m=1$ test has $4.5$--$12.0\%$ error, compared with
$9.5$--$12.1\%$ after $\Gamma(n+b)$ normalization.  It improves the
$n_\ell=3,4$ cases and is essentially tied at $n_\ell=5$.  The exploratory
$m=2$ column happens to improve the first two targets further, but there is
only one withheld order and no independent validation of its extra mode; it
must not be selected after seeing $r_4$.  The $m=3$ column is worse throughout
and uses all three known coefficients to determine three amplitudes.  Thus
supplying the known leading position and exponent helps, but does not yield
uniform percent-level accuracy.

\subsection{Which modes?  The order of a moment renormalon}
\label{sec:order-rule}

Nothing above tells us which $(u_j,\gamma_j)$ to use.  For $\tau$ moments the
answer is fixed by the weight, and the rule of Sec.~\ref{sec:moments} extends
from the presence of a singularity to its order.

Let $\mathcal N_p[F]$ denote the order of the pole of $B[F]$ at $u=p$.  From
Eq.~\eqref{eq:Jwu}, $\sin(\pi u)$ has a simple zero at every integer, while
$\sum_mW_m/(u-m)$ has a simple pole at $u=p$ only when $0\leq p\leq M$ and
$W_p\neq0$.  Hence $K_W$ is finite and nonzero at $u=p$ if $W_p\neq0$ and has
a simple zero otherwise, giving
\begin{equation}
 \mathcal N_p\big[\delta_W^{(0)}\big]
 =\mathcal N_p\big[\Dhat\big]-1+\theta(W_p\neq0),
 \qquad p\in\mathbb Z,
 \label{eq:order-rule}
\end{equation}
with the convention $W_p=0$ for $p<0$ or $p>M$.  The statement of
Sec.~\ref{sec:moments}, that the $u=2$ singularity survives iff $W_2\neq0$, is
the special case $\mathcal N_2[\Dhat]=1$.

In the large-$\beta_0$ limit Eq.~\eqref{eq:large-beta-borel} has a simple pole
at $u=2$ and double poles at every other nonzero integer except $u=1$, so
Eq.~\eqref{eq:order-rule} fixes the entire singularity structure of every
moment without further calculation.  Table~\ref{tab:order-rule} lists the
resulting orders; they agree with a direct numerical determination of the
scaling of $B(p+\epsilon)$ for every entry.

\begin{table*}[!tbp]
\centering
\caption{Order $\mathcal N_p$ of the large-$\beta_0$ Borel singularity at
$u=p$, from Eq.~\eqref{eq:order-rule}.  A zero entry means the singularity is
absent.  Every entry was confirmed numerically from the scaling of
$B(p+\epsilon)$.}
\label{tab:order-rule}
\begin{tabular}{lrrrrrrr}
\toprule
observable & $u=-3$ & $u=-2$ & $u=-1$ & $u=2$ & $u=3$ & $u=4$ & $u=5$\\
\midrule
$\Dhat$ & 2 & 2 & 2 & 1 & 2 & 2 & 2\\
$W_\tau$ & 1 & 1 & 1 & 0 & 2 & 2 & 1\\
$W_{\rm GC}$ & 1 & 1 & 1 & 1 & 2 & 1 & 1\\
$W_{P2}$ & 1 & 1 & 1 & 1 & 2 & 1 & 1\\
\bottomrule
\end{tabular}
\end{table*}

Two structural consequences follow.  First, a pole of order two at $u=p$ is
\emph{exactly} two annihilator modes at the same position,
$(p,\gamma=1)$ and $(p,\gamma=0)$, since
$B\supset K/(p-u)^2+L/(p-u)$ contributes
$\pi(\beta_0/p)^n[K(n+1)!/p^2+Ln!/p]$ to $d_{n+1}$.  Second, because
$B[\Dhat]$ is meromorphic and its growth on $|u|=R$ is bounded by
$e^{5R/3}/R^2$, pushing the Cauchy contour outward gives
\begin{equation}
 b_n=-\sum_{|u_j|\leq R}\mathrm{Res}\,\frac{B(u)}{u^{n+1}}
 +\mathcal O\!\left[\left(\frac{5e}{3n}\right)^{\!n}\right],
\end{equation}
so in the large-$\beta_0$ limit the pole part is an infinite superposition of
modes with $\gamma\in\{0,1\}$ plus a supergeometrically small analytic
remainder.  Truncating a nested mode list at $|u|\leq U$ leaves, up to powers of
$N$ and residue-dependent constants, a relative contribution governed by the
nearest omitted singularity.  This supplies an asymptotic rate estimate for
the declared meromorphic model; it is not a finite-order error bound and has no
direct analogue for a generic full-QCD branch-point spectrum.

We accordingly adopt a mode rule fixed before any target is examined:
\begin{quote}
\textbf{Rule.} Include every Borel singularity with $|u|\leq U$, with the
confluent multiplicity dictated by Eq.~\eqref{eq:order-rule} and
Table~\ref{tab:order-rule}, ordered by $|u|$.
\end{quote}
This gives $m=5,9,13$ modes for the Adler function and $m=2,5,8$
($W_\tau$) or $m=3,6,8$ ($W_{\rm GC}$, $W_{P2}$) at $U=2,3,4$.  No mode list
is altered after inspecting a withheld coefficient.

\section{All-order validation of the annihilator}
\label{sec:annihilator-validation}
The algebra only guarantees cancellation when the assumed modes are the
actual remainders.  We therefore test coefficient forecasts and fixed-coupling
sums against independently generated large-$\beta_0$ targets, keeping the mode
rule fixed before each withheld order is inspected.

The large-$\beta_0$ limit supplies exact coefficients and an exact
principal-value sum, so the annihilator can be tested with nothing withheld
except the target.  Two independent checks anchor the implementation used
here: it reproduces the balanced Borel--Wynn errors of
Table~\ref{tab:large-beta-errors} and the $\theta_2$ errors summarized in
App.~\ref{sec:local-reg-summary} to every digit generated by the accompanying
code, and its Borel
coefficients agree between the Cauchy radii $r=0.50$ and $0.62$ to better than
$7\times10^{-66}$ because $B[\Dhat]$ is evaluated in closed form,
\begin{align}
 \sum_{k\geq2}\frac{(-1)^kk}{[k^2-v^2]^2}
 &={}\frac{1}{16v}\Big[
 \psi'\!\big(\tfrac{2-v}{2}\big)-\psi'\!\big(\tfrac{3-v}{2}\big)
 \nonumber\\[-1mm]
 &\hspace{13mm}{}-\psi'\!\big(\tfrac{2+v}{2}\big)
 +\psi'\!\big(\tfrac{3+v}{2}\big)\Big],\nonumber\\
 &\hspace{30mm}v=1-u,
 \label{eq:trigamma-closed}
\end{align}
rather than by truncating the sum in Eq.~\eqref{eq:large-beta-borel}.

\subsection{Why the exact mode is necessary}

Table~\ref{tab:mode-comparison} compares the two mode functions in the
fixed-coupling test at $a=0.1$.  With the leading estimate
$\Gamma(N+1+\gamma)(\beta_0/u)^Na^{N+1}$ the transformation does not converge:
supplying more coefficients does not systematically reduce the error.  The
prescription-matched mode is substantially better at most displayed orders,
but the residual is not monotone for every observable and the improvement in
this three-mode illustration is typically one to three orders of magnitude,
not four to five.  The justified conclusion is narrower: the factorially
normalized first omitted term is not the tail once $Nz\sim1$, and using the
correct pure-mode tail is necessary but not sufficient for convergence of a
truncated multi-mode model.

\begin{table*}[!tbp]
\centering
\caption{Absolute error (percent) of the principal-value sum at $a=0.1$
using the first three nonzero modes of the declared ordering:
$\{(-1,1),(-1,0),(2,0)\}$ for the Adler function,
$\{(-1,0),(-2,0),(3,1)\}$ for $W_\tau$, and
$\{(-1,0),(2,0),(-2,0)\}$ for $W_{\rm GC}$ and $W_{P2}$.  The leading
remainder estimate is compared with Eq.~\eqref{eq:exact-mode}; $N_{\rm tot}$
is the number of coefficients used.}
\label{tab:mode-comparison}
\begin{tabular}{lrrrrrr}
\toprule
& \multicolumn{2}{c}{$N_{\rm tot}=10$} & \multicolumn{2}{c}{$N_{\rm tot}=14$}
& \multicolumn{2}{c}{$N_{\rm tot}=18$}\\
observable & leading & exact & leading & exact & leading & exact\\
\midrule
Adler & $52.62$ & $1.164$ & $0.8083$ & $0.6579$ & $1.106$ & $0.3893$\\
$W_\tau$ & $0.4896$ & $0.01012$ & $17.63$ & $0.05753$ & $19.76$ & $0.03029$\\
$W_{\rm GC}$ & $304.6$ & $1.162$ & $0.9309$ & $0.6146$ & $11.14$ & $0.2831$\\
$W_{P2}$ & $114.2$ & $1.530$ & $0.4563$ & $0.8163$ & $22.59$ & $0.4160$\\
\bottomrule
\end{tabular}
\end{table*}

\subsection{Withheld coefficients}

\begin{figure*}[!tbp]
 \centering
 \includegraphics[width=\linewidth]{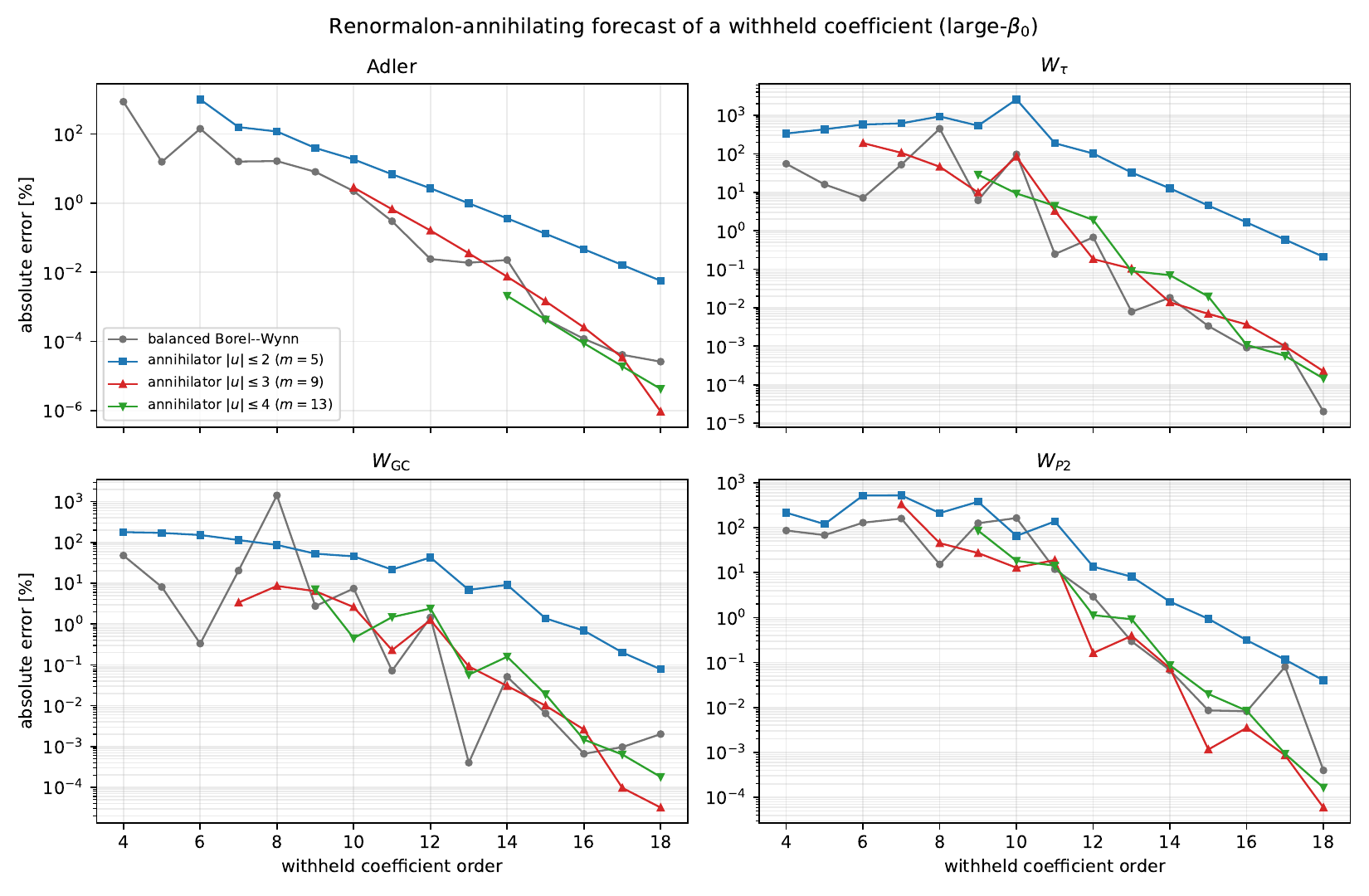}
 \caption{Withheld-coefficient error for the renormalon annihilator against
 the balanced Borel--Wynn construction.  The grey curves reproduce the spikes
 documented in Fig.~\ref{fig:large-beta-errors}, which occur where the dominant
 renormalon changes.  The annihilator, which is told where the renormalons
 are, does not spike.}
 \label{fig:annih-coeff}
\end{figure*}

Table~\ref{tab:annih-coeff} and Fig.~\ref{fig:annih-coeff} give the
coefficient-space test of App.~\ref{sec:coefficient-dual} against the balanced
Borel--Wynn construction of App.~\ref{sec:largebeta}, on identical data and
with the same withheld-target protocol.

\begin{table*}[!tbp]
\centering
\caption{Absolute error (percent) on a withheld large-$\beta_0$ coefficient.
``balanced'' is the near-diagonal Borel--Wynn construction of
Table~\ref{tab:large-beta-errors}; $|u|\leq U$ is the annihilator with the mode
rule of App.~\ref{sec:order-rule}.  Dashes indicate that the mode list is
larger than the number of available coefficients.}
\label{tab:annih-coeff}
\setlength{\tabcolsep}{7pt}
\begin{tabular}{llrrrr}
\toprule
observable & method & order 8 & order 12 & order 16 & order 18\\
\midrule
\multirow{3}{*}{Adler}
& balanced & $16.485$ & $0.02424$ & $1.19\times10^{-4}$ & $2.60\times10^{-5}$\\
& $|u|\leq2$ ($m=5$) & $118.3$ & $2.690$ & $0.04634$ & $5.69\times10^{-3}$\\
& $|u|\leq3$ ($m=9$) & -- & $0.1614$ & $2.54\times10^{-4}$ & $9.34\times10^{-7}$\\
\midrule
\multirow{3}{*}{$W_\tau$}
& balanced & $447.998$ & $0.6740$ & $9.17\times10^{-4}$ & $2.00\times10^{-5}$\\
& $|u|\leq3$ ($m=5$) & $46.05$ & $0.1838$ & $3.66\times10^{-3}$ & $2.24\times10^{-4}$\\
& $|u|\leq4$ ($m=8$) & -- & $1.913$ & $1.08\times10^{-3}$ & $1.45\times10^{-4}$\\
\midrule
\multirow{3}{*}{$W_{\rm GC}$}
& balanced & $1423.44$ & $1.455$ & $6.63\times10^{-4}$ & $2.02\times10^{-3}$\\
& $|u|\leq3$ ($m=6$) & $8.585$ & $1.252$ & $2.61\times10^{-3}$ & $3.19\times10^{-5}$\\
& $|u|\leq4$ ($m=8$) & -- & $2.399$ & $1.48\times10^{-3}$ & $1.80\times10^{-4}$\\
\midrule
\multirow{3}{*}{$W_{P2}$}
& balanced & $15.207$ & $2.898$ & $8.21\times10^{-3}$ & $3.99\times10^{-4}$\\
& $|u|\leq3$ ($m=6$) & $45.09$ & $0.1626$ & $3.52\times10^{-3}$ & $5.93\times10^{-5}$\\
& $|u|\leq4$ ($m=8$) & -- & $1.128$ & $8.46\times10^{-3}$ & $1.65\times10^{-4}$\\
\bottomrule
\end{tabular}
\end{table*}

The comparison is not a uniform victory, and we do not present it as one.  At
the highest orders the two constructions are comparable: at order 18 the
annihilator is better for the Adler function ($9.3\times10^{-7}\%$ versus
$2.6\times10^{-5}\%$), for $W_{\rm GC}$ ($3.2\times10^{-5}\%$ versus
$2.0\times10^{-3}\%$) and for $W_{P2}$, and slightly worse for $W_\tau$.  Nor
does adding modes always help: the $|u|\leq4$ list is worse than $|u|\leq3$ at
orders 12--16 for three of the four observables, because a longer mode list
consumes more coefficients and the discarded singularities are already
suppressed by $4^{-N}$.

What is systematic is the behavior at low and intermediate order, which is
where the balanced constructions of App.~\ref{sec:largebeta} fail.  At withheld order 8 the balanced
construction misses $W_{\rm GC}$ by $1423\%$ and $W_\tau$ by $448\%$; the
annihilator misses them by $8.6\%$ and $46\%$.  Those large balanced errors
are not noise --- they are finite-order rational defects that appear when the
$u=2$ and $u=-1$ contributions exchange dominance, and the annihilator avoids
them for the obvious reason that it is not trying to infer that exchange from
the data.  The residual is also much smoother in $N$, which matters if a
forecast is to be used rather than merely reported.

\subsection{Fixed-coupling summation}

\begin{figure*}[!tbp]
 \centering
 \includegraphics[width=0.9\linewidth]{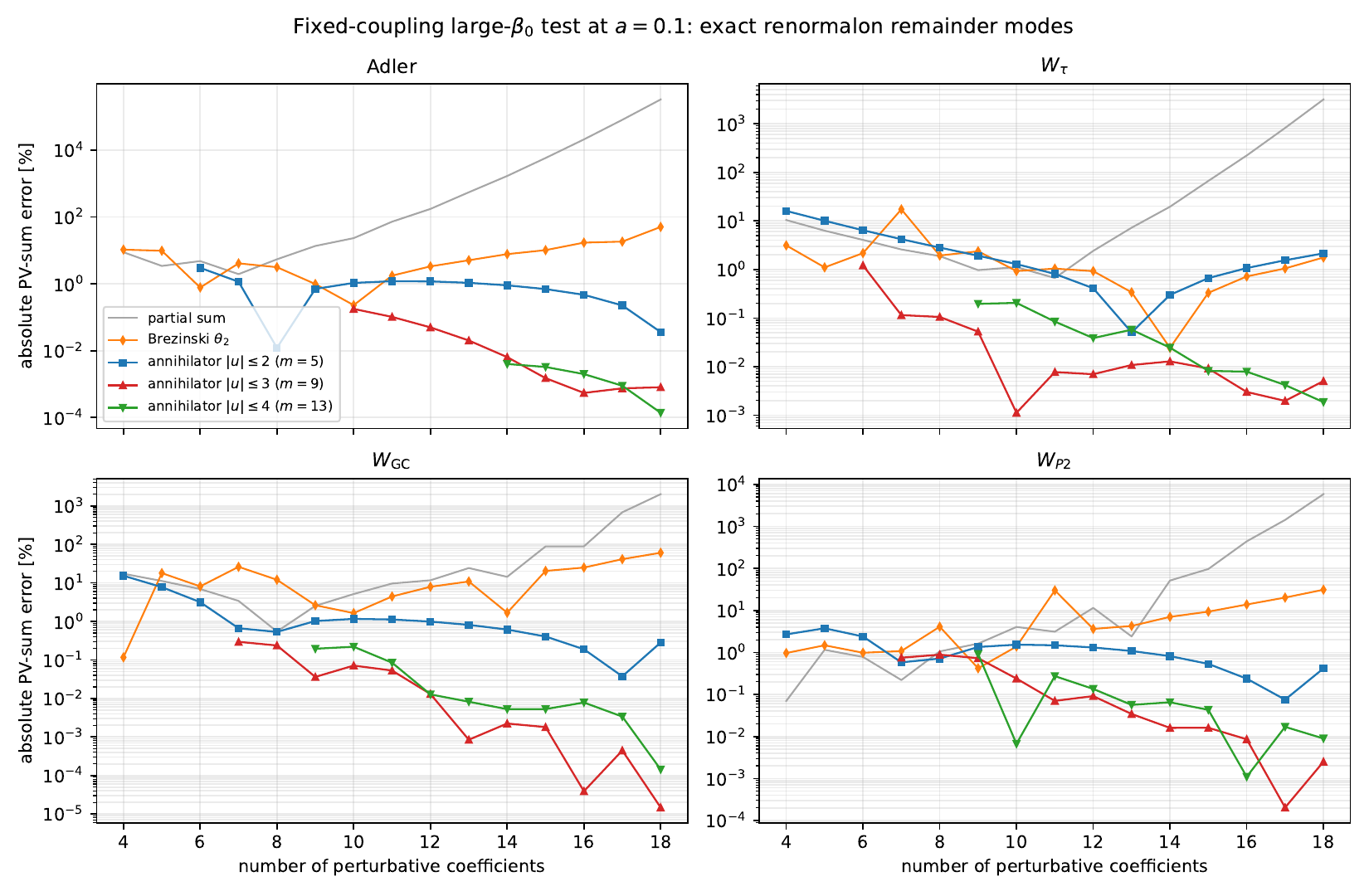}
 \caption{Fixed-coupling large-$\beta_0$ test at $a=0.1$.  Raw partial sums
 diverge past the minimal term.  The fixed $|u|\leq3$ mode model remains much
 more accurate, although its finite-order error is not monotone for every
 observable.}
 \label{fig:annih-pv}
\end{figure*}

Table~\ref{tab:annih-pv} and Fig.~\ref{fig:annih-pv} repeat the fixed-coupling
test of App.~\ref{sec:local-reg-summary} using Eq.~\eqref{eq:R-m-def} with the exact
modes.  Here the improvement is unambiguous, because the competing local
transformations have no access to the renormalon positions at all.

\begin{table*}[!tbp]
\centering
\caption{Absolute error (percent) relative to the principal-value
large-$\beta_0$ sum at $a=0.1$, as a function of the number of perturbative
coefficients supplied.}
\label{tab:annih-pv}
\setlength{\tabcolsep}{6pt}
\begin{tabular}{llrrrr}
\toprule
observable & method & $N_{\rm tot}=8$ & $12$ & $16$ & $18$\\
\midrule
\multirow{3}{*}{Adler}
& partial sum & $5.403$ & $173.4$ & $2.08\times10^{4}$ & $3.25\times10^{5}$\\
& $\theta_2$ & $3.140$ & $3.317$ & $17.03$ & $50.10$\\
& annihilator $|u|\leq3$ & -- & $0.04980$ & $5.45\times10^{-4}$ & $8.06\times10^{-4}$\\
\midrule
\multirow{3}{*}{$W_\tau$}
& partial sum & $1.882$ & $2.415$ & $222.1$ & $3161$\\
& $\theta_2$ & $1.937$ & $0.9244$ & $0.7180$ & $1.760$\\
& annihilator $|u|\leq3$ & $0.1055$ & $6.99\times10^{-3}$ & $3.01\times10^{-3}$ & $5.04\times10^{-3}$\\
\midrule
\multirow{3}{*}{$W_{\rm GC}$}
& partial sum & $0.5461$ & $11.70$ & $88.98$ & $2014$\\
& $\theta_2$ & $12.08$ & $7.930$ & $25.03$ & $60.60$\\
& annihilator $|u|\leq3$ & $0.2373$ & $0.01302$ & $3.89\times10^{-5}$ & $1.47\times10^{-5}$\\
\midrule
\multirow{3}{*}{$W_{P2}$}
& partial sum & $1.058$ & $11.46$ & $439.5$ & $5831$\\
& $\theta_2$ & $4.087$ & $3.625$ & $13.81$ & $31.00$\\
& annihilator $|u|\leq3$ & $0.8838$ & $0.09130$ & $8.57\times10^{-3}$ & $2.51\times10^{-3}$\\
\bottomrule
\end{tabular}
\end{table*}

At $16$ coefficients the annihilator is between roughly two and six orders of
magnitude closer to the principal-value sum than $\theta_2$.  The error can
rise again when another mode or conditioning effect becomes relevant, as the
Adler and $W_\tau$ entries at order 18 show.  Thus the test validates
prescription-matched mode subtraction in this meromorphic laboratory; it does
not make optimal truncation irrelevant in an incompletely specified theory.

\section{Misspecification diagnostics and supervised position scans}
\label{sec:falsifiability}
Convergence with the correct mode list is not enough: a structurally useful
model should also reveal how errors respond to wrong positions, omitted modes,
and spurious modes.  These tests use known large-$\beta_0$ targets and are
therefore supervised validation diagnostics, not a procedure for locating an
unknown QCD renormalon from four coefficients.

Because the sum-space weights depend only on the assumed singularity data, a
wrong assumption can produce a characteristic failure.  Out-of-sample Pad\'e
and Shanks tests are also falsifiable; the additional value here is that the
source of misspecification is explicitly parameterized.

\begin{figure*}[!tbp]
 \centering
 \includegraphics[width=\linewidth]{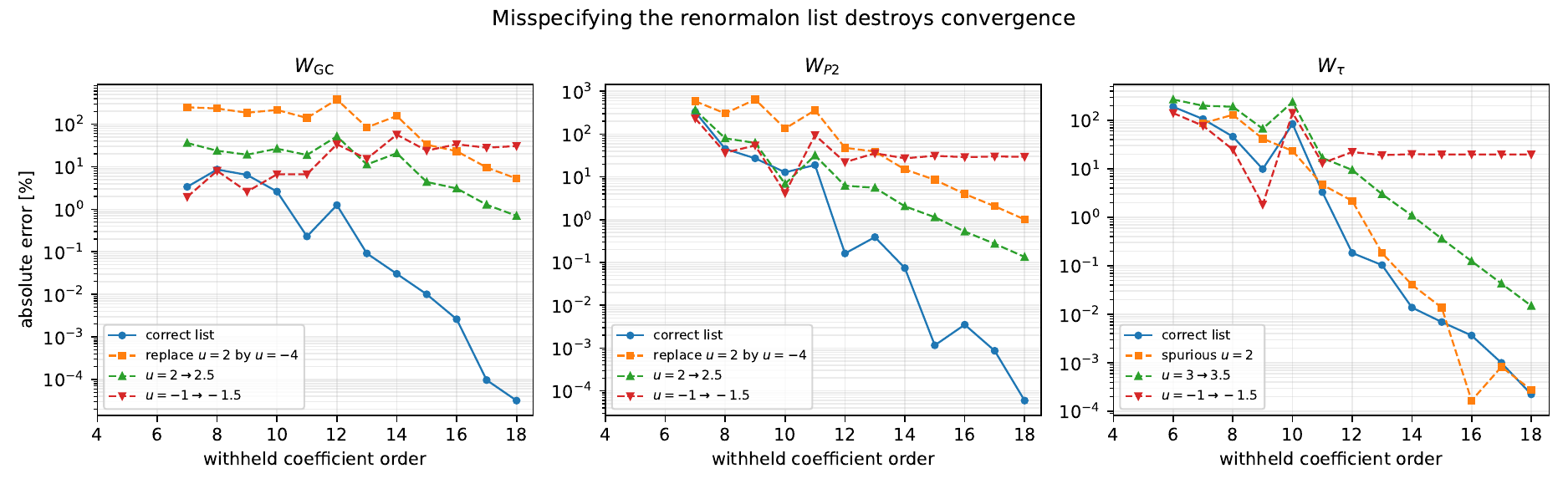}
 \caption{Effect of an incorrect renormalon list.  Replacing $u=2$ by a
 same-dimension spurious $u=-4$ mode keeps the number of fitted amplitudes
 fixed.  A displaced leading ultraviolet position produces a persistent
 error over the displayed range.}
 \label{fig:annih-misspec}
\end{figure*}

Table~\ref{tab:misspec} and Fig.~\ref{fig:annih-misspec} alter the $|u|\leq3$
mode list in four ways while leaving everything else fixed.

\begin{table*}[!tbp]
\centering
\caption{Absolute error (percent) on a withheld coefficient when the
$|u|\leq3$ mode list is misspecified.  Where $u=2$ is absent from the true
structure ($W_\tau$) a spurious mode is added instead of removed.}
\label{tab:misspec}
\setlength{\tabcolsep}{7pt}
\begin{tabular}{llrrr}
\toprule
observable & mode list & order 12 & order 16 & order 18\\
\midrule
\multirow{4}{*}{$W_{\rm GC}$}
& correct & $1.252$ & $2.61\times10^{-3}$ & $3.19\times10^{-5}$\\
& replace $u=2$ by $u=-4$ & $379.9$ & $23.18$ & $5.255$\\
& $u=2\to2.5$ & $51.75$ & $3.089$ & $0.7007$\\
& $u=-1\to-1.5$ & $33.70$ & $33.70$ & $30.56$\\
\midrule
\multirow{4}{*}{$W_{P2}$}
& correct & $0.1626$ & $3.52\times10^{-3}$ & $5.93\times10^{-5}$\\
& replace $u=2$ by $u=-4$ & $47.82$ & $3.987$ & $1.013$\\
& $u=2\to2.5$ & $6.235$ & $0.5285$ & $0.1350$\\
& $u=-1\to-1.5$ & $22.09$ & $28.94$ & $29.45$\\
\midrule
\multirow{4}{*}{$W_\tau$}
& correct & $0.1838$ & $3.66\times10^{-3}$ & $2.24\times10^{-4}$\\
& spurious $u=2$ & $2.182$ & $1.67\times10^{-4}$ & $2.76\times10^{-4}$\\
& $u=3\to3.5$ & $9.460$ & $0.1246$ & $0.01496$\\
& $u=-1\to-1.5$ & $22.17$ & $19.79$ & $19.76$\\
\bottomrule
\end{tabular}
\end{table*}

Three patterns are visible and all of them are informative.

Displacing the leading ultraviolet singularity from $u=-1$ to $u=-1.5$ stops
convergence altogether: the error freezes near $20$--$31\%$ and is essentially
independent of how many coefficients are supplied.  This is the signature of a
mode that is never removed, and it is the cleanest available demonstration
that the method is not fitting the data.

Replacing a genuine singularity is less catastrophic but still decisive.
Replacing the $u=2$ mode by $u=-4$ for $W_{\rm GC}$ raises the order-18 error from
$3.2\times10^{-5}\%$ to $5.3\%$, a factor of $1.6\times10^{5}$; for $W_{P2}$,
whose $u=2$ residue has the opposite sign and smaller magnitude, the
degradation is a factor of $1.7\times10^{4}$.  Convergence continues, but at
the rate set by the omitted mode rather than by the first genuinely neglected
one.

Adding a mode that is not there is inexpensive only in this high-order test.
For $W_\tau$, whose
$x^2$ coefficient vanishes, annihilating a nonexistent $u=2$ renormalon costs
little at order 16--18, but the enlarged fit also uses one additional known
coefficient and is not a fixed-budget comparison.  It therefore cannot support
a general rule to include speculative modes at physical order.

\subsection{Supervised sensitivity to a renormalon position}
\label{sec:supervised-scan}

\begin{figure*}[!tbp]
 \centering
 \includegraphics[width=0.98\linewidth]{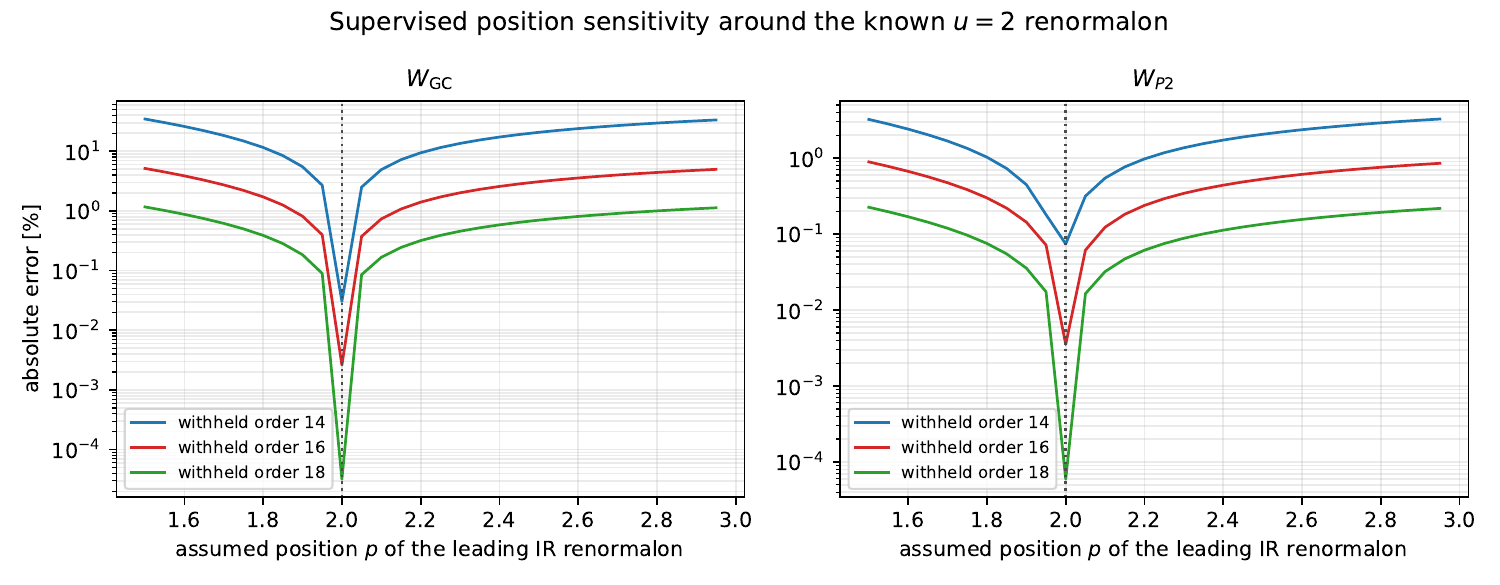}
 \caption{Withheld-coefficient error as a function of the assumed position of
 the leading infrared renormalon, all other modes held at their correct values.
 The minimum sits at $u=2$ and sharpens with the number of coefficients
 supplied.}
 \label{fig:annih-scan}
\end{figure*}

Because a wrong position degrades the forecast smoothly, the position itself
can be scanned.  Figure~\ref{fig:annih-scan} fixes the mode list
$\{-1,p,-2,3,-3\}$, with the $u=3$ pair confluent, and varies the assumed
position $p$ of the leading infrared singularity.

The error minimum reproduces the gluon-condensate position when a withheld
exact coefficient is supplied as the scoring target.  For
$W_{\rm GC}$ at withheld order 18 the error is $3.2\times10^{-5}\%$ at $p=2.0$
and rises to $0.18\%$ at $p=1.9$ and $0.17\%$ at $p=2.1$ --- a contrast of
nearly four orders of magnitude across a $5\%$ displacement --- and the minimum
deepens monotonically with order.  We stress the logical status of this
observation: it is a demonstration in a limit where the answer is known, and it
quantifies the sensitivity available in this supervised experiment.  It is
not an unsupervised locator and does not claim that $u=2$ can be inferred from
the four known QCD coefficients.  Appendix
\ref{sec:threshold} shows that it cannot.

\subsection{A subdominant mode}
\label{sec:u-minus-2}

Two further supervised experiments bound the method from opposite sides: a
subdominant mode, treated here, and a single dominant mode,
App.~\ref{sec:u2-only}.  Both are run in large-$\beta_0$ with every mode other
than the one under test held at its correct value, and both are reported in
summary; the tables and figures behind them are reproduced by the accompanying
code.

\paragraph{A subdominant renormalon: how late it becomes visible.}
The second ultraviolet singularity at $u=-2$ sits behind the leading one at
$u=-1$, and term by term its contribution is suppressed by
$|T_n^{(-2)}/T_n^{(-1)}|\sim2^{-n}\Gamma(n+1+\gamma_{-2})/\Gamma(n+1+\gamma_{-1})$,
so at order $16$ the geometric factor alone is $2^{-16}\approx1.5\times10^{-5}$.
Adding the $u=-2$ mode to a list that already contains $u=-1$ and $u=-3$ has
\emph{negative} marginal value below a threshold order --- it consumes
coefficients while its contribution is still beneath the truncation floor ---
and only at order $17$--$18$ does including it help, improving the Adler
withheld error by a factor of $11$ and the $W_{\rm GC}$ fixed-coupling sum by a
factor of $38$.  A supervised position scan locates it only weakly, late, and
with a systematic outward bias.  The crossover order is the quantitative
fingerprint of a subdominant renormalon: for a dominant mode such as $u=2$ or
$u=-1$ the marginal value is positive at every order tested.  The practical
statement is that resolving a renormalon which is merely subdominant already
requires of order $16$--$18$ coefficients.

\subsection{Annihilating only the \texorpdfstring{$u=2$}{u=2} renormalon}
\label{sec:u2-only}

The opposite limit shortens the mode list to its shortest nontrivial value:
retain only the gluon-condensate mode $u=2$ and discard every other renormalon.
For $m=1$ the construction is elementary --- in coefficient space
Eq.~\eqref{eq:nu-closed} collapses to the single-renormalon ratio
\begin{equation}
 d_{n+1}^{\rm pred}=(n+\gamma)\,\frac{\beta_0}{u}\,d_n
 =1.125\,n\,d_n
 \qquad(u=2,\ \gamma=0),
 \label{eq:u2-single-ratio}
\end{equation}
in the large-$\beta_0$ simple-pole convention.  The all-order behavior is
decisive and is the reason the one-mode continuation is carried only as a
diagnostic in the physical-order survey above: a single $u=2$ mode works during the finite window in
which that mode dominates the coefficient growth, and then fails once the
untouched $u=-1$ ultraviolet renormalon takes over.  It subtracts one component
of the divergence but not the component that controls the ultimate growth.
This is the converse of the subdominant case above: there a mode was too weak
to detect, here the dominant infrared mode is detected but is not by itself
enough.

\section{Required order for renormalon extrapolation}
\label{sec:threshold}
The physical application has only four nontrivial coefficients.  To calibrate
that sparsity, we record when one fixed, externally informed mode model reaches
persistent accuracy in the exactly solvable laboratory.

Suppose the large-$\beta_0$ singularity structure is supplied to the declared
$|u|\leq3$ transformation.  How many coefficients are required before its
forecast remains below a tolerance through the tested range?  The answer is a
benchmark for this construction, not a universal lower bound for every method.

Table~\ref{tab:threshold} reports, for each large-$\beta_0$ observable, the
smallest order beyond which the $|u|\leq3$ annihilator error remains below a
given tolerance for every subsequent order tested.  Requiring persistence
rather than a single crossing avoids rewarding an accidental zero.

\begin{table}[t]
\centering
\caption{Smallest withheld order $T$ beyond which the $|u|\leq3$ annihilator
error stays below the stated tolerance for all orders up to 20, in the exact
large-$\beta_0$ limit.}
\label{tab:threshold}
\begin{tabular}{lrrrrr}
\toprule
observable & $m$ & $<50\%$ & $<10\%$ & $<1\%$ & $<0.1\%$\\
\midrule
Adler & 9 & 10 & 10 & 11 & 13\\
$W_\tau$ & 5 & 11 & 11 & 12 & 14\\
$W_{\rm GC}$ & 6 & 7 & 7 & 13 & 13\\
$W_{P2}$ & 6 & 8 & 12 & 12 & 14\\
\bottomrule
\end{tabular}
\end{table}

Within the tested range through order 20, persistent percent-level control
begins at eleven to thirteen coefficients, and tenth-of-a-percent control at
thirteen to fourteen, even with exact renormalon input, exact coefficients and
no experimental uncertainty.  The $W_{\rm GC}$
entry illustrates why the persistence requirement matters: its error passes
below $10\%$ already at order 7 but does not settle below $1\%$ until order 13,
the intervening range containing the region where the $u=2$ and $u=-1$
contributions are comparable.

Four coefficients are not in the same empirical regime for this model.  This
supports a conservative physical forecast without becoming an impossibility
claim about unrelated algorithms or additional analytic input.

\section{Adler-projected renormalon-transported Shanks and adaptive defect transport}
\label{sec:defect-transport}
The direct Shanks/Pad\'e forecast fails because its implicit geometric
remainder model is too rigid, whereas a full renormalon-mode fit has more
unknown amplitudes than four QCD coefficients can determine stably.  This
section develops an intermediate construction: retain the local Pad\'e
forecast as a baseline, model only its observed one-step defect, and activate
a multi-mode correction only when predictions of earlier, already known
defects have passed an order-local certificate.  The construction is intended
to improve finite-order forecasting while retaining an explicit abstention
state.

\subsection{The all-weight construction: ARTS}
\label{sec:arts-definition}

The modification that cures the observed weight pathology is not another
denominator deformation applied independently to every moment.  It is a
change in the order of operations.  We call it the
\emph{Adler-projected renormalon-transported Shanks} transformation (ARTS):
\begin{enumerate}[leftmargin=2.2em]
 \item invert the known triangular FOPT relations to identify the single
 common Adler coefficient sequence $c_{n,1}$;
 \item form the ordinary Shanks one-step baseline
 $c_N^2/c_{N-1}$ and its observed defects
 $\Delta_n^{\rm P}=c_n-c_{n-1}^2/c_{n-2}$;
 \item transport the last defect with the declared nearest-renormalon
 position $u_\star$, without fitting its exponent;
 \item reconstruct every spectral moment using its exact RG contribution
 $R_{N+1}[W]$.
\end{enumerate}
Thus ARTS is the composition
\begin{equation}
 \mathfrak A_{N+1}[W]
 =\mathcal T_W\circ\mathcal P_{u_\star}\circ\mathcal I
 \label{eq:arts-composition}
\end{equation}
where $\mathcal I$ denotes the common-Adler projection,
$\mathcal P_{u_\star}$ the renormalon transport of the Shanks defect, and
$\mathcal T_W$ the known triangular Adler-to-moment map.  The noncommuting
alternative $\mathcal P_{u_\star}\circ\mathcal T_W$, which transforms each
moment separately, is precisely what fails for $W_{P2}$.

Any common Adler forecast therefore obeys the transfer identity
\eqref{eq:arts-error-transfer} of Sec.~\ref{sec:weight-map}: the
\emph{absolute} forecast errors are equal across weights after reconstruction,
while the relative errors are not, because a moment such as $W_{P2}$ can be
small through cancellation.  Consequently
ARTS can pass an all-weight relative-error test only when the common Adler
error is sufficiently small compared with the smallest tested moment.

The name also makes its relation to the two parent methods explicit.  Setting
the transported defect to zero returns the Shanks/$[1/1]$ coefficient
forecast.  For a pure factorial renormalon mode the first-order difference
operator
\begin{equation}
 \left[\mathsf E-q_n-\frac{\beta_0}{u_\star}\right]
 \Delta_n^{\rm P}=0
 \label{eq:arts-defect-annihilator}
\end{equation}
annihilates the defect exactly.  ARTS is therefore a renormalon annihilator
for the \emph{Shanks remainder defect}, rather than an overparameterized fit
of several raw moment modes.

\subsection{Exact single-mode identity and the index shift}

Suppose first that a coefficient sequence is a single factorial mode,
\begin{equation}
 c_n=K\,\Gamma(n+\gamma)A^{n-1},\qquad
 A=\frac{\beta_0}{u},\qquad n\geq1.
 \label{eq:port-pure-mode}
\end{equation}
The direct one-step geometric forecast formed at order $n-1$ is
$c_{n-1}^2/c_{n-2}$, and its observed defect is $\Delta_n^{\rm P}$, the
quantity \eqref{eq:pade-defect} of Sec.~\ref{sec:weight-map} evaluated on the
pure-mode sequence.  Using $\Gamma(z+1)=z\Gamma(z)$ gives
\begin{align}
 \frac{c_{n-1}^2}{c_{n-2}}
 &=K\Gamma(n-1+\gamma)(n-2+\gamma)A^{n-1},
 \nonumber\\
 \Delta_n^{\rm P}
 &=K\Gamma(n-1+\gamma)A^{n-1}.
 \label{eq:pade-defect-pure}
\end{align}
The Gamma-function argument in the defect is shifted down by one.  Thus the
transport factor is
\begin{equation}
 \frac{\Delta_{n+1}^{\rm P}}{\Delta_n^{\rm P}}
 =A(n-1+\gamma),
 \label{eq:defect-transport-factor}
\end{equation}
not $A(n+\gamma)$.  This off-by-one correction is numerically important.

More usefully, define $q_n=\Delta_n^{\rm P}/\Delta_{n-1}^{\rm P}$.  The
unknown exponent cancels between two successive ratios:
\begin{equation}
 q_n=A(n-2+\gamma),\qquad q_{n+1}=q_n+A.
 \label{eq:port-affine}
\end{equation}
This yields the ARTS/position-only ratio-transport (PORT) forecast
\begin{equation}
 \widehat c_{N+1}^{\rm ARTS}
 =\frac{c_N^2}{c_{N-1}}
 +\Delta_N^{\rm P}
 \left(\frac{\Delta_N^{\rm P}}{\Delta_{N-1}^{\rm P}}
       +\frac{\beta_0}{u_\star}\right).
 \label{eq:port-forecast}
\end{equation}
It requires two known defects.  The sign of the last observed ratio selects
the nearest declared positive- or negative-axis mode; no target coefficient
enters this choice.  Equation~\eqref{eq:port-forecast} is exact for
Eq.~\eqref{eq:port-pure-mode}.  It is not exact for a sum of modes.

\subsection{Why the physical projection must precede the weight map}
\label{sec:arts-order}
\label{sec:port-weights}

The FOPT moment relation is triangular with unit diagonal,
Eq.~\eqref{eq:triangular-weight-map}.  Applying a nonlinear ratio independently to each $d_n[W]$ allows
weight-dependent cancellations to masquerade as information about
$c_{N,1}$.  We instead apply ARTS/PORT once to the common Adler sequence and then
use Eq.~\eqref{eq:triangular-weight-map} for every weight.

With
\begin{equation}
 (c_0,c_1,c_2,c_3)=(1,1,1.639821204,6.371014),
\end{equation}
the two known defects, their ratio, and the direct baseline are
\begin{align}
 \Delta_2^{\rm P}&=0.639821204,
 &\Delta_3^{\rm P}&=3.682000419,\nonumber\\
 q_3&=5.754733347,
 &\frac{c_3^2}{c_2}&=24.75258845.
\end{align}
Choosing the leading positive Adler singularity $u_\star=2$ gives
$\beta_0/u_\star=1.125$ and
\begin{equation}
 \widehat c_{4,1}^{\rm ARTS}=50.08376951
 \qquad(c_{4,1}^{\rm exact}=49.07570).
 \label{eq:port-c4}
\end{equation}
Table~\ref{tab:port-weight-test} reconstructs the withheld fourth-order
moments.  The \emph{direct projected} column is the same Adler-space baseline
$c_3^2/c_2$ followed by the weight map; it is deliberately distinct from the
moment-by-moment Pad\'e forecast of Table~\ref{tab:pade-all-weight}.

\begin{table*}[!tbp]
\centering
\caption{Blind full-QCD test of ARTS (Adler-space PORT) followed by the known
FOPT weight map.  The final column uses a predeclared ten-percent criterion.}
\label{tab:port-weight-test}
\begin{tabular}{lrrrrr}
\toprule
weight & $W_2$ & exact $d_4$ & direct projected & ARTS & ARTS error\\
\midrule
$W_\tau$ & $0$  & $127.0786$ & $102.7555$ & $128.0867$ & $+0.793\%$\\
$W_{\rm GC}$ & $3$ & $214.1398$ & $189.8167$ & $215.1479$ & $+0.471\%$\\
$W_{P2}$ & $-1$ & $21.5087$ & $-2.8144$ & $22.5168$ & $+4.687\%$\\
\bottomrule
\end{tabular}
\end{table*}

ARTS/PORT therefore passes this particular all-weight withheld-$d_4$ test,
including both nonzero-$W_2$ moments.  It is tested on three of the six
leave-one-order-out entries of Table~\ref{tab:qcd-loo}: because
Eq.~\eqref{eq:port-forecast} needs two Pad\'e defects, and hence three prior
coefficients, ARTS is structurally undefined on the withheld-$d_3$ fold and no
claim is made there.  The weights do not supply three
independent determinations: after the common Adler projection they share the
same $c_{4,1}$ error, while their denominators make the relative moment errors
different.  The result is a transfer test of the projection rule, not three
statistically independent successes.

The qualifier ``this particular'' is essential.  The successful certificate
is the blind three-input prediction of the known fourth order for the declared
set $\{W_\tau,W_{\rm GC},W_{P2}\}$ and the predeclared ten-percent tolerance.
It is not a proof that ARTS passes every polynomial weight, because one can
construct a weight for which the exact moment is arbitrarily close to zero,
making any nonzero common Adler error arbitrarily large in relative terms.
Nor does it certify the unknown physical $c_{5,1}$: the order-adaptive
multi-mode extension below requires an additional prior defect and therefore
abstains at the currently available physical order.

\subsection{Nearby-renormalon test}

For the pole--$\MSbar$ mass relation the leading singularity is $u_\star=1/2$
and $\beta_0=(11-2n_\ell/3)/4$.  Table~\ref{tab:port-pole} uses
$r_0=1$ and the known $r_1,r_2,r_3$ to withhold $r_4$.  The same formula,
without an exponent fit, reduces the direct errors from about $22\%$ to below
$4\%$ for all three flavor numbers.
\begin{table}[t]
\centering
\caption{Position-only defect transport for the withheld four-loop
pole--$\MSbar$ coefficient.}
\label{tab:port-pole}
\begin{tabular}{crrrr}
\toprule
$n_\ell$ & exact $r_4$ & direct & PORT & PORT error\\
\midrule
3 & $1702.7109$ & $1315.3269$ & $1637.3588$ & $-3.838\%$\\
4 & $1235.6343$ & $960.8596$ & $1204.1354$ & $-2.549\%$\\
5 & $839.1793$  & $658.3411$ & $825.7565$  & $-1.600\%$\\
\bottomrule
\end{tabular}
\end{table}

\subsection{From one mode to an effective multi-mode defect model}
\label{sec:oamdt-model}

The nonlinearity in Eq.~\eqref{eq:pade-defect} must not be hidden.  If
$c_n=\sum_j K_j\phi_j(n)$, its Hankel numerator is
\begin{align}
 H_n\equiv c_nc_{n-2}-c_{n-1}^2
 &={}\sum_{j,k}K_jK_k\phi_j(n)\phi_k(n-2)\nonumber\\
 &\quad-\sum_{j,k}K_jK_k\phi_j(n-1)\phi_k(n-1),\nonumber\\
 \Delta_n^{\rm P}&=\frac{H_n}{c_{n-2}}.
 \label{eq:multimode-hankel}
\end{align}
Consequently the defect of a sum is not the sum of the pure-mode defects.
For a dominant mode $C_n$ and $c_n=C_n(1+\epsilon_n)$, expansion to first
order gives
\begin{align}
 \Delta_n^{\rm P}
 ={}&\Delta_n^{(C)}
 +C_n\epsilon_n
 -\frac{C_{n-1}^2}{C_{n-2}}
  (2\epsilon_{n-1}-\epsilon_{n-2})
 +\cO(\epsilon^2).
 \label{eq:defect-linearized}
\end{align}
The subleading shapes are therefore mixed modes.  A linear multi-mode defect
fit is an effective local approximation, not an algebraic consequence of a
multi-renormalon coefficient model.

We define the order-adaptive multi-mode defect transport (OAMDT) ansatz
\begin{equation}
 \Delta_n^{\rm P}\simeq
 \sum_{j=1}^{m}A_j\psi_j(n),\qquad
 \psi_j(n)=\Gamma(n-1+\gamma_j)
 \left(\frac{\beta_0}{u_j}\right)^{n-1}.
 \label{eq:oamdt-ansatz}
\end{equation}
The modes $(u_j,\gamma_j)$ are declared from the singularity structure before
the target is inspected.  To avoid overflow and artificial column scaling,
the design matrix at forecast origin $N$ uses
\begin{equation}
 X_{nj}=\frac{\psi_j(n)}{\psi_j(N)},\qquad
 \frac{\psi_j(n+1)}{\psi_j(n)}
 =\frac{\beta_0}{u_j}(n-1+\gamma_j),
 \label{eq:oamdt-normalization}
\end{equation}
and its columns are normalized to unit Euclidean norm before an SVD
least-squares solution.  The fitted amplitudes are transformed back before
forming $\widehat\Delta_{N+1}^{(M)}$ for candidate mode set $M$.

\subsection{Order-adaptive certificate and abstention}
\label{sec:oamdt-certificate}

Every decision uses only defects known before the target:
\begin{enumerate}[label=(\arabic*),leftmargin=2em]
 \item A model with $m$ modes must have at least $m+3$ known defects: one
 overdetermining datum and at least two expanding-window rolling-origin
 predictions.
 \item Its rolling skill relative to a zero-correction baseline is
 \begin{equation}
 Q_M=1-\frac{\sum_r
  (\Delta_{r+1}^{\rm P}-\widehat\Delta_{r+1|r}^{(M)})^2}
 {\sum_r(\Delta_{r+1}^{\rm P})^2}.
 \label{eq:oamdt-skill}
 \end{equation}
 We require $Q_M\geq0.25$.
 \item The column-normalized design and every rolling fit must have
 $\kappa_2\leq10^6$.  The forecast leverage is divided by the largest
 declared one-step modal gain in Eq.~\eqref{eq:oamdt-normalization}; the
 resulting dimensionless leverage must not exceed 2.
 \item At least two independently declared mode sets must pass.  Their
 forecast defects must have one sign and relative full spread no larger than
 $0.50$.
\end{enumerate}
Accepted corrections are skill-weighted and shrunk by the skill-weighted mean
$\overline Q$:
\begin{equation}
 \widehat c_{N+1}^{\rm OAMDT}
 =\frac{c_N^2}{c_{N-1}}
 +\operatorname{clip}(\overline Q,0,1)
 \frac{\sum_{M\in\mathcal A}Q_M\widehat\Delta_{N+1}^{(M)}}
      {\sum_{M\in\mathcal A}Q_M}.
 \label{eq:oamdt-forecast}
\end{equation}
If any consensus condition fails, the algorithm returns the direct baseline
and records \texttt{abstain}; it does not silently select the model closest to
the withheld target.  The round thresholds above define the reported
developmental certificate.  They are not probability levels.

\subsection{Large-\texorpdfstring{$\beta_0$}{beta0} validation}

The candidate sets use the known large-$\beta_0$ pole multiplicities:
$\{(-1,1),(-1,0),(2,0)\}$ for the Adler function,
$\{(-1,0),(3,1),(3,0)\}$ for $W_\tau$, and
$\{(-1,0),(2,0),(3,1),(3,0)\}$ for each nonzero-$W_2$ weight.  The displayed
order is part of the declaration.  The candidate family consists of every
single mode together with the prefixes of lengths two and three of that
ordered list; no candidate is added after the withheld coefficient is seen.
At every order $6\leq N\leq18$, $c_N$ is withheld and all
certification is recomputed from $c_0,\ldots,c_{N-1}$.

The purpose of this scan is diagnostic rather than predictive.  The
certificate frequently abstains, and an accepted correction is not guaranteed
to improve the direct baseline.  The coverage also changes when the declared
skill and consensus thresholds are varied.  These observations are sufficient
for the conclusion used here: the multi-mode extension has not acquired a
uniform all-order certificate and supplies no physical coefficient forecast.

\subsection{Physical status}

Using all four known Adler coefficients in the single-mode PORT formula gives
the diagnostic continuation $\widehat c_{5,1}=566.07$.  We do not promote
this number to a physical prediction: only three Pad\'e defects are known,
whereas even a one-mode OAMDT certificate requires four to provide two
rolling origins.  Thus the certified order-adaptive method abstains at the
physical $c_{5,1}$ order.  The successful $c_{4,1}$ and pole-mass tests show
that transporting the local defect can cure the specific weight pathology;
the large-$\beta_0$ record shows that it has not yet earned a universal
forecasting claim.  The reported physical $c_{5,1}$ and $c_{6,1}$ envelopes
therefore remain those in Tables~\ref{tab:conditional-forecast} and
\ref{tab:sequence-forecast}.

\section{D-log Pad\'e approximants and the all-weight test}
\label{sec:dlog-pade}
A branch point of a function becomes a simple pole of its logarithmic
derivative.  D-log Pad\'e approximation can therefore encode a singularity
position and exponent more economically than a direct rational approximation
to the original function.  This potential advantage is relevant to
renormalons, but it must be tested at the actual number of known coefficients:
with only three inputs in a withheld-$d_4$ test, the logarithmic derivative
contains only two known terms and can represent only one effective
singularity.

\subsection{Definition and one-step forecast}

Let
\begin{align}
 F(z)&=\sum_{n\geq0}f_nz^n,\qquad f_0\neq0,\nonumber\\
 G(z)&\equiv\frac{F'(z)}{F(z)}
 =\sum_{n\geq0}g_nz^n .
 \label{eq:dlog-definition}
\end{align}
The identity $F'=GF$ gives the triangular recursion
\begin{equation}
 (n+1)f_{n+1}=\sum_{k=0}^{n}g_k f_{n-k}.
 \label{eq:dlog-recursion}
\end{equation}
Consequently $f_0,\ldots,f_N$ determine
$g_0,\ldots,g_{N-1}$.  A Pad\'e approximant $[L/M]_G$ with
$L+M=N-1$ predicts $g_N$ by re-expansion, after which
Eq.~\eqref{eq:dlog-recursion} predicts $f_{N+1}$.  This is the D-log Pad\'e
coefficient forecast used below.

For a moment series
\begin{align}
 \delta_W^{(0)}(a)&=d_1a\,F_W(a),\nonumber\\
 F_W(a)&=1+d_2a+d_3a^2+d_4a^3+\cdots,
 \label{eq:dlog-moment-factor}
\end{align}
where $d_1=1$, the three-input withheld-$d_4$ problem gives
\begin{equation}
 g_0=d_2,\qquad
 g_1=2d_3-d_2^2,\qquad
 g_2=3d_4-3d_2d_3+d_2^3.
 \label{eq:dlog-g012}
\end{equation}
The only nontrivial denominator approximant is
\begin{equation}
 [0/1]_G(z)=
 \frac{g_0}{1-(g_1/g_0)z},
 \qquad
 \widehat g_2=\frac{g_1^2}{g_0},
 \label{eq:dlog-01}
\end{equation}
and hence
\begin{equation}
 \widehat d_4^{D\log[0/1]}=
 \frac{1}{3}\left[
 \frac{(2d_3-d_2^2)^2}{d_2}
 +3d_2d_3-d_2^3\right].
 \label{eq:dlog-d4}
\end{equation}
The polynomial alternative $[1/0]_G=g_0+g_1z$ sets
$\widehat g_2=0$.  It is retained as the other accuracy-through-input-order
choice, not as an independent measurement.

For the Borel version we apply the same construction to
\begin{equation}
 B_W(u)=\sum_{n\geq0}\frac{d_{n+1}}{n!}u^n
 \label{eq:dlog-borel}
\end{equation}
and multiply the predicted $u^3$ coefficient by $3!$.  This normalization
removes the leading factorial growth before the logarithmic derivative is
formed.

\subsection{Blind full-QCD weight test}
\label{sec:dlog-weight-test}

Table~\ref{tab:dlog-weight-test} gives signed errors for the original
$W_\tau$ moment and both nonzero-$W_2$ weights.  Every construction uses only
$d_1,d_2,d_3$ and withholds the known $d_4$.
\begin{table*}[!tbp]
\centering
\caption{Blind three-input D-log Pad\'e prediction of the physical full-QCD
$d_4$.  Entries are signed relative errors in percent.}
\label{tab:dlog-weight-test}
\begin{tabular}{lrrr}
\toprule
construction & $W_\tau$ & $W_{\rm GC}$ & $W_{P2}$\\
\midrule
direct D-log $[0/1]$ & $+4.224$ & $-2.506$ & $+110.587$\\
direct D-log $[1/0]$ & $-28.995$ & $-36.988$ & $+67.750$\\
Borel D-log $[0/1]$  & $+2.367$ & $-0.177$ & $+71.896$\\
Borel D-log $[1/0]$  & $+2.220$ & $-0.430$ & $+42.599$\\
\bottomrule
\end{tabular}
\end{table*}

Thus no standard three-input D-log realization passes all three weights at
the declared ten-percent level.  Borel normalization substantially improves
$W_\tau$ and $W_{\rm GC}$ but still misses $W_{P2}$ by at least $42.6\%$.
The failure is not caused by a tunable parameter: two known coefficients of
$G$ fix $[0/1]_G$ uniquely.

The conclusion also survives the common-Adler projection used successfully
for PORT.  Applying Borel D-log $[0/1]$ to
$c_{1,1},c_{2,1},c_{3,1}$ gives
\begin{equation}
 \widehat c_{4,1}^{\,B D\log}=39.05785304,
 \label{eq:dlog-adler-c4}
\end{equation}
and reconstruction through Eq.~\eqref{eq:triangular-weight-map} gives errors
$-7.883\%$, $-4.678\%$, and $-46.576\%$ for
$W_\tau$, $W_{\rm GC}$, and $W_{P2}$, respectively.  The first two pass, but
the cancellation-sensitive $W_{P2}$ moment does not.

\subsection{Why the low-order D-log approximation fails}

If
\begin{equation}
 F(z)\sim\frac{A(z)}{(1-z/z_0)^{1+\gamma}},
 \qquad A(z_0)\neq0,
 \label{eq:dlog-single-branch}
\end{equation}
then
\begin{equation}
 \frac{F'(z)}{F(z)}
 =\frac{1+\gamma}{z_0-z}+\frac{A'(z)}{A(z)}.
 \label{eq:dlog-pole-map}
\end{equation}
This explains the strength of D-log Pad\'es when one branch point dominates.
At physical order, however, $[0/1]_G$ has only one pole and cannot resolve
the competition between the $u=-1$, $u=2$, and more distant structures.
The $W_{P2}$ coefficients exhibit a strong cancellation between these
contributions; taking a logarithmic derivative does not remove that
cancellation and can amplify it when $F$ is locally small.

A partial D-log Pad\'e could factor a known singularity position or exponent
before approximation.  Such a construction may be useful, but it imports the
same external renormalon information as a mode model and must be validated as
a different method.  It is not evidence that the standard three-input
D-log Pad\'e has passed the test.

\subsection{All-order large-\texorpdfstring{$\beta_0$}{beta0} behavior}
\label{sec:dlog-largebeta}

To determine whether the physical failure is permanent or only finite-order,
we construct at every withheld order the balanced D-log approximant to the
exact large-$\beta_0$ series.  For $K$ known coefficients of $G$, the degrees
satisfy $L+M=K-1$ with $M=\lceil(K-1)/2\rceil$.  Both the direct and
factorially normalized Borel sequences are recomputed without the target.

\begin{table}[t]
\centering
\caption{First withheld order from which the balanced Borel--D-log Pad\'e
error remains below ten percent through order 18 in large-$\beta_0$.}
\label{tab:dlog-thresholds}
\begin{tabular}{lrrrr}
\toprule
observable & Adler & $W_\tau$ & $W_{\rm GC}$ & $W_{P2}$\\
\midrule
persistent order & $6$ & $11$ & $4$ & $12$\\
\bottomrule
\end{tabular}
\end{table}

The Borel--D-log sequence therefore can pass all four large-$\beta_0$
observables once sufficiently many coefficients are supplied.  It does not
pass the physical four-order validation.  This distinction agrees with the
balanced Pad\'e/Wynn and annihilator studies: eventual convergence does not
identify its onset from a short, cancellation-dominated series.

\section{Regularized shifted-Wynn transformations}
\label{sec:regularization}
Small denominators can make a local Wynn element unstable, but a regulator is
useful only if its definition is algebraically nontrivial and its parameter
choice transfers out of sample.  We therefore analyze the two proposed forms
exactly and test the complete predeclared grid across all weights.

The parameter grid is fixed in advance,
\begin{equation}
 k,\alpha\in\{0.01,0.02,0.03,0.04,0.05\}.
 \label{eq:fixed-reg-grid}
\end{equation}
No value is selected or discarded after a withheld target is examined.  We
apply the transformations both to the original coefficients $d_n$ and to the
Borel coefficients $d_n/(n-1)!$.  This makes the comparison with the
direct continuation and with the validated Borel--Wynn tables unambiguous.

\subsection{The \texorpdfstring{$\eta_n$}{eta-n} modification}

A practical Vanden Broeck--Schwartz-inspired realization of an additive
denominator control is
\begin{equation}
 \widetilde\varepsilon_{2,\eta}^{(n)}=
 S_{n+2}-\frac{(\Delta S_{n+1})^2}{\Delta^2S_n+\eta_n}.
 \label{eq:eta-new-wynn}
\end{equation}
We do not attribute this particular ``$+\eta_n$'' formula to the historical
VBS transformation.  We test the practical choice
\begin{equation}
 \eta_n=k\Delta^2S_n.
 \label{eq:eta-choice}
\end{equation}
Equations~\eqref{eq:eta-new-wynn} and \eqref{eq:eta-choice} show immediately
that the denominator is $(1+k)\Delta^2S_n$.  Therefore this prescription is a
multiplicative rescaling, not a pole constraint: if $\Delta^2S_n=0$, then
$\eta_n=0$ and the singularity remains.  For the perturbative partial sums it
predicts
\begin{equation}
 d_{N+j}^{(\eta)}=\frac{1}{1+k}\frac{d_N^2}{d_{N-1}}
 \left(\frac{d_N}{d_{N-1}}\right)^{j-1},\qquad j\geq1.
 \label{eq:eta-future}
\end{equation}

Table~\ref{tab:eta-old-comparison} gives the four-input $W_\tau$ continuation
from Eq.~\eqref{eq:eta-future}, retaining the full numerical precision of the
known FOPT coefficients.  These are conditional continuations, not blind
tests.
\begin{table}[t]
\centering
\caption{$\eta_n=k\Delta^2S_n$ continuation of the physical $W_\tau$ series.
The unregularized direct result is $c_{5,1}=304.71$,
$c_{6,1}=3171.04$.}
\label{tab:eta-old-comparison}
\begin{tabular}{crrrrr}
\toprule
$k$ & $0.01$ & $0.02$ & $0.03$ & $0.04$ & $0.05$\\
\midrule
$c_{5,1}$ & $298.65$ & $292.70$ & $286.87$ & $281.15$ & $275.54$\\
$c_{6,1}$ & $3249.83$ & $3327.08$ & $3402.82$ & $3477.11$ & $3549.99$\\
\bottomrule
\end{tabular}
\end{table}

The decisive check is out of sample.  Table~\ref{tab:regularized-qcd-loo}
withholds $d_4$ for the original and both renormalon-sensitive weights.  The
$W_\tau$ improvement at $k=0.05$ is what a $W_\tau$-only optimization would
select.  It does not transfer: every positive $k$ worsens
$W_{\rm GC}$, and the second nonzero-$W_2$ moment remains wrong by more than
$118\%$.

\begin{table*}[!tbp]
\centering
\caption{Direct, three-input prediction of the withheld full-QCD $d_4$.
Entries are signed relative errors in percent.  The interval is the complete
fixed grid in Eq.~\eqref{eq:fixed-reg-grid}.}
\label{tab:regularized-qcd-loo}
\begin{tabular}{lrrrr}
\toprule
weight & original shifted Wynn & $\eta$ grid & $\alpha$ grid & $\theta_2$\\
\midrule
$W_\tau$ & $+5.152$ & $[+0.145,+4.111]$ & $+5.152$ & $+2.367$\\
$W_{\rm GC}$ & $-3.671$ & $[-8.258,-4.625]$ & $-3.671$ & $-0.177$\\
$W_{P2}$ & $+129.932$ & $[+118.983,+127.655]$ & $+129.932$ & $+71.896$\\
\bottomrule
\end{tabular}
\end{table*}

Borel normalization does not rescue the four-order test.  For
$(W_\tau,W_{\rm GC},W_{P2})$, respectively, the absolute errors are
\begin{align*}
 \text{unregularized: }&(57.7,44.5,244.9)\%,\\
 \eta\text{ grid: }&(50.2\text{--}56.2,\ 37.6\text{--}43.1,\
 228.5\text{--}241.5)\%.
\end{align*}
This is consistent with the
earlier conclusion that the physical coefficients are not yet in the
Borel-geometric regime.

\subsection{The \texorpdfstring{$\alpha$}{alpha} modification}
\label{sec:alpha}

The second denominator is
\begin{equation}
 \widetilde\varepsilon_{2,\alpha}^{(n)}=
 S_{n+2}-\frac{(\Delta S_{n+1})^2}
 {\Delta^2S_n+\alpha\Delta S_{n+1}}.
 \label{eq:alpha-new-wynn}
\end{equation}
Writing the last two perturbative increments explicitly gives the exact
future-coefficient formula
\begin{equation}
 d_{N+j}^{(\alpha)}=\frac{d_N^2}{d_{N-1}}
 \left[(1+\alpha)\frac{d_N}{d_{N-1}}\right]^{j-1},
 \qquad j\geq1.
 \label{eq:alpha-future}
\end{equation}
The first unseen coefficient ($j=1$) is independent of $\alpha$.  Hence a
flat $c_{5,1}$ row in an $\alpha$ scan is an algebraic identity and cannot be
interpreted as evidence of stability.  The parameter first enters $c_{6,1}$:
Because of Eq.~\eqref{eq:alpha-future}, the $\alpha$ column in the one-step
blind test, Table~\ref{tab:regularized-qcd-loo}, must coincide with the
unregularized result.  Two-step large-$\beta_0$ tests do depend on $\alpha$,
but show no universal improvement.  At target order 12 after Borel
normalization, the absolute errors on the full fixed grid are
$19.5$--$22.6\%$ (Adler), $99.81$--$99.82\%$ ($W_\tau$),
$24.1$--$29.0\%$ ($W_{\rm GC}$), and $640.9$--$670.2\%$ ($W_{P2}$).
The corresponding unregularized two-step errors are $23.4\%$, $99.8\%$,
$22.9\%$, and $633.5\%$.  The small Adler improvement therefore does not
transfer across weights.

\section{Local regularizations: summary of the numerical tests}
\label{sec:local-reg-summary}

The $\theta_2$ and canonical $\rho_2$ remainder models, and the large-$\beta_0$
behavior of the local $\eta_n$ and $\alpha$ regularizations, were tested on the
same four observables used throughout the validation and renormalon-informed appendices.  The outcome is uniform
and negative, and we record it here rather than tabulating it: neither
$\theta_2$ nor canonical $\rho_2$ converges uniformly across the four
observables, and the local regularizations of App.~\ref{sec:regularization}
reproduce the deterioration of the raw partial sums rather than curing it.
The $\theta_2$ continuation therefore enters the global comparison of
Sec.~\ref{sec:global-comparison} as an accuracy-through-order row with no
external certificate; canonical $\rho_2$ does not enter it because it lacks
accuracy through order; and neither is used in any forecast advanced by this
paper.  The full numerical tables and the corresponding large-$\beta_0$ error
curves are reproduced by the accompanying code.

\section{Derivation of the moment Borel factor}
\label{app:contour}
The weight-selection rule is central to every sensitive-moment test, so this
appendix fixes the contour branch and derives the factor rather than treating
it as a convention-dependent assertion.

Let $x=e^{i\phi}$ with $-\pi<\phi<\pi$.  On the standard contour branch,
\begin{equation}
 \ln(-e^{i\phi})=
 \begin{cases}
 i(\phi+\pi),&-\pi<\phi<0,\\
 i(\phi-\pi),&0<\phi<\pi.
 \end{cases}
\end{equation}
For a monomial $x^m$,
\begin{align}
 K_m(u)&=\frac{1}{2\pi i}\oint\frac{dx}{x}x^m(-x)^{-u}\nonumber\\
 &=\frac{1}{2\pi}\left[
 e^{-i\pi u}\int_{-\pi}^{0}d\phi\,e^{i(m-u)\phi}
 \right.\nonumber\\[-1mm]
 &\hspace{21mm}\left.
 +e^{i\pi u}\int_{0}^{\pi}d\phi\,e^{i(m-u)\phi}
 \right].
\end{align}
Because $m$ is an integer, the two elementary integrals combine to
\begin{equation}
 K_m(u)=\frac{\sin(\pi u)}{\pi(u-m)}.
\end{equation}
Linearity then gives Eq.~\eqref{eq:Jwu}.  The logarithmic moments follow from
\begin{equation}
 J_k[W]=(-1)^k\left.\frac{d^k}{du^k}K_W(u)\right|_{u=0}.
\end{equation}

\section{Pad\'e forecast equations}
\label{app:pade-forecast}
The numerical forecast should be reproducible independently of a particular
software implementation.  We therefore record the defining linear system and
recurrence explicitly.

Let $f(z)=\sum_{n=0}^{N-1}f_nz^n$ and approximate it by
\begin{align}
 [L/M](z)&=\frac{P_L(z)}{Q_M(z)},\nonumber\\
 Q_M(z)&=1+\sum_{j=1}^{M}Q_jz^j,
 \qquad L+M=N-1.
\end{align}
Matching through $z^{N-1}$ gives the linear system
\begin{equation}
 f_n+\sum_{j=1}^{M}Q_jf_{n-j}=0,\qquad
 n=L+1,\ldots,L+M.
\end{equation}
Every forecast coefficient then obeys the recurrence
\begin{equation}
 f_n^{\rm pred}=-\sum_{j=1}^{M}Q_jf_{n-j}^{\rm pred},
 \qquad n\geq N.
\end{equation}
For numerical stability, the expansion variable may be rescaled before this
system is solved and the forecast coefficients transformed back afterward;
this changes neither the matched series nor the predicted coefficients.

\section{Exact remainder mode and the annihilator weights}
\label{app:annihilator}
The main construction depends on the distinction between a first omitted term
and a prescription-matched tail.  This appendix supplies the derivation,
states the meromorphic large-$\beta_0$ scope, and documents the conditioning
check.

\subsection{Derivation of the exact-mode equation}

For an isolated Borel singularity
$B_p(u)=K_p(1-u/p)^{-1-\gamma}$, expanding
$(1-u/p)^{-1-\gamma}$ gives
\begin{equation}
 b_n=\frac{K_p}{p^n}\frac{\Gamma(n+1+\gamma)}{n!\,\Gamma(1+\gamma)},
\end{equation}
so that with $d_{n+1}=\pi b_nn!\beta_0^n$ the perturbative term of order $n+1$
is $T_n=\widehat C\,\Gamma(n+1+\gamma)z^n$ with $z=\beta_0a/p$ and
$\widehat C=\pi K_pa/\Gamma(1+\gamma)$, which is
Eq.~\eqref{eq:single-mode-term}.  The remainder after $N$ terms is the
principal value of $\sum_{n\geq N}T_n$.  Writing
$\Gamma(n+1+\gamma)=\int_0^\infty dt\,t^{n+\gamma}e^{-t}$ and interchanging sum
and integral,
\begin{align}
 \sum_{n\geq N}\Gamma(n+1+\gamma)z^n
 &=\int_0^\infty dt\,t^{\gamma}e^{-t}
 \sum_{n\geq N}(zt)^n\nonumber\\
 &=z^N\!\int_0^\infty dt\,
 \frac{t^{N+\gamma}e^{-t}}{1-zt},
\end{align}
the last step being the geometric sum, valid termwise and defining the
divergent series by the principal value of the resulting integral.  This is
Eq.~\eqref{eq:exact-mode}.  Since
$\int_0^\infty dt\,t^{N+\gamma}e^{-t}=\Gamma(N+1+\gamma)$, the $z\to0$ limit
returns the leading estimate.

Because $B[\Dhat](u)$ is meromorphic with all singularities on the real axis,
and the amplitudes multiplying each mode are real, summing the modes with the
same lateral prescription used for the full Borel integral reproduces the
lateral sum term by term; taking real parts then reproduces its principal
value.

\subsection{Order of a \texorpdfstring{large-$\beta_0$}{large-beta0} pole and confluent modes}

Near a pole of order two at $u=p$, $B(u)=K/(p-u)^2+L/(p-u)+\cO(1)$, and
\begin{equation}
 d_{n+1}=\pi\left(\frac{\beta_0}{p}\right)^{\!n}
 \left[\frac{K}{p^2}(n+1)!+\frac{L}{p}\,n!\right],
\end{equation}
which is exactly two coefficient modes $\omega_n^{(p,1)}$ and
$\omega_n^{(p,0)}$ in the sense of Eq.~\eqref{eq:coeff-model}.  A simple pole
gives one mode with $\gamma=0$.  Together with Eq.~\eqref{eq:order-rule} this
fixes the complete mode list of every observable considered here.

\subsection{Growth bound and the truncation error}

On $|u|=R$ the closed form Eq.~\eqref{eq:trigamma-closed} gives
$|B[\Dhat](u)|\lesssim e^{5R/3}/R^{2}$.  Deforming the Cauchy contour outward,
\begin{equation}
 b_n=-\sum_{|u_j|\leq R}\mathrm{Res}\,\frac{B(u)}{u^{n+1}}
 +\frac{1}{2\pi i}\oint_{|u|=R}\frac{B(u)}{u^{n+1}}du,
\end{equation}
and minimizing $e^{5R/3}R^{-n}$ at $R=3n/5$ bounds the last term by
$(5e/3n)^n$ up to a power prefactor.  The pole decomposition of $b_n$ is
therefore asymptotically controlled.  Truncating a nested mode list at
$|u|\leq U$ leaves a contribution governed by the nearest omitted pole,
multiplied by residue-dependent constants and powers of $n$ set by pole
multiplicity.  The simpler $(U+1)^{-n}$ form is only the leading geometric
factor when the nearest included pole has $|u|=1$ and the next pole is at the
integer $U+1$.  This is the scope of the convergence estimate quoted in
App.~\ref{sec:order-rule}; it is not a nonasymptotic error bound.

\subsection{Closed-form weights: numerical check}

For $N=3$, $a=0.1$ and modes $u=2,3,-1$ with $\gamma=0$,
Eq.~\eqref{eq:lambda-closed} gives
\begin{equation}
 \begin{aligned}
 \lambda={}&(0.517388218595,\;-2.29950319375,\\
 &\phantom{(}0.511000709723,\;2.27111426544),
 \end{aligned}
\end{equation}
agreeing with the direct solution of Eq.~\eqref{eq:lambda-system-general} to
$1.1\times10^{-45}$ at $45$-digit working precision.  The weights are large and
of alternating sign, which is why the transformation must be evaluated in
extended precision: for the $|u|\leq3$ lists at order 18 we observe a loss of
roughly $12$ significant digits, comfortably inside the $45$ digits carried.

\FloatBarrier

\end{document}

%% file: generated_data/cipt_diagonal_table.tex
\begin{tabular}{crrrr}
\toprule
$n$ & $a^n$ & $J_n^{\rm CI}[W_\tau]$ & $J_n^{\rm CI}[W_{\rm GC}]$ & $J_n^{\rm CI}[W_{P2}]$\\
\midrule
$1$ & $1.0173184e-01$ & $1.3788269e-01$ & $1.5808453e-01$ & $1.1619587e-01$\\
$2$ & $1.0349367e-02$ & $1.6069745e-02$ & $2.1266127e-02$ & $1.1375563e-02$\\
$3$ & $1.0528602e-03$ & $1.6428537e-03$ & $2.5725404e-03$ & $9.4492804e-04$\\
$4$ & $1.0710940e-04$ & $1.4829589e-04$ & $2.8741467e-04$ & $6.3293960e-05$\\
$5$ & $1.0896436e-05$ & $1.1575980e-05$ & $3.0114506e-05$ & $2.6479781e-06$\\
\bottomrule
\end{tabular}

%% file: generated_data/cipt_fopt_separation_table.tex
\begin{tabular}{ccrrr}
\toprule
$N$ & weight & $\delta^{\rm FOPT}_{W,N}$ & $\delta^{\rm CIPT}_{W,N}$ & $(\mathrm{CIPT}-\mathrm{FOPT})/\mathrm{FOPT}$\\
\midrule
$4$ & $W_{\rm GC}$ & $0.2206375$ & $0.2234519$ & $+1.28\%$\\
$4$ & $W_{P2}$ & $0.1634646$ & $0.1439761$ & $-11.92\%$\\
$5$ & $W_{\rm GC}$ & $0.2362622$ & $0.2317937$ & $-1.89\%$\\
$5$ & $W_{P2}$ & $0.1608974$ & $0.1447096$ & $-10.06\%$\\
\bottomrule
\end{tabular}

%% file: generated_data/cipt_weight_table.tex
\begin{tabular}{c ccc ccc}
\toprule
 & \multicolumn{3}{c}{FOPT} & \multicolumn{3}{c}{CIPT}\\
\cmidrule(lr){2-4}\cmidrule(lr){5-7}
$N$ & $W_\tau$ & $W_{\rm GC}$ & $W_{P2}$ & $W_\tau$ & $W_{\rm GC}$ & $W_{P2}$\\
\midrule
4 & $\pm12.7$ & $\pm21.4$ & $\boldsymbol{\pm2.15}$ & $\pm123$ & $\boldsymbol{\pm77.7}$ & $\pm227$\\
5 & $\pm58.5$ & $\pm143$ & $\boldsymbol{\pm23.6}$ & $\pm1.6e+03$ & $\boldsymbol{\pm770}$ & $\pm5.46e+03$\\
\bottomrule
\end{tabular}

%% file: generated_data/bj_cut_correction_domain.tex
\begin{tabular}{rrrrc}
\toprule
$n$ & $\tilde b_1$ term & $\tilde b_2$ term & bracket & status\\
\midrule
$0$ & $-0.9990$ & $-2.9042$ & $-2.9032$ & \textbf{fitted}\\
$1$ & $-0.6118$ & $-0.6531$ & $-0.2649$ & \textbf{fitted}\\
$2$ & $-0.4410$ & $-0.2883$ & $+0.2708$ & \textbf{fitted}\\
$3$ & $-0.3447$ & $-0.1624$ & $+0.4929$ & \textbf{fitted}\\
$4$ & $-0.2829$ & $-0.1042$ & $+0.6129$ & \\
$5$ & $-0.2399$ & $-0.0725$ & $+0.6876$ & \\
$6$ & $-0.2083$ & $-0.0534$ & $+0.7383$ & \\
$7$ & $-0.1840$ & $-0.0409$ & $+0.7751$ & \\
$8$ & $-0.1648$ & $-0.0324$ & $+0.8028$ & \\
\bottomrule
\end{tabular}

%% file: generated_data/beta_function_nf3_blind_table.tex
\begin{tabular}{lrr}
\toprule
method & $\widehat\beta_4$ & signed error [\%]\\
\midrule
direct Pad\'e [1/2] & 336.983 & +164.7\\
direct Pad\'e [2/1] & 221.721 & +74.1\\
Borel Pad\'e [1/2] & 256.648 & +101.6\\
Borel Pad\'e [2/1] & 295.628 & +132.2\\
direct D-log Pad\'e [1/1] & 275.051 & +116.0\\
Borel D-log Pad\'e [1/1] & 405.020 & +218.1\\
direct shifted Wynn/Shanks & 221.721 & +74.1\\
direct $\eta$, $k=0.03$ & 215.263 & +69.1\\
direct $\alpha=0.03$ & 221.721 & +74.1\\
direct $\theta_2$ tail & 324.664 & +155.0\\
direct Levin-$u$ ATO & 324.664 & +155.0\\
Borel Levin-$u$ ATO & 353.701 & +177.8\\
\midrule
exact five-loop value & 127.322 & $0$\\
\bottomrule
\end{tabular}

%% file: generated_data/beta_function_nf3_forecast_table.tex
\begin{tabular}{lrr}
\toprule
family & $\widehat\beta_5$ & $\widehat\beta_6$\\
\midrule
direct Pad\'e & 561.0 & 1597.1\\
Borel Pad\'e & 3495.9 & -40522.5\\
direct D-log Pad\'e & [327.2,830.0] & [925.9,1603.6]\\
Borel D-log Pad\'e & [953.8,2587.2] & [-20141.3,2384.6]\\
direct shifted Wynn/Shanks & 343.2 & 925.4\\
direct $\eta$ grid & [326.9,339.8] & [881.3,916.2]\\
direct $\alpha$ grid & 343.2 & [934.6,971.6]\\
direct $\theta$ tail & 88.8 & -1132.9\\
Borel shifted Wynn/Shanks & 429.1 & 1735.1\\
Borel $\eta$ grid & [408.6,424.8] & [1652.4,1717.9]\\
Borel $\alpha$ grid & 429.1 & [1752.4,1821.8]\\
Borel $\theta$ tail & -138.1 & -5145.6\\
direct Levin-$u$ ATO & -124.0 & -3101.9\\
Borel Levin-$u$ ATO & -443.3 & -8546.8\\
\bottomrule
\end{tabular}

%% file: generated_data/large_beta0_pade_errors.tex
\begin{tabular}{llrrrr}
\toprule
observable & method & order 4 & order 8 & order 12 & order 16\\
\midrule
Adler & direct & $538.647$ & $787.006$ & $2.13651$ & $0.522109$\\
Adler & Borel & $857.97$ & $16.4851$ & $0.0242352$ & $0.000118606$\\
Adler & $[0/N-1]$ (diag.) & $558.126$ & $352.782$ & $0.507$ & $5.88059$\\
\midrule
$W_\tau$ & direct & $3.26931$ & $66.2282$ & $14.6749$ & $13.0073$\\
$W_\tau$ & Borel & $54.904$ & $447.998$ & $0.674027$ & $0.00091729$\\
$W_\tau$ & $[0/N-1]$ (diag.) & $3.27496$ & $39.6192$ & $586.268$ & $4.20668$\\
\midrule
$W_{\rm GC}$ & direct & $1.4502$ & $1.96778$ & $29.3992$ & $9.56648$\\
$W_{\rm GC}$ & Borel & $47.8247$ & $1423.44$ & $1.45539$ & $0.000662865$\\
$W_{\rm GC}$ & $[0/N-1]$ (diag.) & $1.44549$ & $0.0292154$ & $277.288$ & $211.995$\\
\midrule
$W_{P2}$ & direct & $24.0699$ & $31.9016$ & $16.8875$ & $4.2562$\\
$W_{P2}$ & Borel & $86.1048$ & $15.2073$ & $2.89824$ & $0.0082106$\\
$W_{P2}$ & $[0/N-1]$ (diag.) & $24.0773$ & $34.4191$ & $95.8077$ & $24.0363$\\
\bottomrule
\end{tabular}